\documentclass[12pt,a4paper,twoside,final]{book}

\usepackage{epsfig}
\usepackage{graphicx}
\usepackage{amsfonts}
\usepackage{amssymb}
\usepackage{amsthm}
\usepackage{fancyheadings}
\usepackage{bbm}
\usepackage{citesort}

\def\L{\mathcal L}

\newcounter{numrom}
{\begin{list}{\Roman{numrom}.}
 {\usecounter{numrom}}}{\end{list}}

\newcommand{\wt}{\widetilde}

\begin{document}

\def\a{\alpha}
\def\b{\beta}
\def\c{\chi}
\def\d{\delta}
\def\e{\epsilon}
\def\f{\phi}
\def\g{\gamma}
\def\h{\eta}
\def\i{\iota}
\def\j{\psi}
\def\k{\kappa}
\def\l{\lambda}
\def\m{\mu}
\def\n{\nu}
\def\o{\omega}
\def\p{\pi}
\def\q{\theta}
\def\r{\rho}
\def\s{\sigma}
\def\t{\tau}
\def\u{\upsilon}
\def\x{\xi}
\def\z{\zeta}
\def\D{\Delta}
\def\F{\Phi}
\def\G{\Gamma}
\def\J{\Psi}
\def\L{\Lambda}
\def\O{\Omega}
\def\P{\Pi}
\def\Q{\Theta}
\def\S{\Sigma}
\def\U{\Upsilon}
\def\X{\Xi}

\def\ve{\varepsilon}
\def\vf{\varphi}
\def\vr{\varrho}
\def\vs{\varsigma}
\def\vq{\vartheta}

\def\dg{\dagger}                                     
\def\ddg{\ddagger}                                   
\def\wt#1{\widetilde{#1}}                    
\def\mt{\widetilde{m}_1}
\def\mti{\widetilde{m}_i}
\def\rt{\widetilde{r}_1}
\def\mtt{\widetilde{m}_2}
\def\mttt{\widetilde{m}_3}
\def\rtt{\widetilde{r}_2}
\def\mb{\overline{m}}
\def\VEV#1{\left\langle #1\right\rangle}        

\def\ds{\displaystyle}
\def\ra{\rightarrow}

\def\matm{m_{\rm atm}}
\def\msol{m_{\rm sol}}
\def\HL{$M_1\ll M_2 \ll M_3$}
\def\DL{$M_1\simeq M_2 \simeq M_3$}
\def\CP{$C\!P$}
\def\be{\begin{equation}}
\def\ee{\end{equation}}
\def\bea{\begin{eqnarray}}
\def\eea{\end{eqnarray}}
\def\NO{\nonumber}
\def\Bar#1{\overline{#1}}
\def\gev{~{\rm GeV}}
\def\ev{~{\rm eV}}

\newfont{\headfo}{cmssbx10 scaled 3000}
\pagestyle{empty}
\begin{center}
\vspace{0.5cm} {\Large \sc Dissertation}

\vspace{1.3cm} {\headfo A New Era of Leptogenesis} 

\vspace{1cm} by

\vspace{0.5cm} {\Large \sc Steve Blanchet}

\vspace{1.5cm}
\end{center}

\begin{center}
\begin{tabular}{c}
{\large Technische Universit\"at M\"unchen}\\
{\large Physik Department}\\
{\large Institut f\"ur Theoretische Physik T30e}\\
{\large Univ.-Prof.~Dr.~Michael Ratz}
\end{tabular}

\begin{figure}[h!]
\begin{center}
\includegraphics[width=0.3\textwidth]{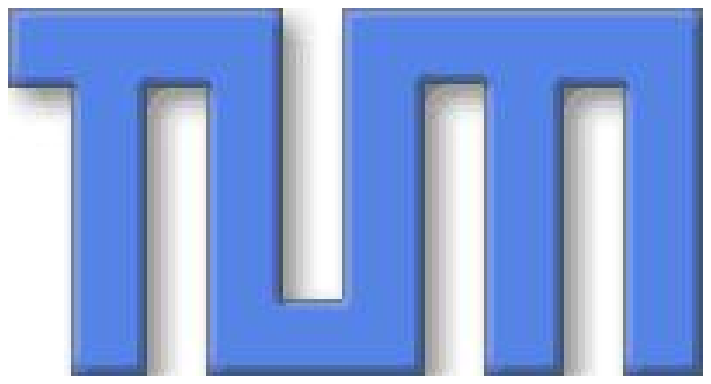}
\end{center}
\end{figure}

\begin{tabular}{c}
angefertigt am\\
Max-Planck-Institut f\"ur Physik,  M\"unchen\\
(Werner-Heisenberg-Institut)\\
unter Betreuung von\\
Dr. Pasquale Di Bari
und Dr. habil. Georg G. Raffelt
\end{tabular}
\end{center}

\begin{figure}[h!]
\begin{center}
\vspace{-0.9cm}
\includegraphics[height=4cm]{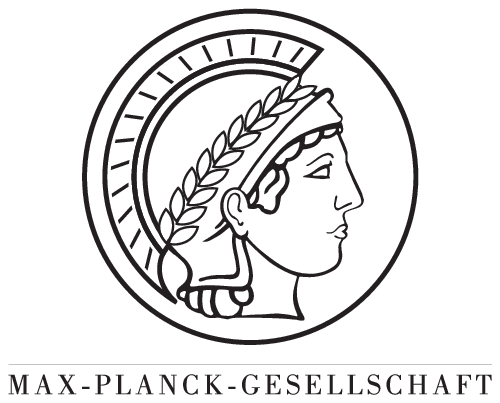}
\end{center}
\end{figure}

\cleardoublepage

\begin{center}

\vspace*{-1cm}
\begin{tabular}{c}
{\large Technische Universit\"at M\"unchen}\\
{\large Physik Department}\\
{\large Institut f\"ur Theoretische Physik T30e}\\
{\large Univ.-Prof.~Dr.~Michael Ratz}
\end{tabular}

\vspace{1.3cm} {\headfo A New Era of Leptogenesis} 
\vspace{3.2cm}

{\Large\sf\sc Steve Blanchet} \vfill

\end{center}
Vollst\"andiger Abdruck der von der Fakult\"at f\"ur Physik der
Technischen Universit\"at M\"unchen zur Erlangung des akademischen
Grades eines
\begin{center}
{\bf \sf Doktors der Naturwissenschaften (Dr. rer. nat.)}
\end{center}
genehmigten Dissertation.

\begin{center}
\begin{tabular}{lll}
Vorsitzender:              &    & Univ.-Prof.~Dr.~L.~Oberauer\\
Pr\"ufer der Dissertation: & 1. & Univ.-Prof.~Dr.~M.~Ratz\\
                           & 2. & Hon.-Prof.~Dr.~W.~F.~L.~Hollik
\end{tabular}
\end{center}

\noindent Die Dissertation wurde am 10.04.2008 bei der Technischen
Universit\"at M\"unchen eingereicht und durch die Fakult\"at f\"ur
Physik am 09.05.2008 angenommen. \cleardoublepage

\frontmatter
\pagenumbering{roman} \pagestyle{plain}


\chapter{Summary}

The present thesis, which is mainly based on my research
papers I--IV listed overleaf, is devoted to some forefront issues 
in the field of leptogenesis. As well as nicely explaining the 
origin of the matter-antimatter asymmetry of the Universe,
this mechanism is intimately related to the nature of neutrino masses
in that it is the cosmological consequence of the see-saw mechanism. 
This connection makes possible to relate parameters
measured or to be measured in neutrino experiments to a
cosmological parameter, the baryon-to-photon ratio, \mbox{$\eta_{B}\simeq
6\times 10^{-10}$}. 

According to the mass $M$ of the heavy neutrino that decays, the 
generation of asymmetry has to be described by different sets of 
classical Boltzmann equations: in the unflavored regime, when
$M \gtrsim 10^{12}\gev$, the lepton flavors are summed over, and
the asymmetry is estimated from two equations, one of which tracks
$N_{B-L}$, the total $B\!-\!L$ number, where $B$ and $L$ denote baryon
and lepton numbers, respectively. In the fully flavored regime, when 
$M \lesssim 10^{12}\gev$, the lepton flavor asymmetries 
$B/3-L_{\a},~\a=e+\m$ or $\t$, must be tracked instead. 
The fact that two descriptions coexist is due to the $\t$-lepton 
Yukawa interactions, which become important roughly when 
the temperature in the primordial plasma drops below $10^{12}\gev$. 
As we pointed out in paper III, the transition between the two regimes 
requires a quantum kinetic 
treatment, where correlations in flavor space and partial losses
of coherence are properly accounted for.

In the unflavored regime, when the heavy neutrino masses are hierarchical, 
one typically obtains a scenario where
only the contribution from the lightest right-handed neutrino, $N_1$, 
matters.
A lower bound on $M_1$ of about $10^{9}\gev$ for successful 
leptogenesis can then be derived. The associated lower bound on 
the reheat temperature $T_{\rm reh}$ of the Universe
after inflation being of the same order, a tension arises
between leptogenesis and 
supergravity theories because of a possible overproduction of gravitinos. 
Furthermore, it is interesting that one finds a stringent upper bound 
on the absolute neutrino mass scale, \mbox{$m_1\lesssim 0.1\ev$}. Finally, it
should be noted that, in the 
unflavored regime, the final asymmetry does not depend on
the lepton mixing matrix, which contains the potentially measurable
Dirac and Majorana \CP-violating phases. The necessary source of
\CP~violation is instead provided by unmeasurable phases in the high-energy
sector.  

In the fully flavored regime, the lowest bounds on $M_1$ and on the
reheat temperature do not change, as we stressed in paper II. On the other hand, 
the parameters accessible in neutrino experiments, notably the angle $\q_{13}$ and
the Dirac and Majorana \CP-violating phases, acquire importance.
According to the value they take, the predictions from leptogenesis
can differ by orders of magnitude, as we showed in paper II. Moreover, 
as investigated in detail in paper IV, the source of 
\CP~violation required for leptogenesis may be exclusively provided 
by the Dirac phase, which is the low-energy 
phase with the best prospects of being measured in the near future.

There are different ways to go beyond the ``vanilla'' picture where
the asymmetry is produced by the lightest right-handed neutrino and the lower
bounds introduced above hold. Following
paper I, three possibilities are discussed. First, a quasi-degenerate 
mass spectrum for the heavy neutrinos is considered, implying that two 
or three heavy neutrinos contribute
to the generation of asymmetry. It is then possible to relax the bounds 
on $M_1$ and $T_{\rm reh}$ to the TeV scale, and 
the upper bound on $m_1$ can be evaded.
Second, specific situations can be found where the asymmetry is produced
by the next-to-lightest right-handed neutrino, $N_2$. The lower bound on $M_1$
is then replaced by a lower bound on $M_2$, still implying a lower
bound on $T_{\rm reh}$ similar to the vanilla case. Finally,
even within the $N_1$-dominated scenario, a special choice of high-energy
parameters, which requires large cancellations, can be chosen in order 
to relax the lower bounds on $M_1$ and $T_{\rm reh}$.

Leptogenesis has recently entered a new era in that some of what has been
discussed above has only been realized in the last couple of years. Notably,
let me mention
the importance of flavor effects, the possible relevance of quantum effects,
as well as the role played by the next-to-lightest right-handed neutrino.

In the future, the case for leptogenesis will be weakened or 
strengthened depending on the outcome of the next-generation neutrinoless 
double-beta decay searches, the planned long-baseline neutrino 
experiments, as well as the LHC.

\newpage

\noindent
The content of this thesis is mainly based on the following works:

\begin{itemize}

\item[ I ]
 S.~Blanchet and P.~Di Bari, \\
\textit{Leptogenesis beyond the limit of hierarchical heavy neutrino masses},\\
JCAP {\bf 0606} (2006) 023 [arXiv:hep-ph/0603107]. 

\item [ II ]
S.~Blanchet and P.~Di Bari, \\
\textit{Flavor effects on leptogenesis predictions},\\
   JCAP {\bf 0703} (2007) 018 [arXiv:hep-ph/0607330].

\item[ III ]
 S.~Blanchet, P.~Di Bari and G.~G.~Raffelt,\\
 \textit{Quantum Zeno effect and the impact of flavor in leptogenesis},\\
 JCAP {\bf 0703} (2007) 012 [arXiv:hep-ph/0611337].

\item[ IV ]  A.~Anisimov, S.~Blanchet and P.~Di Bari,\\
 \textit{Viability of Dirac phase leptogenesis},\\
 JCAP {\bf 0804} (2008) 033 [arXiv:0707.3024].

\end{itemize}

\cleardoublepage

\chapter{Acknowledgments}


It is a very difficult task, if not impossible, to list all people 
who contributed directly or indirectly to the achievement of
my Ph.D. studies. 
Therefore, I shall undertake the more modest task of underlining
the major contributions. 

I would like first to express my gratitude to the 
Max-Planck-Institut f\"ur Physik (MPI) for
providing funding and optimal research possibilities through the 
newly born International Max Planck Research School for
Elementary Particle Physics. Thanks also to the Technische Universit\"at
M\"unchen (TUM), in particular to Prof. Michael Ratz, for providing 
the academic framework to my Ph.D. studies. 

Next, I would like to thank my supervisor within the institute, 
Georg Raffelt, for his kindness, support, 
availability, and the many useful advice he gave me both at the 
scientific and non-scientific levels.

For the research presented in this thesis, I am mostly indebted
to Pasquale Di Bari. I benefited a lot from his expertise 
in the fields of leptogenesis and cosmology in general. Moreover,
I enjoyed the long discussions about our work and beyond.

Then, my thanks go to all members of the ``astroparticle'' 
theory group who overlapped with me since October 
2005 at the MPI for a nice, relaxed and inspiring atmosphere. 
Special thanks to Yvonne Wong for suggestions about English issues
and to Pasquale Serpico for useful information about
the process of terminating a Ph.D. at the MPI and TUM. Thanks also 
to my office mate Stefano 
Pozzorini for the enjoyable ``Swiss'' atmosphere inside the 
office. Finally, thanks to the other IMPRS students, especially
Florian Hahn-W\"ornle, Edoardo Mirabella, and Joseph Pradler.

Last but not least, I am grateful to my parents for the
irreplaceable role they played from the very beginning until the 
present achievement of my studies.

\tableofcontents


\cleardoublepage

\pagestyle{fancyplain} \pagenumbering{arabic}
\renewcommand{\chaptermark}[1]{\markboth{Chap. \thechapter:\ #1}{}}
\renewcommand{\sectionmark}[1]{\markright{\thesection\ #1}}
\lhead[\fancyplain{}{\bfseries\thepage}] %
    {\fancyplain{}{\rightmark}}
\rhead[\fancyplain{}{\bfseries\leftmark}] %
    {\fancyplain{}{\bfseries\thepage}}
\cfoot{}

\mainmatter


\chapter{Introduction}
\label{chap:intro}


\section{The matter-antimatter puzzle}
\label{sec:matter-antimatter}

One can surely say that our understanding of the Universe has made 
a huge leap forward in the last few years. This is partly 
because the amount of data has dramatically increased, so that
an epoch of ``precision cosmology'' has started. But this is also
due to an accumulation of evidence for concepts that
were still considered exotic not so long ago, such as dark matter
and dark energy. Even though their nature is still unknown, at 
least there seems to be a consensus about 
their existence.

\begin{figure}
\begin{minipage}[t]{.46\textwidth}
\hspace{-0.4cm}
\includegraphics[width=1.1\textwidth]{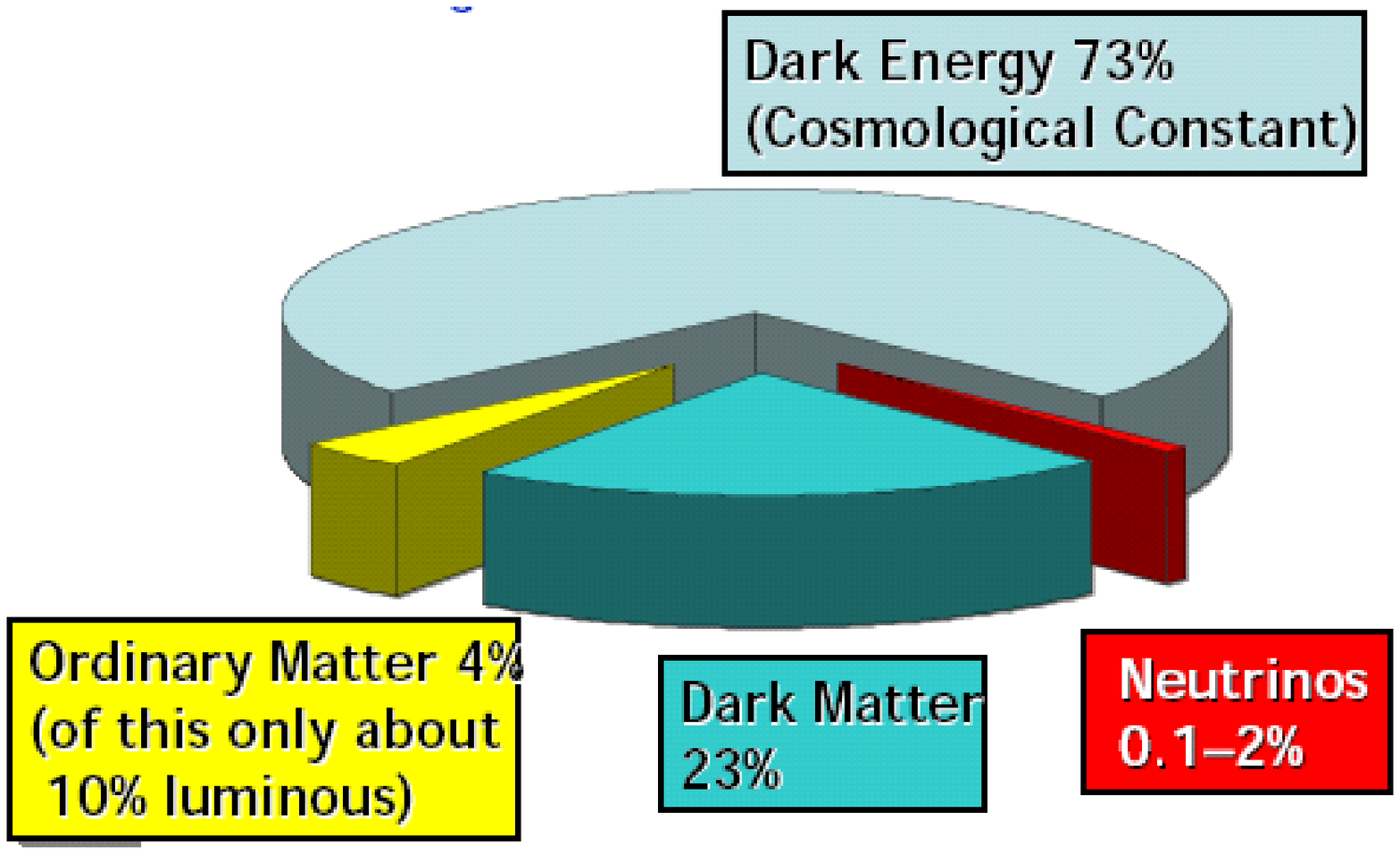}
\caption{The mass-energy budget of the Universe.}
\label{fig:cosmic}
\end{minipage}
\hfill
\begin{minipage}[t]{.46\textwidth}
\hspace{-0.4cm}
\includegraphics[width=1.1\textwidth]{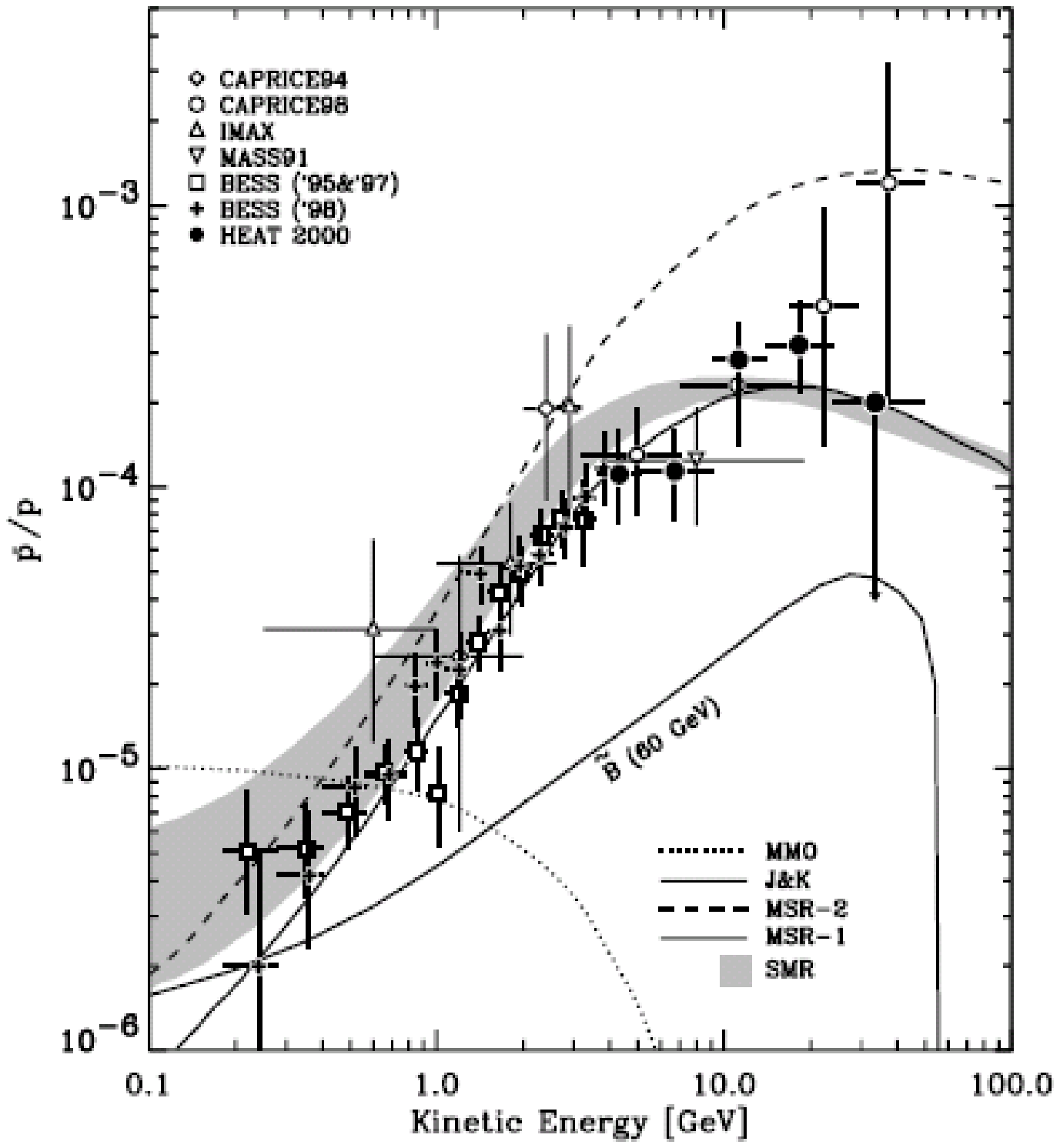}
\caption{The antiproton-to-proton ratio at the top of the 
atmosphere, as observed (points) and predicted from the models 
(lines)~\cite{Beach:2001ub}.}
\label{fig:protons}
\end{minipage}
\end{figure}

Nowadays, one speaks about a  ``Standard Cosmological Model'',
in analogy with its very successful counterpart of particle physics.  
The Standard Cosmological Model tells us that the
Universe is in a phase of accelerated
expansion and that the total energy
in the Universe is shared among at least four components 
(see Fig.~\ref{fig:cosmic}) which sum
to $\O_{\rm tot}\simeq 1$, meaning that the Universe is flat to a 
good precision. The dominant component (about 73\%)
is called dark energy, dark matter makes about 23\%, ordinary
matter (both luminous and dark) only 4\% and neutrinos 0.2--2\%,
the uncertainty here stemming from the unknown absolute neutrino mass
scale, as we shall see in Section~\ref{sec:absolute}.    

It is well known that the nature of dark matter is still mysterious. The particle
interpretation seems to be widely supported, and the candidates are
numerous: axion, lightest supersymmetric particle (neutralino,
gravitino), sterile neutrino, and many others. 

Dark energy is probably a much bigger issue still than dark matter.
It is supposed to drive the accelerated expansion, but its nature
is very unclear. It also raises a fundamental question about how
to treat the vacuum energy in quantum field theory.

Ordinary matter, which constitutes our bodies
as well as the Earth and the stars, does not seem at first to 
introduce any challenge to our understanding. However, this naive
perception is wrong because two very puzzling questions remain:
\begin{enumerate}
\item[1)]
Why is antimatter essentially absent in the observable Universe?
\item[2)] 
Why is the number density of baryons so small compared
to photons or neutrinos?
\end{enumerate}
These two questions are puzzling because, according to the 
Standard Big-Bang Theory, matter and antimatter evolved 
in the same way in the early Universe. On the other hand,
today the observable Universe is composed almost exclusively of
matter. Antimatter is only seen in particle physics accelerators
and in cosmic rays. Moreover, the rates observed in cosmic rays
are consistent with the secondary emission of antiprotons,
$n_{\bar{p}}/n_{p}\sim 10^{-4}$ (see Fig.~\ref{fig:protons}).

Ordinary matter is composed of baryons (protons, neutrons) and leptons
(electrons). One can assign an experimentally conserved number
to baryons and leptons. Baryons and leptons carry one unit of
these numbers, whereas antibaryons and antileptons carry one 
negative unit. In this way one can say that the predominance
of matter over antimatter is equivalent to the existence of a 
net baryon number. 

Following the Standard Big-Bang Theory and relying on the
Standard Model (SM) of particle physics, the relic density 
of baryons, i.e. nucleons here, can be easily estimated. One has 
the usual Boltzmann
equation for the number density of baryons $n_B$ (or antibaryons)
\begin{equation}\label{boltzbaryon}
{{\rm d}n_B\over {\rm d}z}+3Hn_B=- \langle \s_A |v|\rangle\left[n_B^2-\left(n_B^{\rm eq}
\right)^2\right],
\end{equation}
where $z=M_B/T$, $M_B$ is the baryon mass,
$H$ the Hubble expansion rate, $n_B^{\rm eq}$ the equilibrium
number density of baryons, and $T$ the cosmic temperature. 
The collision term on the right-hand side is given in terms
of an thermally-averaged annihilation cross-section 
$\langle \s_A |v|\rangle$, whose definition can be found
in \cite{kolbturner} for example. Eq.~(\ref{boltzbaryon}) can be 
conveniently rewritten in terms of the variable
$n_B/n_{\g}$, where $n_{\g}$ is the number density of photons, allowing one 
to factor out the effects of the expansion of the Universe.
One obtains
\begin{equation}
{{\rm d}(n_B/n_{\g})\over {\rm d}z}=- {n_{\g} z \over H(M_B)}\langle \s_A |v|\rangle
\left[\left({n_B\over n_{\g}}\right)^2-\left({n_B^{\rm eq}\over n_{\g}}\right)^2\right].
\end{equation}

In our case, the important annihilation
channel is into pions (e.g. \mbox{$p+\bar{p}\to \p^++\p^-$}). 
Taking the averaged cross-section to be $\langle \s |v|\rangle=
C_1 M_{\p}^{-2}$, with \mbox{$M_{\p}\simeq 135$ MeV} and where $C_1$ is a 
numerical factor of order unity, the freeze-out occurs at $T\sim 22$~MeV. 
Neglecting the entropy production
in $e^+e^-$-annihilations, one finds for today's abundance (subscript `0')~\cite{kolbturner} 
\begin{equation}\label{classicalcomp}
\left. {n_B\over n_{\g}}\right|_0=\left.{n_{\bar B}\over n_{\g}}\right|_0
\simeq 7\times 10^{-20}C_1^{-1}.
\end{equation}
Note that the ratio of baryon number density to photon number density today
is usually referred to as the \emph{baryon-to-photon ratio}, 
\be
\eta_B\equiv \left. {n_B\over n_{\g}}\right|_0.
\ee
The result of our simple computation, Eq.~(\ref{classicalcomp}), is
clearly a small number, which would perhaps explain the question 2) above. But
one notices immediately a first
problem, namely because the abundances of baryons and antibaryons
are predicted in this way to be the same. Baryons and antibaryons
do not evolve in distinctive ways so that one expects today the same
amount of each of them. So, the argument we have just described
leaves open the question 1) above. But let us for the moment ignore 
this point and
try to see if the abundance of baryons matches observation.

The photon density follows directly from the measurement of the 
Cosmic Microwave Background (CMB) temperature and from Bose-Einstein
statistics: $n_{\g}\sim T^3$. Determining the baryon content of the 
Universe is more difficult. Direct
measurements are not accurate, because only a small fraction of baryons
formed stars and other luminous objects (see Fig.~\ref{fig:cosmic}). 
However, we can rely on two different indirect probes.

\begin{figure}
\begin{center}
\includegraphics[width=0.57\textwidth]{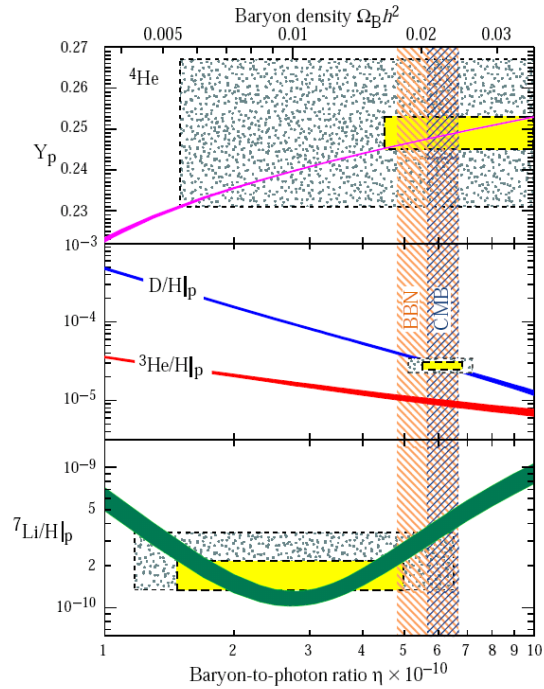}
\caption{
The observed abundances of light elements compared to the standard BBN
predictions~\cite{PDBook}. The smaller boxes indicate $2\s$
statistical errors, the larger ones $\pm 2\s$ statistical and systematic
errors.}
\label{fig:BBN}
\end{center}
\end{figure}

The first probe is Big-Bang Nucleosynthesis (BBN). 
The abundances of light elements such as ${}^4{\rm He}$, D, ${}^3{\rm He}$ 
and ${}^7{\rm Li}$ predicted by the standard theory of BBN crucially depend 
on $\eta_B$. 
Comparing predictions with observations, as shown in Fig.~\ref{fig:BBN},
the following baryon-to-photon ratio is inferred~\cite{Fields:2006ga}:
\begin{equation}
\eta_B\simeq (5.5\pm1.0)\times 10^{-10}.
\end{equation}

\begin{figure}
\begin{center}
\includegraphics[width=0.6\textwidth]{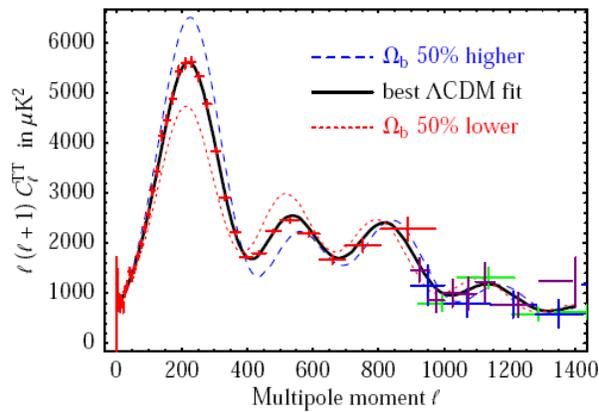}
\caption{The dependence of CMB temperature anisotropies on the 
baryon abundance $\O_b$ (or $\eta_B$), compared with data~\cite{Strumia:2006qk}.}
\label{fig:CMB}
\end{center}
\end{figure}

The CMB temperature anisotropies, very well measured by the WMAP satellite,
offer the second probe. These anisotropies reflect the acoustic oscillations 
of the baryon-photon fluid
which happened around photon decoupling. A precise computation
can be done evolving Boltzmann equations for anisotropies, assuming
that they are generated by quantum fluctuations during inflation.
Fig.~\ref{fig:CMB} illustrates how the amount of anisotropies
with angular scale $\sim 1/\ell$ depends on $\eta_B$. The baryon-to-photon
ratio obtained from 3 years of WMAP data is~\cite{WMAP3}
\begin{equation}\label{WMAP}
\eta_B \simeq (6.1\pm 0.2)\times 10^{-10}.
\end{equation}


The synthesis of light elements occurred during 
the first 3 minutes in the history of the Universe, whereas 
the photon decoupling occurred when the Universe was
400 thousand years old. The fact that these two completely 
different probes of the baryon content of the Universe give compatible
results is one of the great successes of modern cosmology.\footnote{
Throughout the thesis, we shall exlusively use 
the value of $\eta_B$ obtained from CMB temperature anisotropies, 
Eq.~(\ref{WMAP}), which has much smaller errors.}

It should now be clear that the result from the ``classical'' 
computation of the baryon density given in Eq.~(\ref{classicalcomp})
is at odds with the observed value, Eq.~(\ref{WMAP}). In order to avoid the ``baryon
annihilation catastrophe'' leading to the value in Eq. 
(\ref{classicalcomp}), one has to generate a primordial asymmetry 
between baryons and antibaryons. This small asymmetry, 
\emph{at the
level of one part in one billion}, would imply that after the 
annihilation process has occurred at full strength, one remains
with the small excess of baryons over antibaryons. The problem
of generating this small excess of baryons over antibaryons is 
often called the \emph{baryogenesis problem}.

The solution to the baryogenesis problem requires the generation of a small
baryon asymmetry primordially. Sakharov, in 1967, enunciated the
three necessary conditions for such a process to be possible at some
stage in the history of the Universe~\cite{Sakharov:1967dj}:
\begin{enumerate}
\item Baryon number violation.

\item $C$ and \CP~violation.

\item Departure from thermal equilibrium.
\end{enumerate}

In principle, the SM contains all these ingredients. Indeed,
\begin{enumerate}
\item Due to the chiral nature of weak interactions,
$B$ and $L$ are not conserved~\cite{'tHooft:1976up}. At zero
temperature, this has no observable effect due to the smallness
of the weak coupling. However, as the temperature
reaches the critical temperature $T_{\rm EW}$ of the electroweak
phase transition, $B$ and $L$ violating processes
come into thermal equilibrium~\cite{Kuzmin:1985mm,Arnold:1987mh}. 
The rate of these processes is related to the free energy of
the sphaleron-type field configurations which carry 
topological charge~\cite{Manton:1983nd}. In the SM they lead
to an effective interaction operator of all left-handed fermions, 
\mbox{$O_{B+L}
=\prod_i Q_{Li}Q_{Li}Q_{Li}\ell_{Li}$}, which violates baryon and
lepton number by three units. On the other hand, $B\!-\!L$ remains 
conserved.
 
\item The weak interactions of the SM violate $C$ maximally
and violate \CP~via the Kobayashi-Maskawa mechanism~\cite{Kobayashi:1973fv}.
The latter originates from the quark mixing matrix, 
often called the Cabibbo-Kobayashi-Maskawa (CKM) matrix, 
which contains one \CP-violating phase. This phase is known
to be non-zero since \CP~violation has been observed
in the $K$ and $B$ mesons systems (see e.g. the review on 
\CP~violation in~\cite{PDBook}).

\item A strongly 
first-order electroweak phase transition
in the early Universe could provide the out-of-equilibrium condition. A
first-order phase transition proceeds via nucleation and growth of
bubbles~\cite{Bernreuther:2002uj}.
\end{enumerate}
This scenario is called \emph{electroweak baryogenesis in the Standard
Model}. However, it fails for two reasons. First,
it turns out that, for the electroweak phase transition to be
strongly first order, the mass of the Higgs particle should be
smaller than about 45~GeV~\cite{Jansen:1995yg} 
(see also \cite{Shaposhnikov:1987tw}). 
However, LEP~II gives the well-known bound 
$M_h > 114$ GeV~\cite{PDBook}.
Second, the source of \CP~violation in the quark sector is far
too small, due to the smallness of some of the quark masses~\cite{Gavela:1993ts}.

In conclusion, successful baryogenesis requires
physics beyond the SM, just as the dark matter and dark energy problems!
One intriguing solution to the problem of baryogenesis is deeply connected
with the neutrino sector and in particular with neutrino masses. This is
the topic of the next section.

\section{The puzzle of neutrino masses}

\subsection{Theory of neutrino oscillations}
\label{sec:theoryneut}

Even with tiny masses, massive neutrinos can behave very differently
from massless ones. In particular, massive neutrinos naturally lead to 
neutrino mixing and to neutrino oscillations, which have been recently 
observed, as we shall see below. Let us sketch
how this happens.



Consider a neutrino beam created in a charged current
interaction along with the antilepton $\a^{+},~\a=e,\m,\t$. By definition
the neutrino created is called $\n_{\a}$. In general, this is not
a physical particle, but rather a superposition of physical fields
$\n_{i}$ with different masses $m_i$:
\be\label{neutbeam}
|\n_{\a}\rangle=\sum_{i} U^{\star}_{\a i}\,|\n_i\rangle,
\ee
where $U$ is the lepton mixing matrix, also known as the 
Pontecorvo-Maki-Nakagawa-Sakata (PMNS) matrix~\cite{Pontecorvo:1957qd,Pontecorvo:1957cp,Maki:1962mu}. 
By analogy with the CKM matrix in the quark
sector, the lepton mixing matrix can be conveniently parametrized as
\begin{equation}\label{PMNSmatrix2}
U=V \times {\rm diag}\,({\rm e}^{{\rm i}\,{\Phi_1\over 2}}, {\rm e}^{{\rm i}\,{\Phi_2\over 2}}, 1),
\end{equation}
\begin{displaymath}
V=\left( \begin{array}{ccc}
1&0&0\\
0& c_{23} & s_{23}\\
0& -s_{23} &c_{23}
\end{array}\right)\left( \begin{array}{ccc}
c_{13}&0&s_{13}\,{\rm e}^{-{\rm i}\,\d}\\
0& 1 & 0\\
-s_{13}\,{\rm e}^{{\rm i}\,\d}& 0 &c_{13}
\end{array}\right)\left( \begin{array}{ccc}
c_{12}&s_{12}&0\\
-s_{12}& c_{12}&0\\
0&0& 1
\end{array}\right),
\end{displaymath}
where $s_{ij}\equiv \sin \theta_{ij}$, $c_{ij}\equiv \cos \theta_{ij}$,
$\d$ and $\F_{1,2}$ are the Dirac and Majorana \CP-violating
phases, respectively. The Majorana phases, which are not present 
in the quark sector, are related to the possible Majorana nature 
of neutrinos (see Section~\ref{sec:solution}). 

For a simple-minded approach to the propagation of the state $|\n_{\a}\rangle$,
 we assume
that the 3-momentum $\mathbf{p}$ of the different components in the beam are the
same. However, since their masses are different, the energies of all
these components cannot be equal. Rather, for the component $\n_i$,
the energy is given by the relativistic energy-momentum relation
$E_i=\sqrt{\mathbf{p}^2+m_i^2}$.
After a time $t$, the evolution of the initial beam, Eq.~(\ref{neutbeam}),
assuming that neutrinos are stable particles, gives
\be\label{propneut}
|\n_{\a}(t)\rangle=\sum_{i} e^{-{\rm i} E_{i}t} U^{\star}_{\a i}|\n_{i}\rangle.
\ee
Since all $E_{i}$'s are not equal if the masses are not, Eq.~(\ref{propneut})
represents a different superposition of the physical eigenstates $\n_i$
compared to Eq.~(\ref{neutbeam}). In general, the state in Eq.~(\ref{propneut})
can therefore show properties of other flavor states. The amplitude
of finding a $\n_{\a '}$ in the original $\n_{\a}$ beam is
\be
\langle\nu_{\a '}|\n_{\a}(t)\rangle=\sum_{i} e^{-{\rm i} E_i t} 
U^{\star}_{\a i} U_{\a ' i},
\ee
using the fact that $\langle \n_i |\n_j\rangle =\d_{ij}$.
The probability of finding a $\n_{\a '}$ in the original $\n_{\a}$ beam 
at any time $t$ is then the modulus squared of the amplitude,
$P_{\n_{\a}\to \n_{\a'}}(t)=|\langle\nu_{\a '}|\n_{\a}(t)\rangle|^2$.
In all practical situations, neutrinos are extremely relativistic, so that
one can approximate the energy-momentum relation as $E_i\simeq |\mathbf{p}|+m_i^2/
(2|\mathbf{p}|)$ and replace $t$ by the distance $x$ traveled by the beam. After
a few manipulations one can finally write the vacuum oscillation probability as
\bea\label{oscprob}
P_{\n_{\a}\to \n_{\a'}}(x)&=&\d_{\a \a'}-4\sum_{i>j} {\rm Re}\left( 
U^{\star}_{\a i} U_{\a' i} U_{\a j}
U^{\star}_{\a' j}\right) \sin^2\left({\D m_{ij}^2\over 4 E} x\right)\nonumber\\
&& \quad +2\sum_{i>j} {\rm Im}\left( 
U^{\star}_{\a i} U_{\a' i} U_{\a j}
U^{\star}_{\a' j}\right) \sin\left({\D m_{ij}^2\over 4 E} x\right),
\eea
where $E\simeq |\mathbf{p}|$ and $\D m_{ij}^2\equiv m_i^2-m_j^2$. From this
formula it is apparent that neutrino 
oscillations require non-zero neutrino masses and mixings.

\subsection{Experimental evidence for neutrino oscillations}
\label{sec:expev}

The last 10 years have been extremely successful for the
field of neutrino physics. In 1998, the Super-Kamiokande experiment
in Japan reported the first compelling evidence for
neutrino oscillations as a way to explain the anomaly in atmospheric
neutrinos. 
Super-Kamiokande not only confirmed the previously found deficit in 
$\nu_{\mu}$-type events but also measured a zenith angle dependent 
$\nu_{\m}$-deficit which was inconsistent with expectations based
on calculations of the atmospheric neutrino flux. The neutrino 
oscillation explanation $\nu_{\mu}\to \nu_{\t}$ with a quasi-maximal 
mixing angle~\cite{SuperK1} appeared therefore as the most convincing
one. 

\begin{figure}
\begin{minipage}[t]{.47\textwidth}
\hspace{-0.7cm}
\includegraphics[width=1.1\textwidth]{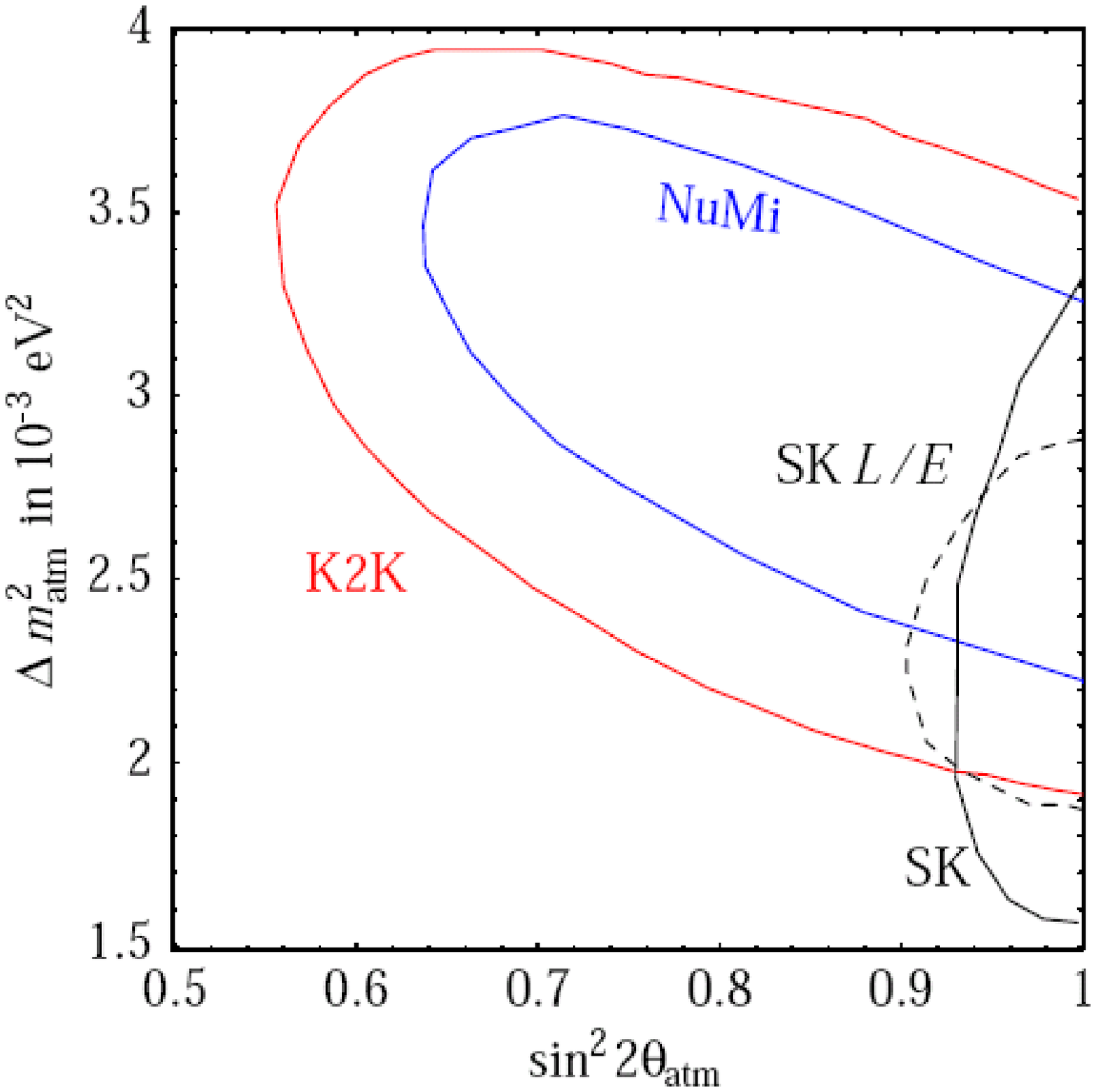}
\caption{Best fit regions at 90\% C.L. 
for atmospheric and accelerator neutrinos~\cite{Strumia:2006db}.} 
\label{fig:atm}
\end{minipage}
\hfill
\begin{minipage}[t]{.47\textwidth}
\hspace{-0.7cm}
\includegraphics[width=1.1\textwidth]{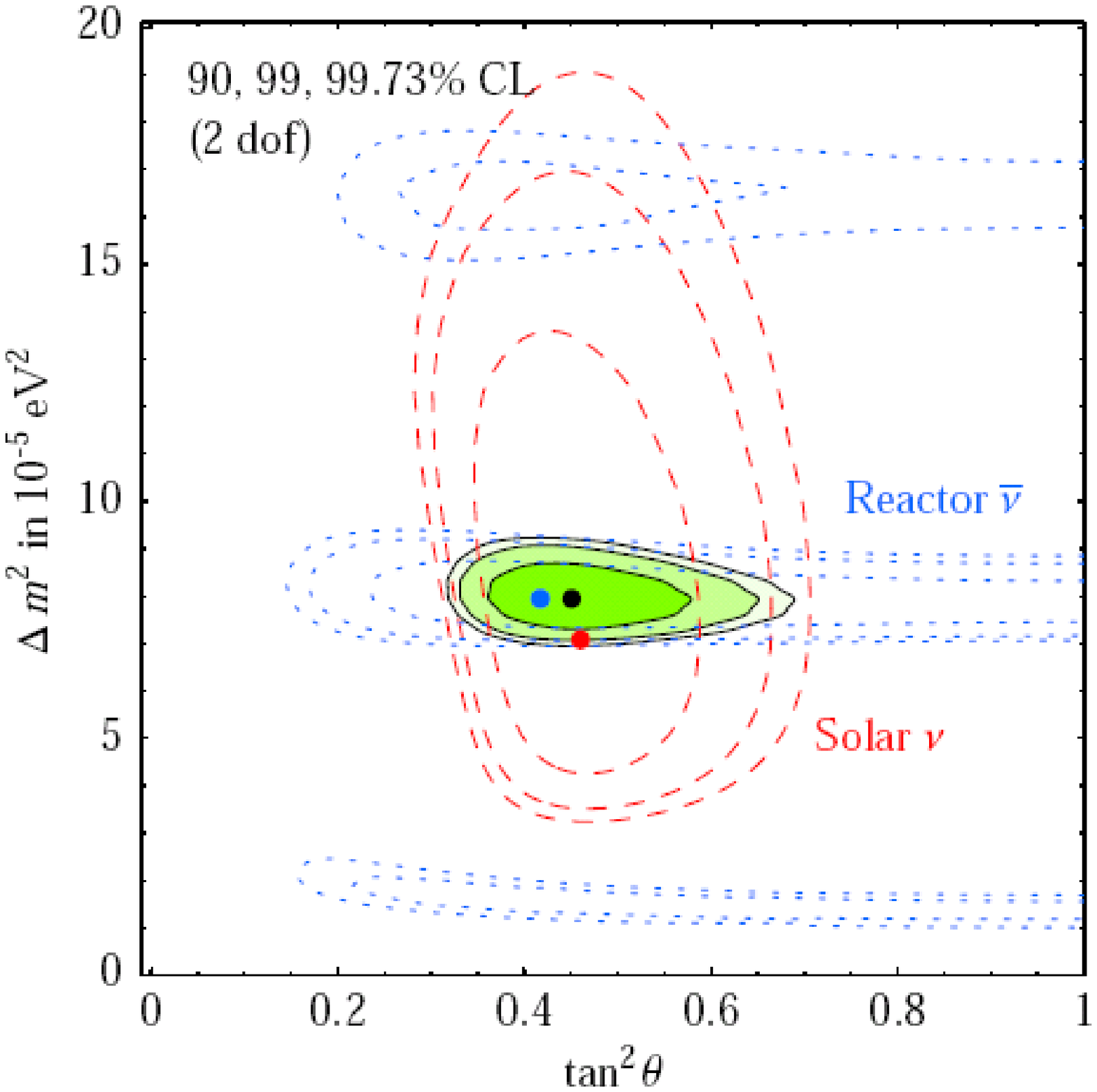}
\caption{Best fit regions at 90, 99 and 99.73\% C.L. for 
solar and reactor neutrinos~\cite{Strumia:2006db}.}
\label{fig:sol}
\end{minipage}
\end{figure}

Two dedicated laboratory experiments have been conceived 
in order to check this picture. 
The experiment K2K in Japan, which collected data until November 2004,
used a pulsed beam of muon-neutrinos produced at KEK and detected at
Super-Kamiokande (distance of 250 km). The currently running MINOS experiment
uses a pulsed beam of muon-neutrinos produced at NuMI (Fermilab), and the far 
detector is located at a distance of 735 km in the Soudan mine, Minnesota. 
Both experiments
point to a neutrino oscillation interpretation of their data,
with mixing parameters compatible with those explaining the atmospheric 
anomaly \cite{K2K,MINOS}. A summarizing
plot for ``atmospheric'' neutrinos can be found in Fig.~\ref{fig:atm}. 
The best-fit parameters are~\cite{Strumia:2006db}:
\begin{eqnarray}
|\Delta m^2_{32}|\equiv \Delta m^2_{\rm atm}&=&(2.5\pm 0.2)\times 10^{-3} 
{\rm eV}^2, \label{atmmass}\\
\sin^2 2\theta_{23}&=&1.02\pm 0.04,\label{atmparam}
\end{eqnarray} 
where the ranges indicated are at $1\s$.

Neutrinos from the Sun were first studied by Ray Davis in the 
Homestake mine in South Dakota in the 1960's~\cite{Davis:1964hf}. This
pioneering work made use of the radiochemical technique,
$\nu_e+{}^{37}{\rm Cl}\to e^- +{}^{37}{\rm Ar}$. He quickly found that the
observed flux was smaller than the one predicted by Bahcall and
collaborators from the Standard Solar Model~\cite{Bahcall:2000nu}. 
The \emph{solar neutrino puzzle} was born. Later,
other experiments were conceived to check this deficit in solar
neutrinos: Kamiokande~\cite{Kamiokande} and later 
Super-Kamiokande~\cite{SuperKsol}, which
used the water Cherenkov technique, also found a deficit, as 
well as SAGE~\cite{SAGE} and GALLEX/GNO~\cite{GALLEX,GNO}, which used the
radiochemical method with Gallium. 

It should be noted that the deficits found in the
different experiments were different, highlighting the energy dependence,
as well as the fact that the experiments were based on different techniques.
The radiochemical method allows only for the measurement
of charged-current (CC) events, which is exclusively sensitive to electron
neutrinos. Experiments relying
on water Cherenkov detection register only elastic scattering (ES) events,
$\n+e\to \n+e$. This technique can measure in principle all neutrino
flavors, but the efficiency in the $\m$ and $\t$ channels is much
lower, due to the lower cross-sections.

Only the SNO experiment in Sudbury, Canada, used
a heavy water detection which allowed for the detection via three channels
simultaneously: CC, ES as well as neutral current (NC) thanks to the 
reaction $\nu +d\to \n +n+p$, where $d$ denotes deuteron. This
made possible a reliable measurement of the total flux of neutrinos
from the Sun. Hence, in addition
to confirming the deficit in solar neutrinos in the CC and ES 
detection channels \cite{SNO1}, this experiment
allowed for the first direct test of the Standard Solar
Model. In 2002, the data were found to be actually consistent with the
prediction \cite{SNO2}, finally resolving the longstanding solar
neutrino puzzle. This can be seen as a huge success in the history of
neutrino physics. The Borexino experiment, located in the
Gran Sasso laboratory, has recently added another milestone to 
the understanding of
neutrinos from the Sun by detecting in real time a flux of ${}^7{\rm Be}$
neutrinos consistent with the Standard Solar Model prediction~\cite{Collaboration:2007xf}. 

The neutrino oscillation explanation $\n_e\to\n_{\m,\t}$ 
to the deficits in neutrinos from the Sun in different 
experiments was popular from the beginning, because it appeared as the simplest and
most elegant solution. But the first experiment which really left
neutrino oscillations as the only possible explanation for
the solar neutrino puzzle is KamLAND, rejecting, for instance \emph{spin
flavor oscillations} \cite{Okun:1986na,Okun:1986hi}, which 
were still allowed after SNO. 
KamLAND is a scintillation detector located in Japan which observed the $\bar{\n}_e$ 
from neighboring nuclear reactors thanks to the process $\bar{\n}_e+p\to n+e^+$. 
This experiment saw a $6\sigma$ evidence
for the disappearance of $\bar{\n}_e$~\cite{KamLAND1,KamLAND2}. A summarizing
plot of the solar neutrino and KamLAND data can be found in Fig.~\ref{fig:sol}. 
The best-fit parameters are~\cite{Strumia:2006db}:
\begin{eqnarray}
|\Delta m^2_{21}|\equiv \Delta m^2_{\rm sol}&=&(8.0\pm 0.3)\times 10^{-5} 
{\rm eV}^2, \label{solmass}\\
\tan^2 \theta_{12}&=&0.45\pm 0.05,\label{solparam}
\end{eqnarray} 
where the ranges indicated are at $1\s$.


For completeness, we should add that a short-baseline reactor
neutrino experiment named CHOOZ, in France, was performed to 
get a handle on the third mixing angle,
$\theta_{13}$ \cite{CHOOZ}. They report no signal, which, together
with atmospheric and K2K data, yields the $1\s$ upper 
bound~\cite{Strumia:2006db}
\begin{equation}\label{theta13}
\sin^2 2\theta_{13}<0.05.
\end{equation}

\subsection{Absolute neutrino mass scale}
\label{sec:absolute}

The oscillation formula Eq.~(\ref{oscprob}) shows that 
neutrino oscillation experiments
are only able to provide information on neutrino mass-squared \emph{differences}.
They are insensitive to the absolute neutrino mass scale. Fortunately,
there are other ways to probe it.

First, the so-called ``direct'' measurement makes use of the very 
precise determination of the upper end of the spectral distribution
of the electron in the tritium $\b$-decay, \mbox{
${}^3{\rm H}\to {}^3{\rm He} +\bar{\n}_e+e^-$}. The Mainz
experiment obtained the bound $m_{\n_e}<2.2\ev$~\cite{Mainz} 
and the Troitsk experiment $m_{\n_e}<2.5\ev$~\cite{Lobashev:1999tp}, both
at 95\% C.L., where
\begin{equation}\label{mne}
m_{\n_e}\equiv \sqrt{\sum_{i=1}^3 |U_{ei}|^2m_i^2},
\end{equation}
with $U$ denoting the lepton mixing matrix we introduced in 
Section~\ref{sec:theoryneut} [cf.~Eq.~(\ref{PMNSmatrix2})]. The upcoming KATRIN 
experiment~\cite{Osipowicz:2001sq},
which is under construction at the moment, is expected to have a discovery 
potential down to $m_{\n_e}\simeq 0.35\ev$. 

Second, if neutrinos are Majorana particles (see next section), then 
it might be possible 
to observe neutrinoless double-beta ($0\n\b\b$) decay for nuclei like ${}^{76}
{\rm Ge}$. 
Double-$\b$ decay, $(A,Z)\to (A,Z+2)+2e^-+2\bar{\n}_e$, 
has been already observed, even though the rate is of second order in
the Fermi coupling constant $G_{\rm F}$, implying for instance a lifetime 
of about $10^{21}$~years for ${}^{76}{\rm Ge}$.
The $0\n\b\b$-decay rate is even more suppressed, and depends
on yet another combination of the neutrino masses as in Eq.~(\ref{mne})
~\cite{Bilenky:1987ty},
\begin{equation}\label{effmass}
|\langle m\rangle|\equiv \left|\sum_{i=1}^3 U_{ei}^2 m_i\right|.
\end{equation} 
Up to now, no compelling signal\footnote{Part of the Heidelberg-Moscow
collaboration claims an evidence for $0\n\b\b$-decay, leading
to the measurement $|\langle m\rangle |=0.2\textrm{--}0.6~{\rm eV}$ 
(90\% C.L.)~\cite{KlapdorKleingrothaus:2004wj,KlapdorKleingrothaus:2001ke},
where the uncertainty is due to the poorly known nuclear matrix 
elements.}  
for $0\n\b\b$-decay has been
observed, and the currently running CUORICINO experiment, which uses
${}^{130}{\rm Te}$,
obtains now a bound~\cite{Arnaboldi:2008ds}
\begin{equation}\label{effneutCUOR}
|\langle m\rangle |<0.19\textrm{--}0.68~{\rm eV}\quad (90\%~{\rm C.L.})
\end{equation}
which is slightly stronger than the one obtained in 
the previous Heidelberg-Moscow~\cite{HM} and
IGEX experiments~\cite{Aalseth:1999ji,Aalseth:2002rf}, where ${}^{76}{\rm Ge}$ was
used.
Note that the range shown in Eq.~(\ref{effneutCUOR}) stems from the uncertainty in
the nuclear matrix elements. Future
planned experiments such as GERDA~\cite{Abt:2004yk}, MAJORANA~\cite{Aalseth:2004yt}
or CUORE~\cite{Ardito:2005ar} aim at a sensitivity in the 
50~meV range.

There is a third probe of the absolute neutrino mass scale, and this
is cosmology. More precisely, the CMB data and the data from the large
scale structure (LSS) have a sensitivity to the sum of the neutrino
masses $\sum_i m_i$. Looking at the literature in the last 3 years that
dealt with this issue, it is clear that the bound on the sum of the neutrino
masses depends on which data sets are included. We simply show here 
the result from a somewhat 
conservative calculation in \cite{Hannestad:2006mi}:
\begin{equation}
\sum_{i=1}^3 m_i<0.6~{\rm eV}\quad (95\%~{\rm C.L.}).
\end{equation}

\begin{figure}
\begin{center}
\includegraphics[width=0.75\textwidth]{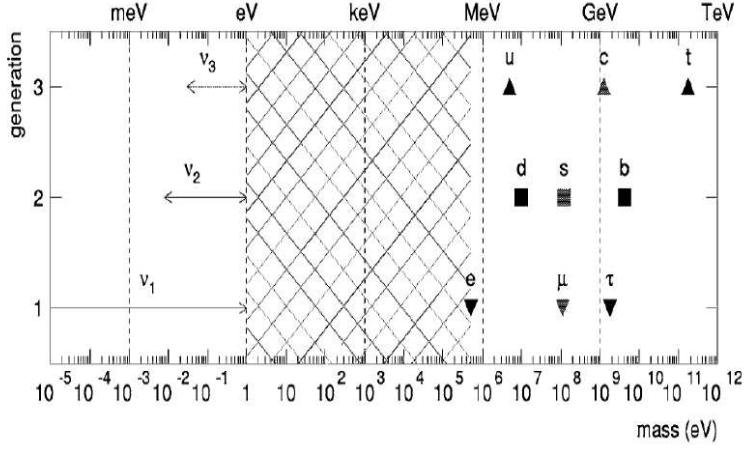}
\caption{Mass spectrum of fermions.}
\label{fig:NeutrinoGap}
\end{center}
\end{figure}

The three probes we have presented involve different combinations of the neutrino
masses, where even cancellations are possible in the case of $0\n\b\b$-decay
[cf. Eq.~(\ref{effmass})]. But the bottom line here 
is that all three probes tend to the same conclusion: \emph{the neutrino mass scale 
has to be in the sub-eV range}. In this sense, this makes neutrinos a very peculiar
particle in the SM, with a mass that is six orders of magnitude smaller
than the electron! Fig.~\ref{fig:NeutrinoGap} illustrates nicely this gap 
between neutrinos
and the other SM particles. Clearly, such a gap must be explained on 
theoretical grounds, and this is the topic of the next section.

\section[The see-saw mechanism and leptogenesis]{An elegant solution: \\
the see-saw mechanism and leptogenesis}
\label{sec:solution}

As we have just seen, neutrinos are very special
particles due to their extreme lightness. This might suggest that 
neutrinos have a unique mass generation mechanism that leads to 
small masses in a natural way.
One of the first attempts in the late 1970's and early 1980's
was the use of the see-saw mechanism~\cite{Minkowski:1977sc,Yanagida,Gell-Mann,Glashow,Barbieri:1979ag,Mohapatra:1980yp}, which we shortly describe now.

Let us imagine that we would like to write down the most simple 
and general extension
of the Standard Model that accomodates neutrino masses. For simplicity
we consider here only one generation of neutrinos. In the SM,
in order to generate massive fermions, one introduces a right-handed (RH)
component for each particle. So let us introduce one RH neutrino $N_R$
in a usual Dirac-type term $m_{\rm D} \overline{\n_{L}} N_{R}+h.c.$ Contrary
to the other SM particles, this is not the only term allowed for neutrinos. 
Since neutrinos are neutral particles, the
RH component can only be a singlet of all SM gauge interactions,
including hypercharge. This unique property of neutrinos implies that
the Majorana mass term ${1\over 2}M_{\rm M}\overline{(N_R)^c} N_R+h.c.$
is allowed, where the superscript $c$ denotes charge conjugation. 
Actually, in general there is no reason why this term should be absent. 
The mass term for neutrinos can then be conveniently written in a
 matrix form as 
\begin{equation}\label{massmatrix1gen}
\mathcal{L}_{\rm mass}=-{1\over 2} \left(\begin{array}{cc}
\overline{\n_L} & 
\overline{(N_R)^c}\end{array}\right)
\left(\begin{array}{cc} 0& m_{\rm D}\\
m_{\rm D} & M_{\rm M} \end{array}\right)\left(\begin{array}{c}
(\nu_L)^c \\ N_R\end{array}\right)+h.c.
\end{equation}
Note that upper left component
of the mass matrix is zero because it is not possible to write 
down a Majorana mass term
for left-handed fields without breaking gauge invariance.

The see-saw mechanism then assumes that the Majorana mass is 
much larger than the Dirac mass, i.e. $M_{\rm M}\gg m_{\rm D}$.
Diagonalizing the mass matrix with this assumption yields
the two mass eigenvalues 
\be\label{mM}
m\simeq {m_{\rm D}^2 \over M_{\rm M}},\quad M\simeq M_{\rm M},
\ee 
where the first eigenvalue corresponds to the mass of a light 
Majorana neutrino, and the second eigenvalue corresponds to the
mass of a heavy Majorana neutrino. It is interesting to notice
that plugging a Dirac mass at the electroweak scale,  
$m_{\rm D}=100\gev$, with a heavy neutrino mass of 
$M_{\rm M}=10^{14}\gev$ maybe related to the underlying Grand Unified
Theory (GUT), one obtains a light neutrino mass $m=0.1\ev$. This
can be seen as a strong support for this model. For the generalization
of the previous discussion to three massive neutrinos and more details, 
see Appendix~\ref{chap:see-saw}.

The version of the see-saw mechanism we have just described, with the
addition of RH neutrinos to the SM, is nowadays 
referred to as the \mbox{type-I} see-saw mechanism. For completeness, let us mention
that other see-saw mechanisms were proposed to explain the smallness 
of neutrino masses:
the addition of $SU(2)$ triplet Higgses to the SM 
leads to the type-II mechanism~\cite{Magg:1980ut,Lazarides:1980nt,Mohapatra:1980yp}, 
and the addition of $SU(2)$ triplet fermions leads to the type-III
mechanism~\cite{Foot:1988aq,Ma:1998dn}.
Throughout the thesis, we shall exclusively deal with the type-I mechanism.

As we have just shown, the see-saw mechanism elegantly solves the problem 
of generating
small neutrino masses in a natural way. But this is not the only
virtue of the model: its 
cosmological consequence is indeed that it can also 
solve the puzzle of the matter-antimatter
asymmetry of the Universe through \emph{thermal leptogenesis}, as first proposed
by Fukugita and Yanagida 20 years ago~\cite{Fukugita:1986hr}. 
We saw in 
Section~\ref{sec:matter-antimatter} that one needs to satisfy all 
three Sakharov's conditions in order to solve the baryogenesis 
problem. Let us see how they are
satisfied in the case of thermal leptogenesis.
\begin{enumerate}
\item Lepton number is violated in the decays of the heavy neutrinos into 
lepton doublets and Higgs doublets. Additionally, $B$ and $L$ are violated 
by the non-perturbative
sphaleron processes, which however conserve $B-L$. This is thus the 
same source of $B$ violation
as the one described in the case of electroweak baryogenesis (see
Section~\ref{sec:matter-antimatter}). The difference is that
the sphaleron transitions are here transferring $L$ number produced
in the decays of the heavy neutrinos into the required $B$ number. 
These transitions should be in
equilibrium for temperatures~\cite{Bodeker:1999gx,Arnold:1996dy,Moore:2000mx}
\be
T_{\rm EW}\sim 100~{\rm GeV}<T\lesssim 10^{12}~{\rm GeV}.
\ee

\item \CP~violation occurs in the decays of the heavy
neutrinos. The relevant quantity to evaluate is the \CP~asymmetry parameter,
defined as
\be\label{CPasym}
\ve_i\equiv -{\G_i-\overline{\G}_i
\over \G_i+\overline{\G}_i},
\ee
where $\G_i=\sum_{\a}\G(N_i\to \ell_{\a}\F^{\dagger})$ and 
$\overline{\G}_i=\sum_{\a}\G(N_i\to \bar{\ell}_{\a}\F)$.
As shown in Fig.~\ref{fig:CPasym}, one can calculate 
such an asymmetry by computing the interference between 
the tree-level diagram and the two relevant one-loop diagrams, namely 
the self-energy diagram and the vertex correction. We will show
the result of such a calculation in the next chapter. 
 This source of \CP~violation is typically sufficient to explain the 
matter-antimatter problem.

\begin{figure}
\begin{center}
\includegraphics[width=1.\textwidth]{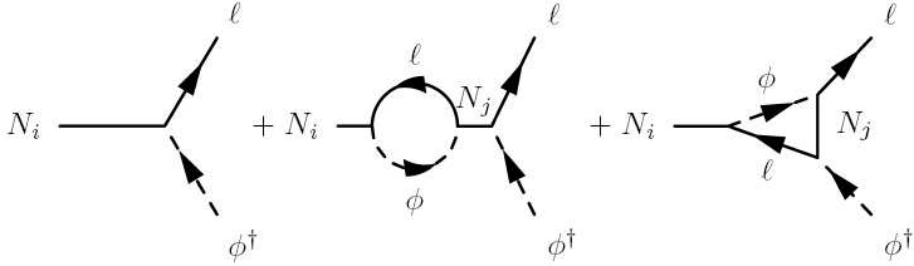}
\caption{The diagrams necessary to compute the \CP~asymmetry
in leptogenesis.}
\label{fig:CPasym}
\end{center}
\end{figure}

\item The out-of-equilibrium condition will be satisfied thanks
to the expansion of the Universe. Convenient quantities to describe
when the decays of the heavy neutrinos freeze out are the
decay parameters, $K_i$,
defined as the ratio of the total decay width of the heavy
neutrino $N_i$ to the expansion rate at $T=M_i$,
\be\label{decpar}
K_i\equiv {\G_{{\rm D},i}(T=0)\over H(T=M_i)},
\ee
where $\G_{{\rm D},i}\equiv \G_i+\overline{\G}_i$.
This is the key quantity for the thermodynamical
description of the decays of heavy particles in the early Universe
\cite{kolbturner}. In leptogenesis it can be conveniently expressed
in terms of the \emph{effective neutrino mass}~\cite{Plumacher:1996kc} 
\be\label{effneut}
\mti\equiv {(m_{\rm D}^{\dagger}m_{\rm D})_{ii}\over M_i}=K_i\,m_{\star}\, ,
\ee
where $m_{\star}$ 
is the \emph{equilibrium neutrino mass}, given by
\be\label{mstar}
m_{\star}\simeq 1.08\times 10^{-3}~{\rm eV}.
\ee

\end{enumerate}

In this thesis,
when referring to ``leptogenesis'' we will exclusively mean the production
of a lepton asymmetry by the decays of heavy singlet neutrinos. Other
``leptogenesis'' scenarios, such as the Affleck-Dine mechanism in supersymmetric
theories~\cite{Affleck:1984fy,Dine:1995kz,Hamaguchi:2002vc} or leptogenesis 
via neutrino oscillations~\cite{Akhmedov:1998qx,Asaka:2005pn} 
will not be discussed here.


After the clear evidence for non-zero neutrino masses appeared in 1998,
the see-saw mechanism, and hence leptogenesis,
became very attractive, solving simultaneously two important puzzles 
of modern physics 
within a simple and minimal extension of the SM. Another
nice feature of the model is that the heavy fermionic singlets 
(RH neutrinos) necessary for the (type-I) see-saw mechanism have a 
natural connection 
with grand unification; e.g. $SO(10)$~\cite{Fritzsch:1974nn} predicts 
the existence of such fermionic singlets.
It is therefore not a surprise if a huge activity was registered in this field
since 1998, with an ever growing precision in the computations.
 
We present in Chapter~\ref{chap:unflavored} the picture of 
leptogenesis that emerges when the flavor content of the leptons coming
from the decays of the heavy neutrinos
is ignored. This is what we call the ``unflavored'' treatment.
In Chapter~\ref{chap:flavor} we discuss why flavor effects
are important in a considerable part of the parameter space, and what their
main implications are.
This will allow us to build up a ``flavored'' picture of leptogenesis.
In Chapter~\ref{chap:densmat} the validity of 
the classical computation of the final baryon asymmetry in the case
of flavored leptogenesis is analyzed. We shall argue that the use of a 
more complicated density 
matrix equation might be relevant in some region of the parameter space. 
Then, in Chapter~\ref{chap:beyond}, an overview of
different ways to go beyond the conventional picture of leptogenesis,
where only the lightest RH neutrino is taken into account and the mass
spectrum of the heavy neutrinos is very hierarchical, is given.
In Chapter~\ref{chap:OmReal} we consider a by-product of 
flavor effects, namely the fact that the low-energy phases in the
PMNS matrix have a more important role than previously thought. In particular, 
the role played by the Dirac phase, source of \CP~violation in neutrino
oscillation experiments, will be thoroughly studied. Finally, we conclude 
with a summary of the main results and a discussion about what one can 
expect from the experimental side in the next few
years regarding a possible test of leptogenesis.



\chapter{Vanilla leptogenesis}
\label{chap:unflavored}


In this chapter, following a purely unflavored\footnote{In the literature
the expressions ``flavor-independent'' or ``one-flavor'' are often used to
describe the same concept.} treatment, 
we first introduce the most general Boltzmann
equations which need to be solved in order to estimate the baryon
asymmetry of the Universe produced through leptogenesis. This will allow
us to introduce very useful quantities and definitions which will
be used in the remainder of the thesis. Then, we describe how 
the general picture simplifies considerably
under natural assumptions, such as a hierarchical mass
spectrum for the heavy neutrinos. This will enable us to 
arrive at a ``vanilla'' leptogenesis, i.e. a typical scenario of
leptogenesis, where the lightest RH neutrino, $N_1$, provides the main 
contribution. We will then derive two important constraints on 
the parameters of 
the model: a lower bound on the mass of the lightest RH neutrino,
$M_1$, and an upper bound on the absolute neutrino mass scale, $m_1$.
Finally, we shortly comment on leptogenesis within a supersymmetric
framework and discuss the so-called \emph{gravitino problem}.

\section{General scenarios of unflavored leptogenesis}
\label{sec:general}

We would like here to describe how to calculate the baryon asymmetry
produced through leptogenesis in full generality, i.e.
for an arbitrary RH neutrino mass spectrum and no restrictions on the parameters
of the model.
 
The Lagrangian in Eq.~(\ref{lagrangian}) contains all the terms relevant
for leptogenesis. Since we are concerned here with a purely 
unflavored analysis, it will prove useful to define the states
\begin{equation}\label{state}
|\ell_i\rangle \equiv {1\over \sqrt{(h^{\dagger}h)_{ii}}}
\sum_{\a}h_{\a i}|\ell_{\a}\rangle
\end{equation}
and
\begin{equation}
|\bar{\ell}'_i\rangle \equiv {1\over \sqrt{(h^{\dagger}h)_{ii}}}\sum_{\a}
h^{\star}_{\a i}|\bar{\ell}_{\a}\rangle ,
\end{equation}
which can be thought of as the states produced in decays
and inverse decays (and $\D L=1$ scattering processes, see below) in 
the process of leptogenesis.
So, to each heavy neutrino $N_i$ corresponds one lepton state 
$|\ell_i\rangle$ and one
anti-lepton state $|\bar{\ell}'_i\rangle$. Note that $|\bar{\ell}'_i\rangle $ 
is not the \CP~conjugate of $|\ell_i\rangle$, although  $|\bar{\ell}_{\a}\rangle$ is 
the \CP~conjugate of $|\ell_{\a}\rangle$. 
We shall discuss this point in more detail in the next
chapter and see which important consequences follow.

Let us
list now the processes that are relevant for the computation
of the baryon asymmetry through 
leptogenesis~\cite{Luty:1992un,Plumacher:1996kc,Giudice:2003jh,Pilaftsis:2003gt}:
\begin{itemize}
\item Decays and inverse-decays: $N_i\leftrightarrow \ell_i\F^{\dagger}$
and the conjugate processes $N_i\leftrightarrow \bar{\ell}_i
\F$.

\item $\D L=1$ Higgs-mediated scattering processes, such
as the $s$-channel $\ell_i N_i 
\leftrightarrow Q_3 \bar{t}$ and the $t$-channels 
$\ell_i \bar{Q}_3 \leftrightarrow N_i \bar{t}$, and 
$\ell_i t \leftrightarrow N_i Q_3$,
where $Q_3$ and $t$ are the third generation quark doublet and the
top $SU(2)$ singlet, respectively, as well as those involving
gauge bosons, such as $\ell_i N_i\to \F^{\dagger} A$ (with $A=W^{3,\pm}$ or $B$).

\item The $s$-channel $\D L=2$ scattering processes 
$\ell_i \F\leftrightarrow \bar{\ell}'_i \F^{\dagger}$ with on-shell $N_i$
are already accounted for by decays and inverse decays.
However, the off-shell $s$-channel contribution, as well as the
$u$-channel and $t$-channel scatterings 
($\ell \ell\leftrightarrow \F^{\dagger}\F^{\dagger}$), must be included. 

\end{itemize}

For simplicity, we shall neglect in the following the
so-called \emph{spectator 
processes}~\cite{Buchmuller:2001sr,Nardi:2005hs}, which
are processes that are not directly related to the generation
of the asymmetry but that are fast and could affect it 
indirectly. For example,
the sphaleron transitions belong to spectator processes, since
they are supposed to be in equilibrium during the leptogenesis
process and since they are responsible for transmitting the $L$ 
asymmetry produced in the decays into a $B\!-\!L$ asymmetry in a 
non-trivial way. For the time
being, we neglect spectator processes, and simply assume
that the asymmetry is directly produced in $B\!-\!L$ instead of $L$.
In any case, spectator processes altogether are not
expected to induce corrections larger than 30\%~\cite{Nardi:2005hs}. 
We shall discuss in detail spectator processes
in the next chapter (Section~\ref{sec:spectator}). 

Thermal corrections will
also be neglected. They may lead to relevant corrections, though with big 
theoretical uncertainties, in the weak washout regime ($K_i\lesssim 5$), 
but negligible corrections are expected 
in the more relevant strong washout regime ($K_i\gtrsim 5$)~\cite{Giudice:2003jh}.

Following an  unflavored treatment,
the set of Boltzmann equations can be worked out in the
form~\cite{Luty:1992un,Plumacher:1996kc,Branco:2002xf}
\begin{eqnarray}
{{\rm d}N_{N_i}\over {\rm d}z} & = & -(D_i+S_i)\,(N_{N_i}-N_{N_i}^{\rm eq}) \;,
\hspace{10mm} i=1,2,3 \label{dlg1} \\
{{\rm d}N_{B-L}\over {\rm d}z} & = &
\sum_{i=1}^3\,\varepsilon_i (D_i+S_i)\,(N_{N_i}-N_{N_i}^{\rm eq})-N_{B-L}\,W \;,
\label{dlg2}
\end{eqnarray}
where $z\equiv M_1/T$. With $N_X$ we denote the abundance of
$X$ per RH neutrino in ultra-relativistic
thermal equilibrium. Defining $x_i\equiv M_i^2/M_1^2$
and $z_i\equiv z\,\sqrt{x_i}$, the decay factors are given by
\be
D_i \equiv {\G_{{\rm D},i}\over H\,z}=K_i\,x_i\,z\,
\left\langle {1\over\gamma_i} \right\rangle   \, ,
\ee
where $H$ is the expansion rate and the decay parameters $K_i$ 
were introduced in Eq.~(\ref{decpar}). The total decay rates,
$\G_{{\rm D},i}\equiv \G_i+\overline{\G}_i$,
are the product of the decay widths times the
thermally averaged dilation factors
$\langle 1/\gamma_i\rangle$, given by the ratio
${\cal K}_1(z_i)/ {\cal K}_2(z_i)$ of the modified
Bessel functions.

As we have seen, the $\D L=1$ scatterings $S_i$ and the related washout contribution
$W^{\D L=1}_i$ arise from two different classes of Higgs- and lepton-mediated
inelastic scatterings involving the top quark ($Q_3$) and gauge bosons
($A$). Their main effect is to enhance the heavy neutrino production and thus the
efficiency factor in the weak washout regime, and, since they also contribute
to the washout, they lead to a correction of the efficiency factor
in the strong washout regime as well. Top quark and gauge boson scattering 
terms are expected to be of similar size. However, the reactions densities
for the gauge boson processes are presently 
controversial~\cite{Pilaftsis:2003gt,Giudice:2003jh}. Therefore,
we will not discuss these processes here. A simple analytic 
approximation for the sum $D_1+S_1$ was 
obtained in~\cite{Buchmuller:2004nz},
\be\label{D+S}
D_1+S_1\simeq 0.1\, K_1\left[1+\ln\left({M_1\over M_h}\right)z^2 \ln\left(1+{a\over z}\right)
\right],
\ee
where
\be
a\simeq {10\over\ln(M_1/M_h)} .
\ee
The Higgs mass $M_h$ was introduced to cut off the infrared divergence of 
the $t$-channel process. It turns out that $D_1+S_1\simeq D_1$ in the
strong washout regime. The generalization of 
Eq.~(\ref{D+S}) for $i\neq 1$
can be derived from~\cite{Pilaftsis:2003gt}.

The equilibrium abundances of the heavy neutrinos and their rates are 
also expressed through the modified Bessel functions,
\be
N_{N_i}^{\rm eq}(z_i)= {1\over 2}\,z_i^2\,{\cal K}_2 (z_i) \;\; ,
\hspace{10mm}
{{\rm d}N_{N_i}^{\rm eq}\over {\rm d}z_i} =
-{1\over 2}\,z_i^2\,{\cal K}_1 (z_i) \, .
\ee
The RH neutrinos can be produced by inverse decays and $\D L=1$
scatterings. Nevertheless, in the relevant strong washout regime,
inverse decays alone are sufficient to make
the RH neutrino abundance reach its thermal equilibrium 
value prior to the decays which produce the asymmetry. 
The asymmetry generated together with the RH neutrino 
production will then be efficiently washed out, so that the details of the 
RH neutrino production will not affect the final
asymmetry, thus greatly reducing theoretical uncertainties. 
This is one of the nice features of the
strong washout regime on which we shall focus. 
For this reason, the $\D L=1$ scattering terms $S_i$ will
play a subdominant role.

The washout factor $W$ in Eq.~(\ref{dlg2}) can be written as the sum
of two contributions~\cite{Buchmuller:2002rq},
\be\label{W}
W=\sum_i\,W_i(K_i)+\D\, W \, .
\ee
The first term is the sum of the contributions from inverse decays 
and $\D L=1$ scatterings,
\be
W_i=W^{\rm ID}_i + W^{\D L =1}_i.
\ee
For the lightest RH neutrino, $N_1$, it was shown in~\cite{Buchmuller:2004nz} 
that $W_1$ can be conveniently rewritten as 
$W_1(z)= j(z)W^{\rm ID}_1(z)$, where
\be\label{jz}
j(z)=0.1\left(1+{15\over 8z}\right)\left[z\ln\left(M_1\over M_h\right)
\ln\left(1+{a\over z}\right)+{\mu\over z} \right],
\ee
where $\mu=1~(2/3)$ in the strong (weak) washout regime. The 
corresponding results for $i\neq 1$ can again be obtained 
from~\cite{Pilaftsis:2003gt}. 

In the strong washout regime, inverse decays,
where the resonant \mbox{$\D L=2$} contribution has to be
properly subtracted \cite{Kolb:1979qa,Giudice:2003jh},
dominate $\D L=1$ scatterings~\cite{Giudice:2003jh,Buchmuller:2004nz},
so that
\be\label{WID}
W_i(z)\simeq W_i^{\rm ID}(z) =
{1\over 4}\,K_i\,\sqrt{x_i}\,{\cal K}_1(z_i)\,z_i^3 \, .
\ee

The second term in Eq.~(\ref{W}) arises from the non-resonant $\D L=2$ processes
and gives typically a non-negligible contribution only in the 
non-relativistic limit, for $z\gg 1$~\cite{Luty:1992un,Buchmuller:2002rq,Buchmuller:2004nz}.
For hierarchical light neutrinos, it can be safely
neglected for reasonable values $M_1\ll 10^{14}\,{\rm GeV}$. We shall
come back to this contribution when discussing the upper bound
on the absolute neutrino mass scale at the end of Section~\ref{sec:N1DS}.

The effects of production and washout
are simultaneously accounted for by the efficiency factors
$\k_i$ associated with each $N_i$.
Let us indicate with $N_{B-L}^{\rm in}$ a possible pre-existing
asymmetry at the initial temperature of leptogenesis $T_{\rm in}$.
The final $B\!-\!L$ asymmetry can then be written as~\cite{kolbturner,Buchmuller:2004nz},
\be\label{integral}
N_{B-L}^{\rm f}=N_{B-L}^{\rm in}\exp\left(-\sum_i \,\int\,{\rm d}z'\,W_{i}(z')\right)+
\sum_i\,\ve_i\,\k_{i}^{\rm f}\,,
\ee
with the final efficiency factors $\k_i^{\rm f}\equiv \k_i(z\to \infty)$ given by
\begin{equation}\label{ki}
\k_i^{\rm f}= - \int_{z_{\rm in}}^{\infty}\, {\rm d}z'\
{{\rm d}N_{N_i}\over {\rm d}z'}\
\exp\left(-\sum_i\,\int_{z'}^{z}\ {\rm d}z''\, W_{i}(z'')\right)\,,
\end{equation}
where we defined $z_{\rm in}\equiv M_1/T_{\rm in}$.
Notice that each efficiency factor depends in general on 
all decay parameters,
i.e. $\k_{i}^{\rm f}=\k_i^{\rm f}(K_1,K_2,K_3)$.

The baryon-to-photon ratio at recombination can then
be calculated as
\be\label{asymgeneral}
\eta_B=a_{\rm sph} {N_{B-L}^{\rm f}\over N_{\g}^{\rm rec}}\simeq 0.96\times
10^{-2} N_{B-L}^{\rm f}\,,
\ee
where $N_{\g}^{\rm rec}\simeq 37$, and $a_{\rm sph}=n_B/ n_{B-L}$ 
accounts for the 
sphaleron conversion of a $B\!-\!L$ asymmetry into a $B$ asymmetry.
If the electroweak sphalerons go out of equilibrium  before the 
electroweak phase transition, one has
\mbox{$a_{\rm sph}=28/79$}~\cite{Harvey:1990qw}. If, instead, electroweak 
sphalerons remain
in equilibrium until slightly after the electroweak phase transition
(as would be the case if, as presently believed, the electroweak 
phase transition was not strongly first order), this factor
would be $a_{\rm sph}=12/37$~\cite{Laine:1999wv}. Both coefficients
are of order 1/3 and, for definiteness, we shall use 
$a_{\rm sph}=28/79$ throughout the thesis.

\begin{figure}
\centerline{\psfig{file=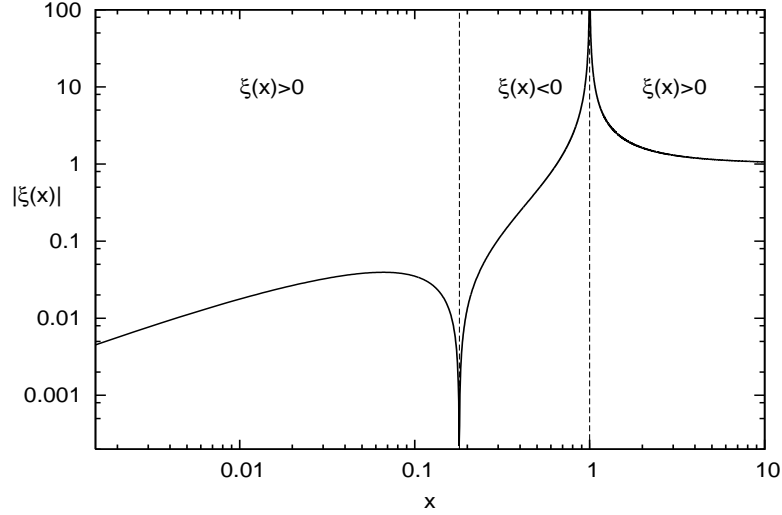,height=7cm,width=11cm}}
\caption{The function $\xi(x)$ defined in Eq. (\ref{xi}).}
\label{fig:xi}
\end{figure}

If the mass differences satisfy the condition for the
applicability of perturbation theory, $|M_j-M_i|/M_i \gg {\rm
max}[(h^{\dagger}\,h)_{ij}]/(16\,\pi^2)$ with $j\neq i$
\cite{Anisimov:2005hr}, then a perturbative calculation from the
interference of tree-level with one-loop self-energy and vertex
diagrams (see Fig.~\ref{fig:CPasym}) gives~\cite{Flanz:1994yx,Covi:1996wh}
\be\label{CPas} 
\ve_i =\, {3\over 16\pi}\, \sum_{j\neq i}\,
{{\rm Im}\,\left[(h^{\dagger}\,h)^2_{ij}\right] \over
(h^{\dagger}\,h)_{ii}} \,{\xi(x_j/x_i)\over \sqrt{x_j/x_i}}\, , 
\ee
where the function $\xi(x)$, shown in Fig.~\ref{fig:xi}, is defined as
\cite{Buchmuller:2003gz} \be\label{xi} \xi(x)= {2\over 3}\,x\,
\left[(1+x)\,\ln\left({1+x\over x}\right)-{2-x\over 1-x}\right] \,.
\ee
A particularly useful parametrization of the Yukawa coupling matrix in the
context of leptogenesis involves the orthogonal matrix 
$\O$~\cite{Casas:2001sr}\footnote{Compared to the $R$ matrix 
in~\cite{Casas:2001sr},
one has the simple relation $\O=R^{\dagger}$.},
\be\label{CI}
h=U\,\sqrt{D_m}\,\O\,\sqrt{D_M}/v ,
\ee
where we defined $D_M\equiv {\rm diag}(M_1,M_2,M_3)$ and 
$D_m\equiv {\rm diag}(m_1,m_2,m_3)$. The matrix $U$ diagonalizes
the light neutrino mass matrix $m_{\n}$ given in Eq.~(\ref{mnu}), i.e. \mbox{
$U^{\dagger}\,m_{\nu}\,U^{\star}\equiv D_m$}, and it can be identified
with the lepton mixing matrix in a basis where the charged lepton
mass matrix is diagonal (see Appendix~\ref{chap:param}). 
Moreover, neglecting the effect of the running
of neutrino parameters from high energy to low 
energy~\cite{Babu:1993qv,Antusch:2003kp}, one can assume the
$U$ matrix to be identified with the PMNS matrix, partially measured
in neutrino oscillation experiments. In the following, we shall
refer to the parametrization Eq.~(\ref{CI}) as the ``orthogonal''
or the ``Casas-Ibarra'' parametrization.

The $\O$ matrix can be conveniently parametrized as
\be\label{second}
\O({\o}_{21},{\o}_{31},{\o}_{32})
=R_{12}(\o_{21})\,\,
 R_{13}(\o_{31})\,\,
 R_{23}(\o_{32})\,\, ,
\ee
where
\bea\label{R12}
R_{12}&=&
\left(
\begin{array}{ccc}
 \sqrt{1-{\o}^2_{21}}  &  -{\o}_{21}          & 0 \\
            {\o}_{21}  & \sqrt{1-{\o}^2_{21}} & 0 \\
  0 & 0 & 1
\end{array}
\right) \, , \\ \label{R13}
R_{13}&=&
\left(
\begin{array}{ccc}
 \sqrt{1-{\o}^2_{31}}  & 0 &  - {\o}_{31} \\
    0 & 1 & 0 \\
  {\o}_{31} & 0 & \sqrt{1-{\o}^2_{31}}
\end{array}
\right)  \, ,
\,\,\\ \label{R23}
R_{23}&=&
\left(
\begin{array}{ccc}
  1  &  0   & 0   \\
  0  & \sqrt{1-{\o}^2_{32}} & -{\o}_{32} \\
  0 &  {\o}_{32} & \sqrt{1-{\o}^2_{32}}
\end{array}
\right)\, .
\eea
This parametrization for an orthogonal complex matrix
corresponds to the transposed form of the
CKM  matrix in the quark sector or of the
PMNS matrix in neutrino mixing [cf.~Eq.~(\ref{PMNSmatrix2})], 
with the difference
that here one has complex rotations instead of real ones.

Notice that, using the orthogonal parametrization, Eq.~(\ref{CI}), the
decay parameters $K_i$ can be expressed as linear combinations
of the neutrino masses~\cite{Fujii:2002jw,Buchmuller:2004nz},
\be\label{decparCI}
K_i=\sum_j {m_j \over m_{\star}}|\O_{ji}^2|.
\ee

The parametrization Eq.~(\ref{second}) is especially useful
to understand the general structure of
different scenarios of leptogenesis.
\begin{itemize}
\item For $\O=R_{13}$, one has $\ve_2=0$, while $\ve_1$ is maximal 
if~\cite{DiBari:2005st} 
\be
m_3\,{{\rm Re}(\o^2_{31})/ |\o_{31}^2|}=
 m_1\,{[1-{\rm Re}(\o^2_{31})]/ |1-\o_{31}^2|}.
\ee
In the hierarchical limit for the heavy neutrinos (HL),
i.e. \mbox{\HL}, one obtains the $N_1$-dominated
scenario ($N_1$DS), where the final asymmetry
is the result of only $N_1$-related processes.

\item
For $\O=R_{23}$, one has $\ve_1=0$, while $\ve_2$
is maximal if
\be\label{R23max}
m_2\,{{\rm Re}(\o^2_{32})/|\o_{32}^2|}=
m_3\,{[1-{\rm Re}(\o^2_{32})]/ |1-\o_{32}^2|}.
\ee
At the same time, one has $\mt=m_1$, so that the washout from
$N_1$ can be neglected if $m_1$ is small enough.
Therefore, in the HL and for hierarchical light neutrinos,
one obtains the $N_2$-dominated scenario ($N_2$DS)~\cite{DiBari:2005st},
which will be the topic of Section~\ref{sec:N2}.

\item If $\O=R_{12}$, then $\ve_1$ undergoes
a phase suppression compared to its maximal value but
$|\ve_2|\propto (M_1/M_2)\,|\ve_1|$. This implies that in the
HL one again recovers the $N_1$DS \cite{DiBari:2005st}.
On the other hand,
if $M_1\simeq M_2$, both $N_1$ and $N_2$ should be taken into
account since both \CP~asymmetries are expected to be of the
same order.
\end{itemize}

Before ending this section, we would like to make two
remarks. First, using the Casas-Ibarra parametrization,
Eq.~(\ref{CI}), it can be easily seen that the PMNS
matrix $U$ cancels out in combinations like
$(h^{\dagger}h)_{ij}$ (or $(m_{\rm D}^{\dagger}m_{\rm D})_{ij}$).
This implies that both the \CP~asymmetry,
Eq.~(\ref{CPas}), and the decay parameters or effective neutrino
masses, Eq.~(\ref{effneut}), do not depend on $U$. 
Hence, the final asymmetry is completely independent of the 
PMNS matrix. In particular, if the \CP~asymmetry is insensitive to the
\CP-violating phases in $U$, then the source of \CP~violation 
necessary for leptogenesis must come from the ``high-energy'' 
sector,
i.e. the $\O$ matrix, which is not probed in neutrino
experiments. We shall see in the next chapter that the 
situation is completely different when flavor effects
are taken into account.

Second, the border between the $N_1$DS and the $N_2$DS will 
also be affected by flavor effects.
Furthermore, contrary to the unflavored case, where $\ve_3$ is always
suppressed by factors $M_1/M_{2,3}$, the flavored \CP~asymmetries $\ve_{3\a}$ 
are not suppressed, opening the way to a potential $N_3$-dominated scenario.
We shall nonetheless see, for instance in Section~\ref{sec:DiracHL},
that, even though the \CP~asymmetry may not be suppressed, the
washout from the other two heavy neutrinos is difficult to avoid. 


\section[The $N_1$-dominated scenario]{The \boldmath{$N_1$}-dominated scenario}
\label{sec:N1DS}

Let us now discuss the $N_1$DS, where the asymmetry is dominantly generated
by the lightest RH neutrino, $N_1$. This typically (but not
necessarily) occurs when a hierarchical heavy neutrino spectrum is 
considered, $3\,M_1 \lesssim M_2$~\cite{Blanchet:2006dq}.

The general expression
for the final asymmetry, Eq.~(\ref{asymgeneral}), reduces to
\be\label{etaBunfllightest}
\eta_B\simeq 10^{-2}\,\ve_1\,\k_1^{\rm f},
\ee
where $\k_1^{\rm f}$ can be calculated solving a
system of just two kinetic equations~\cite{kolbturner,Buchmuller:2002rq,Giudice:2003jh},
\bea\label{unflke1}
{{\rm d}N_{N_1}\over {\rm d}z}& = & -(D_1+S_1)\,(N_{N_1}-N_{N_1}^{\rm eq}) 
\, , \\*  \label{unflke2}
{{\rm d}N_{B-L}\over {\rm d}z}& = & \ve_1\,(D_1+S_1)\,(N_{N_1}-N_{N_1}^{\rm eq})
-W_1\,N_{B-L} \, .
\eea
These equations are
obtained from the general set,~Eqs.~(\ref{dlg1}) and (\ref{dlg2}),
neglecting the asymmetry generation and the washout terms
from the two heavier RH neutrinos.

\begin{figure}[t!]
\centerline{\psfig{file=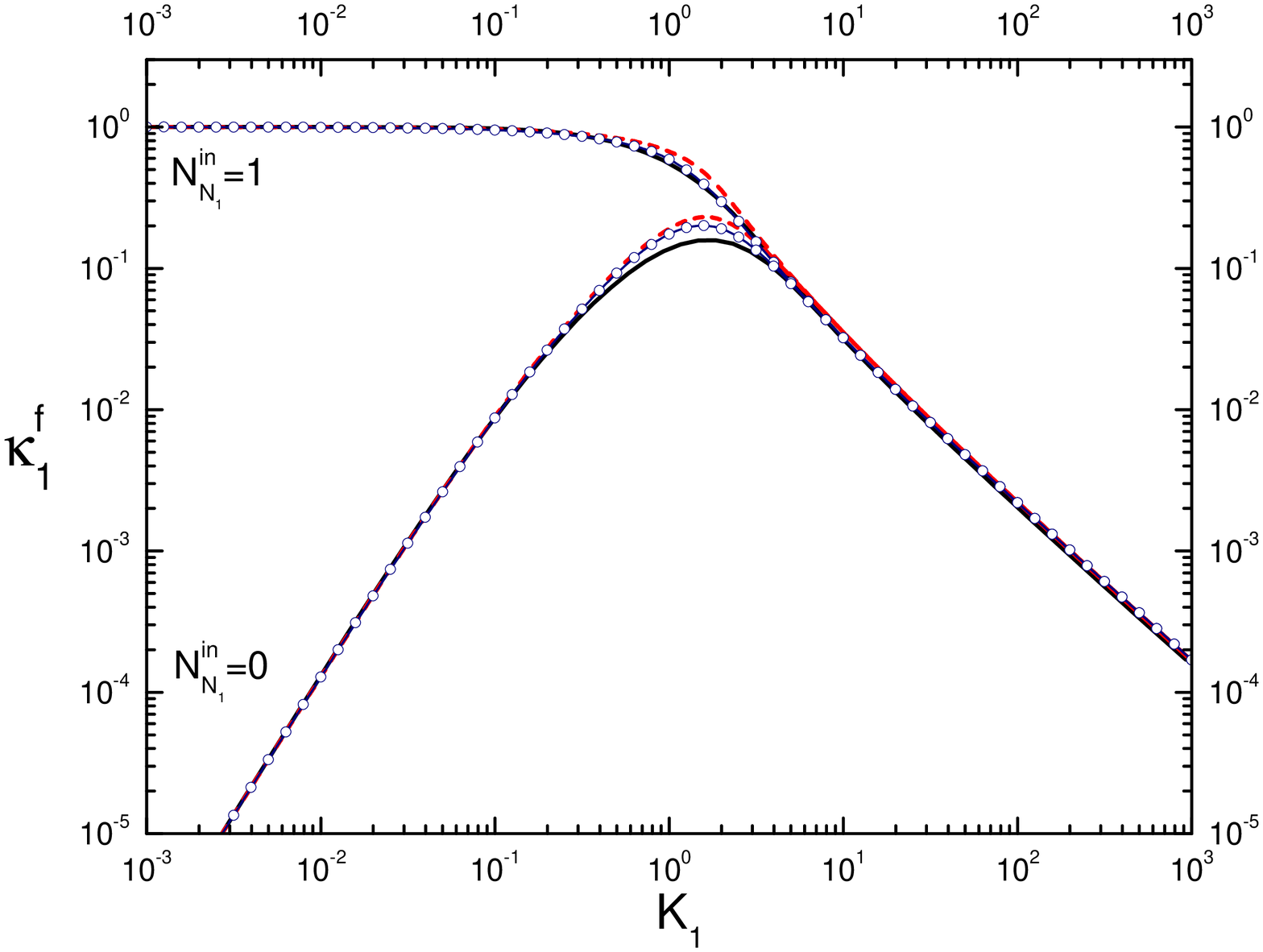,height=8cm,width=14cm}}
\caption{Efficiency factor in the $N_1$DS. The solid lines
are the numerical solutions of Eqs.~(\ref{unflke1}) and
(\ref{unflke2}) with $\D W=0$. The analytical expression~(\ref{k}) 
yields the dashed line if $z_{\rm B}(K_1)$
is given by Eq.~(\ref{zBK}) and the
circled line if it is given by Eq.~(\ref{zB}).}
\label{fig:EffUnfl}
\end{figure}

For $M_1 \ll 10^{14}\,{\rm GeV}\,(m_{\rm atm}^2/\sum\,m_i^2)$,
the term $\D W(z)$ in the washout term [cf.~Eq.~(\ref{W})] is negligible
and the solutions depend just on $K_1$,
since this is the only parameter in the equations.
The $B\!-\!L$ asymmetry can be worked out in an integral form~\cite{kolbturner},
and one obtains a special case of the more general Eq.~(\ref{integral}),
\be\label{NBmLf}
N_{B-L}(z;\bar{z})=\overline{N}_{B-L}\exp\left(-\int_{\bar{z}}^{z}\,{\rm d}z'\,W_{1}(z')\right)
+\ve_1\,\k_1(z;\bar{z}) \, ,
\ee
where now a possible asymmetry produced by the two heavier
RH neutrinos and frozen at $\bar{z}\geq z_{\rm in}$ is included in
$\overline{N}_{B-L}$.
The efficiency factor $\k_1(z;\bar{z})$ can be expressed through
 a Laplace integral,
\bea\label{k1int}
\k_1(z;\bar{z})&=&
-\int_{\bar{z}}^{z}\,{\rm d}z'\,{{\rm d}N_{N_1}\over {\rm d}z'}\,
\exp\left(-\int_{z'}^{\infty}\,{\rm d}z''\,W_{1}(z'')\right)\NO \\
&=&\int_{\bar{z}}^{z}\,{\rm d}z' \exp\left[-\psi(z',z)\right].
\eea
In the strong washout regime we are interested in, for $K_1\gtrsim 5$,
one can safely use the approximations ${\rm d}N_{N_1}/{\rm d}z'\simeq {\rm d}N_{N_1}^{\rm eq}/{\rm d}z'$,
$D_1+S_1\simeq D_1$, \mbox{$W_1(z')\simeq W_1^{\rm ID}(z')$} [cf.~Eq.~(\ref{WID})], 
and one finds that the final value for the efficiency
factor, when $z\rightarrow\infty$, is given by 
\cite{Buchmuller:2004nz}
\be\label{k}
\k_{1}^{\rm f}(K_1)\simeq \k(K_1) \equiv {2\over K_1\,z_{\rm B}(K_1)}\,
\left[1-\exp\left(-{1\over 2}{K_1\,z_{\rm B}(K_1)}\right)\right] ,
\ee
if $\bar{z}\lesssim z_{\rm B}-2$. The value $z'=z_{\rm B}(K_1)$ is where the
quantity $\psi(z',\infty)$ has a minimum and the integral
in Eq.~(\ref{k1int}) receives a dominant contribution from a restricted
$z'$-interval centered around it. In the strong washout regime,
it can be calculated as a solution of
\begin{equation}\label{zBK}
W_{1}^{\rm ID}(z_{\rm B}) =
\left.
{{\rm d}^2 N_{N_1}^{\rm eq}/{\rm d}z^2\over |{\rm d}N_{N_1}^{\rm eq}/{\rm d}z|}
\right|_{z=z_{\rm B}}=
\left\langle {1\over\gamma}\right\rangle^{-1}(z_{\rm B})\,
-\, {3\over z_{\rm B}}\;.
\end{equation}
For very large $K_1$, the right-hand side of this equation
tends to unity and $z_{\rm B}\simeq z_{\rm off}$, the value of $z$
when the washout from inverse decays switches off, i.e. 
$W_{\rm ID}(z>z_{\rm off})<1$.
Fig.~\ref{fig:EffUnfl} shows (dashed lines) that Eq.~(\ref{k}), 
with $z_{\rm B}(K_1)$ given by Eq.~(\ref{zBK}), reproduces the
numerical result (solid line)
within $10\%$ for \mbox{$K_1\gtrsim 3$}~\cite{Buchmuller:2004nz}.

\begin{figure}[!p]
\vspace{40mm}
\begin{center}
\begin{picture}(100,100)
\put(-100,0){\includegraphics{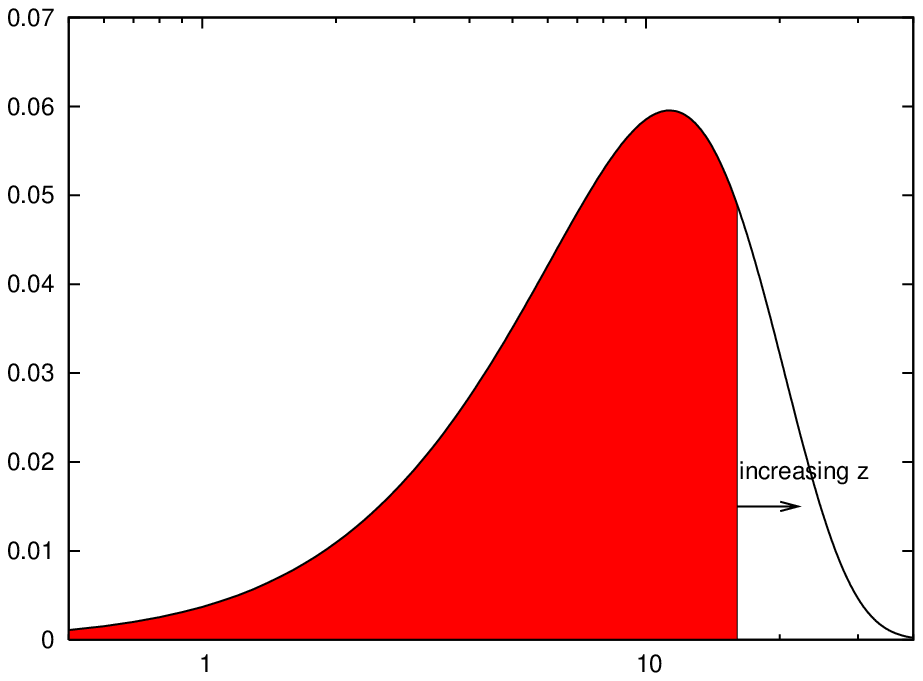}}
\put(-50,160){$K_1=10^{-2}$}
\put(95,200){$z_{\rm max}^{\rm weak}(0.01)\simeq 12$}
\put(101,172){\Huge $\uparrow$}
\put(-35,115){$\left\arrowvert \frac{{\rm d}N_{N_1}}{{\rm d}z'}\right| \simeq
\left\arrowvert\frac{{\rm d}N_{N_1}^{\rm weak}}{{\rm d}z'} \right|$}
\put(15,80){\Huge $\searrow$}
\put(50,-10){$z'$}
\put(-132,100)
{\large $\left.\frac{{\rm d}\kappa_1}{{\rm d}z'}\right|_{z' \leq z}$}
\end{picture}
\end{center}
\vspace{28mm}
\begin{center}
\begin{picture}(100,100)
\put(-100,0){\includegraphics{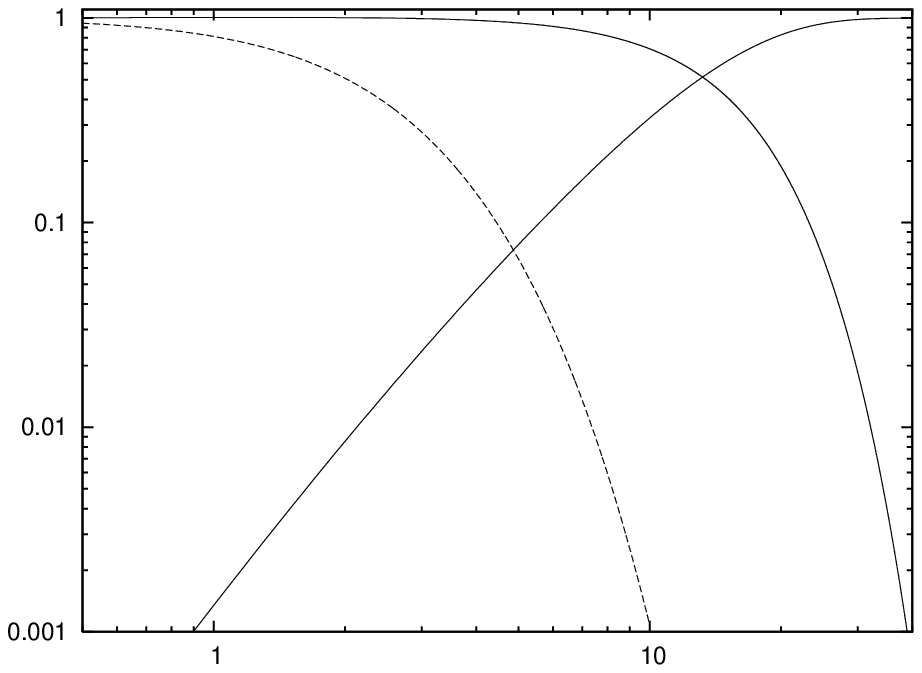}}
\put(-50,160){$K_1=10^{-2}$}
\put(50,-10){$z=\frac{M_1}{T}$}
\put(-25,50){$\kappa_1$}
\put(77,100){$N_{N_1}^{\rm eq}$}
\put(127,120){$N_{N_1}$}
\end{picture}
\end{center}
\caption{Dynamics in the weak washout regime for a
thermal initial $N_1$-abundance ($N_{N_1}^{\rm in}=1$).
Top panel: rates. Bottom panel: efficiency factor $\k_1$ and $N_1$-abundance.
The maximum of the asymmetry production rate occurs at
$z'\simeq z_{\rm max}^{\rm weak}(K_1=0.01)\simeq 12$ [cf.~Eq.~(\ref{zweakmax})].}
\label{fig:DynamicsWeak}
\end{figure}

\begin{figure}[!hp]
\vspace{5cm}
\begin{center}
\begin{picture}(100,100)
\put(-100,0){\includegraphics{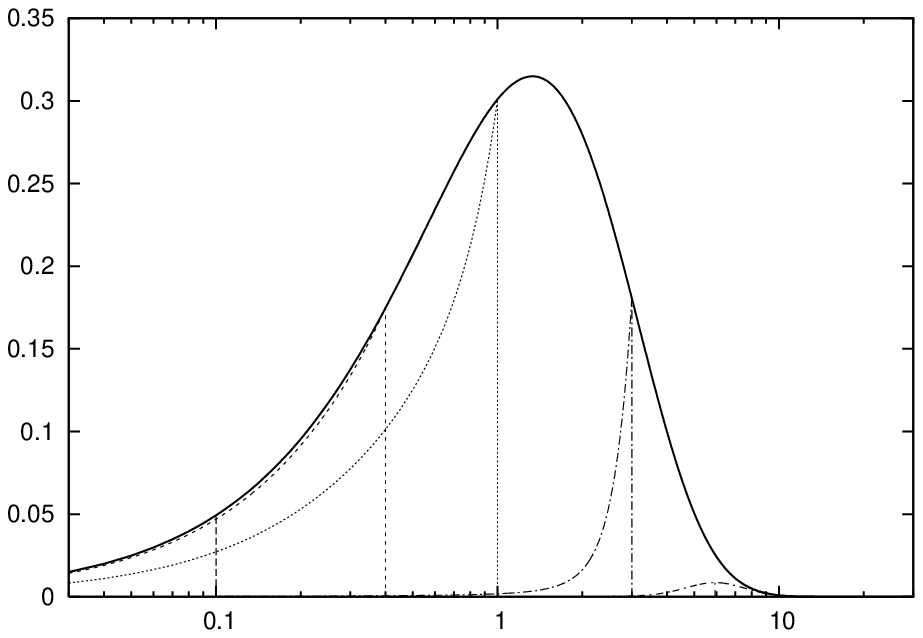}}
\put(-50,160){$K_1=10$}
\put(50,-10){$z'$}
\put(50,80){$z=1$}
\put(63,168){\Large $\uparrow$}
\put(50, 187){$z_{\rm max}^{\rm eq}\simeq 1.33$}
\put(-40,20){$z=0.1$}
\put(10,35){$z=0.4$}
\put(80,45){$z=3$}
\put(115,5){$z_{\rm B}$}
\put(120,20){$\swarrow$}
\put(130,30){$z\gg z_{\rm B}$}
\put(-50,135){$\left\arrowvert \frac{{\rm d}N_{N_1}}{{\rm d}z'}\right| \simeq
\left\arrowvert\frac{{\rm d}N_{N_1}^{\rm eq}}{{\rm d}z'} \right|$}
\put(3,105){\Huge $\searrow$}
\put(-132,100)
{\large $\left.\frac{{\rm d}\kappa_1}{{\rm d}z'}\right|_{z'\leq z}$}
\end{picture}
\end{center}
\vspace{30mm}
\begin{center}
\begin{picture}(100,100)
\put(-100,0){\includegraphics{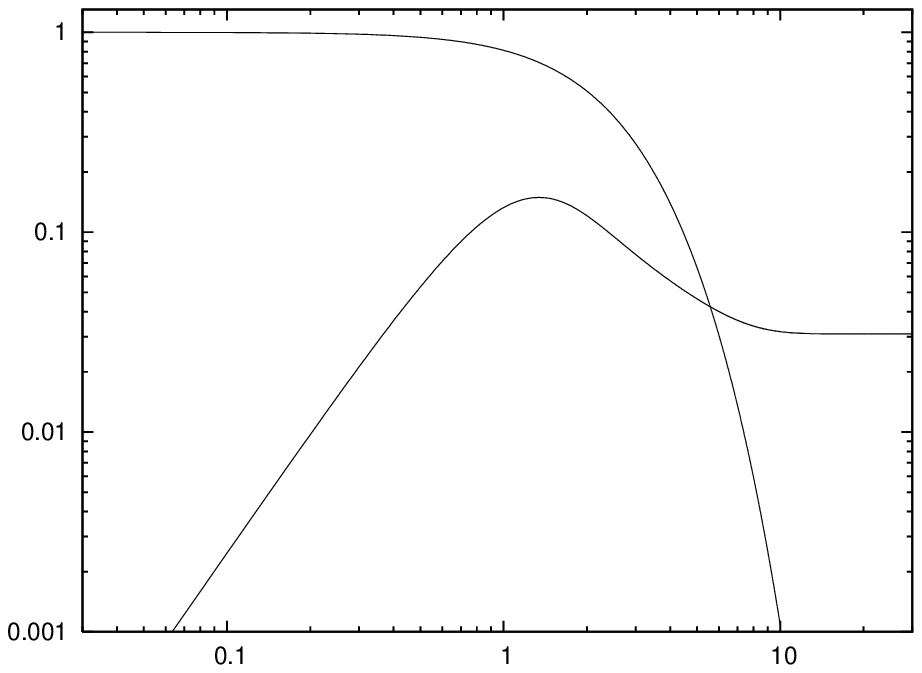}}
\put(-50,160){$K_1=10$}
\put(2,100){$\kappa_1$}
\put(50,-10){$z=\frac{M_1}{T}$}
\put(105,150){$N_{N_1}^{\rm eq} \simeq N_{N_1}$}
\end{picture}
\end{center}
\caption{Dynamics in the strong washout regime.
Top panel: rates. Bottom panel: efficiency factor $\k_1$ and $N_1$-abundance.
The maximum of the final asymmetry production rate
occurs at $z'\simeq z_{\rm B}$.}
\label{fig:DynamicsStrong}
\end{figure}

Even though the approximation ${\rm d}N_{N_1}/{\rm d}z' \simeq {\rm d}N_{N_1}^{\rm eq}/{\rm d}z'$
works rigorously only in the strong washout regime,
Eq.~(\ref{k}) describes also the correct
weak washout regime for a thermal initial $N_1$-abundance ($N^{\rm in}_{N_1}=1$), 
because $\k_1^{\rm f}$ depends only on the value of the initial
abundance and not on the decay rate. However,
in the intermediate regime ($K_1\simeq 1$) the error is about $30\%$. For
$K_1 \lesssim 1$, the approximation ${\rm d}N_{N_1}/{\rm d}z'\simeq {\rm d}N_{N_1}^{\rm eq}/{\rm d}z'$
does not work well and $z_{\rm B}(K_1)$, evaluated with Eq.~(\ref{zBK}),
incorrectly saturates to a constant value $z_{\rm max}^{\rm eq}\simeq 1.33$
in the limit $K_1\to 0$,
corresponding to the maximum of ${\rm d}N_{N_1}^{\rm eq}/{\rm d}z$ (see
upper panel of Fig.~\ref{fig:DynamicsStrong}). As a matter of fact,
in the weak washout regime the maximum of the asymmetry production
does not occur at $z_{\rm max}^{\rm eq}$ but at higher values
\be\label{zweakmax}
z_{\rm max}^{\rm weak}(K_1)\simeq 1/\sqrt{K_1}+15/8,
\ee
roughly when the age of the Universe is equal to the RH neutrino lifetime.
Indeed, in the weak washout regime, the $N_1$'s decay far from equilibrium,
when inverse decays can be neglected. In this case one has approximately
$N_{N_1}(z)\simeq N_{N_1}^{\rm weak}(z)$, where~\cite{Buchmuller:2004nz}
\bea
N_{N_1}^{\rm weak}(z) &\simeq&  N_{N_1}^{\rm in}\,
\exp\left[-\int_{z_{\rm i}}^{z}\,{\rm d}z'\, D_1(z')\right]\NO\\*
&\simeq& N_{N_1}^{\rm in}
\exp\left\{-K_1 \left[{z^2 \over 2} - {15\,z \over 8}
+ \left({15\over 8}\right)^2 \ln\left(1+{8\over 15}z\right)
\right]\right\},\NO
\eea
with $N^{\rm in}_{N_1}=1$ for a thermal initial $N_1$-abundance.
The maximum of ${\rm d}N_{N_1}^{\rm weak}/{\rm d}z$ gives the
value of $z$ where the production of the asymmetry is
maximum in the weak washout regime, and one can easily find that
this value is given approximately by
$z_{\rm max}^{\rm weak}(K_1)$ [cf.~Eq.~(\ref{zweakmax})].
An example is shown in the upper panel of Fig.~\ref{fig:DynamicsWeak}
for $K_1=0.01$.
An improvement of the analytical expression in Eq.~(\ref{k})
is obtained replacing $z_{\rm B}(K_1)$ from Eq.~(\ref{zBK}) with an expression
that coincides with it at large $K_1$ and with $z_{\rm max}\simeq 2$
at $K_1\simeq 1$, such as
\be\label{zB}
z_{\rm B}(K_1) \simeq 2+4\,{K_1}^{0.13}\,\exp\left(-{2.5\over K_1}\right).
\ee
Fig.~\ref{fig:EffUnfl} shows (circles) that this expression, plugged into
Eq.~(\ref{k}), reproduces the numerical solution (solid line)
with a precision always better than 10\%.

Notice that, in the particularly relevant range $5\lesssim K_{1}\lesssim 100$, 
Eq.~(\ref{k}) is well approximated by a power law~\cite{DiBari:2004en},
\be\label{plaw}
\k_{1}^{\rm f}(K_1)\simeq {0.5\over K_{1}^{1.2}} \, .
\ee

In the case of a vanishing initial $N_1$-abundance ($N_{N_1}^{\rm in}=0$),
one has to take into account two different contributions,
a negative and a positive one,
\be
\k_{1}^{\rm f} (K_1)
=\k_{-}^{\rm f}(K_1)+ \k_{+}^{\rm f}(K_1) \, .
\ee
Defining $z_{\rm eq}$ by the condition 
$N_{N_1}(z_{\rm eq})=N_{N_1}^{\rm eq}(z_{\rm eq})$, the negative contribution 
arises from a first stage when
$N_{N_1}\leq N_{N_1}^{\rm eq}$, for $z\leq z_{\rm eq}$,
and is given approximately by~\cite{Buchmuller:2004nz}
\be\label{k-unfl}
\k_{-}^{\rm f}(K_1)\simeq
-{2}\,\exp\left(-{3\,\pi\,K_{1} \over 8}\right)
\left[\exp\left({1\over 2} N_{N_1}(z_{\rm eq})\right) - 1 \right] .
\ee
The $N_1$-abundance at $z_{\rm eq}$ is well approximated by
\begin{equation}\label{nka}
N_{N_1}(z_{\rm eq}) \simeq \overline{N}(K_1)\equiv
{N(K_1)\over\left(1 + \sqrt{N(K_1)}\right)^2}\, ,
\end{equation}
interpolating between the limit $K_1\gg 1$, where $z_{\rm eq}\ll 1$ and
$N_{N_1}(z_{\rm eq})=1$, and the limit $K_1\ll 1$, where
$z_{\rm eq}\gg 1$ and $N_{N_1}(z_{\rm eq})=N(K_1)\equiv 3\p K_1/4$.
The positive contribution arises from a second stage when
$N_{N_1}\geq N_{N_1}^{\rm eq}$, for $z\geq z_{\rm eq}$,
and is approximately given by~\cite{Buchmuller:2004nz}
\be\label{k+unfl}
\k_{+}^{\rm f}(K_1)\simeq
{2\over z_{\rm B}(K_{1})\,K_{1}}
\left[1-\exp\left(-{1\over 2} K_{1} z_{\rm B}(K_{1}) N_{N_1}(z_{\rm eq})\right)\right] \, .
\ee
As can be seen in Fig.~\ref{fig:EffUnfl}, the use of Eq.~(\ref{zB})
yields an improvement also for a vanishing initial $N_1$-abundance.

Figures~\ref{fig:DynamicsWeak} and~\ref{fig:DynamicsStrong} show,
for a thermal initial $N_1$-abundance,
the dynamics of the asymmetry generation,
comparing one example of  weak washout
with one example of strong washout~\cite{Blanchet:2006dq}.
In the top panels we show
the function ${\rm d}\k_1/{\rm d}z'\equiv \exp\left[-\psi(z',z)\right]$, defined for
$z'\leq z$, for different values of $z$.
The difference between the two cases is striking. In the weak washout
regime, each decay contributes to the final asymmetry for any value
of $z'$ at which the asymmetry is produced.
In the strong washout regime, all the asymmetry produced at $z'\lesssim z_{\rm B}-2$
is efficiently washed out by inverse decays, so that only decays
occurring at $z'\sim z_{\rm B}$ give a contribution to the final asymmetry.


After having thoroughly studied the efficiency factor, we would like
to turn now to the other important piece in the calculation of
the baryon asymmetry Eq.~(\ref{etaBunfllightest}), namely
the \CP~asymmetry $\ve_1$. The general expression~(\ref{CPas}) 
for $\ve_1$ can be re-cast through the $\O$ matrix as~\cite{DiBari:2005st}
\be\label{ve1}
\ve_1 = \xi(x_2)\,\ve_1^{\rm HL}(m_1,M_1,\O_{21},\O_{31})+
[\xi(x_3)-\xi(x_2)]\,
\D\ve_1 (m_1,M_1,\O_{21},\O_{31},\O_{22}) \, .
\ee
In the hierarchical limit (HL), for $x_3,x_2\gg 1$, one has
$\xi(x_2)\simeq \xi(x_3) \simeq 1$, yielding 
$\ve_1\simeq \ve_1^{\rm HL}(m_1,M_1,\O_{21},\O_{31})$.
Therefore, the dependence on four of the
see-saw parameters, namely $M_2,M_3$ and $\O_{22}$, disappears
in the HL, and one is left with only six parameters. Notice moreover
that $\k_1^{\rm f}=\k_1^{\rm f}(m_1,M_1,K_1)$,
where $K_1=K_1(m_1,\O_{21},\O_{31})$,
and thus the final asymmetry depends on the same six 
parameters. In particular, let us emphasize once more that the
final asymmetry is independent of the PMNS matrix,
which contains three mixing angles and three phases [cf.~Eq.~(\ref{PMNSmatrix})].

Let us define
\be
\overline{\ve}(M_1)\,
\equiv {3\over 16\,\pi}\,{M_1\,m_{\rm atm}\over v^2}\,
\simeq 10^{-6}\,\left({M_1\over 10^{10}\,{\rm GeV}}\right)
\,\left({m_{\rm atm}\over 0.05\,{\rm eV}}\right), 
\ee
and
\be
\b(m_1,\O_{21},\O_{31})\equiv
{\sum_j\,m^2_j~{\rm Im}[\O^2_{j1}]
\over m_{\rm atm}\,\sum_j\,m_j\,|\O_{j1}|^2} \, .
\ee
The HL for $\ve_1$ can then be written as \cite{Davidson:2002qv}
\be
\ve_1^{\rm HL}(m_1,M_1,\O_{21},\O_{31}) \equiv \,
\overline{\ve}(M_1)\,\b(m_1,\O_{21},\O_{31}).
\ee 
It is interesting that $\b(m_1,\O_{21},\O_{31})\leq 1$,
so that in the HL one has the upper bound
$|\ve_1^{\rm HL}|\leq\overline{\ve}(M_1)$ \cite{Asaka:1999jb}.
A more precise bound, which is known as the Davidson-Ibarra
bound, was later derived~\cite{Davidson:2002qv},
\be\label{DIbound}
|\ve_1^{\rm HL}|\leq \overline{\ve}(M_1){m_{\rm atm}\over m_1+m_3},
\ee
where we used the fact that $(m_3-m_1)/\matm=\matm/(m_1+m_3)$.
Further refinements were added, leading to the even more accurate
$\mt$-dependent bound, 
\be\label{CPbound}
|\ve_1^{\rm HL}|\leq \overline{\ve}(M_1){m_{\rm atm}\over m_1+m_3}
f(m_1,\mt),
\ee
where~\cite{Buchmuller:2003gz,Hambye:2003rt}
\be\label{fm1}
f(m_1,\mt)\simeq \left\{\begin{array}{cc} 1-m_1/\mt &{\rm if}~m_1\ll m_3,\\
\sqrt{1-(m_1/\mt)^2} & {\rm if}~m_1\simeq m_3,
\end{array}\right.
\ee
obtained by maximizing over the $\O$-parameters at fixed $\mt$.
From this bound one gets that $f(m_1,\mt)$ and hence the \CP~asymmetry
are maximal in the limit $m_1\to 0$. One notices as well that the 
\CP~asymmetry vanishes when $\mt=m_1$.

The presence of the bound on the \CP~asymmetry Eq.~(\ref{CPbound})
 motivates the introduction of the so-called ``effective leptogenesis phase'' 
$\d_{\rm L}$ as
\be\label{sindelta}
\b(m_1,\O_{21},\O_{31})=\b_{\rm max}(m_1,\mt)~
\sin\d_{\rm L}(m_1,\O_{21},\O_{31}) \, ,
\ee
where
\be\label{bmax}
\b_{\rm max}(m_1,\mt)={m_{\rm atm}\over m_1+m_3}\,f(m_1,\mt),
\ee
such that the upper bound in Eq.~(\ref{CPbound})
corresponds to \mbox{$\sin\d_{\rm L}=1$}.

It is also useful to express the function $f(m_1,\mt)$ as~\cite{DiBari:2005st}
\be
f(m_1,\mt)={m_1+m_3\over \mt}\,Y_{\rm max}(m_1,\mt) \, ,
\ee
where $Y_{\rm max}(m_1,\mt)$ represents the configuration of $\O$ parameters
which maximizes the \CP~asymmetry at fixed $\mt$. It turns out that 
such a configuration always occurs for $\O_{21}=0$~\cite{DiBari:2005st},
and hence $Y_{\rm max}$ is the maximum of ${\rm Im}[\O^2_{31}]$ at fixed $\mt$.
In other words, for $\O_{21}=0$ and 
$Y_{\rm max}(m_1,\mt)={\rm Im}[\O^2_{31}]$, the phase $\d_{\rm L}$ is maximal, while
for a generic choice of $\O$,
the $C\!P$ asymmetry undergoes a phase suppression
\bea
\sin\d_{\rm L}(m_1,\O_{21},\O_{31})&=&{m_1+m_3\over \mt\,f(m_1,\mt)}\,
({\rm Im}[\O^2_{31}]+\sigma^2 {\rm Im}[\O^2_{21}])\nonumber \\
&=&{{\rm Im}[\O^2_{31}]+\sigma^2 
{\rm Im}[\O^2_{21}]\over Y_{\rm max}(m_1,\mt)}\, ,\label{sind}
\eea
where $\s \equiv \sqrt{m_2^2-m_1^2}/m_{\rm atm}$. It can be readily seen 
that $\sin\d_{\rm L}=1$ for ${\rm Im}[\O^2_{21}]=0$ and 
${\rm Im}[\O^2_{31}]=Y_{\rm max}$.

\begin{figure}
\begin{center}
\includegraphics[angle=-90,width=0.8\textwidth]{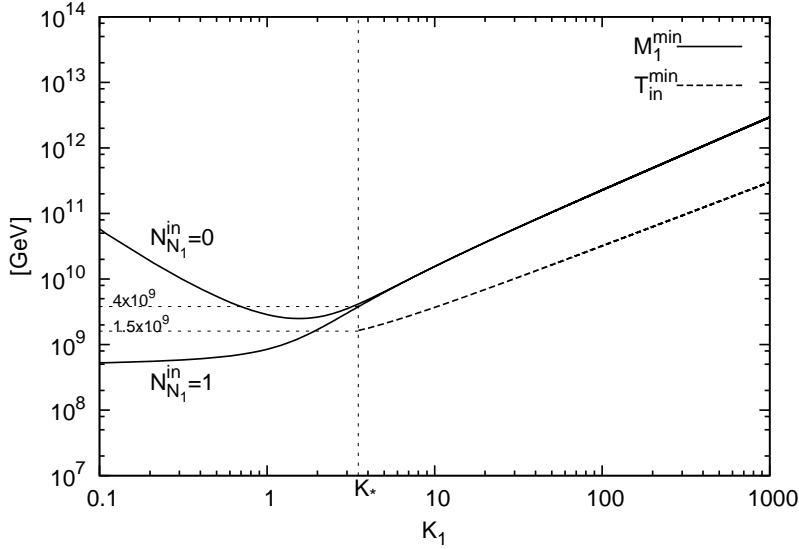}
\caption{Lower bounds on $M_1$ and $T_{\rm in}$ vs.
$K_1$ [cf. Eqs. (\ref{M1minMbar}) and (\ref{lbTin})] in the case
of maximal phase ($\sin\d_{\rm L}=1$) and for $m_1=0$.}
\label{fig:M1minUnfl}
\end{center}
\end{figure}

It is instructive to calculate $\sin\d_{\rm L}$ for each of the three
elementary complex rotations that can be used to parametrize
$\O$ [cf. Eq.~(\ref{second})]:
\begin{itemize}
\item For $\O=R_{13}$, one has $\sin\d_{\rm L}={\rm Im}[\O^2_{31}]/Y_{\rm max}$
and the phase is maximal if
${\rm Im}[\O^2_{31}]=Y_{\rm max}=\mt/m_{\rm atm}$; notice that
there is no difference between normal and inverted hierarchy.

\item For $\O=R_{12}$, one has
$\sin\d_{\rm L}=\s^2{\rm Im}[\O^2_{21}]/Y_{\rm max}\leq \s$, larger
for inverted hierarchy than for normal; however, for fully
hierarchical light neutrinos ($m_1\ll \msol$), one has that
$K_1=K_{\rm sol}\,|\O_{21}^2|$ for normal hierarchy
and $K_1\simeq K_{\rm atm}\,|\O_{21}^2|$ for inverted hierarchy,
and since $\k_1^{\rm f}\propto K_1^{-1.2}$ [cf.~Eq.~(\ref{plaw})],
the final asymmetry is slightly higher for normal
hierarchy compared to inverted at fixed 
$|\O_{21}^2|$~\cite{DiBari:2005st}.

\item For $\O=R_{23}$, one has $\sin\d_{\rm L}=\ve_1=0$; one can
check that $\ve_1=0$ applies independently of
$M_2$ and $M_3$ and therefore not only in the HL. Notice that
the conclusions in the previous two cases are still valid if
one multiplies $R_{13}$ or $R_{12}$ with $R_{23}$ respectively,
since it does not affect $\sin\d_{\rm L}$.
\end{itemize}
Interesting constraints follow
if one imposes that the asymmetry produced from leptogenesis
explains the value of the baryon-to-photon ratio inferred 
from CMB observations, 
Eq.~(\ref{WMAP}),~\cite{Buchmuller:2002rq,WMAP3}
\be
\eta_B(m_1,M_1,\O_{21},\O_{31})=
\eta_B^{\rm CMB} = (6.1 \pm 0.2)\times 10^{-10} \, .
\ee
If $M_1 \ll 10^{14}\,{\rm GeV}\,(m_{\rm atm}^2/\sum_i\,m_i^2)$, then
\bea\label{M1minMbar}
M_1 &=&{\overline{M}_1\,\over
\k(K_1)\,
\b_{\rm max}(m_1,K_1)\,\sin\d_{\rm L}(\O_{21},\O_{31})} \NO\\
&\geq& M_1^{\rm min}(K_1)\equiv {\overline{M}_1\,\over
\k(K_1)\,\b_{\rm max}(m_1,K_1)}   \, ,
\eea
where we introduced the quantity
\be\label{barM1}
\overline{M}_1 \equiv {16\,\pi\over 3}\,
{N_{\g}^{\rm rec}\,v^2\over a_{\rm sph}}\,
{\eta_B^{\rm CMB}\over m_{\rm atm}}
=(6.25\pm 0.4)\times 10^8\,{\rm GeV} \gtrsim 5\times 10^8{\rm GeV} \, .
\ee
The last inequality gives the $3\s$ value that we used to obtain
all the results shown in the figures.\footnote{Notice that $\overline{M}_1$ 
gives the lower bound on $M_1$
for a thermal initial $N_1$-abundance in the limit 
$K_1\rightarrow 0$.}
Eq.~(\ref{M1minMbar}) is quite general and shows the effect of the
phase suppression \cite{DiBari:2005st} and of a higher
absolute neutrino mass scale \cite{DiBari:2004en}
in making $M_1$ higher.
In Fig.~\ref{fig:M1minUnfl} we show $M_1^{\rm min}$ (thick solid line)
for fully hierarchical light neutrinos ($m_1=0$)
and maximal phase ($\sin\d_{\rm L}=1$).
It is convenient to introduce a value $K_{\star}$ such that,
for $K_1\geq K_{\star}$, the final asymmetry calculated for a
thermal initial $N_1$-abundance ($N^{\rm in}_{N_1}=1$) differs from the one 
calculated for a vanishing
initial $N_1$-abundance ($N^{\rm in}_{N_1}=0$) by less than some quantity
$\d$. When $\d=10\%$, $K_{\star}\simeq 3.5$ 
and one obtains the lowest value \cite{DiBari:2005st},
\be\label{lbM1min}
M_1\gtrsim 4\times 10^{9}\,{\rm GeV} \, .
\ee
The lower bound on $M_1$ also translates into a lower bound on $T_{\rm in}$,
the initial temperature of leptogenesis,
\be\label{lbTin}
T_{\rm in}\geq (T_{\rm in}^{\rm min})_{\rm HL}\simeq
{(M_1^{\rm min})_{\rm HL}\over z_{\rm B}(K_1)-2}
\gtrsim 1.5\,\times 10^{9}\,{\rm GeV}
\hspace{8mm} (K_1\gtrsim 3.5) \, .
\ee
A plot of this lower bound is shown in Fig.~\ref{fig:M1minUnfl} 
(thick dashed line).
The relation between $M_1^{\rm min}$ and $T_{\rm in}^{\rm min}$
can be understood from the top panel of Fig.~\ref{fig:DynamicsStrong},
which shows that the final asymmetry is the result of the decays
occurring just around $z_{\rm B}$, when inverse decays switch off,
whereas all the asymmetry produced before is efficiently washed out.

Assuming a period of inflation at early stages, 
the minimal initial temperature $T_{\rm in}^{\rm min}$ that allows for successful
leptogenesis can be identified
with the minimal reheat temperature $T_{\rm reh}^{\rm min}$ after
inflation. In locally supersymmetric 
theories, the high temperature required poses a problem known as 
the \emph{gravitino problem}, as we shall see in the next
section.

For increasing values of the absolute neutrino mass scale, $m_1$,
there is a joint effect of the suppression of the
\CP~asymmetry, more specifically of $\b_{\rm max}(m_1,\mt)$
[cf.~Eq.~(\ref{bmax})],
plus the extra washout from $\D L=2$ processes mediated
by off-shell RH neutrinos, which
can be conveniently factored out of the efficiency factor
in the following way~\cite{Buchmuller:2004nz}:
\be
\bar{\k}^{\rm f} (\mt, M_1 \overline{m}^2)=\k^{\rm f}(\mt)
\exp \left\{-{\o \over z_{\rm B}}\left(M_1\over 10^{10}~{\rm GeV}\right)
\left(\overline{m}\over {\rm eV}\right)^2\right\},
\ee
where $\o\simeq 0.186$ and $\overline{m}^2\equiv m_1^2+m_2^2+m_3^2$. 
This suppression leads to a stringent upper bound on the lightest neutrino mass, 
\be\label{m1max}
m_1 \leq 0.12\,{\rm eV},
\ee
as derived analytically in \cite{Buchmuller:2004nz} and numerically in
\cite{Buchmuller:2002jk,Buchmuller:2003gz,Giudice:2003jh}. 
The upper bound can be seen in the upper part of Fig.~\ref{fig:mMplot},
corresponding to the unflavored regime. It is obtained for high values 
\mbox{$M_1\sim 10^{13}$~GeV}. 

There have been earlier attempts to get a bound on the absolute neutrino mass
scale
simply from the fact that $m_1\leq \mt$ and  then imposing an upper limit
\mbox{$\mt\lesssim 10^{-3}\ev$}, justified as a generic `out-of-equilibrium'
condition. However, it turns out that such a restrictive upper limit
on $\mt$ does not hold for reasons that are clear from Fig.~\ref{fig:DynamicsStrong}:
the `out-of-equilibrium' condition is realized also in the strong 
washout regime when $z\gtrsim z_{\rm off}\simeq z_{\rm B}$. Therefore,
within the $N_1$DS, it is not a problem to have $\mt$ as large as
$1\,{\rm eV}$. Incidentally, the upper bound can only be understood 
when the washout
from $\D L=2$ processes and the upper bound on the $C\!P$ asymmetry are
jointly taken into account~\cite{Buchmuller:2002jk}.

We would like now to summarize the results of this section. As we have seen, 
in the limit where the heavy neutrino mass spectrum is hierarchical
(HL), \mbox{$M_1\ll M_2\ll M_3$}, the picture of leptogenesis in the
unflavored case simplifies considerably. A typical scenario of leptogenesis, which we
call ``vanilla leptogenesis'', emerges and it turns out that
one has to study only the production of asymmetry by the lightest RH neutrino,
 $N_1$, reducing the number of parameters from 10 to 6. This is possible because the 
theoretically favored range for $K_1$ lies between \mbox{$K_{\rm sol}\equiv
m_{\rm sol}/m_{\star}=8.2$}, given by the solar scale, 
and $K_{\rm atm}\equiv m_{\rm atm}/m_{\star}= 48$, given by the atmospheric scale. 
In this range, the
washout from the lightest RH neutrino is strong enough to make
the contributions from the heavier two RH neutrinos negligible, and
the final result is independent of the initial number of 
$N_1$~\cite{Buchmuller:2003gz,Buchmuller:2004nz,DiBari:2005st}. This
is an extremely nice feature of the strong washout regime.
Note that the $N_1$DS can actually be ensured by setting
$R_{23}=\mathbbm{1}$, which exludes a large contribution from $N_2$~\cite{DiBari:2005st}.
Within vanilla leptogenesis, it is possible to derive from successful
leptogenesis a lower bound on $M_1$ and on the initial temperature $T_{\rm in}$, 
as shown in Eqs.~(\ref{lbM1min}) and (\ref{lbTin}). Moreover,
there is a stringent upper bound on the absolute neutrino
mass scale, as shown in Eq.~(\ref{m1max}).

We shall discuss the implications of going beyond the assumptions
leading to vanilla leptogenesis in Chapter~\ref{chap:beyond}.

\section{Leptogenesis and supersymmetry}
\label{sec:lepsuper}

In the Minimal Supersymmetric Standard Model (MSSM) complemented with three RH 
neutrinos and the corresponding superpartners, the picture of leptogenesis
is qualitatively quite different from the non-supersymmetric case, but it turns out that,
quantitatively, they are very similar. 

The interactions of the heavy
(s)neutrino field can be derived from the leptonic superpotential
\be
W={1\over 2} M_i N_i N_i +h_{\a i} L_{\a} H_{u} N_i+f'_{\a} L_{\a} H_{d} E_{\a},
\ee
where $L_{\a}$ and $E_{\a}$ are the $SU(2)$ lepton doublets and singlets 
chiral superfields,
respectively, and $H_u$ and $H_d$ are the Higgs chiral superfields. The scalar 
components
of both Higgs bosons, which we denote $\F_{u,d}$, have vacuum expectation values: 
$\langle \F_u\rangle\equiv v_u= v \sin\b$ and $\langle \F_d\rangle\equiv v_d= v \cos\b$,
where $v=174\gev$. Their ratio is then given by
\be
\tan \b\equiv {v_u\over v_d}.
\ee
When flavor effects (see next chapter) are included in supersymmetric leptogenesis,
the value of $\tan \b$ is relevant because \mbox{$f_{\a}'^2= (1+\tan^2\b)f_{\a}^2$}, 
where $f_{\a}$ is the SM Yukawa coupling.

Typically, supersymmetry breaking terms are of no relevance for 
the mechanism
of lepton number generation, and we are left with the following
trilinear couplings in the Lagrangian, written in terms
of four-component spinors,
\be
-h_{\a i}\left[ M_i\tilde{L}^{\dg}_{\a}\tilde{N}_i \F_u+\bar{\ell}_{\a} P_R
N_i\F_u +\bar{\ell}_{\a}P_R \tilde{\f}^c\tilde{N}_i +
\tilde{L}^{\dg}_{\a}P_R \tilde{\f}^c N_i\right] +h.c.,
\ee 
where $\tilde{L}$, $\f$ and $\tilde{N}$ denote sleptons, higgsinos and
singlet sneutrinos, respectively.

From these couplings one obtains the tree-level relations
\be
\G_{N_i \ell}+\G_{N_i \bar{\ell}}=\G_{N_i\tilde{L}}+
\G_{N_i\tilde{L}^{\dg}}=\G_{\tilde{N}_i^{\star} \ell}=\G_{\tilde{N}_i
\tilde{L}}={\left(h^{\dg}h\right)_{ii}\over 8\p} M_i.
\ee

There are now new diagrams contributing to the \CP~asymmetry.
On top of the usual contributions shown in Fig.~\ref{fig:CPasym}, 
there are three additional sources coming from the decay of
the heavy neutrinos into sleptons, from the decay of RH sneutrinos
into leptons and from the decay of RH sneutrinos into sleptons. One
can then define a \CP~asymmetry for the decay of RH neutrinos
into leptons and sleptons, and another one for the decay
of RH sneutrinos into leptons and sleptons, as follows: 
\bea
\tilde{\ve}_{N}&\equiv& -{(\G_{N\tilde{L}}+\G_{N \ell})
-(\G_{N\tilde{L}^{\dg}}+\G_{N \bar{\ell}})
\over \G_{N}},\\
\tilde{\ve}_{\tilde{N}}&\equiv& -{(\G_{\tilde{N}^{\star}\ell}+
\G_{\tilde{N}\tilde{L}})-(\G_{\tilde{N}\bar{\ell}}+
\G_{\tilde{N}^{\star}\tilde{L}^{\dg}})
\over \G_{\tilde{N}}},
\eea
where $\G_N$ and $\G_{\tilde{N}}$ denote the total decay rate of RH
neutrinos and RH sneutrinos, respectively.

These \CP~asymmetries were computed in~\cite{Covi:1996wh} to be
\be
\tilde{\ve}_N=\tilde{\ve}_{\tilde{N}}={1\over 8 \p}{1\over (h^{\dg}h)_{11}}
\sum_{j\neq 1} {\rm Im}\left[\left(h^{\dg}h\right)_{1j}^2
\right]g(x_j),
\ee 
where we recall that $x_j={M_j^2/ M_1^2}$, and 
\be
g(x)=\sqrt{x}\left[{2\over x-1}+\ln\left({1+x\over x}\right)\right]
\stackrel{x\gg 1}{\longrightarrow} {3\over \sqrt{x}}.
\ee
In the HL ($x_j\gg 1$), the \CP~asymmetry in the MSSM is therefore twice as large
as the one in the SM.

Then, since in the MSSM there are two Higgses, the coefficient ${a_{\rm sph}}$ is
slightly different from its value in the SM, and one obtains for the
baryon-to-photon ratio
\be
\eta_B\simeq 1.03\times
10^{-2} \tilde{\ve}_N \k_1^{\rm f}.
\ee

As we have seen, there are new decay channels in the MSSM, which yield an enhancement
of the \CP~asymmetry by a factor 2. On the other hand, there is also an
enhancement of the washout by a factor of 2, which implies that the constraints
on $M_1$, $T_{\rm in}$ and $m_1$ derived in the last section remain 
essentially unchanged~\cite{Giudice:2003jh}.

Assuming a period of inflation and reheating before
leptogenesis occurs, the lower bound on the initial temperature of leptogenesis 
$T_{\rm in}$ can be identified with a lower bound on the
reheat temperature $T_{\rm reh}$ of the Universe after inflation.
Within locally supersymmetric theories, it is well known that gravitinos are
produced during the reheating phase. The point is actually that they may
be overproduced, i.e. their abundance may overclose the Universe, 
leading to the so-called \emph{gravitino problem} (for early discussions, see
\cite{Pagels:1981ke,Weinberg:1982id,Khlopov:1984pf,Ellis:1984eq}). There
are two situations to be distinguished: stable gravitinos and unstable
ones. 

If the gravitino is the lightest supersymmetric particle, it is stable
and therefore represents a good dark matter candidate. In order for gravitinos not 
to exceed the dark matter abundance,
the reheat temperature has to satisfy~\cite{Moroi:1993mb,Bolz:2000fu,Pradler:2006qh,Pradler:2006hh}
\be
T_{\rm reh}\lesssim 10^7\textrm{--}10^9~{\rm GeV}.
\ee

On the other hand, if gravitinos are unstable, they may lead 
to a large entropy production when they decay during or after big-bang 
nucleosynthesis, spoiling the nice agreement between
theory and observations (see Fig.~\ref{fig:BBN}). This leads to
the bound (see \cite{Kohri:2005wn} and references therein)
\be
T_{\rm reh}\lesssim 10^6~{\rm GeV},
\ee
unless the gravitino mass is larger than about 20~TeV.

Consequently, whatever specific scenario of supergravity one considers, 
there is a clear tension with the lower bound from leptogenesis given in~Eq.~(\ref{lbTin}). 
Different ways to
relax this tension have been proposed in the literature. Let us give three well-known
examples.

One possibility is to produce
RH neutrinos non-thermally in the decays of the 
inflaton~\cite{Lazarides:1991wu,Murayama:1992ua,Asaka:1999yd,Asaka:1999yd,Jeannerot:2001qu,Hamaguchi:2001gw,Giudice:2003jh}. A recent 
study~\cite{HahnWoernle:2008pq} shows that the lower bound on $T_{\rm reh}$ 
from leptogenesis can be relaxed in this way by two orders of magnitude. 

Another possibility is provided by 
``soft leptogenesis''~\cite{Grossman:2003jv,D'Ambrosio:2003wy}, which
is a supersymmetric scenario which requires only one heavy RH neutrino. The
interference between the \CP-odd and \CP-even states of the heavy
scalar neutrino resembles very much the neutral kaon system. The mass splitting
as well as the required \CP~violation in the heavy sneutrino system comes from
the soft supersymmetry breaking $A$ and $B$ terms, associated with the Yukawa coupling
and mass term of $N_1$, respectively. The lower bound on the reheat temperature
in this scenario can go as low as $10^6~{\rm GeV}$~\cite{Giudice:2003jh}.

Finally, it is also possible to use the enhancement of the \CP~asymmetry
for quasi-degenerate heavy neutrinos $M_1\simeq M_2 \simeq M_3$~\cite{Covi:1996wh} (see
Fig.~\ref{fig:xi}) in order to relax the lower bound on the reheat temperature. 
This inspired the scenario of 
``resonant leptogenesis''~\cite{Pilaftsis:1997jf,Pilaftsis:1998pd,Pilaftsis:2003gt}, 
in which case the scale can be lowered to TeV. We shall come 
back to quasi-degenerate heavy neutrinos in Section~\ref{sec:beyondDL}.



\chapter{Adding flavor to vanilla leptogenesis}
\label{chap:flavor}


We described in the last chapter a picture of leptogenesis where
the flavor content of the lepton doublets produced in the decays of the 
heavy neutrinos is neglected. This is done
by summing over the flavor in the \CP~asymmetry parameter 
[cf.~Eq.~(\ref{CPasym})], in the decay, inverse-decay and scattering rates, 
and the properly normalized number density
in the Boltzmann equation~(\ref{unflke2}) is $N_{B-L}=\sum_{\a}N_{B/3 -L_{\a}}$. 
Summing over the flavor implies that all quantities in 
unflavored leptogenesis depend on some combination of 
$(h^{\dagger}h)_{ij}$, where the sum over $\a$ was explicitly carried out. 
Thus, even though the Yukawa coupling in the Lagrangian has a flavor 
index $\a$ [cf.~Eq.~(\ref{lagrangian})], 
the latter never shows up in the calculation.

This picture of leptogenesis was thought to be the correct one, or at least
a good approximation of it, until the beginning of 2006, when
two groups published independently results that showed the importance of 
flavor effects in leptogenesis~\cite{Nardi:2006fx,Abada:2006fw}. In this
chapter we would like to explain in detail why flavor effects may be 
important, when they are expected to matter and how they modify in practice
the results presented in the previous chapter. 

\section{When does flavor matter and why?}
\label{sec:flavor}

The SM Lagrangian contains a term that gives rise to the masses
of the charged leptons. Before spontaneous symmetry breaking,
this term can be written as $f_{\a}\bar{\ell}_{L\a}e_{R\a}\F +h.c.$, where
$e_{R\a}$ denote the right-handed lepton fields, which are singlets
under $SU(2)$, but with hypercharge -2. Note that we chose a lepton basis
such that this term is flavor diagonal. 

In the early Universe 
the Yukawa coupling $f_{\a}$ can be strong enough to maintain
processes like $\ell_{\a}\bar{e}_{\a}\leftrightarrow \F^{\dagger}$ 
or $\ell_{\a}\bar{e}_{\a}\leftrightarrow \F^{\dagger} A$, where $A=
W^{3,\pm},B$ are the $SU(2)\times U(1)$ gauge bosons, in 
equilibrium, i.e. $\G_{\a}\gtrsim H$. With a vacuum Higgs mass of $120\gev$, the 
interaction rate was estimated in~\cite{Cline:1993bd} to be
\be\label{rateflav}
\G_{\a}\simeq 5\times 10^{-3}f_{\a}^2 T.
\ee
Obviously, when the temperature drops due to the expansion 
of the Universe, the first 
interactions that will enter equilibrium are the ones 
involving the $\t$-lepton, simply because
its Yukawa coupling is larger. It turns out that the temperature
at which the $\t$-lepton Yukawa interactions enter equilibrium is
$\sim 10^{12}$~GeV. For the muon, this will happen at 
$T\sim 10^9$~GeV, and for the electron, at still much lower
temperature.\footnote{In supersymmetric leptogenesis, the rate in 
Eq.~(\ref{rateflav}) is multiplied by $(1+\tan \b)^2$ (see Section~\ref{sec:lepsuper}),
so that flavor effects can start to matter at higher temperatures~\cite{Antusch:2006gy}.}

If the charged-lepton Yukawa interactions are in equilibrium
($\G_{\a}> H$) and faster than inverse decays (see Section~\ref{sec:maxflavored}),
\be\label{condition}
\G_{\a}\gtrsim \sum_{i}\,\G_{\rm ID}^i \, ,
\ee
during the relevant period of the asymmetry generation,
then the lepton quantum states $|\ell_i\rangle$ [cf.~Eq.(\ref{state})]
lose coherence between the production
at decay and the subsequent absorption in inverse processes.
When the loss of coherence is complete, the Higgs bosons 
will interact with incoherent lepton flavor eigenstates instead
of the coherent superposition $|\ell_i\rangle$ produced in the
decays. 
In the limit where the quantum state becomes completely incoherent
and is fully projected onto one of the flavor eigenstates, 
each lepton flavor
can be treated as a statistically independent particle species.
This is what we call the ``fully flavored regime''.
One has to distinguish a two-flavor regime, for
$10^{9}~{\rm GeV}\lesssim M_1\lesssim 10^{\rm 12}\,{\rm GeV}$, such that
the condition Eq.~(\ref{condition}) is satisfied
only for $\a=\t$, and a three-flavor regime, for
$M_1\lesssim 10^9~{\rm GeV}$, where the
condition Eq.~(\ref{condition}) applies also to $\a=\m$.

In the fully flavored regime, there are two new effects
compared  to the unflavored regime \cite{Nardi:2006fx}. These can
be understood introducing the projectors and writing them
as the sum of two terms,
\bea\label{defproj}
P_{i\alpha} & \equiv  &
|\langle \ell_{i}|\ell_{\alpha}\rangle |^2  =
P_{i\alpha}^0 + {\Delta P^0_{i\alpha}\over 2} \\ \label{defantiproj}
\overline{P}_{i\alpha}& \equiv &
|\langle \bar{\ell}'_{i}|\bar{\ell}_{\alpha}\rangle |^2  =
P_{i\alpha}^0 - {\Delta P^0_{i\alpha}\over 2} \, .
\eea
The first effect is a reduction of the washout compared to the
unflavored regime and is described by the tree-level contribution
$P_{i\a}^0=(P_{i\a}+\overline{P}_{i\a})/2$, which sets
the fraction of the total asymmetry produced in $N_i$-decays
that goes into each single flavor $\a$. In the fully flavored regime,
the Higgs will make inverse decays on flavor eigenstates 
$|\ell_{\a}\rangle$, instead of the linear superposition $|\ell_i\rangle$,
and hence the washout rate is reduced by the projector $P_{i\a}^0$.

The second effect is an additional \CP-violating 
contribution due to
a different flavor composition between
$|\ell_i\rangle$ and $C\!P |\bar{\ell}_i'\rangle$.
This can be described in terms of the projector differences
$ \D\,P_{i\a}\equiv P_{i\alpha}-\overline{P}_{i\alpha}$,
such that \mbox{$\sum_{\a}\,\D\,P_{i\a}= 0$}.
Indeed, defining the flavored \CP~asymmetries, 
\be
\ve_{i\a}\equiv -{\G_{i\a}-\overline{\G}_{i\a}
\over \G_i+\overline{\G}_i},
\ee
where $\G_{i\a}\equiv P_{i\a}\G_{i}$ and $\overline{\G}_{i\a}\equiv
\overline{P}_{i\a}\overline{\G}_{i}$ and the total decay rates $\G_i$ and 
$\overline{\G}_i$ were introduced after Eq.~(\ref{CPasym}), these can be 
now expressed as
\be\label{veiaDP}
\ve_{i\a}=\ve_i\,P^{0}_{i\a}+ {\D\,P_{i\a}\over 2} \, ,
\ee
showing that the first term is the usual contribution due to
a different decay rate into leptons and antileptons, and
the second is the additional contribution
due to a possible different flavor composition
between $|\ell_i\rangle$ and $C\!P |\bar{\ell}_i'\rangle$.
Note that when the flavored \CP~asymmetries are summed 
over the flavor, one recovers the total \CP~asymmetry used 
in the unflavored regime, i.e.
\mbox{$\sum_{\a}\ve_{i\a}=\ve_{i}$}. 

It is interesting to notice at this point that
one can imagine a scenario where the total
\CP~asymmetry $\ve_i$ is 0, i.e. no asymmetry would
have been produced in the unflavored regime, but, thanks
to the $\D P$ contribution, the flavored \CP~asymmetries
$\ve_{i\a}$ do not vanish. More specifically, one can
imagine that the only source of \CP~violation comes
from low-energy phases in the PMNS matrix. Since the
total \CP~asymmetry is insensitive to the PMNS
matrix, $\ve_i=0$ in this case. However, $\D P_{i\a}$ and
$\ve_{i\a}$ do explicitly depend on the PMNS matrix [see for instance
Eq.~(\ref{e1alOm})], so that they
can be non-zero only thanks to the low-energy phases. 
In great contrast to the unflavored picture, fully flavored 
leptogenesis can then be successful solely thanks to the \CP~violation coming
from low-energy phases. This
represents a very nice possibility which will be the 
subject of Chapter~\ref{chap:OmReal}.

\section{Flavored Boltzmann equations and spectator processes}
\label{sec:spectator}

We follow here the approach presented in~\cite{Nardi:2005hs,Nardi:2006fx}.
Heavy neutrino decays produce lepton flavor asymmetries, $N_{L_{\a}}$, 
that are computed using Boltzmann equations similar to 
Eqs.~(\ref{unflke1}) and (\ref{unflke2}).
Rigorously, one should include the effect of electroweak sphalerons,
which constitute an additional source of lepton flavor violation.
This can be symbolically done by adding a washout term ${\rm d} N_{L_{\a}}
^{\rm EW}/{\rm d}z$. Then, for consistency, we also need to add the equation
${\rm d}N_B/{\rm d}z={\rm d}N_B^{\rm EW}/{\rm d}z$ to account for baryon number violation by the
sphaleron processes. Given that the sphaleron interactions preserve
the three charges $\D_{\a}\equiv B/3-L_{\a}$ associated to 
anomaly-free currents, it follows that \mbox{$N_B^{\rm EW}/3=N_{L_{\a}}^{\rm EW}$}.
By subtracting the equations for the lepton flavor densities from the 
equation for baryon number weighted by a suitable factor 1/3, one
obtains the following network of flavored Boltzmann equations
~\cite{Nardi:2006fx}:
 \bea\label{NN1}
{{\rm d}N_{N_1}\over {\rm d}z} & = & -D_1\,(N_{N_1}-N_{N_1}^{\rm eq})\\ \label{Eq1spec}
{{\rm d}N_{\D_{\a}}\over {\rm d}z} & = &
\ve_{1\a}\,D_1\,(N_{N_1}-N_{N_1}^{\rm eq})
-P_{1\a}^{0}\, W_1^{\rm ID}\,(N_{\ell_{\a}}+N_{\F}) \,
,
\eea
where, $\a=e,\m,\t$ and, for simplicity, we included 
only decays and inverse decays with proper subtraction of the
resonant contribution from $\D L=2$ and $\D L=0$ processes,
and only focused on the lightest RH neutrino $N_1$. Notice that
$N_{\D_{\a}}$ is the $\D_{\a}$ number density, properly normalized,
and \mbox{$N_{L_{\a}}=N_{\ell_{\a}}+N_{e_{\a}}$}, where $e_{\a}$ denote
the RH lepton fields. 

We are neglecting non-resonant $\D L=2$ and $\D L=0$ processes, 
a good approximation for $M_1\ll 10^{14}~{\rm GeV}$, as we will
always consider. We are also neglecting $\D L=1$ scatterings, 
which give a correction to a level less than $\sim 10\%$ in
the most interesting strong washout regime~\cite{Blanchet:2006be},
and thermal corrections.


The number density asymmetries for the particles $X$ entering
in Eq.~(\ref{Eq1spec}) are related to the corresponding
chemical potentials through
\be
n_X-n_{\bar{X}}={g_X T^3\over 6}\left\{\begin{array}{l}
\m_X/T\quad {\rm fermions},\\
2\m_X/T\quad {\rm bosons},
\end{array}\right.
\ee
where $g_X$ is the number of degrees of freedom of $X$. For any given
temperature regime, the specific set of reactions that are in 
chemical equilibrium enforce algebraic relations between different
chemical potentials~\cite{Harvey:1990qw}. In the entire range of 
temperatures relevant for leptogenesis, the interactions mediated by
the top-quark Yukawa coupling $h_t$, and by the gauge interactions, 
are always in equilibrium. Moreover, at the intermediate-low temperatures
where flavor effects can be important, strong QCD 
sphalerons~\cite{McLerran:1990de} are also
in equilibrium. This situation has the following consequences:
\begin{itemize}
\item Equilibration of the chemical potentials for the different quark
colors is guaranteed because the chemical potentials of the gluons
vanish, $\m_g=0$.

\item Equilibration of the chemical potentials for the two members of a
$SU(2)$ doublet is guaranteed by the fact that $\m_{W^+}=-\m_{W^-}=0$ 
above the electroweak phase transition. This condition was implicitly
implemented in Eq.~(\ref{Eq1spec}), where we used $\m_{\ell}\equiv
\m_{e_L}=\m_{\n_L}$ and $\m_{\F}\equiv \m_{\f^+}=\m_{\f^0}$ to write the
particle number asymmetries directly in terms of the number densities
of the $SU(2)$ doublets.

\item Hypercharge neutrality implies
\be
\sum_i(\m_{Q_i}+2\m_{u_i}-\m_{d_i}-\m_{\ell_i}-\m_{e_i})+2\m_{\F}=0,
\ee
where $u_i$, $d_i$ and $e_i$ denote the $SU(2)$ singlet fermions of the 
$i$-th generation.

\item The equilibration condition for the Yukawa interactions of the 
  top quark $\m_t=\m_{Q_3}+\m_{\F}$.

\item Because of their larger rates, QCD sphalerons equilibration occurs
at higher temperatures than for the corresponding electroweak processes,
presumably around $T_s\sim 10^{13}~{\rm GeV}$~\cite{Bento:2003jv,Moore:1997im},
and in any case long before
equilibrium is reached for the $\t$-Yukawa processes. This implies the 
additional constraint
\be
\sum_i (2\m_{Q_i}-\m_{u_i}-\m_{d_i})=0.
\ee

\end{itemize}

To express the asymmetries $N_{\ell_{\a}}$ and $N_{\F}$ in terms of the
$N_{\D_{\a}}$, we define two matrices, $C^{\ell}$ and $C^{\F}$, through
the relations~\cite{Nardi:2006fx,Barbieri:1999ma}:
\be
N_{\ell_{\a}}=-\sum_{\b}C^{\ell}_{\a\b}N_{\D_{\b}},\qquad 
N_{\F}=-\sum_{\b}C^{\F}_{\b}N_{\D_{\b}},
\ee
so that Eq.~(\ref{Eq1spec}) can be now rewritten as follows:
\be\label{kematrix}
{{\rm d}N_{\D_{\a}}\over {\rm d}z}  = 
\ve_{1\a}\,D_1\,(N_{N_1}-N_{N_1}^{\rm eq})
-P_{1\a}^{0}\, W_1^{\rm ID}\,\sum_{\b}(C^{\ell}_{\a\b}+C^{\F}_{\b})
N_{\D_{\b}}.
\ee
The numerical values of the entries in $C^{\ell}$ and $C^{\F}$ are 
determined by the constraints among the various chemical potentials 
enforced by the fast reactions that are in equilibrium in the temperature
range ($T\sim M_1$) where the $\D_{\a}$ asymmetries are produced.

In the temperature range $10^9~{\rm GeV} \lesssim T\lesssim 
10^{12}~{\rm GeV}$, we consider the bottom-, tau- and charm-Yukawa interactions 
to be in equilibrium, implying that the asymmetries in the $SU(2)$ singlets
$b$, $e_{\t}$ and $c$ degrees of freedom are populated. The corresponding
chemical potentials obey the equilibrium constraints $\m_{b}=\m_{Q_3}-\m_{\F}$,
$\m_{c}=\m_{Q_2}+\m_{\F}$ and $\m_{\t}=\m_{\ell_{\t}}-\m_{\F}$. 
Moreover, the electroweak sphaleron processes are also in equilibrium, implying
\be
\sum_i (3\m_{Q_i}+\m_{\ell_i})=0.
\ee
As concerns lepton number, each electroweak sphaleron transition creates
all the doublets of the three generations, implying that individual
lepton flavor numbers are no longer conserved. An asymmetry will then be
generated along the $\ell_{\t}$ and $\ell_{e+\m}$ directions in flavor space,
Even though the electroweak sphalerons induce $L_{\perp}\neq0$, where by the 
subscript $\perp$ we mean
the direction in flavor space perpendicular to $\ell_{\t}$ and $\ell_{e+\m}$, the 
condition $\D_{\perp}=0$ is not violated, and hence
Eq.~(\ref{kematrix}) consists of just two equations for $N_{\D_{\t}}$
and  $N_{\D_{e+\m}}$. This is what we call the \emph{two-flavor regime}.
As concerns baryon number, electroweak sphalerons are the only source
of $B$ violation, implying that baryon number is equally distributed among
the three quark generations, i.e. $B/3$ in each generation. This modifies
the detailed equilibrium conditions for the quark chemical potentials.
Solving the corresponding set of linear equations for the chemical potentials,
one obtains~\cite{Blanchet:2008pw}
\be\label{Cl}
C^{\F}={1\over 158}(41,56),\qquad C^{\ell}={1\over 316}\left(\begin{array}{cc}
270&-32\\ -17&208 \end{array}\right).
\ee

In the temperature range $T\lesssim 10^9~{\rm GeV}$, one has to include the equilibration
constraints from the strange-quark Yukawa interactions and, more 
importantly, from the muon-Yukawa interactions.
Given that the electron remains the only lepton with a negligible Yukawa
coupling, the Yukawa interactions completely define the flavor basis for the 
leptons as well as for the antileptons (that are now the \CP-conjugate
states of the leptons). Correspondingly, the lepton asymmetries are also
completely defined in the flavor basis. In this regime the coefficients
$C^{\ell}_{\a\b}$ and $C_{\b}^{\F}$ are given by~\cite{Nardi:2006fx,Abada:2006ea}
\be\label{3flav}
C^{\F}={1\over 179}(37,52,52),\qquad C^{\ell}={1\over 1074}\left(
\begin{array}{ccc}
906&-120&-120\\
-75&688&-28\\
-75&-28&688\end{array}\right).
\ee

\section{In practice, what changes?}
\label{sec:practice}

We have seen in the last section that, if the relevant temperature
for leptogenesis is below roughly $10^{12}$~GeV, a rigorous description of 
the asymmetry evolution has to be performed
in terms of the individual flavor asymmetries $\D_{\a}\equiv B/3-L_{\a}$
rather than in terms of the total asymmetry $N_{B-L}=\sum_{\a}N_{\D_{\a}}$,
as usually done in the unflavored treatment. 

We would like to make two remarks:
\begin{itemize}
\item
From Eq.~(\ref{kematrix}) it can be seen that the two diagonal entries
of $C^{\ell}$ sum up with the entries in $C^{\F}$. Actually,
this sum turns out to be close to~1 both using Eq.~(\ref{Cl}) and 
Eq.~(\ref{3flav}); moreover, the addition of the
coefficients in $C^{\F}$ tends to equalize the flavored diagonal 
elements~\cite{Blanchet:2008pw}.

\item It was shown in~\cite{JosseMichaux:2007zj} that the off-diagonal
elements in the matrix $C^{\ell}$ have an effect smaller than 
40\% on the final asymmetry. Taking into account the matrix $C^{\F}$,
this effect is even reduced.
\end{itemize}

Therefore, for simplicity, we shall use in the following $C^{\ell}+C^{\F}=\mathbbm{1}$,
which takes into account spectator processes in an approximate way. 
The generalization of the flavored Eqs.~(\ref{NN1})
and (\ref{Eq1spec}) to three RH neutrinos ($i=1,2,3$) is then given by
\cite{Endoh:2003mz,Pilaftsis:2005rv,Nardi:2006fx,Abada:2006fw}
\bea\label{flke}
{{\rm d}N_{N_i}\over {\rm d}z} & = & -D_i\,(N_{N_i}-N_{N_i}^{\rm eq})\\ 
{{\rm d}N_{\D_{\a}}\over {\rm d}z} & = &
\sum_i\,\ve_{i\a}\,D_i\,(N_{N_i}-N_{N_i}^{\rm eq})
-\sum_i\,P_{i\a}^{0}\, W_i^{\rm ID}\,N_{\D_{\a}}.\label{flke2}
\eea
Notice that, in the two-flavor regime, $\D_{\a}=B/2-L_{\a}$, where
$\a=e+\m$ or $\t$, and the two equations for the individual
electron and muon asymmetries are replaced by one kinetic equation
for the sum $N_{\D_{e+\m}}$,
where the individual flavored $C\!P$ asymmetries and projectors
have to be replaced by their sum, namely
$\ve_{1\,e+\m}\equiv \ve_{1\m}+\ve_{1e}$ and $P^0_{1\,e+\m}\equiv
P^0_{1\m}+P^0_{1 e}$~\cite{Abada:2006ea}. The total asymmetry
is then given by $N_{B-L}=N_{\D_{e+\m}}+N_{\D_{\t}}$.
The calculation is therefore intermediate between the
unflavored case and the three-flavor regime, though the results
are very similar to the three-flavor regime~\cite{Blanchet:2006be}.

The evolution of the $N_{\D\a}$'s can be worked out in an integral form,
\be\label{NDeltaInt}
N_{\D\a}(z)=N_{\D\a}^{\rm in}\,
\exp\left(-\sum_i\,P_{i\a}^0\,\int_{z_{\rm in}}^z\,{\rm d}z'\,W_i^{\rm ID}(z')\right)
+\sum_i\,\ve_{i\a}\,\k_{i{\a}}(z) \,  ,
\ee
with  the 6, in the two-flavor case, or 9, in the
three-flavor case, efficiency factors given by
\be\label{ef}
\hspace{-0.3cm}
\k_{i\a}(z;K_i,P^{0}_{i\a})=-\int_{z_{\rm in}}^z\,{\rm d}z'\,{{\rm d}N_{N_i}\over {\rm d}z'}\,
\exp\left(-\sum_i\,P_{i\a}^0\,\int_{z'}^z\,{\rm d}z''\,W_i^{\rm ID}(z'')\right).
\ee
The final $B\!-\!L$ asymmetry is then given by
\be
N_{B-L}^{\rm f}=\sum_{\a}\,N_{\D_\a}^{\rm f},
\ee 
from which one obtains the baryon-to-photon ratio $\eta_B$ 
using Eq.~(\ref{asymgeneral}).

Using the convenient Casas-Ibarra
parametrization, Eq.~(\ref{CI}), the tree-level projectors can be written
as
\be\label{proj}
P^0_{i\a}={|\sum_j\,\sqrt{m_j}\,U_{\a j}\,\O_{j i}|^2
\over \sum_j\,m_j\,|\O^2_{ji}|} \, .
\ee
It will also prove useful to introduce the flavored decay 
parameters,
\be\label{Kialpha}
K_{i\a}\equiv P_{i\a}^{0}\,K_i=\left|\sum_j\,\sqrt{{m_j\over m_{\star}}}
\,U_{\a j}\,\O_{j i}\right|^2,
\ee
obtained using Eqs.~(\ref{proj}) and (\ref{decparCI}).

The flavored $C\!P$ asymmetries are given by the following 
expression~\cite{Covi:1996wh}:
\bea\label{veia}
\ve_{i\a}&=&
\frac{3}{16 \p (h^{\dag}h)_{ii}} \sum_{j\neq i} \left\{ {\rm Im}\left[h_{\a i}^{\star}
h_{\a j}(h^{\dag}h)_{i j}\right] \frac{\x(x_j/x_i)}{\sqrt{x_j/x_i}}\right.\NO\\
&&\hspace{2.5cm} \left.+\frac{2}{3(x_j/x_i-1)}{\rm Im}
\left[h_{\a i}^{\star}h_{\a j}(h^{\dag}h)_{j i}\right]\right\} \, ,
\eea
where $\xi(x)$ was defined in Eq.~(\ref{xi}), and we recall that
$x_i\equiv M_i^2/M_1^2$.

In general, the final asymmetry will depend on all 18 see-saw
parameters. As explained in the introduction (Section~\ref{sec:expev}),
until now we have only measured two mass-squared differences and
two mixing angles in neutrino oscillation experiments. This is essentially
the only information on the 18 see-saw parameters we have.
Thus, we can write that \mbox{$\eta_B=\eta_B(m_1,U,M_i,\o_{21},\o_{31},
\o_{32})$}.
It is interesting that including flavor effects there is a
potential dependence
of the final asymmetry also on the unknown parameters contained
in the PMNS mixing matrix $U$~\cite{Endoh:2003mz,Nardi:2006fx},
namely the mixing angle $\q_{13}$ and the \CP-violating phases
$\d$ and $\F_{1,2}$ [cf.~Eq.~(\ref{PMNSmatrix})].

Let us assume that the mass spectrum of the heavy neutrinos is
hierarchical, \HL.
In this case the general expression Eq. (\ref{veia}) for
the $C\!P$ asymmetries $\ve_{i\a}$ reduces to
\begin{eqnarray} 
\label{ve1a}
\ve_{1\alpha}&\simeq& \frac{3}{16 \p (h^{\dag}h)_{11}}
\sum_{j\neq 1} \frac{M_1}{M_j} {\rm Im}
\left[h_{\a 1}^{\star} h_{\a j}(h^{\dag}h)_{1 j}\right],\\   \label{ve2a}
\ve_{2\alpha}&\simeq&
\frac{3}{16 \p (h^{\dag}h)_{22}}
\left\{\frac{M_2}{M_3}{\rm Im}
\left[h_{\a 2}^{\star} h_{\a 3}(h^{\dag}h)_{2 3}\right]\right.\NO\\
&&\hspace{3cm}\left.-\frac{2}{3} {\rm Im}\left[h_{\a 2}^{\star} h_{\a 1}(h^{\dag}h)_{1 2}\right]
\right\} ,\\
\ve_{3\alpha}&\simeq& -\frac{1}{8\,\p (h^{\dag}h)_{33}}
\sum_{j\neq 3}
{\rm Im}\left[h_{\a 3}^{\star} h_{\a j}(h^{\dag}h)_{j 3}\right].
\end{eqnarray}
Expressing the flavored \CP~asymmetries $\ve_{1\a}$ 
in terms of the orthogonal parametrization, one obtains~\cite{Abada:2006ea}
\be\label{e1alOm}
\ve_{1\a}=-\,\frac{3 M_1}{16 \p v^2}\sum_{h,l}\,
{m_l\,\sqrt{m_l\,m_h}\over \mt}
\,{\rm Im}[U_{\a h}\,U_{\a l}^{\star}\,\O_{h1}\,\O_{l1}]
\, ,
\ee
where the explicit dependence on the PMNS matrix can be seen. Similar
expressions can be found for $\ve_{2\a}$ and $\ve_{3\a}$, but as they are
slightly more complicated, we do not show them here.

In the following, we shall assume no rotation in the plane 
23, i.e. \mbox{$R_{23}=\mathbbm{1}$} [cf.~Eq.~(\ref{R23})].
Under these conditions,
both total $C\!P$ asymmetries $\ve_2$ and $\ve_3$
are suppressed like $\sim M_1/M_{2,3}$. On the other
hand, it is interesting to notice
that the $\ve_{2\a}$'s and the
$\ve_{3\a}$'s  are not necessarily suppressed.
This can potentially lead to a scenario where the
final asymmetry is produced by the decays of the two heavier
RH neutrinos, provided $M_{2,3}\lesssim 10^{12}\,{\rm GeV}$
in order for the flavored regime to apply.
Here we do not pursue this possibility (see, however,
Sections~\ref{sec:N2} and \ref{sec:DiracLep}) and 
focus on a typical $N_1$-dominated scenario where
the dominant contribution to the final asymmetry comes
from the decays of the lightest RH neutrino, so that
\be\label{N1DS}
N^{\rm f}_{B-L}\simeq \left. N_{B-L}^{\rm f} \right|_{N_1}
\equiv \sum_{\a}\,\ve_{1\a}\,\k_{1\a}\, .
\ee
It will prove important for our discussion that both the total
$C\!P$ asymmetry $\ve_1$ and the flavored ones $\ve_{1\a}$
cannot be arbitrarily large. The
total $C\!P$ asymmetry is indeed upper bounded 
by Eq.~(\ref{CPbound}), and
each flavored \CP~asymmetry $\ve_{1\a}$ is bounded by
\cite{Abada:2006ea}
\be\label{bound}
|\ve_{1\a}|< \overline{\ve}(M_1)\,\sqrt{P_{1\a}^0}\,
{m_3\over m_{\rm atm}}\,
{\rm max}_{\rm j}\,\,[|U_{\a j}|] \,,
\ee
where we recall that $\overline{\ve}(M_1) 
\equiv 3\,M_1\,m_{\rm atm}/(16\,\pi\,v^2)$. Therefore, while the 
total $C\!P$ asymmetry is suppressed when
$m_1$ increases, the single-flavor \CP~asymmetries can be enhanced.
The existence of an upper bound on the quantity
\be\label{r1a}
r_{1\a}\equiv \ve_{1\a}/\overline{\ve}(M_1)
\ee
independent of $M_1$,
implies, as in the unflavored analysis,
Eq.~(\ref{M1minMbar}),
the existence of a lower bound on $M_1$ given by
\be\label{lbM1}
M_1  \geq M_1^{\rm min}(K_1)={\overline{M}_1 \over \k_1^{\rm f}(K_1)\,
\xi_1^{\rm max}(K_1)} \, ,
\ee
where we will always use the 3$\s$ lower value for $\overline{M}_1$ 
[cf.~Eq.~(\ref{barM1})], and we 
indicated with $\k^{\rm f}_1(K_1)$ the efficiency factor
in the unflavored case, corresponding to
$\k_{1\a}^{\rm f}$ with $P^0_{1\a}=1$. We also defined
\be\label{xi1a}
\xi_1 \equiv \sum_{\a}\,\xi_{1\a}\, ,
\hspace{10mm}\mbox{with}\hspace{12mm}
\xi_{1\a}\equiv {r_{1\a}\,\k_{1\a}^{\rm f}(K_{1\a})
                      \over \k_1^{\rm f}(K_1)}\, .
\ee
This quantity represents the deviation introduced by flavor
effects compared to the unflavored treatment in the
hierarchical light neutrino case.
We could then use $r_{1\a}\leq\sqrt{P^0_{1\a}}\,m_3/m_{\rm atm}$
to maximize $\xi_1$.
Notice however that, first, the $r_{1\a}$'s cannot
be simultaneously equal to $\sqrt{P^0_{1\a}}$ because of the bound on the
total asymmetry, and,
second, there are sign cancellations in $\xi_1$.
Therefore, the bound~(\ref{lbM1}) is more restrictive
than this possible estimation, and we prefer to keep it in this form,
maximizing $\xi_1$ in each particular situation.

As usual, the lower bound on $M_1$ implies an associated lower bound
on the initial temperature $T_{\rm in}$ of leptogenesis and hence on the
reheat temperature $T_{\rm reh}$.

\section{Dependence on the initial conditions and lower bounds}
\label{sec:dependence}

From Eq.~(\ref{ef}), extending an analytic procedure
derived within the unflavored treatment \cite{Buchmuller:2004nz},
one can obtain simple expressions for the flavored efficiency factors 
$\k_{1\a}^{\rm f}$.
In the case of a thermal initial $N_1$-abundance ($N_{N_1}^{\rm in}=1$), one has
\be\label{k1a}
\k_{1\a}^{\rm f} \simeq \k(K_{1\a}) \, ,
\ee
where the function $\k(x)$ was given in Eq.~(\ref{k}).
Notice that, in the particularly relevant range
$5\lesssim K_{1\a}\lesssim 100$, this expression is well
approximated by Eq.~(\ref{plaw}) replacing $K_1\to K_{1\a}$.

In the case of vanishing initial $N_1$-abundance ($N_{N_1}^{\rm in}=0$),
one has to take into account two different contributions,
a negative and a positive one,
\be
\k_{1\a}^{\rm f}
=\k_{-}^{\rm f}(K_1,P_{1\a}^{0})+
 \k_{+}^{\rm f}(K_1,P_{1\a}^{0}) \, .
\ee
The negative contribution arises from a first stage when
$N_{N_1}\leq N_{N_1}^{\rm eq}$, for $z\leq z_{\rm eq}$,
and is given approximately by
\be\label{k-}
\k_{-}^{\rm f}(K_1,P_{1\a}^{0})\simeq
-{2\over P_{1\a}^{0}} \exp\left(-{3\,\pi\,K_{1\a} \over 8}\right)
\left[\exp\left({P_{1\a}^{0}\over 2}\,N_{N_1}(z_{\rm eq})\right) - 1 \right],
\ee
where $N_{N_1}(z_{\rm eq})$ was defined in Eq.~(\ref{nka}).

The positive contribution arises from a second stage when
$N_{N_1}\geq N_{N_1}^{\rm eq}$, for $z\geq z_{\rm eq}$,
and is approximately given by
\be\label{k+}
\hspace{-0.3cm}
\k_{+}^{\rm f}(K_1,P_{1\a}^{0})\simeq
{2\over z_{\rm B}(K_{1\a})\,K_{1\a}}
\left[1-\exp\left(-{1\over 2} {K_{1\a}\,z_{\rm B}(K_{1\a})\,N_{N_1}
(z_{\rm eq})}\right)\right].
\ee
It is interesting to notice that $N_{N_1}(z_{\rm eq})$ is 
still regulated by $K_1$ [cf.~Eq.~(\ref{nka})], since
the RH neutrino production is not affected by flavor
effects, contrarily to the washout, which is reduced because
regulated by $K_{1\a}$.

These analytic expressions make transparent the two conditions
to have independence from the initial conditions.
The first is the thermalization of the $N_1$-abundance,
such that, for an arbitrary initial $N_1$-abundance,
one has $N_{N_1}(z_{\rm eq})=1$.
The second is that the asymmetry produced
during the non-thermal stage, for $z\leq z_{\rm eq}$,
has to be efficiently washed out, leading to
\mbox{$|\k_{-}^{\rm f}| \ll \k_{+}^{\rm f}$}. They
are both realized for large values $K_1\gg 1$.
More quantitatively, we introduced in Section~\ref{sec:N1DS} 
the quantity  $K_{\star}$ such that, for
$K_1\geq K_{\star}$, the final asymmetry calculated
for a thermal initial $N_1$-abundance differs from the one
calculated for a vanishing initial abundance by 
less than some quantity
$\d$. This can be used as a precise definition of
the strong washout regime.
Let us consider some particular cases, showing
how flavor effects tend to enlarge the domain of the
weak washout at the expense of the strong washout regime.

\subsection{Alignment}

The simplest situation is the alignment case,
realized when the $N_1$-decays are just into one flavor
$\a$, so that $P_{1\a}=\overline{P}_{1\a}=1$ and
$P_{1\b\neq\a}=\overline{P}_{1\b\neq\a}=0$, implying
$\ve_{1\a}=\ve_1$. Notice that we do not have to worry about
the fact that
the lightest RH neutrino inverse decays might not be able
to wash out the asymmetry generated from the decays of the
two heavier neutrinos, since we are assuming negligible
$\ve_{2\b}$ and $\ve_{3\b}$ anyway.
In this case the general set of kinetic equations~(\ref{flke}) and
(\ref{flke2}) reduces to the usual unflavored equations and 
all results coincide
with those in the unflavored analysis~\cite{Nardi:2006fx}.
In particular one has $N_{B-L}^{\rm f}=\ve_1\,\k_{1\a}^{\rm f}$.

\begin{figure}[h!]
\begin{center}
\includegraphics[angle=-90,width=0.7\textwidth]{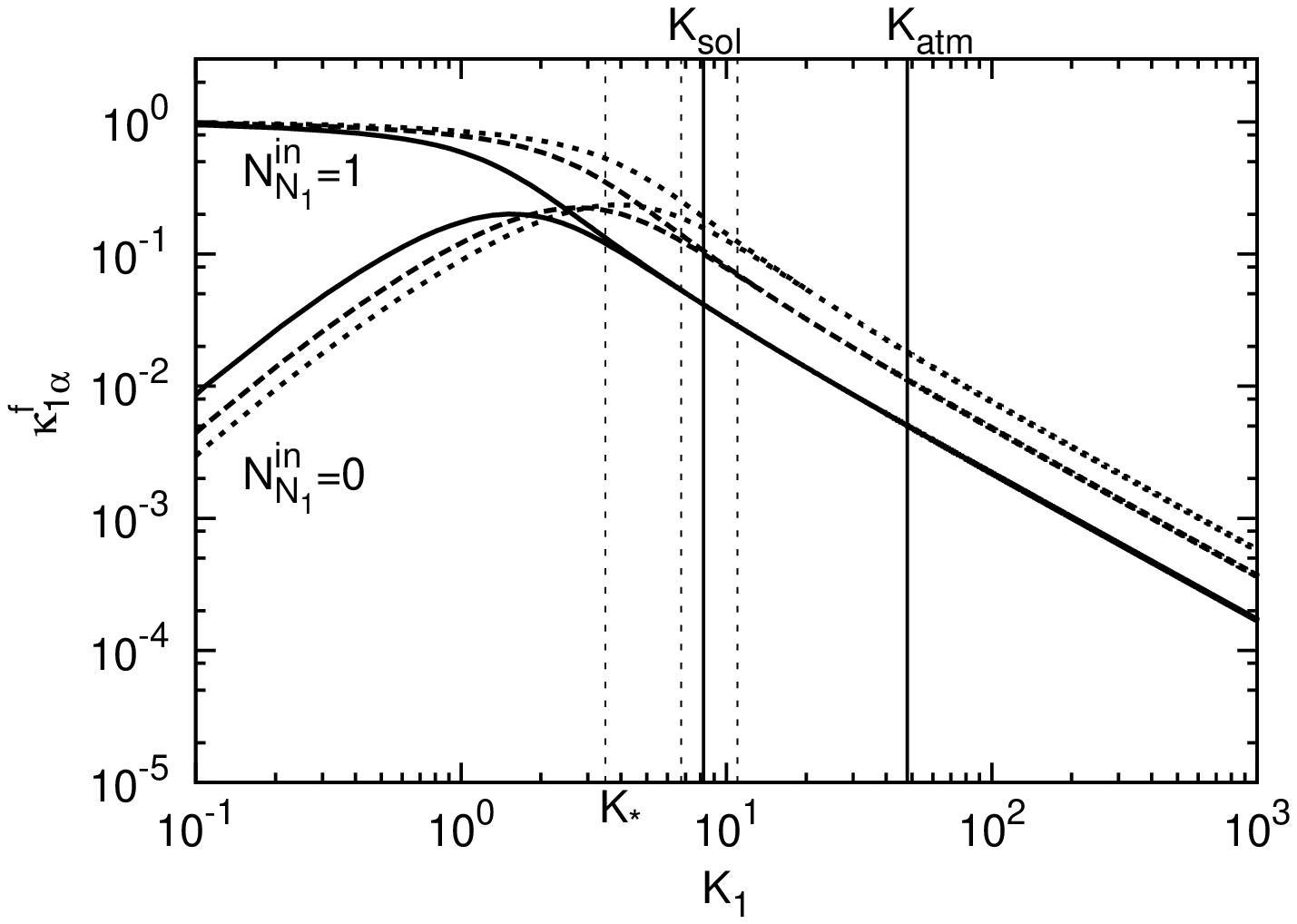}
\includegraphics[angle=-90,width=0.7\textwidth]{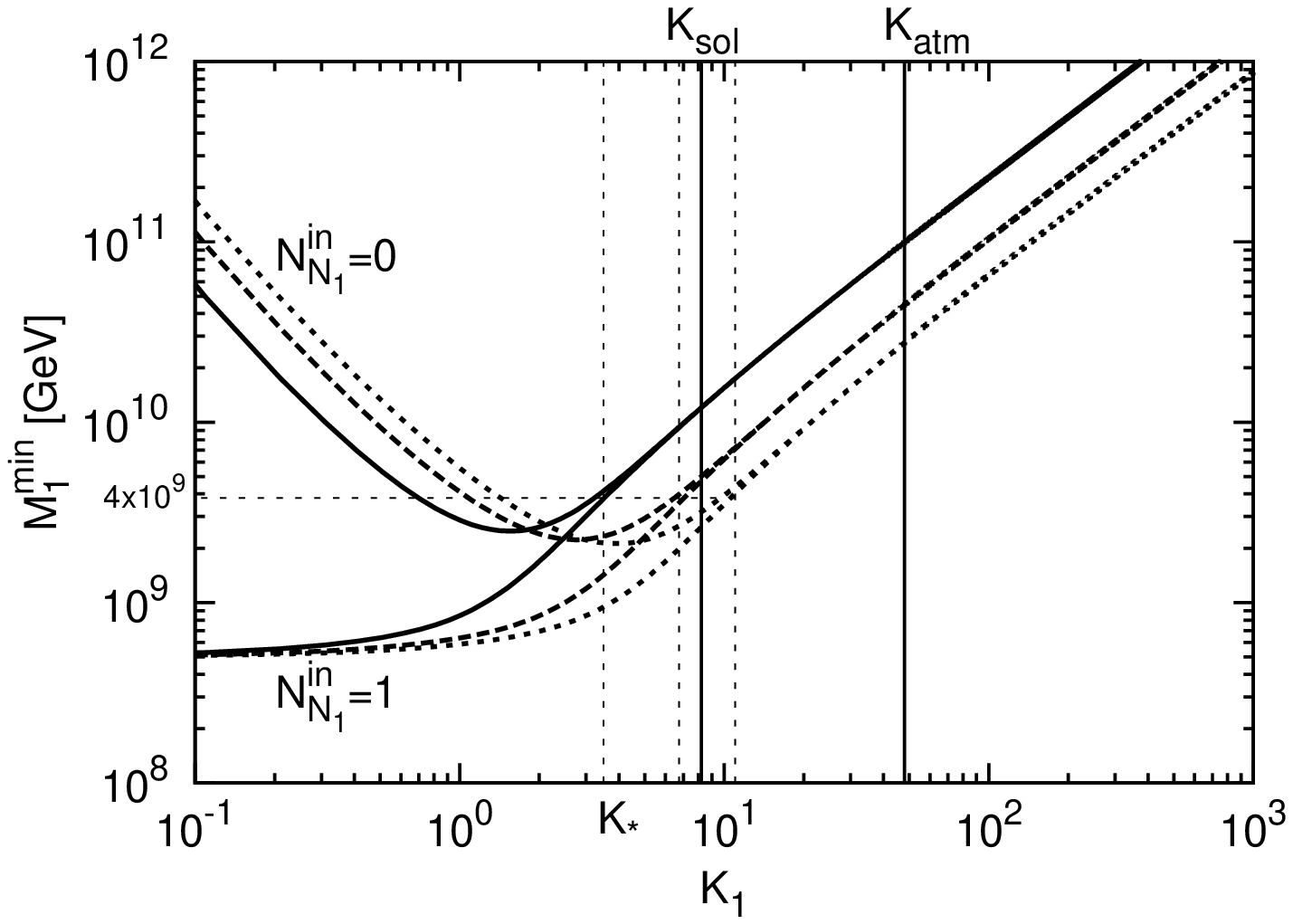}
\caption{Efficiency factors (upper panel) and lower bounds on
$M_1$  (lower panel) in the alignment (solid lines),
semi-democratic (dashed lines) and democratic (short-dashed lines) cases.}
\label{fig:EffandLowerBound}
\end{center}
\end{figure}

In the case of alignment we are considering, we obtain
$K_{\star}\simeq 3.5$ for $\d=0.1$, as shown in 
Fig.~\ref{fig:EffandLowerBound}.
The value of $K_{\star}$ plays a relevant role since
only for $K_1\gtrsim K_{\star}$ one has 
predictions from leptogenesis on the final baryon asymmetry 
resulting from
a self-contained set of assumptions. On the other hand,
for $K_1\lesssim K_{\star}$ leptogenesis has to be complemented
with a model for the initial conditions.
Additionally, the calculation of the final 
asymmetry in the weak washout regime requires a precise description of
the RH neutrino production, potentially sensitive to many poorly known
effects. It is then interesting that
current neutrino mixing data favor $K_1$ to be in the range
$K_{\rm sol}\simeq 8.2 \lesssim K_1 \lesssim 48 \simeq K_{\rm atm}$
\cite{Buchmuller:2003gz,Buchmuller:2004nz,DiBari:2005st}, 
where one can have a mild washout
assuring full independence from the initial conditions, as one can
see in Fig.~\ref{fig:EffandLowerBound}, but still successful leptogenesis.
In this case one can place constraints
on the see-saw parameters which do not depend on specific assumptions
for the initial conditions and 
with reduced theoretical uncertainties.

Since one has $\xi_1=1$ here [cf.~Eq.~(\ref{xi1a})], 
the general lower bound on $M_1$ in~Eq.~(\ref{lbM1}), like all other quantities,
becomes the usual lower bound derived from an unflavored treatment.
In the lower panel of Fig.~\ref{fig:EffandLowerBound}, we have plotted it
both for $N_{N_1}^{\rm in}=1$ and $N_{N_1}^{\rm in}=0$ (solid lines).
This corresponds precisely to the lower
bound in the unflavored treatment shown in Fig.~\ref{fig:M1minUnfl}.
One can see 
that the dependence on the initial conditions in $\k_{1\a}^{\rm f}$
translates into a dependence on the initial conditions in $M_1^{\rm min}$.
The lowest model-independent values are then obtained
for $K_1= K_{\star} \simeq 3.5$ and are given by
\be\label{M1Treh}
M_1\gtrsim 4\times 10^{9}\,{\rm GeV}
\hspace{10mm}\mbox{and}\hspace{10mm}
T_{\rm reh}\gtrsim 1.5\times 10^{9}\,{\rm GeV}\, .
\ee
We did not show the lower bound on $T_{\rm reh}$ in 
Fig.~\ref{fig:EffandLowerBound}, not to overload the
plot, but it is precisely given by the dashed line in
Fig.~\ref{fig:M1minUnfl}.
Another typically quoted lower bound on $M_1$
is the one obtained for a thermal initial $N_1$-abundance
in the limit $K_1\rightarrow 0$, given
by $M_1\gtrsim 5\times 10^8\,{\rm GeV}$~\cite{Buchmuller:2002rq}.

\subsection{Democratic and semi-democratic cases}
\label{sec:democratic}

Let us now discuss another possibility. For definiteness,
we assume a three-flavor regime; the extension of the results
to the two-flavor regime is straightforward.
Let us assume a democratic situation where $P_{1\a}=\overline{P}_{1\a}=1/3$ for
any $\a$ and consequently $\D\,P_{1\a}=0$.
This case was also considered in~\cite{Abada:2006fw}.
From Eq.~(\ref{ef}) it follows that the three efficiency factors
$\k_{1\a}^{\rm f}$, like the three \CP~asymmetries
$\ve_{1\alpha}$, are all equal and thus  Eq.~(\ref{N1DS}) 
simplifies into $N_{B-L}^{\rm f}=\ve_1\,\k_{1\a}^{\rm f}$,
as in the usual unflavored treatment.
However, the washout is now reduced by the
presence of the projector, so that
$K_1\rightarrow P^{0}_{1\a}\,K_1=K_1/3$.
The result is that, in the case of a thermal initial $N_1$-abundance,
the efficiency factor, as a function of $K_1$, is simply shifted.
The same happens for vanishing initial $N_1$-abundance in the strong
washout regime. However, in the weak washout regime,
there is not only a simple shift,
since the RH neutrino production is still depending on $K_1$.
A plot of $\k_{1\a}^{\rm f}$ is shown in the upper panel of 
Fig.~\ref{fig:EffandLowerBound}
(short-dashed lines).
One can see how the reduced washout increases
the value of $K_{\star}$ to $\sim 10$, approximately
$1/P^0_{1\a}\simeq 3$ larger, thus compensating almost
exactly the washout reduction by
a factor $\sim 3^{1.2}$ [cf.~Eq.~(\ref{plaw})].

In this way the lowest bound on $M_1$ in the strong washout regime, 
at $K_1=K_{\star}$, is almost unchanged. On the other hand,
for a given value $K_1\gg K_{\star}$, the lower bound gets approximately
relaxed by a factor $3$~\cite{Abada:2006fw}. The lower bound for
the democratic case is shown in the lower panel of 
Fig.~\ref{fig:EffandLowerBound} (short-dashed lines).
It is apparent that the lower bound for 
\mbox{$K_1\rightarrow 0$} and thermal initial $N_1$-abundance does 
not change with respect to the alignment case (or the unflavored
treatment). One can then say that
flavor effects simply induce a shift
of the dependence of the lower bound on $K_1$.

The semi-democratic case is intermediate
between the democratic and the alignment cases. It
is obtained when one projector vanishes, for example $P_{1\b}=0$, and
the other two are $1/2$. In this case,  $K_{\star}\sim 7$.
The corresponding plots of the efficiency factor and of the lower bound
on $M_1$ are also shown in Fig.~\ref{fig:EffandLowerBound} 
(dashed lines). The semi-democratic case can actually be identified with
the two-flavor regime, where the two projectors are equal, 
\mbox{$P^0_{\t}=P^0_{e+\mu}=1/2$}.

\subsection{One-flavor dominance}
\label{sec:oneflav}

There is another potentially interesting situation that motivates
an extension of the previous results to arbitrarily
small values of $P_{1\a}^{0}$. This occurs when
the final asymmetry is dominated by one flavor $\a$,
and Eq.~(\ref{N1DS}) can be further simplified into
\be
N_{B-L}^{\rm f}\simeq \ve_{1\a}\,\k_{1\a}^{\rm f} \, ,
\ee
analogously to the alignment case but with $P^0_{1\a}\ll 1$.
Notice that this cannot happen due to a dominance of one of the
$C\!P$ asymmetries, for example with $\ve_{1\a}$ being close to
its maximum value, Eq.~(\ref{bound}), much larger than the other two
that are strongly suppressed, simply because one has $\sum_{\a}\,\D P_{1\a}=0$.
One has then to imagine a situation where the $C\!P$ asymmetry
$\ve_{1\a}$ is comparable to the sum of the other two, but
$K_{1\b\neq\a}\gg K_{1\a}\gtrsim 1$, so that
$\k_{1\a}^{\rm f}\gg \k_{1\b}^{\rm f}$. The dominance
is then a result of the much weaker washout.

\begin{figure}
\begin{center}
\includegraphics[width=0.7\textwidth]{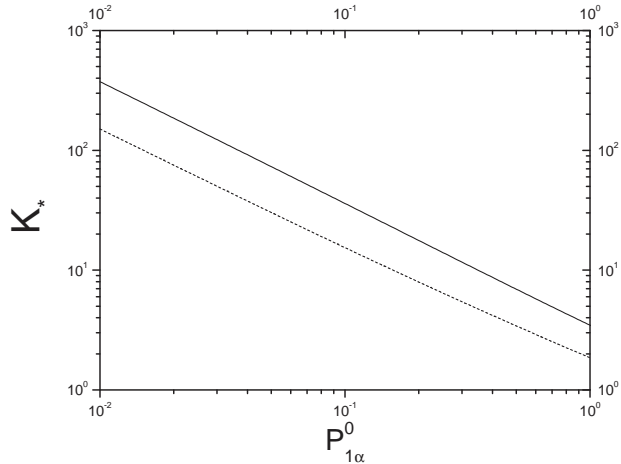}
\caption{Values of $K_{\star}$
defining the strong washout regime, for $\d=10\%$ (solid line)
and $\d=50\%$ (dashed line).}
\label{fig:Kstar}
\end{center}
\end{figure}

\begin{figure}
\hspace*{-10mm}
\psfig{file=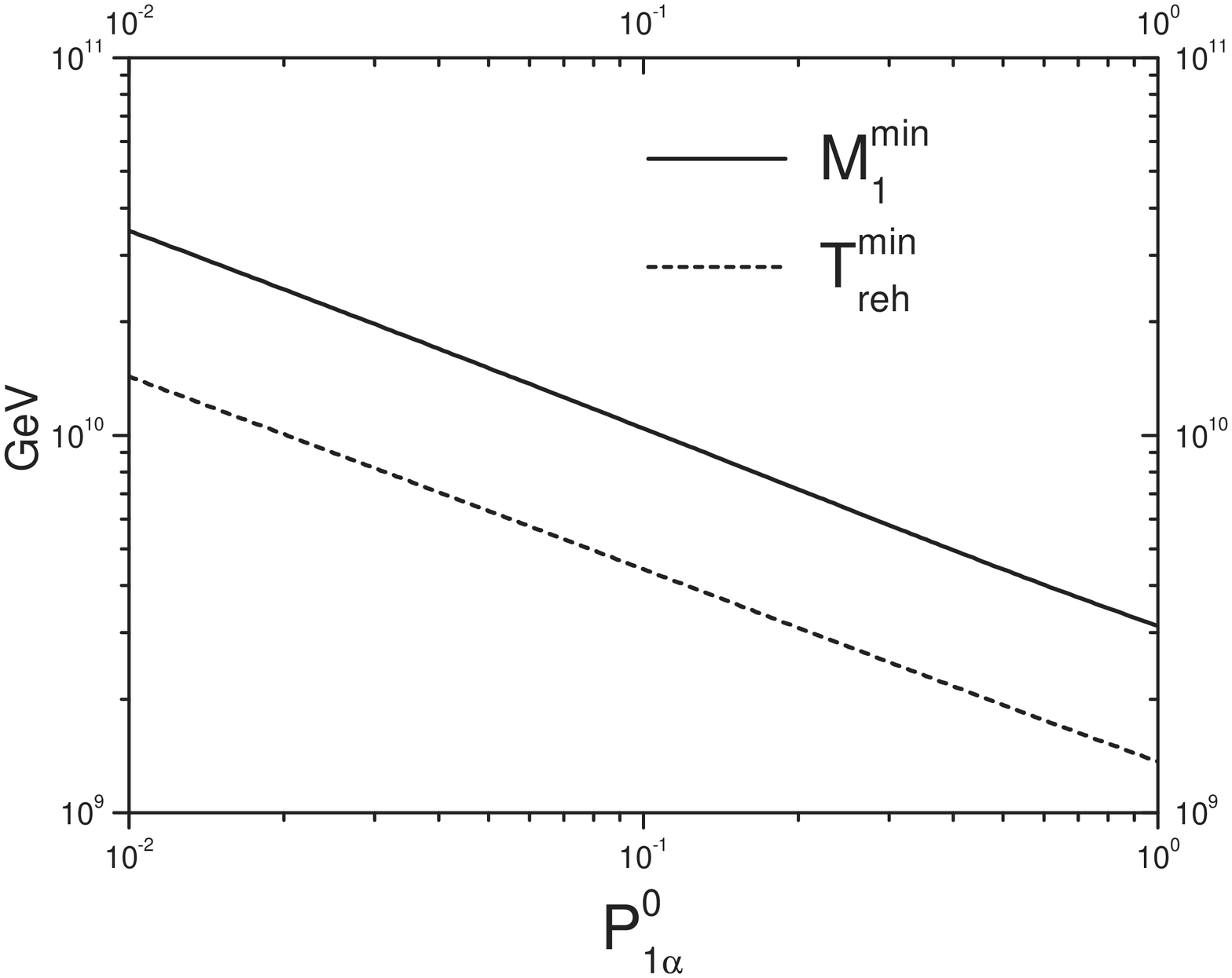,height=7cm,width=8cm}
\hspace{-10mm}
\psfig{file=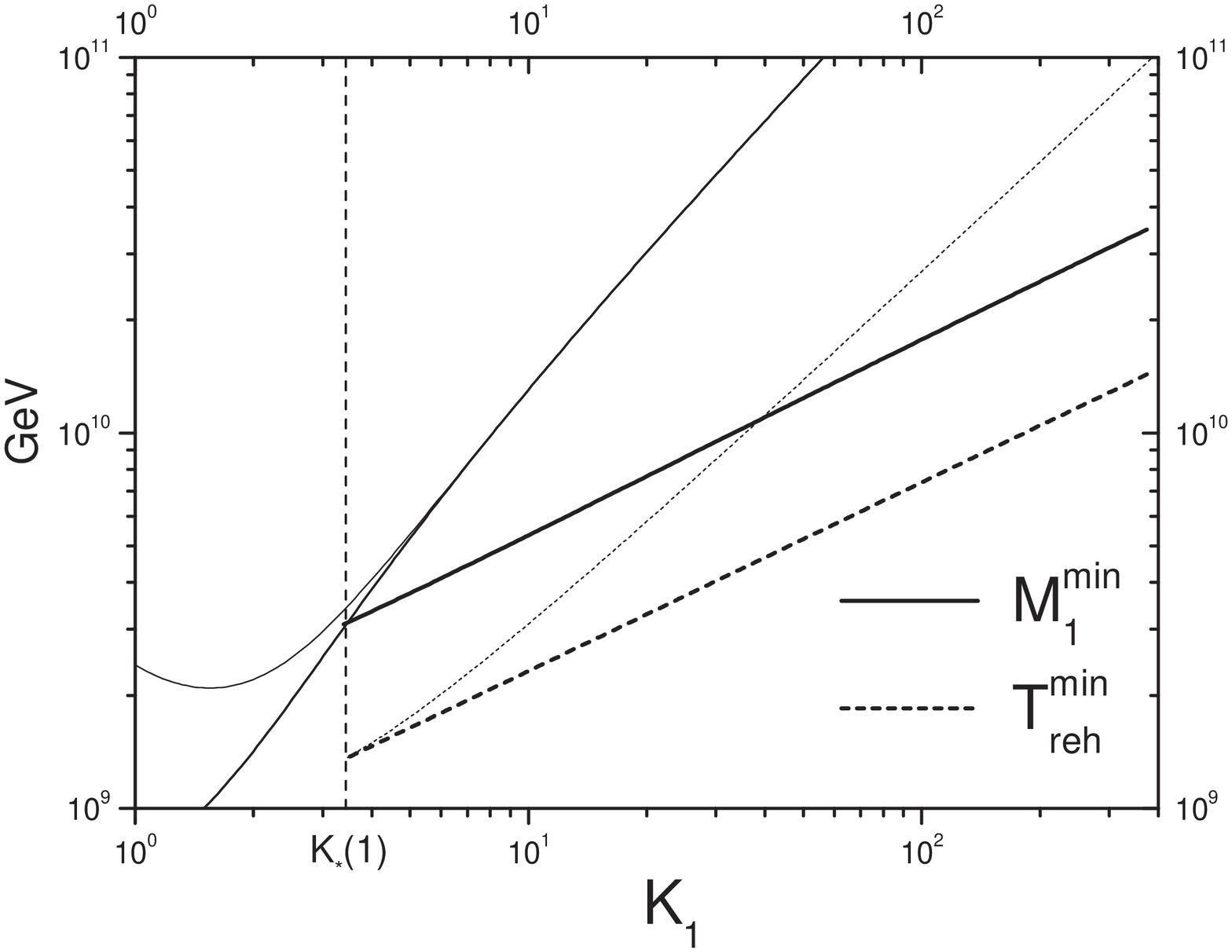,height=7cm,width=8cm}
\caption{Lower bounds on $M_1$ and $T_{\rm reh}$ calculated
choosing $P^0_{1\a}=K_{\star}(P^0_{1\a}=1)/K_1$ such that
$K_1=K_{\star}(P^0_{1\a})$ (thick lines)
and compared with the usual bounds for $P^0_{1\a}=1$ (thin lines). In the left
panel they are plotted as a function of $P^0_{1\a}$, while in the
right panel as a function of $K_1$.}
\label{fig:LowerBoundKstar}
\end{figure}

The analysis of the dependence on the initial conditions can then be
performed as in the previous cases calculating the value of $K_{\star}$ for
any value of $P_{1\a}^{0}$. The result is shown in Fig.~\ref{fig:Kstar}.
The alignment case corresponds to $P_{1\a}^{0}=1$, the semi-democratic
case to $P_{1\a}^{0}=1/2$ and the democratic case to $P_{1\a}^{0}=1/3$.
Notice that the result is very close to the simple estimation
 $K_\star(P^0_{1\a})=K_\star(1)/P^0_{1\a}$ which
 would follow if $\k_{1\a}^{\rm f}$ were just depending on $K_{1\a}$.
In Fig.~\ref{fig:LowerBoundKstar} we have plotted the values of the 
lower bounds
on $M_1$ and $T_{\rm reh}$ for hierarchical light neutrinos,
implying $m_3=m_{\rm atm}$ in Eq.~(\ref{bound}).
These can be obtained plugging
$\xi_1=\sqrt{P^0_{1\a}}\,\k_{1\a}^{\rm f}(K_{1\a})/\k_1^{\rm f}(K_1)\leq 1$
in Eq.~(\ref{lbM1}).
They correspond to the lowest values in the strong
washout regime, when $K_1\geq K_{\star}$.

There are two possible ways to look at the results.
On the one hand, flavor effects can relax the lower bounds
for fixed values of $K_1\gg 1$. Indeed, for each value $K_1\gg 1$,
one can choose $P^0_{1\a}=K_{\star}(P^0_{1\a}=1)/K_1$ such that
$K_{1}=K_{\star}(P^0_{1\a})$, thus obtaining the highest possible
relaxation in the strong washout regime.
This is shown in the right panel of Fig.~\ref{fig:LowerBoundKstar}.
It is  important to say that this relaxation is potential.
A direct inspection is indeed necessary to determine
whether it is really possible to achieve at the same time not only
small values of $P^0_{1\a}$ but also a single-flavor \CP~asymmetry
$\ve_{1\a}$ that is not suppressed compared to $\ve_{1\b\neq\a}$.

On the other hand, as a function of $P_{1\a}^0$
the bounds get more stringent when $P^0_{1\a}$ decreases, so that
the minimum is obtained in the alignment case, corresponding to
the unflavored case. This is clearly visible in the left panel of 
Fig.~\ref{fig:LowerBoundKstar}.
Therefore, it is important to emphasize here that
flavor effects cannot help to alleviate the conflict of the lower bound 
on the reheat temperature $T_{\rm reh}$ from successful leptogenesis with the upper bound
on $T_{\rm reh}$ in order not to overproduce gravitinos 
(see Section~\ref{sec:lepsuper} for a discussion). 
In particular, the bounds on $M_1$ and $T_{\rm reh}$ that are usually
quoted in the literature [cf. Eq.~(\ref{M1Treh})] are not
changed by flavor effects.

Notice that together with the one-flavor dominance case,
one can also envisage, in the three-flavor regime,
a two-flavor dominance case, where
two projectors are equally small and the third is necessarily close
to one, while all the three flavored \CP~asymmetries are
comparable.

In the next section we consider a specific example
that illustrates what we have discussed on general grounds.
At the same time, it will help to understand which are
realistic values for
the projectors and their differences, given a specific set
of see-saw parameters and using the information on the PMNS mixing
matrix we have from neutrino oscillation experiments.

\section{Study of a specific example}
\label{sec:specificexample}

The previous results have been obtained assuming no restrictions
on the projectors. Moreover, in the one-flavor dominance case,
where there can be a relevant relaxation of the
usual lower bounds derived following an unflavored treatment,
we have assumed that the upper bound on $\ve_{1\a}$, Eq.~(\ref{bound}),
is saturated independently of the value of the projector.

This assumption does not take into account that the values
of the projectors depend on the different see-saw parameters, in particular
on the neutrino mixing parameters, and that severe
restrictions could apply. Let us show a definite example
considering a particular form of the orthogonal matrix,
$\O=R_{13}$ [cf. Eq.~(\ref{R13})].
This case is particularly meaningful,
since it realizes one of the conditions ($\O^2_{21}=0$)
to saturate the bound~(\ref{CPbound}) for $\ve_1$.
Moreover, the decay parameter is given by
\be
K_1=K_{\rm min}\,|1-\o^2_{31}|+K_{\rm atm}\,|\o_{31}^2| \, ,
\ee
where $K_{\rm min}\equiv m_1/m_{\star}$, and we recall that 
$K_{\rm atm}\equiv m_{\rm atm}/m_{\star}$.
The expression~(\ref{proj}) for the projector
gets then specialized as
\be
\hspace{-0.5cm}
P^0_{1\alpha}= {m_1\,|U_{\a 1}|^2\,|1-\o_{31}^2|
+m_3\,|U_{\a 3}|^2\,|\o_{31}^2| +2\,\sqrt{m_1\,m_3}\,{\rm Re}
[U_{\a 1}\,U_{\a 3}^{\star}\sqrt{1-\o_{31}^2}\,\o_{31}^{\star}]
\over m_1\,|1-\o^2_{31}|+m_3\,|\o_{31}^2|} \, ,
\ee
while, specializing Eq.~(\ref{e1alOm}) for $\ve_{1\a}$,
one obtains
\begin{eqnarray}\label{full}
r_{1\a} & = &
{Y_3}\,{m_{\rm atm}\over K_1\,m_{\star}}\,\left[
|U_{\a 3}|^2+{m_1^2\over m_{\rm atm}^2}\,
(|U_{\a 3}|^2-|U_{\a 1}|^2)\right] \\ \nonumber
 &-& {m_{\rm atm}\over K_1\,m_{\star}}\,\sqrt{{m_1\over m_{\rm atm}}\,
 {m_3\over m_{\rm atm}}}
 \left[\left({m_1+m_3\over m_{\rm atm}}\right)\,
 {\rm Im}\left[\o_{31}\,\sqrt{1-\o^2_{31}}\right]
 \,{\rm Re}[U^{\star}_{\a1}U_{\a3}]\right. \\ \nonumber
 &  & \;\;\;\;\;\;\;\;\;\;\;\;\;\;\;\;\;\;\;\;\;\;\;\;
 + \left.\left({m_3-m_1\over m_{\rm atm}}\right)\,
 {\rm Re}\left[\o_{31}\,\sqrt{1-\o^2_{31}}\right]
 \,{\rm Im}[U^{\star}_{\a1}U_{\a3}]
\right] \, ,
\end{eqnarray}
where $m_3/m_{\rm atm}=\sqrt{1+m_1^2/m_{\rm atm}^2}$.

If we first consider the case
of fully hierarchical light neutrinos, $m_1=0$, then
\be
P^0_{1\a}={\ve_{1\a}\over \ve_1}=|U_{\a 3}|^2  \,
\hspace{20mm} \mbox{and} \hspace{20mm}
{\D P^0_{1\a}\over 2\,\ve_1}=0  \, .
\ee
For the PMNS matrix $U$ we adopt the parametrization 
Eq.~(\ref{PMNSmatrix}). One then finds
\be
P^0_{1e}\lesssim 0.03 \, ,\hspace{10mm}
P^0_{1\mu}\simeq P^0_{1\tau}\simeq 1/2 \, .
\ee
The fact that the projector on the electron flavor is very small 
and $\D P^0_{1\a}= 0$ for all flavors implies that the asymmetry generated 
in the electron flavor will be small, while the muon and tauon contributions
will be equal, since the projectors on these flavors are equal.
Summing Eq.~(\ref{flke2}) over $\a$, one then simply recovers the 
unflavored case with the washout reduced by a factor of~$2$.
This is a realization of the semi-democratic case that
 we were envisaging at the end of Section~\ref{sec:democratic} where
$K_{\star}\simeq 7$. In this situation flavor effects do not
produce large modifications to the usual results, essentially
a factor of 2 reduction of the washout in the strong washout regime
with a consequent equal relaxation of the lower bounds (see
dashed lines in Fig.~\ref{fig:EffandLowerBound}). Moreover, there is 
practically no difference
between a calculation in the two- or three-flavor regime.

Let us now consider the effect of a non-vanishing
but small lightest neutrino mass $m_1$,
for example $m_1=0.1\,m_{\rm atm}$. In this case
the results can also depend on the
Majorana and Dirac phases. We will show the results for
$\o^2_{31}$ purely imaginary, the second condition
that maximizes the total $C\!P$ asymmetry if $m_1\ll m_{\rm atm}$
\cite{DiBari:2005st}. Notice that this is not in general the condition
that maximizes $r_{1\a}$ for $m_1\gtrsim \matm$; however,
we will also use it in this case for simplicity.

\begin{figure}
\begin{center}
\includegraphics[width=0.32\textwidth]{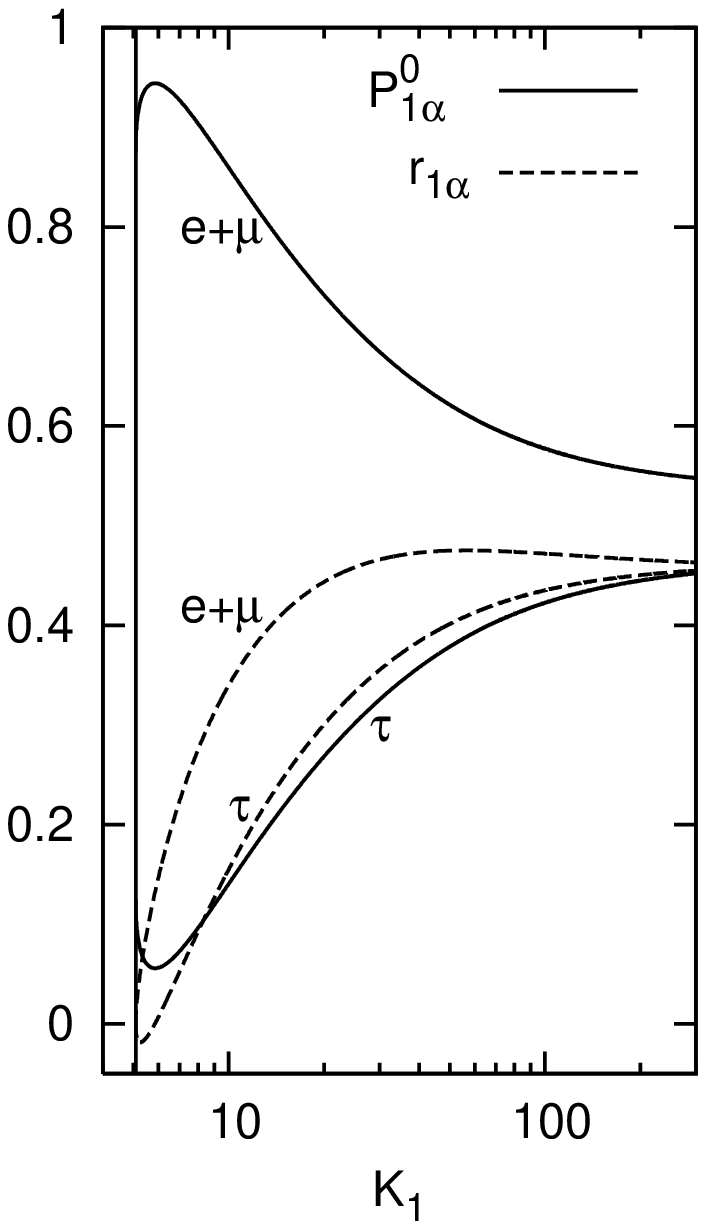}
\includegraphics[width=0.32\textwidth]{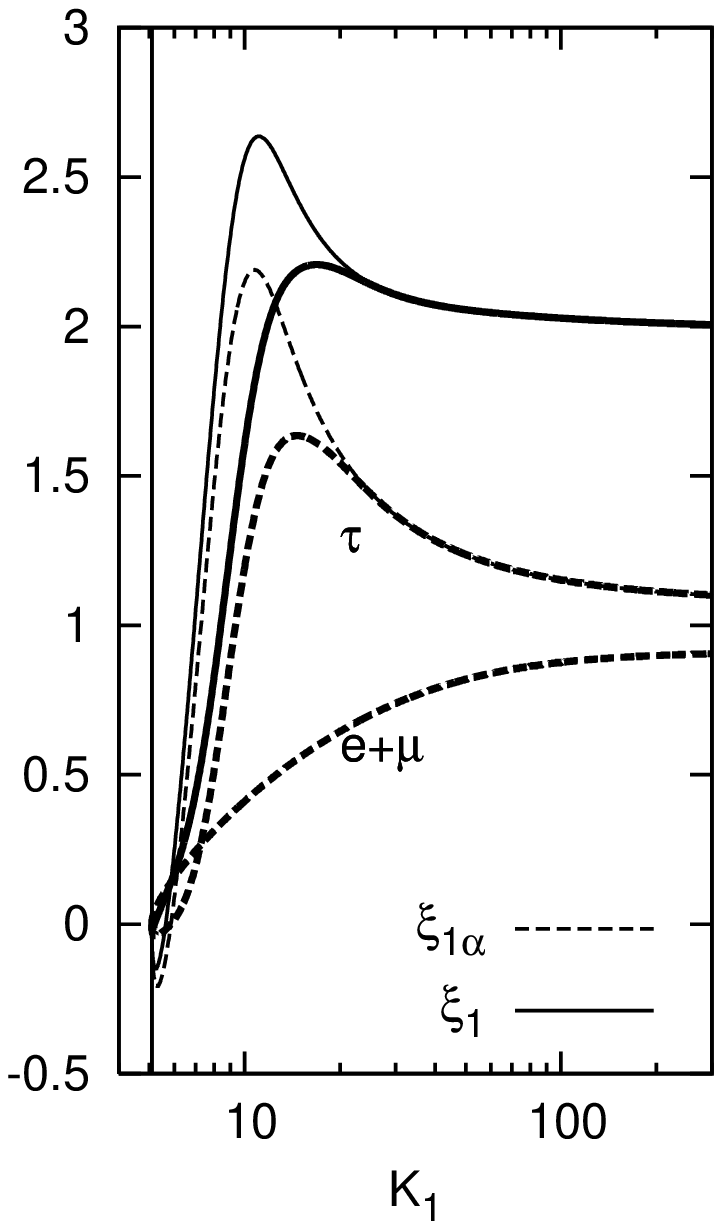}
\includegraphics[width=0.32\textwidth]{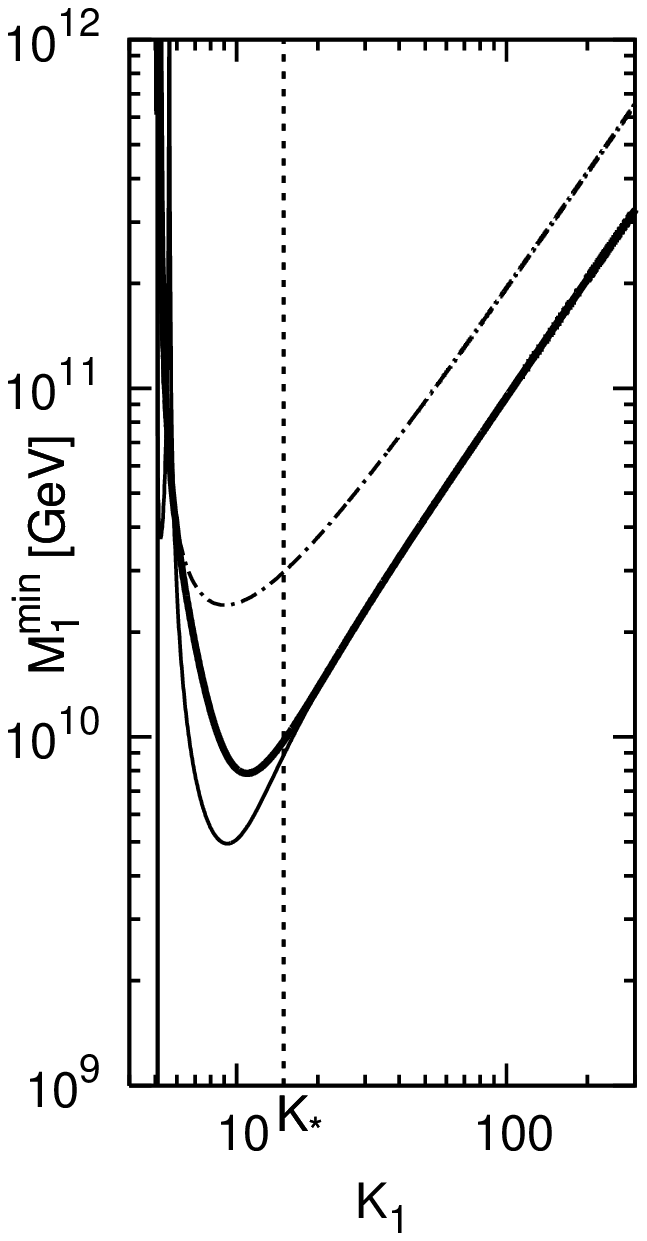}
\caption{Dependence of different quantities on $K_1$ for
$m_1/m_{\rm atm}=0.1$ and real $U$. Left panel: projectors $P^0_{1\a}$
and normalized \CP~asymmetries $r_{1\a}$; central panel: $\xi_{1\a}$ and 
$\xi_1$ as defined in Eq.~(\ref{xi1a}) for thermal (thin)
and vanishing (thick) initial $N_1$-abundances;
right panel: lower bound on $M_1$ for thermal (thin solid) and
vanishing (thick solid) abundances compared with the unflavored
result (dash-dotted line).}
\label{fig:Ex1}
\end{center}
\end{figure}

We first consider
the case of a real $U$. The results are only slightly sensitive
to a variation of $\theta_{13}$ within the experimentally
allowed $3\s$ range \mbox{0--0.2}. Therefore, we shall set $\theta_{13}=0$,
corresponding to $U_{e3}=0$, in all examples.
In the left panel of Fig.~\ref{fig:Ex1}
we show the values of the projectors $P^0_{1\a}$ and
of the normalized \CP~asymmetries $r_{1\a}$ as a function of $K_1$. 
The calculations are
performed in the two-flavor regime since we obtain that
successful leptogenesis is possible only for
$M_1 > 10^9\,{\rm GeV}$, where the two-flavor regime applies.
Now a difference between the tauon and the sum of the
muon and electron projectors and asymmetries arises. On the other hand,
for $K_1\gg 100$, this difference tends to vanish and
the semi-democratic case is again recovered.

In the central panel we show the
quantities $\xi_{1\a}$ [cf. Eq.~(\ref{xi1a})] and their sum
$\xi_1$. We recall that $\xi_1$ gives the deviation
of the total asymmetry from an unflavored calculation 
for hierarchical light neutrinos. One can see how
the contribution to the total asymmetry from the tauon flavor
is now twice as large as from the electron plus muon flavors.
For $K_1\gg 100$, the semi-democratic case is restored,
the two contributions tend to be equal to the unflavored case, 
and the total
final asymmetry is about twice larger. Finally,
in the right panel we show the lower bound on $M_1$, and
compare it with the results from the unflavored
analysis (dash-dotted line). At $K_{\star}\simeq 14$,
the relaxation in the strong washout is maximum, a factor $\sim 3$.
For $K_1\gg K_{\star}$, the relaxation
is reduced to a factor $2$, as in the semi-democratic case.

\begin{figure}
\begin{center}
\includegraphics[width=0.32\textwidth]{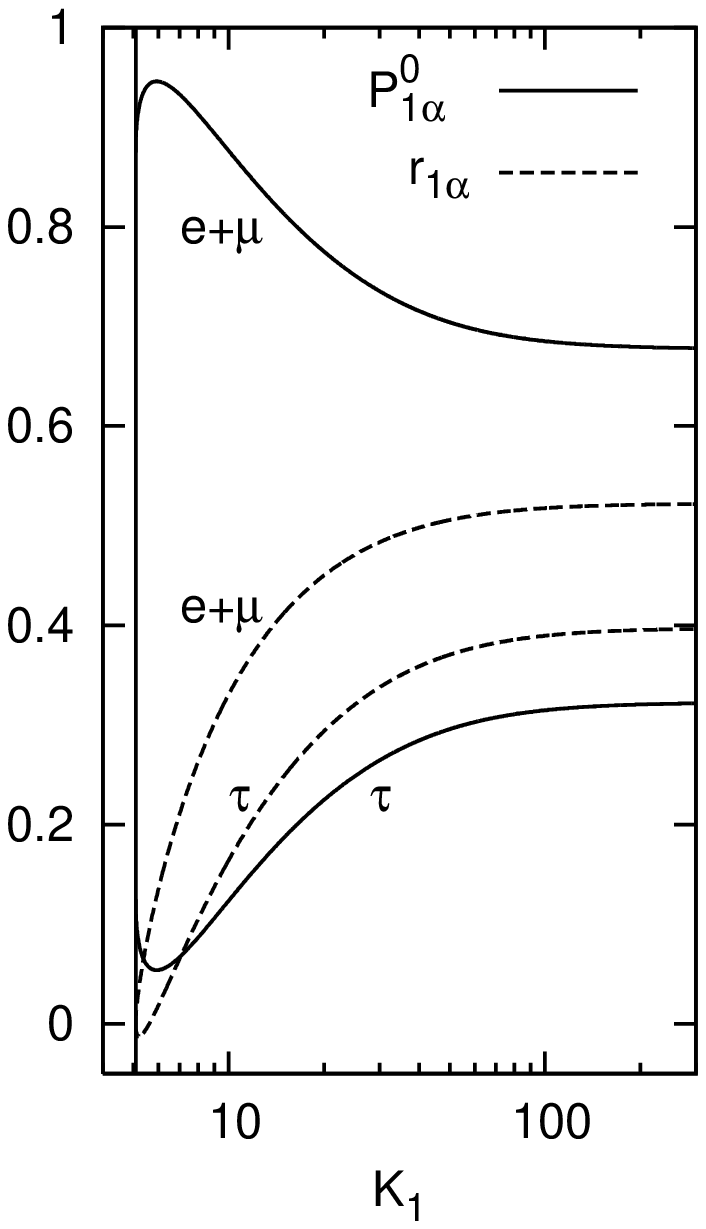}
\includegraphics[width=0.32\textwidth]{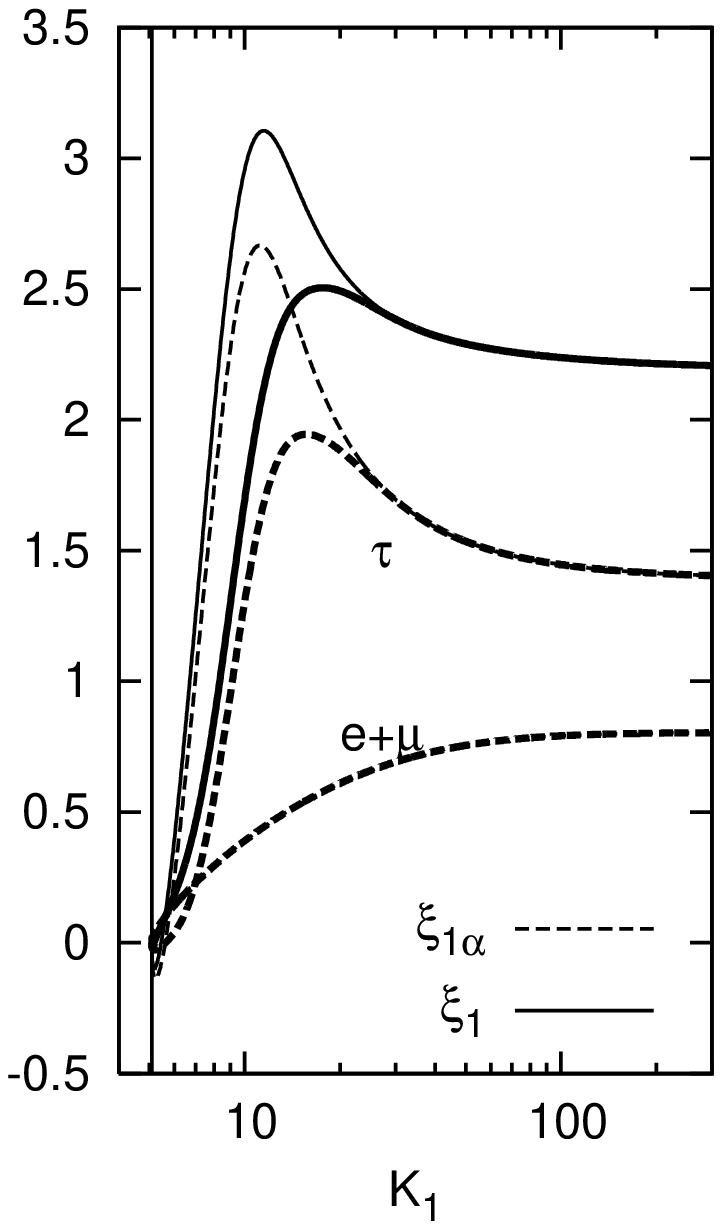}
\includegraphics[width=0.32\textwidth]{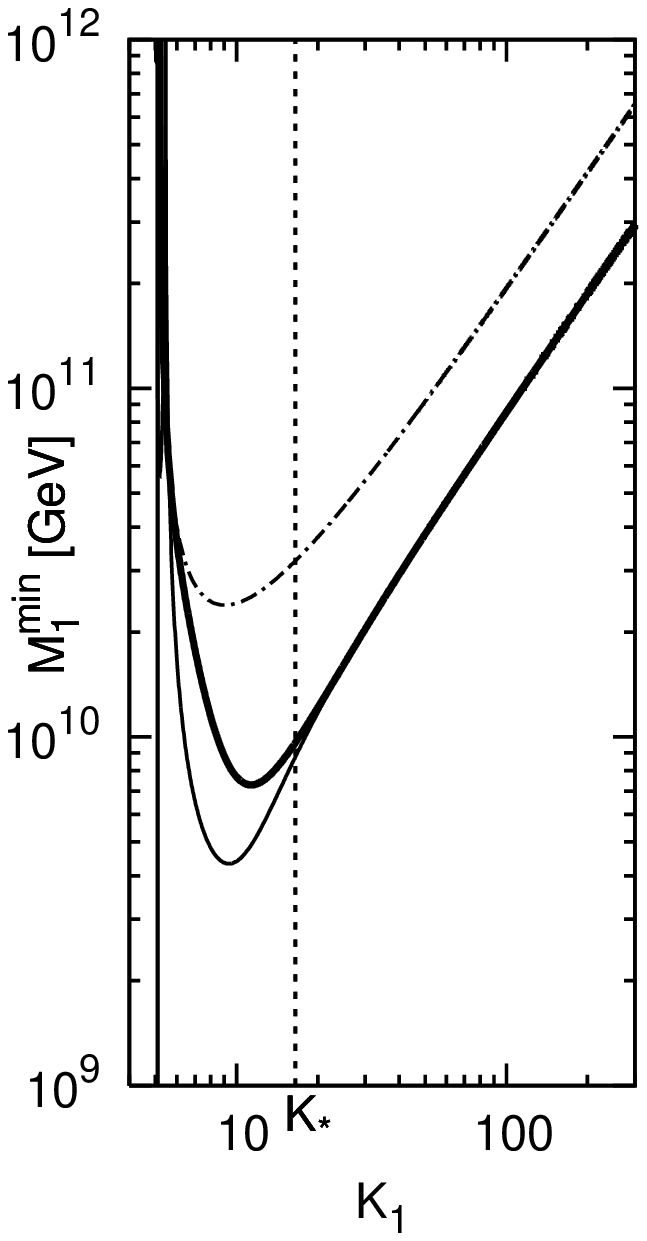}
\end{center}
\caption{Same quantities as in the previous figure but with one
non-vanishing Majorana  phase: $\Phi_1=-\pi$.}
\label{fig:Ex2}
\end{figure}

\begin{figure}
\begin{center}
\includegraphics[width=0.32\textwidth]{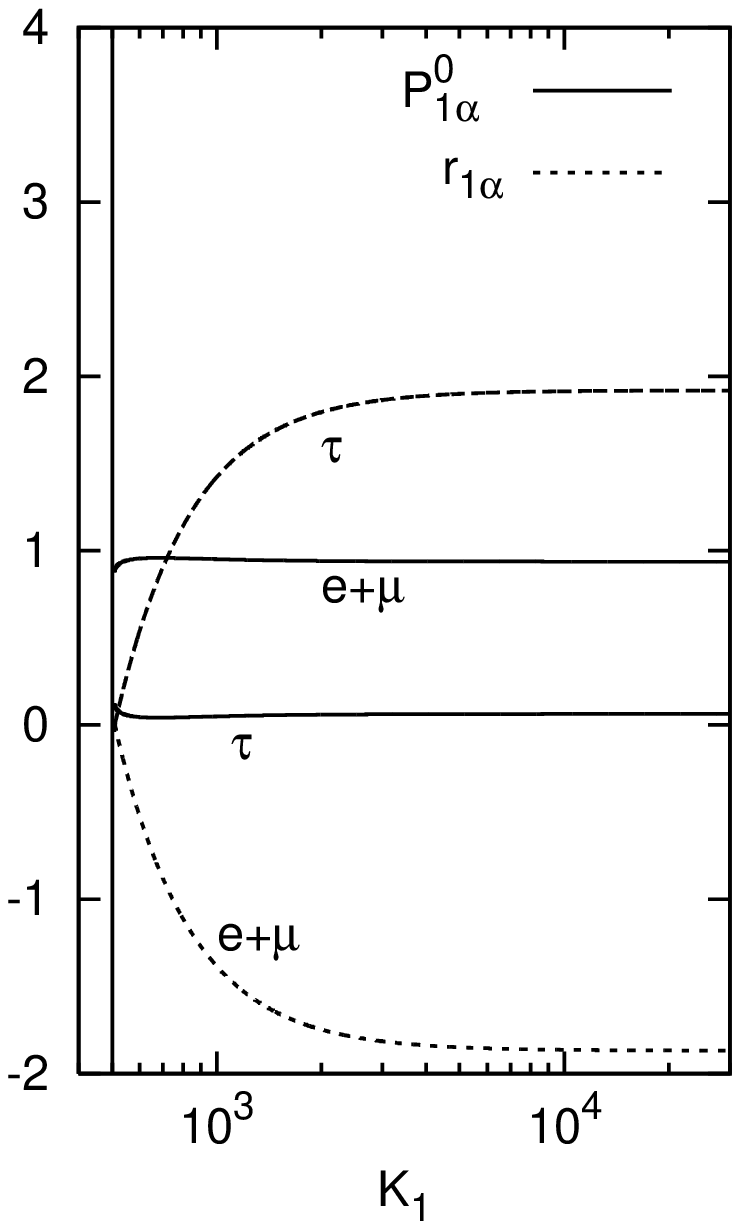}
\includegraphics[width=0.32\textwidth]{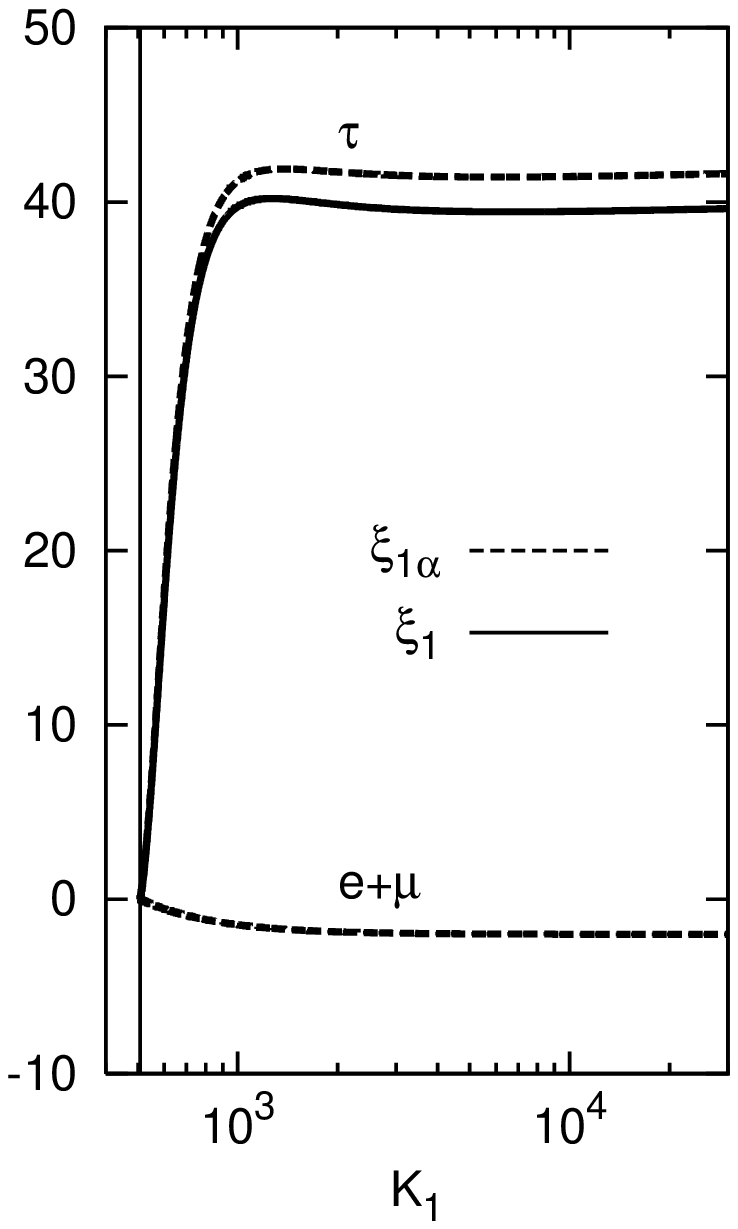}
\includegraphics[width=0.32\textwidth]{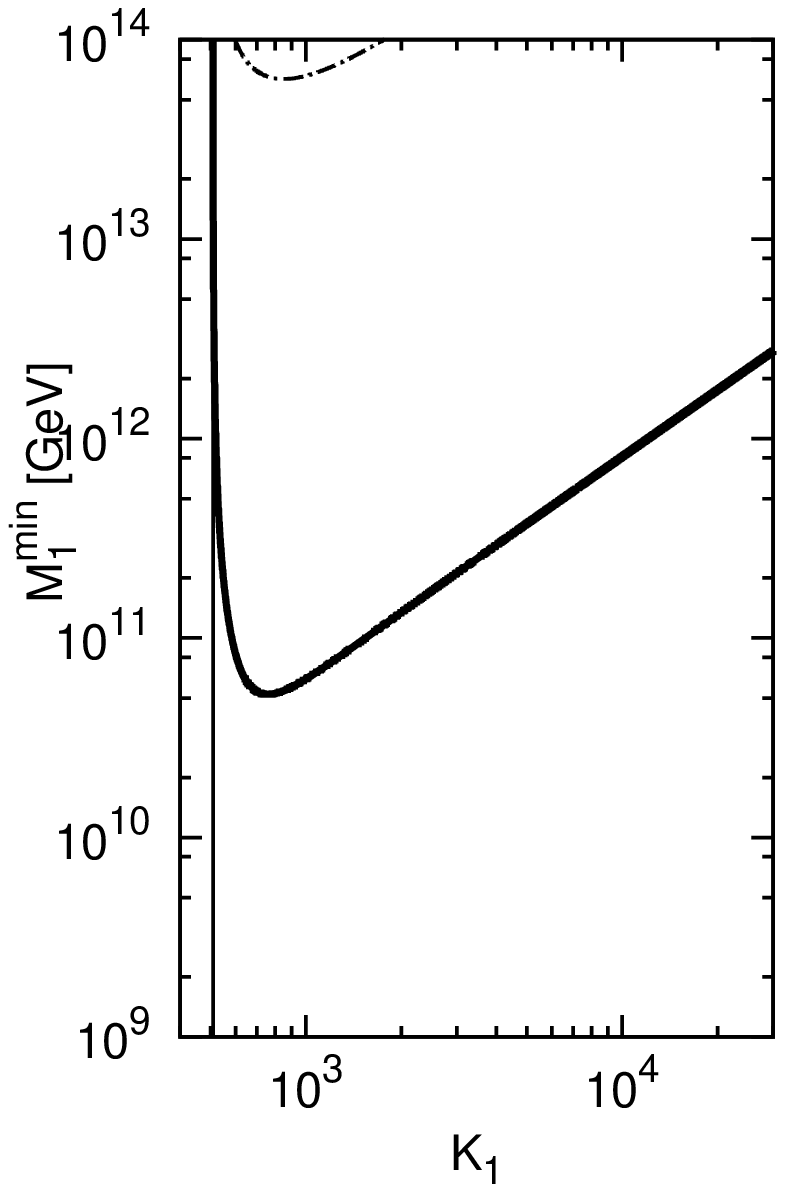}
\end{center}
\caption{Same quantities as in the previous two
figures but with $m_1/m_{\rm atm}=10$
and one non-vanishing Majorana  phase: $\Phi_1=-\pi$.}
\label{fig:Ex3}
\end{figure}

Let us now study the effect of switching on phases
in the $U$ matrix, again for $m_1=0.1\,m_{\rm atm}$.
 The most important effect arises from one of the two
 Majorana phases, namely $\Phi_1$. The other Majorana phase $\F_2$ 
is irrelevant in the particular model we consider, $\O=R_{13}$. As for
the Dirac phase $\d$, it plays a role similar to $\F_1$, but its
effect is suppressed by the small angle $\q_{13}$.
In Fig.~\ref{fig:Ex2} we show, again in three panels, the same
quantities as in Fig.~\ref{fig:Ex1} for $\Phi_1=-\pi$.
One can see how this further increases  the
difference between the $e+\m$ and the $\t$ contributions
and further relaxes the lower bound on $M_1$.
The effect is small for the considered value
$m_1/m_{\rm atm}=0.1$. However, considering
a much larger neutrino mass $m_1$ while keeping
$\Phi_1=-\pi$, the effect becomes dramatically bigger.
In Fig.~\ref{fig:Ex3} we show the same quantities
as in Fig.~\ref{fig:Ex1} and Fig.~\ref{fig:Ex2} for $m_1/m_{\rm atm}=10$.
In the left panel one can see that now, for $K_1\gg K_{\rm min}$,
$|r_{1e+\mu}| \simeq |r_{1\t}| \simeq 2$, much larger than
in the previous case. This means that the dominant
contribution to the flavored $C\!P$ asymmetries comes now
from the $\D P_{1\a}$ term [cf.~Eq.~(\ref{veiaDP})]. At the same time, very importantly,
$P^0_{1\t}\ll P^0_{1e+\m}$ and in this way,
as one can see in the central panel, the dominant contribution to $\xi_1$
is given by $\xi_{1\t}$. This case thus finally realizes a one-flavor
dominance. The final effect is that the lower bound on $M_1$ is about
three orders of magnitude relaxed compared to the unflavored case.

\begin{figure}
\begin{center}
\includegraphics[angle=-90,width=0.7\textwidth]{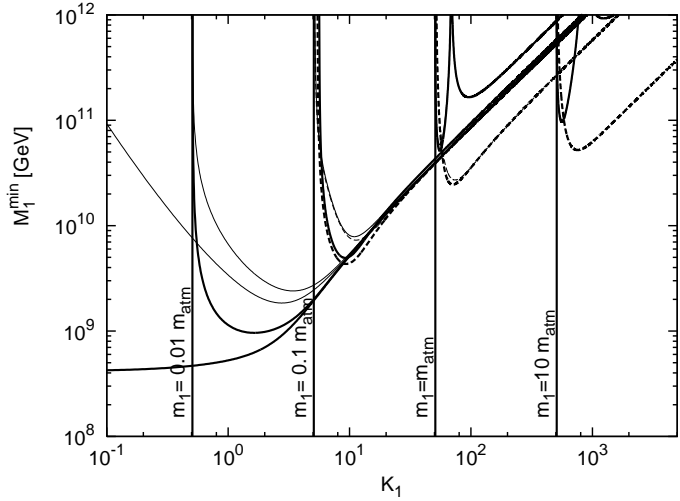}
\end{center}
\caption{Lower bound on $M_1$. The solid lines are for vanishing phase,
while the dashed lines are for $\Phi_1=-\pi$. The results for both vanishing 
(thick lines) and thermal (thin lines) initial abundances are presented.}
\label{fig:Comp}
\end{figure}

In Fig.~\ref{fig:Comp} the dependence on $m_1/m_{\rm atm}$
of the lower bound on $M_1$ is summarized, showing both the case with zero Majorana
phase and the case with $\F_1=-\p$. One can notice how the effect of 
the phase in relaxing the lower bound increases with $m_1/m_{\rm atm}$.

\begin{figure}
\begin{center}
\includegraphics[width=0.32\textwidth]{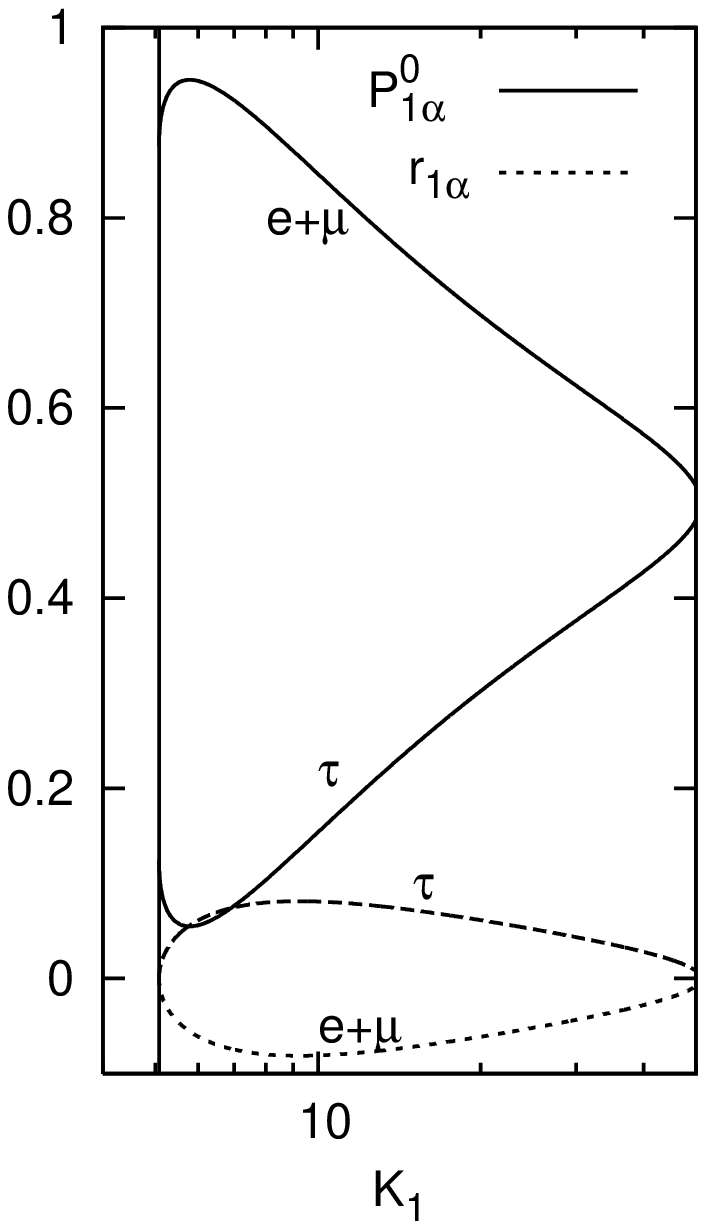}
\includegraphics[width=0.32\textwidth]{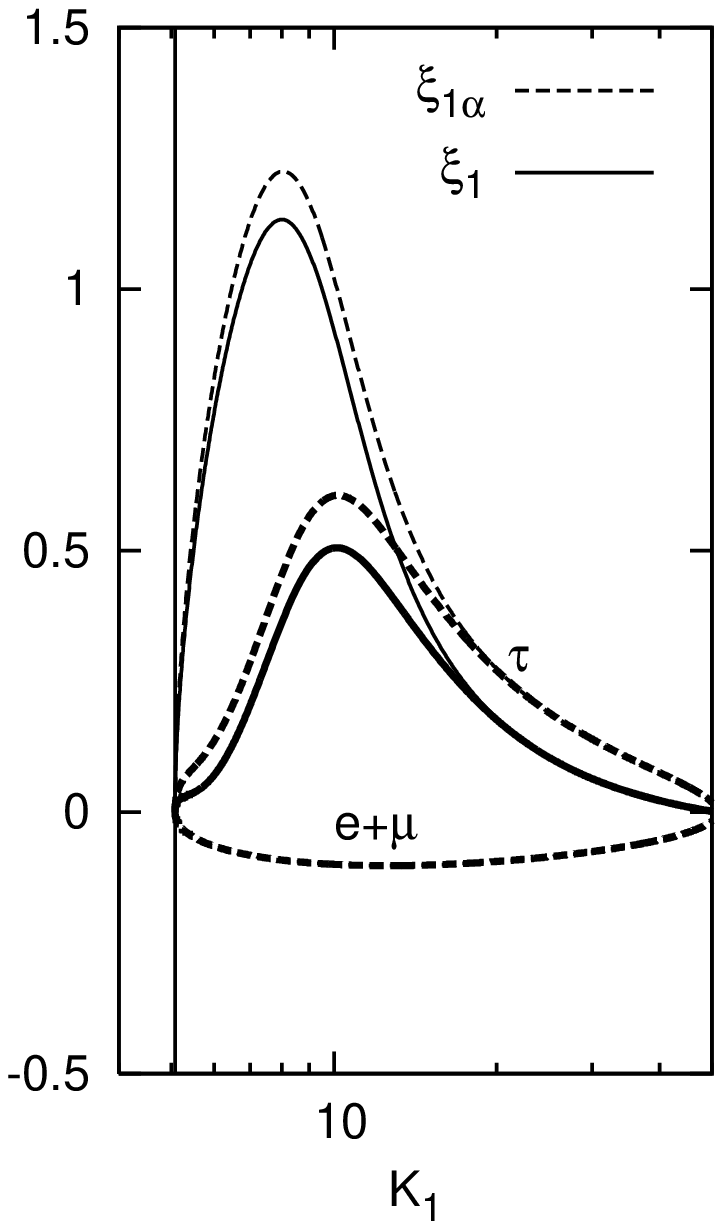}
\includegraphics[width=0.32\textwidth]{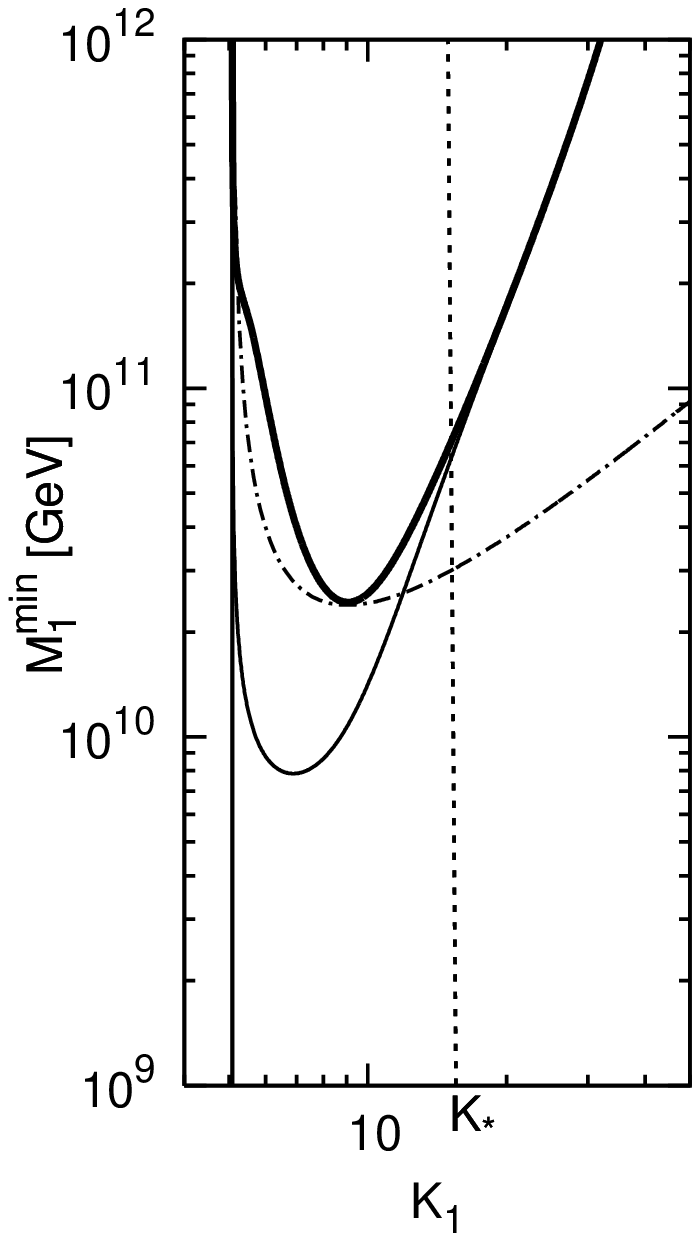}
\end{center}
\caption{Same quantities as in Figs.~\ref{fig:Ex1}, \ref{fig:Ex2} and
\ref{fig:Ex3} in the case of real $\O$ for $m_1/m_{\rm atm}=0.1$
and $\Phi_1=\pi/2$. The dot-dashed line still refers to the unflavored
case for $\O=R_{13}$ and purely imaginary $\o_{31}^2$.}
\label{fig:OmReal}
\end{figure}

Finally, we want to study the interesting case of a real orthogonal 
matrix $\O$, implying $\ve_1=0$. In the particular model we are 
considering, namely $\O=R_{13}$, there is no asymmetry produced
if $m_1=0$, since $\ve_{1\a}\propto \ve_1 =0$ [cf. Eq.~(\ref{full})]. For
a non-vanishing $m_1$ and a non-real $U$, we have that $\D P_{1\a}\neq 0$ 
and consequently $\ve_{1\a}\neq 0$. In Fig.~\ref{fig:OmReal} we show
the results for $m_1/m_{\rm atm}=0.1$ and $\Phi_1=\pi/2$. We present again
the same quantities as in Figs.~\ref{fig:Ex1}, \ref{fig:Ex2} and
\ref{fig:Ex3}. One can see that an asymmetry can still be produced,
as envisaged in~\cite{Nardi:2006fx}. However, successful leptogenesis is 
possible almost only in the weak washout regime. In the strong washout
regime, for values $M_1\lesssim 10^{12}\gev$, there is a small allowed
region only for $K_{\star}\simeq 14 \leq K_1\lesssim 30$.

Let us summarize the main results of this section, where we assumed 
the $\O$ matrix to have the simple form $\O=R_{13}$.

We have seen that the relaxation of the lower bounds compared to
the unflavored case is only of order one for $m_1\ll m_{\rm atm}$, but
grows to several orders of magnitude for $m_1\gtrsim 10\,m_{\rm atm}\simeq
0.5\ev$, i.e. in the quasi-degenerate limit. The reason is that for quasi-degenerate
neutrinos ($m_1\simeq m_2\simeq m_3$) a partial cancellation in one of the projectors, 
Eq.~(\ref{proj}), can be more easily obtained, simply because all terms in the numerator 
are of the same order. When a partial cancellation occurs, one projector will
be suppressed, leading to a one-flavor dominance and thus to a large relaxation 
of the lower bound at large values of $K_1$. This is exactly what happened in
the situation depicted in Fig.~\ref{fig:Ex3}.

The role of the phases in the $U$ matrix to achieve a cancellation
in one of the projectors is crucial. For $\O=R_{13}$, only one Majorana 
phase, $\F_1$, is relevant, and the specific value it takes leads to 
very different results. In Fig.~\ref{fig:Ex3} we used the value $\F_1=-\p$
which approximately maximizes the asymmetry. The Dirac phase $\d$ can also 
play a similar role
to $\F_1$ for $\O=R_{13}$, but its effect is always suppressed by the small
angle $\q_{13}$. It should be noted that for more general cases
like $\O=R_{12}R_{13}$, the cancellation in the projector leading to a one-flavor
dominance is also possible when $m_1=0$, because there is more freedom 
in the $\O$ matrix to achieve a cancellation. Moreover, both Majorana 
phases in $U$ can play a role in principle, not only $\F_1$.

It seems therefore that when flavor effects are taken into account, one 
obtains an opposite result
compared to the unflavored analysis we performed in the previous chapter,
where one had a suppression of the
final asymmetry for growing absolute neutrino mass scale, leading to 
the stringent upper bound Eq.~(\ref{m1max}). This upper bound seems now
to hold only for $M_1\gtrsim 10^{12}\gev$~\cite{Abada:2006fw}. However,
as we shall argue in the next chapter, this issue requires a full quantum
kinetic calculation, in order to keep track of the correlations in flavor
space and of partial losses of coherence~\cite{Blanchet:2006ch}.

We wish also to emphasize that, when flavor effects
are included, leptogenesis is an interesting example of phenomenology,
beyond neutrinoless double beta decay, where Majorana phases
play an important role. The Dirac phase is also relevant, although it appears 
in combination with the small angle $\q_{13}$.
The importance of the \CP-violating phases in $U$ shows up at two levels. First, as 
we have discussed above, they are crucial to achieve a cancellation in one of the
projectors, leading to a one-flavor dominance. Second, they can provide
the unique source of \CP~violation required for successful leptogenesis,
as illustrated in Fig.~\ref{fig:OmReal}. We shall come back to this interesting
possibility in Chapter~\ref{chap:OmReal}.



\chapter{From classical to quantum kinetic equations}
\label{chap:densmat}

We saw in the last two chapters that two quite different pictures 
of leptogenesis are valid in two different temperature regimes. Roughly
speaking, when the temperature relevant for leptogenesis is above 
$10^{12}$~GeV, the unflavored picture should hold, whereas at lower
temperatures, $T\lesssim 10^{12}$~GeV, the fully flavored regime
should apply, either with two independent flavors 
($10^{9}~{\rm GeV}\lesssim T\lesssim 10^{12}$~GeV) or with three 
($T\lesssim 10^{9}$~GeV). 

The final baryon asymmetry was estimated
in both regimes, unflavored and fully flavored, by means of classical
Boltzmann equations. However, one could expect that quantum effects, such as
correlations in flavor space and partial losses of coherence, might play 
some role in the intermediate
regime around $10^{12}$~GeV. In such a case, the classical Boltzmann equation
for the $\D_{\a}$ asymmetry given in Eq.~(\ref{flke2})
is not expected to describe correctly the generation of asymmetry.
One should indeed turn to a density matrix equation, which provides
the right formalism to keep track of correlations in flavor space and
partial losses of coherence. 

In this chapter we re-visit the conditions of validity of 
both the unflavored and the fully flavored pictures. We shall give
physical arguments why one actually expects to have a substantial 
part of the parameter space where quantum effects are important. 
Interestingly, this part of the parameter space is exactly the one relevant
for a discussion about the upper bound on the absolute neutrino mass scale,
which is evaded if one believes a fully flavored analysis. We shall
argue that only solving the relevant density matrix equation will 
give a definite answer to that question. In the final section we
discuss the density matrix formalism and give tentative equations for 
leptogenesis in the transition region.

\section{Validity of the different pictures}

Let us first discuss in some detail the quantity that will be the 
angular stone of the subsequent analysis, namely the washout
rate by inverse decays [cf.~Eq.~(\ref{WID})]. As usual, we
restrict our analysis to the lightest RH neutrino $N_1$.

It was shown in~\cite{Buchmuller:2004nz}
that the inverse-decay rate reaches a maximum $W_1^{\rm
ID}(z_{\rm max})\simeq 0.3\,K_1$ at $z_{\rm max}\simeq 2.4$. In
the weak washout regime, when $K_1\lesssim 3.3$, one has
$W_1^{\rm ID}(z)< 1$ for any value of $z$. In this case the washout 
is negligible, and the final asymmetry depends on the initial
conditions. In the strong washout regime, when $K_1\gtrsim 3.3$,
there is an interval $[z_{\rm on},z_{\rm off}]$ where $W_{1}^{\rm
ID}\geq 1$. The asymmetry produced at $z\lesssim z_{\rm off}$ is
very efficiently washed out, and thus the final asymmetry is
essentially what is produced around $z_{\rm B}\simeq z_{\rm off}$ by the
out-of-equilibrium decays of the residual RH neutrinos, whose
number corresponds approximately to the final value of the
efficiency factor.

\begin{figure}[ht]
\begin{center}
\includegraphics[width=0.6\textwidth,angle=-90]{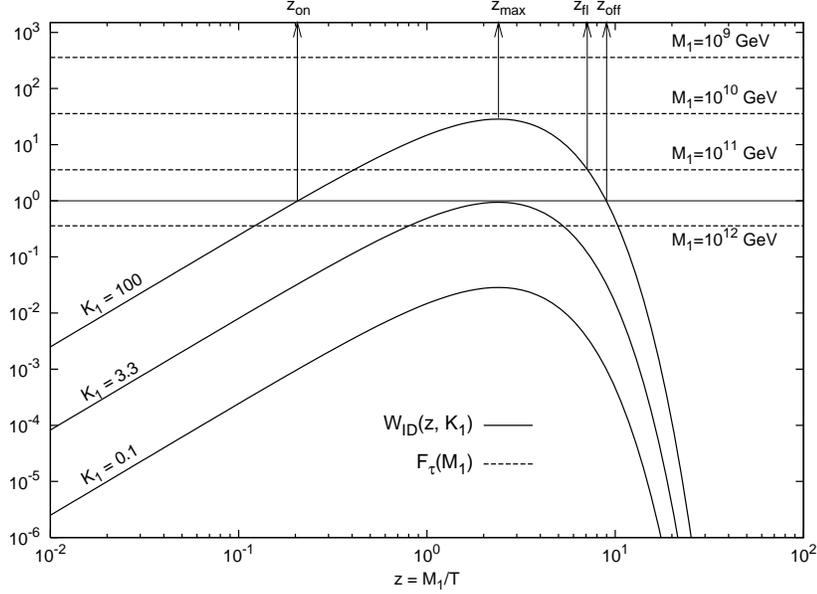}
\caption{Comparison between the washout term $W_1^{\rm ID}(z)$
(thick solid lines), defined in Eq.~(\ref{WID}) and plotted for the
three indicated values of $K_1$, and the charged-lepton Yukawa
interaction term $F_{\tau}$, defined in Eq.~(\ref{Ftau}) and plotted
for the indicated values of $M_1$.}\label{fig:washoutterm}
\end{center}
\end{figure}

In Fig.~\ref{fig:washoutterm} we show the washout term from inverse
decays for three different values of $K_1$. For $K_1=100$, we show
the interval $[z_{\rm on},z_{\rm off}]$. The  maximum value,
$W_1^{\rm ID}(z_{\rm max})\simeq 33$, is reached at $z_{\rm
max}\simeq 2.4$. For $K_1\simeq 3.3$, one has $z_{\rm on}\simeq
z_{\rm max}\simeq z_{\rm off}$. This can be taken as the threshold
value distinguishing between the strong and the weak washout
regimes. For $K_1=10^{-1}$, one has $W_1^{\rm ID}\ll 1$ for any value
of $z$. Notice however that even in this case the weak washout can
be important for successful leptogenesis if the initial $N_1$-abundance 
is zero, since it prevents a full cancellation between two
different sign contributions to the final
asymmetry~\cite{Buchmuller:2004nz}.

\subsection{When are flavor effects important?}

As discussed in Section~\ref{sec:flavor}, flavor 
effects are caused by charged-lepton Yukawa
interactions~\cite{Barbieri:1999ma} that occur with the
rate given in Eq.~(\ref{rateflav}). The largest rate is for $\a=\t$ where
\begin{equation}
 {\G_{\t}\over H}\simeq \left(10^{12}~{\rm GeV}\over T\right)\,.
\end{equation}
Therefore, if $T\gtrsim 10^{12}\,{\rm GeV}$, charged-lepton Yukawa
interactions  are not effective, and all processes in the early
Universe are flavor blind, justifying the unflavored treatment. For
$T \lesssim 10^{12}\,{\rm GeV}$, the $\tau$-Yukawa coupling is
strong enough that the scatterings $\ell_{\t}\,\bar{e}_{\t}\leftrightarrow
\Phi^{\dagger}$ are in equilibrium. However, this condition is not 
necessarily sufficient for
important flavor effects to occur, because we need to compare the
speed of the Yukawa interactions with that of the RH neutrino decays
and inverse decays. To this end, we study the weak and strong
washout regimes separately and consider only a two-flavor case,
because the $\tau$-lepton Yukawa coupling causes the main
modification.

In the weak washout regime, assuming a vanishing initial $N_1$-abundance,
the production of RH neutrinos by inverse decays occurs around
$T\sim M_1$. At this epoch, inverse decays are by definition slower
than the expansion rate. Therefore, the condition $T\lesssim
10^{12}\,{\rm GeV}$ is sufficient to conclude that the
charged-lepton Yukawa interactions are faster than the inverse decay
rate. This translates into the condition $M_1\lesssim 10^{12}\,{\rm
GeV}$ because the RH neutrino production occurs at $T\sim M_1$, in
agreement with the previous
literature~\cite{Nardi:2006fx,Abada:2006fw}.

However, this condition does not guarantee that flavor effects
indeed have an impact on the final asymmetry, because this impact
depends on washout playing some role. For a vanishing initial
$N_1$-abundance, this is the case in that washout effects prevent a full
sign cancellation between the asymmetry produced when
$N_{N_1}<N_{N_1}^{\rm eq}$ and the asymmetry produced later on. On
the other hand, for a thermal initial $N_1$-abundance, no such effect
arises from the weak washout and flavor effects do not modify the
final asymmetry. We will come back to this point later on.

In the strong washout regime, the situation is very different. The
rate of RH neutrino inverse decays at $T\sim  M_1$ is larger than
the expansion rate. Therefore, we need to compare the charged-lepton
Yukawa rate $\Gamma_{\t}$ with the RH neutrino inverse-decay rate
$\Gamma^{\rm ID}_1$. For the unflavored treatment to be valid for
$z\lesssim z_{\rm fl}\leq z_{\rm B}$ then requires
\begin{equation}\label{M1}
 M_1\gtrsim {10^{12}\,{\rm GeV}\over 2\,W_1^{\rm ID}(z_{\rm fl})}\,,
\end{equation}
where $z_{\rm fl}$ is that value of $z$ where the two rates are
equal, i.e. $\Gamma^{\rm ID}_1(z_{\rm fl})=\Gamma_{\t}(z_{\rm fl})$. 
This condition guarantees that at temperatures $T>T_{\rm
fl}=M_1/z_{\rm fl}$ flavor effects will not be able to break the
coherent propagation of the lepton states. The final asymmetry is
dominantly produced around $z\sim z_{\rm B}$. Therefore, the condition for
flavor effects to be negligible is
\begin{equation}\label{unflavored}
M_1\gtrsim 5\times 10^{11}\,{\rm GeV} \, ,
\end{equation}
similar to the weak washout regime. However, the
corresponding condition on the temperature
\begin{equation}
 T \gtrsim {10^{12}\,{\rm GeV}\over 2\,z_{\rm B}(K_1)}
\end{equation}
is now less restrictive.

If one starts with a non-vanishing initial $N_1$-abundance, then the final
asymmetry is also determined by how efficiently the initial value is
washed out; this is described by the integral in Eq.~(\ref{NBmLf}).
In this case even a value of $z_{\rm fl}< z_{\rm B}$ would be important
to determine the final asymmetry since washout in the unflavored
regime would be effective for $z\lesssim z_{\rm fl}< z_{\rm B}$, while a
reduced washout would apply in the flavored regime at lower
temperatures such that $z_{\rm fl}\lesssim z\lesssim z_{\rm B}$.

We conclude that the condition~(\ref{M1}) obeys the intuitive
expectation that there is always a threshold value for $K_1$ above
which the unflavored case is recovered. In this case the temperature
below which flavor effects play a role indeed becomes smaller and
smaller. The situation is illustrated in Fig.~\ref{fig:washoutterm} 
where we compare $W_{\rm ID}$ with
\begin{equation}\label{Ftau}
F_{\t}\equiv {1\over 2}\,{\G_{\t}\over H\,z}
\simeq {5\times 10^{11}\,{\rm GeV}\over M_1} \, ,
\end{equation}
the analogous quantity for the charged-lepton Yukawa interactions.
For any value of $M_1$ and $K_1$, there is a value $z_{\rm fl}$ such
that $F_{\t}\gtrsim W_{\rm ID}$ for $z>z_{\rm fl}$. If $M_1\lesssim
2\times 10^{12}\,{\rm GeV}/K_1$ and $K_1\gtrsim 3.3$, corresponding
to $F_{\tau}\gtrsim W_{\rm ID}(z_{\rm max})$ in the strong washout
regime, then $z_{\rm fl}=0$, meaning that flavor effects are
important during the entire thermal history. On the other hand, for
a fixed value of $M_1$, one has $z_{\rm fl}\rightarrow \infty$ for
$K_1\rightarrow \infty$, implying that flavor effects tend to
disappear for sufficiently large values of $K_1$. Notice however
that if $M_1 \gtrsim 5\times 10^{11}\,{\rm GeV}$, then $z_{\rm
fl}\gtrsim z_{\rm off}\simeq z_{\rm B}$ for any value of $K_1$. This
confirms that only for $M_1\gtrsim 5\times 10^{11}\,{\rm GeV} $
flavor effects can be neglected and the unflavored regime is
recovered.

\subsection{Maximum flavor effects}               \label{sec:maxflavored}

We now turn to the opposite extreme case when flavor effects are
maximal, the ``fully flavored regime.'' In other words, the
charged-lepton Yukawa interactions are now taken to be so fast that
the lepton flavor content produced in $N\to\ell+\Phi$ on average
fully collapses before the inverse reaction can take place, i.e.,
the $\ell$ density matrix in flavor space is to be taken diagonal in
the charged-lepton Yukawa basis. In this case each single-flavor
asymmetry has to be calculated separately because generally the 
washout by inverse decays is different for each flavor. Moreover, the
single-flavor \CP~asymmetries now have an additional
contribution compared to the total, as shown 
in~Eq.~(\ref{veiaDP})~\cite{Nardi:2006fx,Abada:2006fw}.
Finally, the inverse decay involving a lepton in the flavor $\alpha$
does not wash out as much asymmetry as the one produced by one RH
neutrino decay. The reduction is quantified by the probability
$P^0_{i\a}$, averaged over leptons and antileptons, that the lepton
$\ell_i$ produced in the decay of $N_i$ collapses into the flavor
eigenstate $\ell_{\alpha}$. Focusing on the lightest RH neutrino
$N_1$, the fully flavored Boltzmann equations are given by 
Eqs.~(\ref{flke}) and (\ref{flke2}), with $i=1$. Since we are dealing 
with the two-flavor case, here $\a=\t$ or $e+\m$
where the latter stands for a suitable superposition of the $e$ and
$\m$ flavors, as explained after Eq.~(\ref{flke2}). 


As in the unflavored case, we next identify the condition for the
fully flavored approximation to hold. The final asymmetry in the
flavor $\a$ is dominantly produced at $z\simeq z_{{\rm B}\a}\equiv
z_{\rm B}(K_{1\a})$, where $K_{1\a}\equiv P^0_{1\a}\,K_1$. Therefore, one
must require that $\G_{\a}\gtrsim \G_1^{\rm ID}$ holds already at
$z\sim z_{{\rm B}\a}$, or else the washout reduction takes place too late. We
stress that flavor effects modify the final asymmetry only if the
flavor projection takes place before the washout by inverse decays
freezes out. Otherwise the washout epoch is over and the unflavored
behavior is recovered. It is easy to verify that if the projectors
are set to unity and the equations are summed over flavors, the
kinetic equation~(\ref{unflke2}) for $N_{B-L}$ holding in the unflavored 
regime is recovered. Therefore, we require
\begin{equation}\label{M1fl}
 M_1\lesssim {10^{12}\,{\rm GeV}\over 2\,W_1^{\rm ID}(z_{{\rm B}\a})}
\end{equation}
as an approximate condition for the fully flavored behavior.

\begin{figure}[t]
\begin{center}
\includegraphics[width=0.6\textwidth,angle=-90]{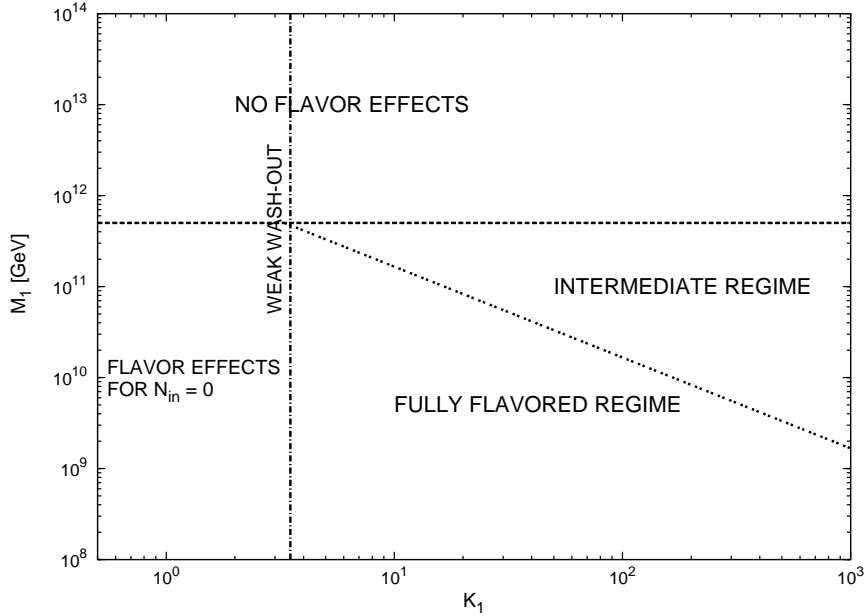}
\caption{Relevance of flavor effects in schematic regions of
parameters $K_1$ and $M_1$. The region above the
horizontal dashed line corresponds to the condition
(\ref{unflavored}).
The vertical dot-dashed line is the border
between the weak and the strong washout regime. The region below
the inclined dotted line corresponds to the condition
(\ref{M1fl}) for $z_{{\rm B}\a}=z_{\rm max}$.} \label{fig:flavorornot}
\end{center}
\end{figure}

In Fig.~\ref{fig:flavorornot} we summarize the different possible
cases in the plane of parameters $K_1$ and $M_1$. For $M_1\gtrsim
5\times 10^{11}\,{\rm GeV}$, above the dashed line, flavor
effects are not important independently of~$K_1$. The condition~(\ref{M1fl}), 
in the most restrictive case when $z_{{\rm B}\a}=z_{\rm
max}$ and $W_1^{\rm ID}\simeq 0.3\,K_1$, is satisfied below the
inclined dotted line. This case typically occurs in a one-flavor
dominated scenario, as we explain below. The vertical dot-dashed line is the
border that separates the weak from the strong washout regime in the
unflavored case. In the flavored case the condition $K_1\lesssim 3.3$
still implies a weak washout regime because flavor
effects can only reduce the washout. However, the
condition for the strong washout regime can be more
restrictive than $K_1\gtrsim 3.3$, as discussed in
\cite{Blanchet:2006be}.
For $K_1\lesssim 3.3$, flavor effects
modify the final asymmetry only marginally and more
specifically only if the initial $N_1$-abundance vanishes, as indicated in
Fig.~\ref{fig:flavorornot}. On the other hand, for $K_1\gtrsim
3.3$ and below the diagonal line, flavor modifications of the final
asymmetry can be large, especially in the one-flavor dominated
scenario.

There is a region in parameter space where neither condition~(\ref{M1}) 
nor~(\ref{M1fl}) holds. This intermediate regime can
become very large in the case of a one-flavor dominated scenario,
where several orders of magnitude enhancement of the final asymmetry 
compared to the unflavored calculation are possible (see Fig.~\ref{fig:Ex3}). 
In this case one of the two projectors is very small
compared to the other, and so the washout is very asymmetric in the
two flavors. On the other hand, if the two flavored \CP~asymmetries 
are comparable, then the final asymmetry is dominantly
produced into one flavor and deviations from the unflavored regime
can become very large. This scenario is realized, in particular,
when the absolute neutrino mass scale increases, relaxing the
traditional neutrino mass bound.

However, the condition~(\ref{M1fl}) strongly restricts the
applicability of the one-flavor dominated scenario. Even though the
reduction of the washout is driven by $K_{1\a}\ll K_1$, implying
$z_{{\rm B}\a}\ll z_{\rm B}$, the possibility for flavor effects to be relevant
relies on the dominance of the charged-lepton Yukawa interaction
rate compared to the RH neutrino inverse-decay rate, which however is
still driven by $K_1$. Therefore, increasing $K_1$, one can enhance
the asymmetry in the one-flavor dominated scenario compared to the
unflavored case, if smaller and smaller values of the projector
$P^0_{1\a}$ are possible. On the other hand, the inverse-decay rate
increases so that the fully flavored behavior may no longer apply.
In particular notice that the maximum enhancement of the asymmetry is
obtained when $K_1\gg K_{1\a}\simeq 1$, when
$z_{{\rm B}\a}\simeq z_{\rm max}$ and the condition (\ref{M1fl}) is
maximally restrictive.

\section{Neutrino mass bound}
\label{sec:transition}

\begin{figure}[p!]
\begin{center}
\includegraphics[width=0.6\textwidth,angle=-90]{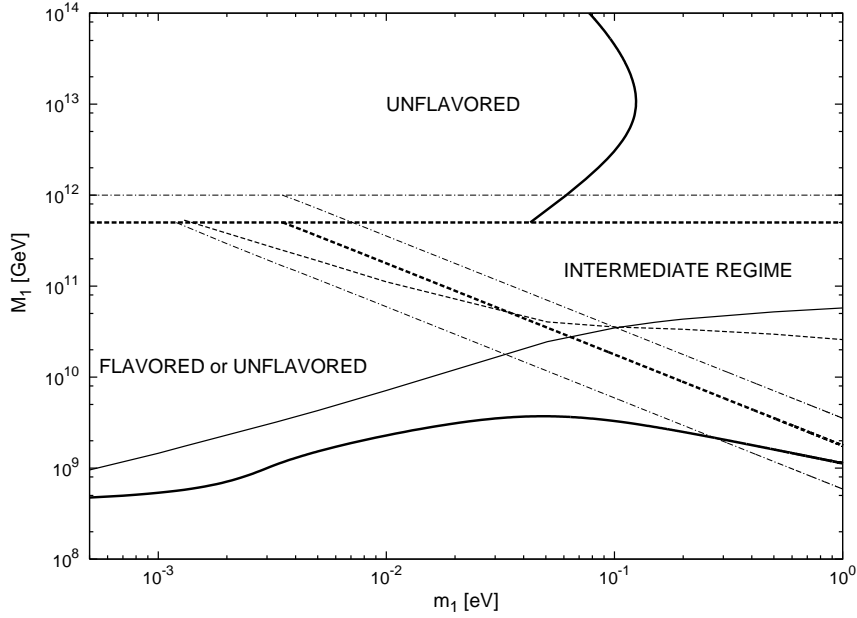}
\caption{Relevance of flavor effects similar to
Fig.~\ref{fig:flavorornot}, now mapped to schematic regions of
parameters $m_1$ and $M_1$. The region above the horizontal dashed
line corresponds to the condition~(\ref{unflavored}) for the
applicability of the unflavored regime. The region below the
inclined thick dashed line corresponds to the condition~(\ref{M1fl})
calculated for that value of $z_{{\rm B}\a}$ that maximizes the final
asymmetry in the one-flavor dominated scenario and for
$K_1=m_1/m_{\star}$. In this same case, the lower
inclined dot-dashed lines includes also the effect of scatterings
in the condition~(\ref{M1fl}), while the upper inclined and the horizontal
dot-dashed lines include the effect of oscillations.
The area between the two inclined dot-dashed lines gives an estimation
of the uncertainty on the condition for the
fully flavored regime to hold. The area between the horizontal
thin dot-dashed line and the horizontal thick dashed line
gives an estimation of the uncertainty on the condition
for the unflavored regime to hold. The two
thick solid lines borders the region where successful
leptogenesis is possible: on the left in the unflavored regime
and above in the fully flavored regime. The thin solid line
is a more restrictive border obtained for a specific choice
of the see-saw orthogonal matrix ($\O=R_{13}$) and the thin dashed line is
the corresponding condition~(\ref{M1fl}). In this case
one has $K_1> m_1/m_{\star}$.
}\label{fig:mMplot}
\end{center}
\end{figure}

One possible consequence of flavor effects is to relax the
traditional upper bound on the neutrino mass that is implied by
successful leptogenesis. In order to explore the impact of our
modified criteria, we first recall the origin of this bound in the
unflavored case. Maximizing the final value of the asymmetry over
all see-saw parameters except $M_1$ and $m_1$
yields~\cite{Buchmuller:2004nz}
\begin{eqnarray}
{\eta_B^{\rm max}(M_1,m_1)\over \eta_B^{\rm CMB}}&\simeq&
 3.8\,\left({m_{\star}\over m_1}\right)^{1.2}\,
 \left({M_1\over 10^{10}\,{\rm GeV}}\right)\,
 {m_{\rm atm}\over m_1+m_3}\,\NO\\ 
&&\times \exp\left[-{\o\over z_{\rm B}}\,\left({M_1\over 10^{10}\,{\rm GeV}}\right)\,
 \left(\overline{m}\over{\rm eV}\right)^2\right]
\geq 1\,,
\end{eqnarray}
where we have approximated $\k(K_1)\simeq 0.5\,K_1^{-1.2}$ [cf.~Eq.~(\ref{plaw})],
and we have neglected the dependence of $z_{\rm B}$ on $K_1$ in the derivative.
This constraint translates into $m_1<m_1^{\rm max}(M_1)$ shown by
the curved solid line in the upper part of Fig.~\ref{fig:mMplot}
where the unflavored behavior obtains. This curve sports an absolute
maximum, $m_1\lesssim 0.12\,{\rm eV}$, for
$M_1\simeq 10^{13}\,{\rm GeV}$.

The possibility that flavor effects could relax this bound is based
on the observation that the flavored \CP~asymmetries, for
$m_1\gtrsim m_{\rm atm}$, are proportional to $m_1$ [cf.~Eq.~(\ref{bound}]. 
On the other hand, the total \CP~asymmetry is suppressed like $m_1^{-1}$,
contributing to the upper bound in the unflavored regime. However,
if $m_1$ increases, then $K_1$ has to increase as well, so that it is
not guaranteed that the fully flavored treatment remains justified.

Quantitatively, the value of $K_1$ is bounded
from below by \cite{Fujii:2002jw}
\begin{equation}
 K_1\geq {m_1\over m_{\star}} \,.
\end{equation}
For $K_1\geq m_1/m_{\star}\gg 1$, the final asymmetry is maximized
when a one-flavor dominated scenario is realized. In this case the
final asymmetry is approximately $N_{B-L}^{\rm f} \simeq
\ve_{1\a}\,\k_{1\a}^{\rm f}$. The bound on the single-flavor \CP~asymmetry 
was given in Eq.~(\ref{bound}).
It is then possible to find the value of $P_{1\a}^0$ that maximizes
the asymmetry as a function of $K_1$ and the corresponding value of
$z_{{\rm B}\a}$. Imposing $\eta_B^{\rm max}\geq \eta_B^{\rm CMB}$ implies
a lower bound on $M_1$ as a function of $K_1$.

This limit can be translated into a lower bound on $M_1$ as a
function of $m_1$ by replacing $K_1$ with its minimum value
$m_1/m_{\star}$. In this way the washout is always minimized and
the final efficiency factor and the final asymmetry are maximized.
Notice that the single-flavor \CP~asymmetries, like the
total, vanish for $K_1=m_1/m_{\star}$. Therefore, this lower bound
cannot be saturated. Notice moreover that for $m_1\lesssim
m_{\star}$ the one-flavor dominated scenario does not necessarily
hold because it is possible that $K_1\lesssim 1$. Actually, for
$K_1\rightarrow 0$, flavor effects disappear and one recovers
the usual asymptotic value of the lower bound obtained in the
unflavored case for thermal initial $N_1$-abundance, $M_1\gtrsim 5\times
10^{8}\,{\rm GeV}$ [cf.~Eq.~(\ref{barM1})]. For intermediate values of $K_1$,
one can use a simple interpolation. The final result is shown in the bottom part
of Fig.~\ref{fig:mMplot} as a thick solid line.

In Fig.~\ref{fig:mMplot} we also show the condition~(\ref{M1fl}) 
calculated for the same value of $z_{{\rm B}\a}$ that
maximizes the final asymmetry , but replacing $K_1$
with its minimum value $m_1/m_{\star}$ (thick dashed line).
Since $W_1^{\rm ID}$
increases with $K_1$, this produces a necessary, but not sufficient,
condition in the $m_1$-$M_1$ plane for the fully flavored behavior.
This condition matches the validity of the unflavored regime at
$m_1\simeq 3\times 10^{-3}\,{\rm eV}$ and the lower bound on $M_1$
at $m_1\simeq 2\,{\rm eV}$. This means that for $m_1\gtrsim
2\,{\rm eV}$ the fully flavored behavior does not obtain. Notice
also that this upper limit is quite conservative because the lower bound
on $M_1$ has been obtained neglecting that, for $K_1=m_1/m_{\star}$,
the flavored \CP~asymmetry vanishes and thus the bound cannot
be saturated. Moreover, we have assumed that $P^0_{1\a}$ can always
assume the value that maximizes the asymmetry.

In Fig.~\ref{fig:mMplot} we also show (thin solid line) the lower
bound $M_1(m_1)$ in the specific scenario considered in 
Section~\ref{sec:specificexample}, namely $\O=R_{13}$ and 
$\o^2_{31}$ is taken purely imaginary [cf.~Eq.~(\ref{R13})].
In this case the value of $z_{{\rm B}\a}$ is not necessarily the same as the
one that maximizes the asymmetry in the one-flavor dominated scenario, and
$K_1> m_1/m_{\star}$. Therefore,
a specific calculation  is  necessary in order to work out correctly
the condition~(\ref{M1fl}).
The result is shown in Fig.~\ref{fig:mMplot} with a thin dashed line.

In this case the upper limit on $m_1$ for
the applicability of the fully flavored regime is much smaller,
$m_1\simeq 0.1\,{\rm eV}$. Allowing for a non-vanishing
real part of $\o^2_{31}$, slightly larger values are possible.
It should be however kept in mind that these values are indicative since they
rely on a condition for the fully flavored regime that comes from a
simple rate comparison.

\section{Limitations of a simple rate comparison}   \label{sec:beyond}

We have exploited a somewhat qualitative rate comparison for the
determination of the region where the fully flavored regime obtains.
While we believe that our approach nicely illustrates the
modifications that derive from our more restrictive criterion for
the significance of flavor effects, there are also important
shortcomings. First, we have simply compared the inverse-decay rate
with the charged-lepton Yukawa interaction rate, ignoring flavor
oscillations caused by the flavor-dependent lepton dispersion
relation in the medium. If the oscillations are much faster than the
inverse-decay rate, they also contribute effectively, together with
inelastic scatterings, to project the lepton state on the flavor
basis. Therefore, including oscillations will tend to enlarge the
region where the fully flavored behavior obtains (the inclined upper
dot-dashed line in Fig.~\ref{fig:mMplot}) and to reduce the one
where the unflavored behavior obtains (the horizontal upper
dot-dashed line in Fig.~\ref{fig:mMplot}). In our case, the
oscillation frequency is comparable to $\G_{\alpha}$, and so the two
estimations are not too far off.

Moreover, we have also neglected $\D L=1$ scatterings. They also
contribute, like inverse decays, both to generate the asymmetry and
to the washout and hence, together with inverse decays,
contribute to preserving the flavor direction of the leptons. At the
relevant $z\sim z_{{\rm B}\a}\sim 2$, the $\D L=1$ scattering rate is
actually larger than the inverse-decay rate and thus tends to reduce
the region where the fully flavored behavior obtains
(the lower inclined dot-dashed line in Fig.~\ref{fig:mMplot}). Therefore, the
effects of oscillations and of $\D L=1$ scatterings may partially
cancel each other. In Fig.~\ref{fig:mMplot} the region between the
two inclined dot-dashed lines gives then an indication of the 
theoretical uncertainty on the
determination of the region where the fully flavored regime
holds. It can be seen that current calculations cannot establish whether
the upper bound holding in the unflavored regime is
nullified, just simply relaxed or still holding, when flavor effects
are included.

Only a full quantum kinetic treatment can give a final verdict on
the effectiveness of flavor effects in leptogenesis and its impact
on the neutrino mass limit. While we have verified that our rate
criteria are borne out by the quantum kinetic equations stated in
Ref.~\cite{Abada:2006fw} (see \cite{DeSimone:2006dd}), these equations 
are not necessarily
complete in that the term describing the generation of
the asymmetry has been added by hand. Moreover, the ``damping
rate'' caused by the flavor-sensitive Yukawa interactions ultimately
derives from a collision term in the kinetic
equation~\cite{Raffelt:1992uj}. Extending the pioneering treatment
of Ref.~\cite{Abada:2006fw} to allow for a complete understanding of
flavor effects remains a challenging~task. In the next section we
show how such a quantum kinetic equation could be derived.

\section{Density matrix equation}

Here we aim to derive a density matrix equation for leptogenesis
following~\cite{Raffelt:1992uj}, where the general formalism was introduced.
We consider exclusively the decay of the lightest RH neutrino, $N_1$,
with a mass between $10^9$~GeV
and $10^{14}$~GeV where only the $\t$-Yukawa interactions may be 
faster than the Hubble rate. We are therefore either in the
unflavored or in the two-flavor regime, the two flavors being $\t$ and 
a combination of $e$ and $\m$ which we call $\b$ in the following.
 
We follow the definition of the matrices of densities for leptons 
and antileptons given in~\cite{Raffelt:1992uj}:
\begin{eqnarray}
\r_l&=&\left(\begin{array}{ll} \r_{\t\t} & \r_{\b \t}\\
\r_{\t \b} & \r_{\b \b}\end{array}\right) \\
\r_{\bar{l}}&=&\left(\begin{array}{ll} \bar{\r}_{\t\t} &
 \bar{\r}_{\t \b}\\
\bar{\r}_{\b \t} & \bar{\r}_{\b \b}\end{array}\right)
\end{eqnarray}
where $\r_{ij}=\langle a^{\dagger}_j(t) a_i(t)\rangle $, $\bar{\r}_{ij}=\langle 
b^{\dagger}_i(t) b_j(t)\rangle$, with $a$ ($a^{\dg}$) denoting the annihilation 
(creation) operator for a lepton and $b$ ($b^{\dg}$) the annihilation (creation)
operator for an antilepton. 
All matrix components are implicitly given in a portion of comoving volume that
would contain one heavy neutrino in ultra-relativistic thermal equilibrium.
Note that we introduce here matrices of number densities,
not occupation numbers as in~\cite{Raffelt:1992uj}. We shall explain later
on how the integration over all modes can be performed here. As usual in 
leptogenesis, one assumes kinetic 
equilibrium, an assumption which should be good at the 10\% level in the strong
washout regime~\cite{Basboll:2006yx}. Following this assumption, all interaction 
rates are thermally averaged.

Let us first study the two simplest -- nevertheless very important -- processes,
namely decays and inverse decays: $N_1 \leftrightarrow \ell_{1}+\Phi^{\dagger}$ and 
$N_1 \leftrightarrow \bar{\ell}_{1}+\Phi$. The state $|\ell_{1}\rangle$ 
was defined in Eq.~(\ref{state}).

The idea is to write the equation for the generation of lepton number
in the basis $\{|\ell_1\rangle,|\ell_{\perp}\rangle\}\equiv \{\ell_1\}$, and 
then rotate it into the flavor basis $\{|\ell_{\t}\rangle,|\ell_{\b}\rangle\}
\equiv \{\ell_{\a}\}$. The unitary
matrix for the change of basis is given by
\begin{equation}\label{transflep}
U (\{\ell_1\} \to \{\ell_{\a}\}) =\frac{1}{\sqrt{(h^{\dagger}h)_{11}}}
\left(\begin{array}{cc} \langle \ell_{\t}|\ell_1\rangle & -\langle \ell_{\b}|\ell_1\rangle \\
\langle \ell_1|\ell_{\b}\rangle & \langle \ell_1|\ell_{\t}\rangle \end{array}\right).
\end{equation}
The same can be done for antileptons, and one gets the following matrix for the change
of basis:
\begin{equation}\label{transfantilep}
U '(\{\bar{\ell}_1'\} \to \{\bar{\ell}_{\a}\}) =\frac{1}{\sqrt{(h^{\dagger}h)_{11}}}
\left(\begin{array}{cc} \langle \bar{\ell}_{\t}|\bar{\ell}_1'\rangle & 
-\langle\bar{\ell}_1'
|\bar{\ell}_{\b}'\rangle \\
\langle\bar{\ell}_{\b}|\bar{\ell}_1'\rangle &
\langle\bar{\ell}_1'|\bar{\ell}_{\t}\rangle\end{array}\right),
\end{equation}
where it should be remembered that $\bar{\ell}'_1$ is not the \CP-conjugate 
of $\ell_1$ at the one-loop level.

Decays and inverse decays
are just source and sink terms, so one can proceed by analogy 
with~\cite{Raffelt:1992uj} with the conventions of~\cite{Buchmuller:2004nz} on 
leptogenesis quantities. 
In the basis $\{\ell_1\}$, the equation tracking the density matrix for leptons is given by
\begin{equation}\label{densunfl1}
{{\rm d}\over {\rm d}z} \r_{\ell} = {1\over 2}(1+\ve_1)D N_{N_1}  \left(\begin{array}{ll}
1 & 0 \\ 
0 & 0 \end{array}\right) -\frac{1}{2}
(1-\ve_1) W_{\rm ID} \left[ \left(\begin{array}{ll}
1 & 0 \\
 0 & 0 \end{array}\right) \r_{\ell} + 
\r_{\ell}\left(\begin{array}{ll}
1 & 0 \\
 0 & 0 \end{array}\right)\right],
\end{equation}
where $z=M_1/T$, and for antileptons:
\begin{equation}\label{densunfl2}
{{\rm d}\over {\rm d}z}\r_{\bar{\ell}} = {1\over 2}(1-\ve_1) D N_{N_1} \left(\begin{array}{ll}
1 & 0 \\ 
0 & 0 \end{array}\right)  -\frac{1}{2}
(1+\ve_1) W_{\rm ID} \left[ \left(\begin{array}{ll}
1 & 0 \\
 0 & 0 \end{array}\right) \r_{\bar{\ell}} + 
\r_{\bar{\ell}}\left(\begin{array}{ll}
1 & 0 \\
 0 & 0 \end{array}\right)\right],
\end{equation}
where both equations are written to first order in the \CP~asymmetry parameter $\ve_1$, 
and we neglected Pauli blocking effects.

Let us now rotate these equations to the flavor basis $\{\ell_{\a}\}$, applying the 
transformations introduced above, Eqs.~(\ref{transflep}) and (\ref{transfantilep}). 
It will prove useful to define the following matrices 
which will appear in the new equations:
\begin{equation}
U \left(\begin{array}{ll} 1 & 0 \\ 0 & 0 \end{array}\right) U^{\dagger} = \left(
\begin{array}{cc} P_{1\t} & \langle \ell_{\t}|\ell_1\rangle \langle \ell_1|\ell_{\b}\rangle \\ 
\langle \ell_1|\ell_{\t}\rangle \langle \ell_{\b}|\ell_1\rangle &
P_{1\b} \end{array}\right) \equiv \mathcal{P},
\end{equation}
and 
\begin{equation}
U' \left(\begin{array}{ll} 1 & 0 \\ 0 & 0 \end{array}\right) U'^{\dagger} = \left(
\begin{array}{cc} \overline{P}_{1\t} & \langle\bar{\ell}_1'|\bar{\ell}_{\t}\rangle \langle 
\bar{\ell}_1'
|\bar{\ell}_{\b}\rangle \\ \langle\bar{\ell}_1'|\bar{\ell}_{\t}\rangle \langle 
\bar{\ell}_1'|\bar{\ell}_{\b}\rangle &
\overline{P}_{1\b} \end{array}\right)\equiv \overline{\mathcal{P}},
\end{equation}
where the projectors are defined as $P_{1\a}=|\langle \ell_{\a}|\ell_1\rangle|^2$ and 
$\overline{P}_{1\a}=|\langle \bar{\ell}_{\a}|\bar{\ell}_1'\rangle|^2$ 
[cf.~Eqs.~(\ref{defproj}) and (\ref{defantiproj})]. At tree level,
the two matrices just defined are equal and given by
\begin{equation}
\mathcal{P}^0={1\over (h^{\dagger}h)_{11}}\left(\begin{array}{cc} |h_{\tau 1}|^2 
& h_{\tau 1}^* h_{\b 1} \\ h_{\tau 1} h_{\b 1}^* & |h_{\b 1}|^2 \end{array}\right).
\end{equation}

In the flavor basis, Eqs.~(\ref{densunfl1}) and (\ref{densunfl2}) 
become
\begin{equation}\label{densflav1}
{{\rm d}\over {\rm d}z} \r_{\ell} = {1\over 2}(1+\ve_1) D N_{N_1} \mathcal{P}  -\frac{1}{2}
(1-\ve_1) W_{\rm ID} \left[ \mathcal{P} \r_{\ell} + 
\r_{\ell} \mathcal{P}\right],
\end{equation}
and 
\begin{equation}\label{densflav2}
{{\rm d}\over {\rm d}z}\r_{\bar{\ell}} = {1\over 2}(1-\ve_1)D N_{N_1}
\overline{\mathcal{P}}  -\frac{1}{2}
(1+\ve_1) W_{\rm ID} \left[ \overline{\mathcal{P}} \r_{\bar{\ell}} + 
\r_{\bar{\ell}} \overline{\mathcal{P}}\right].
\end{equation}

Subtracting the first equation with the second, and defining $\r_{\ell}-
\r_{\bar{\ell}}\equiv \r_{\ell-\bar{\ell}}$,
one obtains
\begin{equation}
{{\rm d}\over {\rm d}z} \r_{\ell-\bar{\ell}} = \ve_1 D \left(N_{N_1}
+N_{N_1}^{\rm eq}\right) \mathcal{P}^0 +\frac{\mathcal{P}- \overline{\mathcal{P}}}{2}D\left(N_{N_1}
- N_{N_1}^{\rm eq}\right)-
\frac{1}{2}W_{\rm ID}\{\mathcal{P}^0,\r_{\ell-\bar{\ell}}\},
\end{equation}
to first order both in $\ve_1$ and $\mathcal{P}- \overline{\mathcal{P}}$. 
Note that we used the fact that
\mbox{$\r_{\ell}+\r_{\bar{\ell}}\simeq 2 N_{\ell}^{\rm eq} \mathbbm{1}$}, where we 
neglected contributions of $\mathcal{O}(\ve_1)$, 
because the whole term is already at first order in $\ve_1$. As usual in leptogenesis, 
one notices that there is an asymmetry production even in thermal
equilibrium~\cite{Kolb:1979qa}. This is due to the missing contribution from the 
on-shell $\D L =2$ processes
$\bar{\ell} \Phi \leftrightarrow \ell \Phi^{\dagger}$. Properly accounting for 
them yields~\cite{Barbieri:1999ma,Abada:2006fw}
\begin{equation}\label{DensEq}
{{\rm d}\over {\rm d}z}\r_{\ell-\bar{\ell}} = \ve_1 D \left(N_{N_1}
-N_{N_1}^{\rm eq} \right) \left(\mathcal{P}^0 +\frac{\mathcal{P}- \overline{\mathcal{P}}}{2\ve_1}\right)-
\frac{1}{2} W_{\rm ID}\{\mathcal{P}^0,\r_{\ell-\bar{\ell}}\}      ,
\end{equation}
 which looks intuitively correct, with both the reduction of the washout by the projector (a
 matrix now!) and the additional contribution to the source term, of the type $\D P$ (also a
 matrix). If the off-diagonal elements are quickly damped (see below), one recovers
 exactly the fully flavored equations on the diagonal entries
[cf.~Eq.~(\ref{flke2}) with $i=1$].

Actually, the matrix of \CP~asymmetries, which can be defined as 
\be
\e\equiv \ve_1 \left(\mathcal{P}^0 +\frac{\mathcal{P}- \overline{\mathcal{P}}}{2\ve_1}\right),
\ee
was proposed in~\cite{Abada:2006fw} for the three-flavor regime to be given by 
\be
\e_{\a\b}={3\over 32 \p}{1\over (h^{\dg}h)_{11}}\sum_{j\neq 1}{M_1\over M_j} {\rm Im}
\left(h^{\star}_{\a 1} (h^{\dg}h)_{1j} h_{\b j}- h_{\b 1}(h^{\dg}h)_{j1} h^{\star}_{\a j}
\right),
\ee
in the HL, $M_1\ll M_2 \ll M_3$, where $\a,\b=e,\m,\t$. It is straightforward to 
check that the diagonal 
elements correspond to the single-flavor \CP~asymmetries shown in Eq.~(\ref{ve1a}).

Let us now examine the effect of the term $f_{\a } \overline{\ell_{L\a}} e_{R\a} \F$, which we
said in Section~\ref{sec:flavor} is responsible for flavor effects to manifest themselves.
This interaction is flavor diagonal, and we expect it to affect Eq.~(\ref{DensEq}) in two
ways. First, it induces a contribution to the refractive index for the lepton doublet
$\ell_{\a}$, and, second, it makes the lepton doublet interact with the lepton 
singlet $e_{\a}$.

\begin{figure}
\begin{center}
\includegraphics{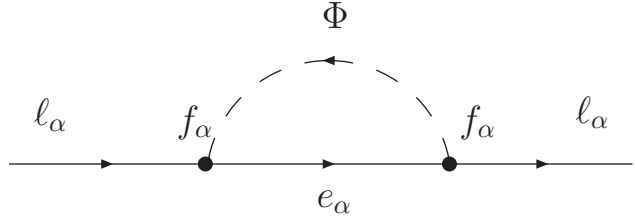}
\end{center}
\caption{Flavor diagonal contribution to the refractive index for the leptons.}
\label{fig:refractive}
\end{figure}

Starting with the index of refraction effect, in perfect analogy 
with~\cite{Raffelt:1992uj}, one has
a contribution in the form of a commutator both for leptons and antileptons:
\begin{eqnarray}
{{\rm d}\over {\rm d}z}\r_{\ell} &=& [\textrm{RHS of Eq.~(\ref{densflav1})}]-i[\L^{\t}_{\o},
\r_{\ell}]\label{eq1} \\ 
{{\rm d}\over {\rm d}z}\r_{\bar \ell} &=& [\textrm{RHS of Eq.~(\ref{densflav2})}] + 
i[\L^{\t}_{\o},\r_{\bar \ell}],\label{eq2}
\end{eqnarray}
where 
\be
\L^{\t}_{\o}\simeq  {1\over Hz}\left(\begin{array}{cc} \langle M^2_{\t}/2p \rangle & 0\\
0& 0\end{array}\right)\simeq {1\over Hz}{T\over 64}
\left(\begin{array}{cc} f_{\t}^2 & 0\\ 0& 0 \end{array}\right) 
\ee
The effective mass $M_{\a}$ aquired by the lepton in the medium due to the Yukawa coupling 
$f_{\a}$ was calculated in~\cite{Weldon:1982bn} from the real part of the diagram
shown in Fig.~\ref{fig:refractive} and used in the second equality. 


It is important to be aware that the
possibility to use a density matrix of number densities, i.e. to have been able to integrate
over the 3-momentum of the leptons, is non-trivial and comes from~\cite{Bell:2000kq}, 
where it was noticed that when gauge interactions are very fast, all modes oscillate
at the same frequency, given by the thermal average $\langle M^2/2p\rangle$ in this case.

Similar to the effect of a vacuum mass, leptons and antileptons oscillate in opposite
directions, so that subtracting Eq.~(\ref{eq2}) to Eq.~(\ref{eq1}) leads to
\begin{equation}\label{DensEqOsc} 
{{\rm d}\over {\rm d}z} \r_{\ell-\bar{\ell}}= [\textrm{RHS of Eq.~(\ref{DensEq})}]-i[\L^{\t}_{\o},
\r_{\ell} +\r_{\bar \ell}]=
[\textrm{RHS of Eq.~(\ref{DensEq})}]-i[\L^{\t}_{\o},\r_-],
\end{equation}
where, in the second equality, we defined 
$\r_-\equiv \r_{\ell} +\r_{\bar \ell}-2N_{\ell}^{\rm eq}\mathbbm{1}$, 
which has all elements of $\mathcal{O}(\ve_1)$ since the diagonal part of 
$\r_{\ell} +\r_{\bar \ell}$ has
a part proportional to $2N_{\ell}^{\rm eq}$ which drops out of the commutator. 
In order to have a closed set of 
equations, we must give the equation of evolution of $\r_-$,
\begin{equation}\label{rho-}
{{\rm d}\over {\rm d}z} \r_- = i[\L^{\t}_{\o},\r_{\ell-\bar{\ell}}].
\end{equation}

As for the new interactions between lepton doublets and singlets, they yield again
two terms, one source and one sink term, which describe how the lepton asymmetry is 
shared between the doublets and the singlets. As we already mentioned, only the 
$\t$-Yukawa coupling is relevant for the range of RH neutrino masses we consider, and thus the 
relevant interaction rate is given by $F_{\t}$, which we defined in Eq.~(\ref{Ftau}).
Assuming a negligible asymmetry in
the Higgs field, the resulting term can be written in the
form
\bea\label{rhoL}
{{\rm d}\over {\rm d}z}\r_{\ell-\bar{\ell}}&\simeq& \e D \left(N_{N_1}
-N_{N_1}^{\rm eq} \right) -{1\over 2}W_{\rm ID}\left\{\mathcal{P}^0,\r_L\right\}
-i[\L^{\t}_{\o},\r_-]\NO\\
&+&2 F_{\t}(N_{e_{\t}}-N_{\bar{e}_{\t}})\left(\begin{array}{ll}
1 & 0 \\ 0 & 0\end{array}\right)
- {1\over 2}F_{\t}\left\{\left(\begin{array}{ll}
1 & 0 \\ 0 & 0\end{array}\right),\r_{\ell-\bar{\ell}}\right\}.
\eea


The asymmetry in the lepton singlets is found by solving the following Boltzmann
equation:
\begin{equation}\label{NE}
{{\rm d}\over {\rm d}z} (N_{e_{\t}} - N_{\bar{e}_{\t}})= F_{\t}\left[\left(N_{\ell_{\t}}-
N_{\bar{\ell}_{\t}}\right) - 2(N_{e_{\t}} -N_{\bar e_{\t}})\right],
\end{equation}
where $N_{\ell_{\t}}-N_{\bar{\ell}_{\t}}$ is the upper left component of the 
density matrix $\r_{\ell-\bar{\ell}}$.
Clearly, this equation will tend to equalize $N_{\ell_{\t}}-N_{\bar{\ell}_{\t}}$ 
and $2(N_{e_{\t}} -N_{\bar e_{\t}})$. It is interesting to notice that when the latter two
quantities are equal, Eq.~(\ref{rhoL}) can be written as
\begin{eqnarray}\label{rhoL2}
{{\rm d}\over {\rm d}z} \r_L &\simeq& \e D \left(N_{N_1}
-N_{N_1}^{\rm eq} \right) -{1\over 2}W_{\rm ID}\left\{\mathcal{P}^0,\r_L\right\}
-i[\L^{\t}_{\o},\r_-]\nonumber\\*
&&
\qquad -{1\over 2}F_{\t} \left[\left(\begin{array}{ll}
1 & 0 \\ 0 & 0\end{array}\right),\left[\left(\begin{array}{ll}
1 & 0 \\ 0 & 0\end{array}\right),\rho_L\right]\right],
\end{eqnarray}
where the total lepton number $N_L$ is defined as $(N_{\ell}-N_{\bar{\ell}})+
(N_{e}-N_{\bar{e}})$. In particular, this means that the upper left component
of $\r_L$ is given by \mbox{$2(N_{\ell}-N_{\bar{\ell}})/3$}. The coefficient
$2/3$ translates the fact that some of the lepton asymmetry, being stored in 
the right-handed fields, escapes the washout by inverse decays. This is nothing
else than an effect caused by one spectator process, as discussed in Section~\ref{sec:spectator},
with the difference that we consider here a simplified situation where the 
asymmetry stored in the Higgs is assumed to be zero.

Eq.~(\ref{rhoL2}) is very illustrative in that it includes all the well-known 
contributions discussed in~\cite{Raffelt:1992uj} in a completely different context, 
namely the production (source term), the washout (sink term), the double commutator,
which damps the off-diagonal elements, and the commutator, which drives the oscillations
in flavor space. 

In the transition region $T\sim 10^{12}\gev$, one cannot assume 
$N_{\ell_{\t}}-N_{\bar{\ell}_{\t}}=2(N_{e_{\t}} -N_{\bar e_{\t}})$, and one has to solve simultaneously 
four coupled equations, 
Eqs.~(\ref{NN1}), (\ref{rho-}), (\ref{rhoL}) and (\ref{NE}). They constitute 
the system of equations to solve in order to have a correct description
of leptogenesis at the transition between the fully flavored and the unflavored
regime, with an approximate treatment of spectator processes. 

The system of equations we propose differ from what can be found in the
present literature~\cite{Abada:2006fw,DeSimone:2006dd} in that
\begin{enumerate}
\item we explicitly
include the effect of oscillations in flavor space, which adds one equation,
since leptons and antileptons oscillate in different directions in flavor
space;
\item we keep track of the asymmetry in $e_{\t}$, which adds another equation, 
as should be done in the transition regime around $T\sim 10^{12}\gev$.
\end{enumerate}
The second point is actually the addition of one spectator process to 
the picture. In principle, all spectator processes should be taken into 
account, and a large number of Boltzmann equations should be solved 
simultaneously. In particular, sphalerons should be included and the final asymmetry
obtained would then be in $\D_{\a}\equiv B/3 -L_{\a}$, as required. We did
not include these effects here for the sake of simplicity and clarity,
as well as because the difference is expected to be of order 20--30\%.

The conditions for the validity of the different pictures of leptogenesis
which we discussed in the previous sections can be put in perspective
by looking at Eq.~(\ref{rhoL}). Roughly speaking, if
$W_{\rm ID}\gg F_{\t}, \L^{\t}_{\o}$, one expects the unflavored regime
to be recovered, whereas in the opposite situation, the fully flavored
regime should hold. To follow up on the comments made in the last section
about the inclusion of $\D L=1$ scatterings, it should be noted that,
even though we did not include them for simplicity in our analysis, their 
addition is straightforward in that one has to replace $D_1 \to D_1+S_1$ and 
$W_1^{\rm ID} \to W_1^{\rm ID}+W_1^{\rm \D L=1}=j(z)W_{\rm ID}$, with the corresponding
analytic expressions given in Eqs.~(\ref{D+S}) and (\ref{jz}), respectively. 

Finally, let us note that there is another quantum limit which can be considered
for the Boltzmann equations relevant for leptogenesis. It relies on the closed
time path formalism for non-equilibrium quantum field theory. Within such an
approach, the so-called ``memory effects'' are for instance accounted for. In the case
of leptogenesis, these effects translate into a time-dependent \CP~asymmetry. 
This quantum limit was studied in~\cite{DeSimone:2007rw,DeSimone:2007pa}, 
and the main conclusion reached there is that memory effects have some impact 
in the case of resonant leptogenesis in the weak washout regime. In all 
other cases, they can be safely neglected.



\chapter{Going beyond vanilla leptogenesis}
\label{chap:beyond}

We have always assumed up to now that the final $B\!-\!L$  
asymmetry was predominantly
produced by the lightest RH neutrino, $N_1$. A $N_1$-dominated scenario 
typically follows from the assumption of hierarchical heavy neutrino 
masses, \HL. 
Actually, it can even be enforced by setting $R_{23}=\mathbbm{1}$, 
since a large contribution from $N_2$ is then unprobable, as we
explained in Section~\ref{sec:general}.
This is ``vanilla'' leptogenesis, where successful leptogenesis 
requires high-scale values of the lightest RH neutrino mass, $M_1$, 
and of the reheat temperature,
$T_{\rm reh}$. Moreover, the upper bound on the absolute neutrino mass
scale implied by successful leptogenesis was discussed in 
Section~\ref{sec:N1DS} within an unflavored
treatment and in Section~\ref{sec:transition} including flavor effects.

In this chapter we would like to go beyond this picture. Actually,
flavor effects themselves allow that to some extent. For example,
they might modify the upper bound on the absolute neutrino mass
scale, and they open new ways to have $N_2$- and $N_3$-dominated
scenarios. First, we want to relax the assumption of a hierarchical
heavy neutrino mass spectrum, and concentrate on quasi-degenerate 
spectra, \DL.
Then, we discuss the implications of having $R_{23}\neq \mathbbm{1}$, opening
the way to the $N_2$-dominated scenario. Actually, we also analyze
this possibility within a flavored perspective and notice that the window
for $N_2$- and $N_3$-leptogenesis opens up compared to the unflavored
case. Finally, we discuss the possible effects coming from
one element of the $\O$ matrix, namely $\O_{22}$.

\section{Degenerate limit for the heavy neutrinos}
\label{sec:beyondDL}

On theoretical grounds it is not difficult to motivate models
with a quasi-degenerate heavy neutrino mass spectrum. For example, a
slightly broken $U(1)_L$ lepton flavor symmetry is enough to 
achieve this goal~\cite{Branco:1988ex,Shaposhnikov:2006nn,Kersten:2007vk}.
Note that this possibility has other implications as well, as we
shall see below.
Such a spectrum can also be motivated in the context of 
``radiative leptogenesis''~\cite{GonzalezFelipe:2003fi,Branco:2005ye}.

In leptogenesis a quasi-degenerate mass spectrum for the heavy neutrinos
leads to important qualitative and quantitative differences with respect to 
the vanilla picture described in Chapters~\ref{chap:unflavored} and \ref{chap:flavor}, 
because all heavy neutrino contributions must be taken into account,
both at the production and washout levels. 

It is straightforward
to generalize, including flavor effects, a result obtained in~\cite{Blanchet:2006dq}
for the efficiency factors within an unflavored treatment in the degenerate 
limit, $(M_3-M_1)/M_1\lesssim 0.1$, and for $K_i\gg 1$, for all $i$. Indeed,
one can now approximate ${\rm d}N_i/{\rm d}z'\simeq {\rm d}N_i^{\rm eq}/{\rm d}z'$ in
Eq.~(\ref{ef}), obtaining that 
\be\label{fullDL}
\k_{i\a}^{\rm f}\simeq \k(K_{1\a}+K_{2\a}+K_{3\a}).
\ee 
The function
$\k(x)$ was defined in Eq.~(\ref{k}), and it
approximates $\k_{1\a}^{\rm f}$ in the hierarchical limit when
$x=K_{1\a}$. In the degenerate limit (DL) one has then only to replace
$K_{1\a}$ with the sum $K_{1\a}+K_{2\a}+K_{3\a}$. The number of
efficiency factors to be calculated reduces from 6 to 2 in the two-flavor
case and from 9 to 3 in the three-flavor case, i.e. one for each flavor,
like in the hierarchical limit.
If, instead of a full degeneracy, one has only a partial degeneracy,
$M_1\simeq M_2\ll M_3$, then 
\be\label{partialDL}
\k_{3\a}^{\rm f}\ll \k_{1\a}^{\rm f}\simeq \k_{2\a}^{\rm f}
\simeq \k(K_{1\a}+K_{2\a}).
\ee

It is interesting to notice that,
as a consequence of the orthogonality of $\O$, one has
\be\label{sumKia}
K_{1\a}+K_{2\a}+K_{3\a}=\sum_k\,{m_k\over m_{\star}}\,|U_{\a k}|^2 \, .
\ee
This means that the sum over the decay parameters will typically
be in the strong washout range, $K_{1\a}+K_{2\a}+K_{3\a}\gtrsim 3$. This is a nice
property of leptogenesis with degenerate heavy neutrinos which still
holds when flavor effects are included. Consequently, we do not need
to introduce analytic expressions for a vanishing initial number of 
heavy neutrinos,
since it will give the same result as $\k(K_{1\a}+K_{2\a}+K_{3\a})$, which
was derived for a thermal initial $N_1$-abundance. 

Concerning the flavored \CP~asymmetries, one can rewrite the general 
expression~(\ref{veia}) in the DL,
\be\label{veiaDL}
\ve_{i\a}=
\frac{1}{16 \p (h^{\dag}h)_{ii}} \sum_{j\neq i} \left\{ {\rm Im}\left[h_{\a i}^{\star}
h_{\a j}(h^{\dag}h)_{i j}\right]+
{\rm Im}
\left[h_{\a i}^{\star}h_{\a j}(h^{\dag}h)_{j i}\right]\right\} \d_{ji}^{-1} \, ,
\ee
where we used the fact that $\xi(x_j/x_i)\simeq 1/(3\d_{ji})$ and 
we defined 
\be\label{deltaji}
\delta_{ji}\equiv {M_j-M_i\over M_i} = \sqrt{x_j\over x_i}-1 \, .
\ee
As it is clear from Eq.~(\ref{veiaDL}), there can be an enhancement of 
the \CP~asymmetry proportional to $\d_{ji}^{-1}$~\cite{Covi:1996wh} for
quasi-degenerate heavy neutrino masses. This enhancement originates
from the one-loop self energy contribution. Note however that an enhancement
of the \CP~asymmetry might require that all three heavy neutrinos
are quasi-degenerate. As a matter of fact, when $\O=R_{13}$, there is
no enhancement when $M_2\to M_1$, but only when 
$M_3 \to M_1$~\cite{Blanchet:2006dq}.

We can now write in general for the final $B\!-\!L$ asymmetry 
\be
N^{\rm f}_{B-L}=\sum_{\a}\left\{ \left(\sum_k \ve_{k \a}\right)\, 
\k\left(\sum_i K_{i\a}\right)\right\},
\ee
where the sums over $i$ and $k$ run from 1 to 3 in the case of full DL and
from 1 to 2 in the case of partial DL.
When computing the final asymmetry and the lower bound on 
$M_1$ in the DL, there will be a competition between two effects,
namely the increase of the washout because of the sum of the decay
parameters in the efficiency factor, and the enhancement of the \CP~asymmetry.
The second effect will become dominant once the first one saturates,
which occurs for $\d_{ji}\sim 0.01$~\cite{Blanchet:2006dq}. From
this point going to higher degeneracies will lower the bounds
proportionally to $\d_{ji}$. Therefore, as we pointed out in 
Section~\ref{sec:lepsuper}, it is possible to relax the lower bound
on the reheat temperature in such a way that the gravitino problem
is avoided. 

One can go even further and look for the most extreme relaxation
one can achieve. This occurs when the resonance in the \CP~asymmetry
is reached, for $M_j-M_i\simeq \G_j/2$~\cite{Pilaftsis:1997jf,Pilaftsis:1998pd}.
We shall use a slightly modified condition,
\be\label{condRL}
\d_{ji}^{\rm res}\simeq d\,\bar{\ve}(M_i)/3,
\ee
where we introduced the uncertainty parameter $d=1\div 10$ because
of the claim in~\cite{Anisimov:2005hr} that it is not possible to
be exactly on resonance without breaking perturbation theory, contrary
to what was stated in~\cite{Pilaftsis:1997jf,Pilaftsis:2003gt}. We do not
aim here at a resolution of this discrepancy and therefore introduced the
parameter $d$.

It is straightforward to check that the resonance condition~(\ref{condRL}) 
implies that \mbox{$\ve_1=1/d$} 
in the unflavored case with maximal phase. In other words, the
resonance condition simply cancels out the dependence on the heavy
neutrino mass scale in the \CP~asymmetry and therefore also
in the final baryon asymmetry. Hence, in resonant leptogenesis there is 
essentially no lower bound on $M_1$ and $T_{\rm reh}$ and one can 
go down to the TeV scale~\cite{Pilaftsis:1997jf,Pilaftsis:2003gt}. The only requirement
is actually to let the heavy neutrinos decay before the freeze-out 
of the sphalerons (at around 100~GeV), in order to produce 
not only lepton number but also baryon number. In~\cite{Pilaftsis:2003gt} 
the authors followed an unflavored treatment where the third heavy neutrino
$N_3$ is decoupled and the other two are quasi-degenerate. In such
a scenario one cannot avoid the fact that the Yukawa coupling must be 
quite small in order to have a successful leptogenesis.

Accounting for flavor effects in the case of resonant leptogenesis, it
was realized in~\cite{Pilaftsis:2004xx,Pilaftsis:2005rv} that an even more 
dramatic situation could be envisaged. If three quasi-degenerate RH neutrinos
are considered, it is possible to have some of the Yukawa couplings of order one, 
leading to phenomenological implications, such as a rate for the lepton
flavor violating decay $\m\to e\g$ within reach of future experiments, as 
well as successful leptogenesis at the TeV scale.
The crucial point there was to have one flavor, the $\t$, that is very weakly
washed out thanks to a very small projector $P_{1\t}$, and one heavy
neutrino, $N_3$, that is weakly coupled. However, the latter condition does not
necessarily imply that the \CP~asymmetry $\ve_3$ is suppressed because it
receives other contributions when  $M_3\simeq M_{1,2}$, as we will explain 
in Section~\ref{sec:O22}. It must be
said, however, that the extreme situation needed seems to occur in a region 
of the parameter space where the condition of validity~(\ref{M1fl}) 
which was derived in the previous chapter is not satisfied.

Actually, strictly speaking, the possibility of explaining the
smallness of neutrino masses with large
Yukawa couplings and relatively light RH neutrino masses 
(TeV scale) does not involve a conventional ``see-saw mechanism'', even though
the same matrix for the light neutrinos applies, Eq.~(\ref{mnu}).
The point is that the neutrino masses are not in this case small because
of the suppression due to the heavy neutrino mass, but because of a
very large cancellation due to the $\O$ matrix. The elements of the 
$\O$ matrix can indeed be
arbitrarily large, as long as they satisfy $\O^T\O=\O\O^T=\mathbbm{1}$. This can be 
theoretically motivated by a slightly broken $U(1)_L$ lepton flavor
symmetry~\cite{Branco:1988ex,Shaposhnikov:2006nn,Kersten:2007vk}. It
is interesting that this type of model may lead to possible 
signatures of heavy neutrinos at future 
colliders~\cite{delAguila:2007em,Kersten:2007vk,deGouvea:2007uz}.

\section[$N_2$-dominated scenario]{\boldmath{$N_2$}-dominated scenario}
\label{sec:N2}

In this section we consider again a hierarchical heavy
neutrino mass spectrum, \HL.
In the unflavored regime, this assumption typically implies
a $N_1$-dominated scenario, where the final asymmetry is
dominated by the contribution from the lightest RH neutrino decays,
Eq.~(\ref{N1DS}). Indeed, in general, in the HL one has two effects.
The first effect is that the asymmetry production from the two heavier RH
neutrinos, $N_2$ and $N_3$, is typically later on washed out by the
$N_1$ inverse processes, implying $\k_3^{\rm f},\k_2^{\rm f}\ll \k_1^{\rm f}$.
 The second effect is a consequence of the fact that the total
$C\!P$ asymmetries vanish in the limit when all particles running in the
loops become massless, and this yields typically
$|\ve_3|\ll |\ve_2| \ll |\ve_1|$.

However, for a particular choice of the see-saw parameters,
$\O = R_{23}$ [cf.~Eq.~(\ref{R23})] and $m_1\ll m_{\star}$ [cf.~Eq.~(\ref{mstar})],
the contribution to the final asymmetry from the next-to-lightest
RH neutrino $N_2$ is not only non-negligible but even dominant, giving rise
to a $N_2$-dominated scenario \cite{DiBari:2005st}. Indeed,
for $\O= R_{23}$, different things happen simultaneously.
First, $N_2$, even though decoupled from $N_1$, is still
coupled to $N_3$ and in the HL the total $C\!P$ asymmetry
$\ve_2$ not only does not vanish, since it receives an unsuppressed contribution
from graphs where $N_3$ runs in the loops, but can even be 
maximal [cf. Eq.~(\ref{R23max})].
On the other hand, one has now $\ve_1=0$, since $N_1$
is essentially decoupled from the other two RH neutrinos. At the same time,
one has $K_1 =m_1/m_{\star}\ll 1$, so that the washout 
from $N_1$ inverse processes is negligible. The final result is that
$|\ve_2\,{\k_2}|\gg |\ve_{i\neq 2}\,\k_{i\neq 2}^{\rm f}|$,
and the final asymmetry is dominantly produced from $N_2$-decays.

A nice feature of this model is that
the lower bound on $M_1$ does not hold anymore, being replaced by a lower
bound on $M_2$ that, however, still implies a lower bound on 
$T_{\rm reh}$~\cite{DiBari:2005st}. If one switches on some small 
$R_{12}$ and $R_{13}$ complex rotations, then the lower bounds on
$M_2$ and $T_{\rm reh}$ become necessarily more stringent. Therefore,
there is a border beyond which this scenario is not viable and one is
forced to go back to the usual $N_1$-dominated scenario for successful leptogenesis.

Within an unflavored treatment and in the HL,
the condition $w_{32}\simeq 0$ in the
$\O$-matrix parametrization [cf. Eq.~(\ref{second})], implying
$R_{23}\simeq \mathbbm{1}$,
is sufficient to have a negligible asymmetry production
from the two heavier RH neutrinos and to guarantee that
the $N_1$-dominated scenario holds. This condition is even not 
necessary for $m_1\gg m_{\star}$, since in this case, due to the
fact that $\mt\geq m_1$, one has necessarily $K_1\gg 1$
and the washout from $N_1$ inverse processes is strong enough to 
suppress a possible contribution to the
final asymmetry from $N_2$-decays.

When flavor effects are taken into account,
the domain of applicability of the $N_1$-dominated
scenario reduces because the importance of $N_2$ and $N_3$
increases. There are two aspects to be considered.

The first aspect is that the washout from $N_1$ inverse processes
becomes less efficient. Indeed, the projectors $P_{1\a}$ can considerably reduce
the washout of the asymmetry produced in the flavor $\a$ from
$N_2$-decays \cite{Vives:2005ra}. This turns the condition
$m_1\gg m_{\star}$ into a looser condition $m_1\gg m_{\star}/P_{1\a}$.
Another effect is that $N_1$ inverse processes can be fast enough
to quickly destroy the coherence of $\ell_2$. Then a statistical
mixture of $\ell_1$ and of the state orthogonal to $\ell_1$ builds
up, and hence part of the asymmetry produced in $N_2$-decays is
protected from the washout from $N_1$ inverse 
processes~\cite{Barbieri:1999ma,Strumia:2006qk,Engelhard:2006yg}. These
effects may occur in the unflavored regime ($M_1\gtrsim 10^{12}\gev$)
or in the two-flavor regime ($10^9\gev \lesssim M_1\lesssim 10^{12}\gev$), 
but disappear in the three-flavor regime, where the full flavor basis
is resolved, since no direction in flavor space is protected from
the $N_1$-washout. Recently, it has been also pointed out that 
the off-diagonal elements in the matrix $C^{\ell}$ introduced in 
Eq.~(\ref{Cl}) and which encodes the effects of all spectator processes
can lead to a reduction of the washout from $N_1$ inverse processes
as well \cite{Shindou:2007se}. One can therefore conclude that
the assumption $\k_{2\a}\ll \k_{1\a}$ is not valid in general, even when
$M_1 \ll M_2$.

The second aspect concerns the flavored \CP~asymmetries.
Even though the results obtained at the end
of Section~\ref{sec:general} in the particular cases $\O=R_{23}$ 
and $\O=R_{13}$ are still valid in the flavored case, i.e. 
$\ve_{1\a}=0$ for $\O=R_{23}$, and
$\ve_{2\a}=0$ for $\O=R_{13}$, there are other effects to keep in mind.
As we shortly mentioned in Section~\ref{sec:practice}, the flavored 
\CP~asymmetries $\ve_{2\a}$ are not necessarily
suppressed by factors $M_1/M_2$ compared to $\ve_{1\a}$,
as it is the case in the unflavored regime for $\ve_2$ compared
to $\ve_1$. This observation \cite{Blanchet:2006be} can also potentially 
contribute to enlarge
the domain of applicability of the $N_2$-dominated scenario when flavor
effects are taken into account. Another related observation is that
the $\ve_{3\a}$'s, contrarily to $\ve_3$, are not suppressed
in the HL. This could open the door even to a $N_3$-dominated
scenario, although  this is possible only for
$M_3\lesssim 10^{12}\,{\rm GeV}$, when flavor effects affect the
generation of asymmetry by $N_3$.

Therefore, when flavor effects are taken into account,
the conditions of applicability
of the $N_1$-dominated scenario become potentially more restrictive
than in the unflavored case.
There is a clear choice of the parameters,
for $\O=R_{13}$ and $M_3\gtrsim 10^{12}\,{\rm GeV}$,
where the $N_1$-dominated scenario holds. Indeed, in this case,
one has that $\ve_{2\a}=0$ and $\ve_3$ is suppressed as $M_1/M_3$. 
This can be considered somehow
opposite to the case $\O=R_{23}$, where the $N_2$-dominated
scenario holds~\cite{DiBari:2005st}.

In general, one can say that the asymmetry produced
from the two heavier RH neutrinos is non-negligible
if two conditions are satisfied:
\begin{itemize}
\item[(i)]
 The asymmetry generated from $N_{2,3}$-decays
at $T\sim M_{2,3}$ is non-negligible compared to the
asymmetry generated at $T\sim M_1$ from $N_1$-decays.
This depends on an evaluation of the $C\!P$ asymmetries
$\ve_{2,3}^{\a}$ and of the washout due to the same
$N_{2,3}$ inverse processes.
\item[(ii)]
The asymmetry produced from $N_{2,3}$-decays is not
afterwards washed out by $N_1$ inverse processes.
Notice that this second condition is subordinate to
the first condition.
\end{itemize}

\section[Effects of $|\O_{22}|$]{Effects of \boldmath{$|\O_{22}|$}}
\label{sec:O22}

The general formula for the total \CP~asymmetry $\ve_1$ was given 
in Eq.~(\ref{CPas}) with $i=1$. It can be rewritten in the illustrative
form Eq.~(\ref{ve1}), where the second term was said to be usually
negligible for hierarchies larger than $3 M_1 \lesssim M_2$.
In fact, there is a situation where the term $[\xi(x_3)-\xi(x_2)]\D \ve_1$ 
in Eq.~(\ref{ve1}) becomes dominant over $\xi(x_2)\ve_1^{\rm HL}$ and one has to require 
a stronger hierarchy than $3M_1 \lesssim M_2$ to 
recover the results presented in Section~\ref{sec:N1DS}~\cite{Hambye:2003rt}.
From another perspective one can say that the second term in Eq.~(\ref{ve1})
offers a way to evade the bound on the \CP~asymmetry Eq.~(\ref{CPbound}) 
for mild hierarchies.

The term proportional to $\D \ve_1$ is maximized when $x_3\gg 1$, implying
$\xi(x_3)\simeq 1$. Moreover, if for definiteness one imposes 
$\O_{21}=0$ and ${\rm Re}[\O_{31}^2]=0$, so that $\sin \d_{\rm L}=1$
[cf.~Eq.(\ref{sindelta})], then
\be
\xi_{\ve_1}\equiv {\ve_1 \over \bar{\ve}(M_1)}=\xi(x_2)+
[1-\xi(x_2)]({\rm Re}[\O_{22}^2]+\mt/m_{\rm atm}\,{\rm Im}[\O_{22}^2]).
\ee 
One can immediately see that when ${\rm Re}[\O_{22}^2]=1$ and 
${\rm Im}[\O_{22}^2]=0$, corresponding to $\O=R_{13}$, one recovers $\xi_{\ve_1}=1$,
independently of the value of $x_2$. Notice also that when $M_2=M_3$, the
$\D \ve_1$ term vanishes. However, one can now perceive
another possibility: if $|\O_{22}|\gg 1$, then the \CP~asymmetry can
be enhanced, i.e. $\xi_{\ve_1}>1$, even when $x_2\gg 1$.

This possibility is interesting because having large $|\O_{22}|$ values
does not imply that the washout is large. As a matter of fact, the
decay parameter $K_1$ does not depend on $\O_{22}$. So there can
be an enhancement of the \CP~asymmetry without any enhancement
of the washout.  This observation was made in~\cite{Hambye:2003rt} and 
supplemented with theoretical motivations for models with large 
$|\O_{22}|$ in~\cite{Raidal:2004vt}. These models make
possible to evade the bound on the \CP~asymmetry Eq.~(\ref{CPbound}) 
without reducing the efficiency factor, thus relaxing the lower
bounds Eqs.~(\ref{lbM1min}) and (\ref{lbTin}). It was
found that values of $M_1\simeq 10^6\gev$ for mild hierarchies
$M_2/M_1\sim 10$ were possible, hence avoiding the gravitino problem.
Note moreover that models with large $|\O_{22}|$
can yield observable signals at experiments searching for 
lepton flavor violating decays $\m \to e\g$, $\t\to \m \g$ and 
$\t\to e\g$~\cite{Hayasaka:2007vc,Aubert:2006cz,PDBook}.

There is another important consequence of the term $\D \ve_1$, which
concerns the upper bound on the absolute neutrino mass scale,
Eq.~(\ref{m1max}). First, when the
absolute neutrino mass scale increases, it is well known that the upper 
bound on the \CP~asymmetry decreases [cf. Eq.~(\ref{CPbound})]. Thus,
a dominance of the $\D \ve_1$ term over the first term in Eq.~(\ref{ve1}) 
can be more easily obtained when $m_1$ is larger, 
even for moderate values $|\O_{22}|\sim 1$. The term $\D \ve_1$ is indeed
not suppressed at large $m_1$. Second, it can be argued
that when $m_1\sim 0.1~\ev$, the light neutrinos are quasi-degenerate,
so that quasi-degenerate heavy neutrinos with a mild degeneracy are more
natural. Under the assumption of `natural' mild degeneracy, the effect 
of the $\D \ve_1$ term has to be included and leads to a relaxed
upper bound $m_1\lesssim 0.6\ev$~\cite{Hambye:2003rt}. Note that 
the washout from the heavier RH neutrino $N_2$ must be
taken into account for degeneracies $\d_{21}\lesssim 0.1$, as emphasized
in Section~\ref{sec:beyondDL}. 


The bottom line is that there are situations where the $\D \ve_1$ term 
in Eq.~(\ref{ve1}) is crucial and hence should not be neglected. The
first situation is when a large $|\O_{22}|$ allows to evade the upper 
bound on the \CP~asymmetry even when $m_1 \to 0$ and for moderate 
hierarchies
$M_2/M_1\simeq 10$, without reducing the efficiency factor. The lower
bounds on $M_1$ and $T_{\rm reh}$ can then be relaxed by a few orders
of magnitude. The second situation is when a quasi-degenerate mass spectrum 
for the heavy neutrinos is considered, $\d_{ji}\lesssim 1$, and 
$|\O_{22}|\sim 1$. Then the contribution from the $\D \ve_1$ term can be
large, especially for quasi-degenerate light neutrinos. 

We think it is judicious to recall that Eq.~(\ref{ve1}) is only a convenient 
rewriting of 
the general expression (\ref{CPas}). All the effects discussed in
this section are of course present in the general expression, but they can
be overseen when the basic assumptions behind some simplified
expressions like $\ve_1^{\rm HL}$ in Eq.~(\ref{ve1}) are forgotten, such as 
a sufficiently large hierachy \HL.



\chapter{Leptogenesis from low-energy 
\textbf{\textit{CP}}-violating phases}
\label{chap:OmReal}


The possibility of relating the \CP~violation required for 
successful leptogenesis with the one that could be seen in future
neutrino experiment is very attractive. However, in the context
of unflavored leptogenesis (see Chapter~\ref{chap:unflavored}), we
saw that the PMNS matrix, which includes the observable \CP-violating
phases [cf.~Eq.~(\ref{PMNSmatrix})], cancels out in the general case. 
Yet, even in this unfavorable situation, 
there have been attempts to relate low-energy \CP-violating
phases with the baryon asymmetry of the Universe produced through
leptogenesis ~\cite{Branco:2001pq,Branco:2002kt,Davidson:2003vc,Rebelo:2002wj,Endoh:2002wm,Frampton:2002qc,Branco:2002xf}. It must be however said that a link could only be 
made for specific textures of the Dirac mass matrix. 

As discussed in Chapter~\ref{chap:flavor}, flavor effects have modified
the conventional picture of leptogenesis in a number of ways. In particular,
the fact that the final asymmetry in the fully flavored regime depends
explicitly on the PMNS mixing matrix is interesting. Thanks to the
new source of \CP~violation implied by flavor effects, namely the $\D P$
contribution, Eq.~(\ref{veiaDP}), one can imagine a situation
where the \emph{only} source of \CP~violation comes from the PMNS 
matrix or, equivalently, where the contributions from the 
unobservable phases in the $\O$ matrix are negligible. This leads to the 
exciting possibility of explaining the baryon asymmetry of the Universe 
thanks to \CP-violating phases that are accessible in neutrino
experiments. This scenario has attracted some attention 
recently~\cite{Blanchet:2006be,Pascoli:2006ie,Branco:2006ce,Pascoli:2006ci,Anisimov:2007mw,Uhlig:2007xe,Molinaro:2007uv}.

In this chapter we first introduce the concepts of low-energy \CP~violation
due to the Dirac phase and to the Majorana phases. Then, we study the 
possibility
that the Dirac phase, which offers the best prospects for a possible
measurement in the next years, provides the unique source of \CP~violation 
required for leptogenesis. This can be considered as the
most conservative case, since we know that the Dirac phase comes
always together with the small angle $\q_{13}$ [cf. Eq.~(\ref{PMNSmatrix2})].
Afterwards, we shortly comment on the role played by the Majorana
phases as sole sources of \CP~violation in leptogenesis. Finally, we
discuss the theoretical relevance of models where $\O$ is real.

\section[\textit{CP} violation in neutrino physics]{\textbf{\textit{CP}} violation in neutrino physics}

\subsection{Neutrino oscillations and the Dirac phase}

Searching for \CP-violating effects in neutrino oscillations is the
only practical way to get information about Dirac \CP~violation
in the lepton sector, associated with the phase $\d$ in the PMNS mixing
matrix $U$ [cf.~Eq.~(\ref{PMNSmatrix})]. A measure of \CP~and $T$ violation
is provided by the asymmetries~\cite{Cabibbo:1977nk,Bilenky:1980cx,Barger:1980jm}
\bea
A^{(\a,\a')}_{\rm \it CP}&=&P(\n_{\a}\to \n_{\a'})-P(\bar{\nu}_{\a}\to \bar{\nu}_{\a'}),\NO\\
A^{(\a,\a')}_{T}&=&P(\n_{\a}\to \n_{\a'})-P(\n_{\a'}\to \n_{\a}),\NO
\eea
where $\a\neq \a'=e,\m,\t$. For three-neutrino oscillations in vacuum, which respect
the $C\!P T$~symmetry, one has~\cite{Krastev:1988yu}
\bea
A^{(e,\m)}_{T}&=&A^{(\m,\t)}_{T}=-A^{(e,\t)}_{T}=J_{\rm \it CP}F^{\rm vac}_{\rm osc},
\quad A^{(\a,\a')}_{\rm \it CP}=A^{(\a,\a')}_T,\NO \\
J_{\rm \it CP}&=&{\rm Im}\left(U_{e1}U_{\m 2}U_{e2}^{\star}U_{\m 1}^{\star}\right)
={1\over 4}\sin 2\q_{12}\sin 2\q_{23} \cos^2\q_{13} \sin\q_{13} \sin\d,\NO\\
F^{\rm vac}_{\rm osc}&=&\sin \left({\D m_{21}^2\over 2E}x\right)+
\sin \left({\D m_{32}^2\over 2E}x\right)+\sin \left({\D m_{13}^2\over 2E}x\right)\NO,
\eea
where $x$ is the distance travelled by the neutrinos, and $E$ is their common energy
[cf.~Eq.~(\ref{oscprob})].
Thus, the magnitude of the \CP-violating effects in neutrino oscillations
is controlled by the rephasing invariant associated with the Dirac phase $\d$,
the so-called \emph{Jarlskog invariant}
$J_{\rm \it CP}$~\cite{Jarlskog:1985ht}. The existence of Dirac \CP~violation in the lepton
sector would be established if, e.g., some of the vacuum oscillation asymmetries
$A^{(\a,\a')}_{\rm \it CP,T}$ are proven experimentally to be non-zero.
This would imply that $J_{\rm \it CP}\neq 0$ and, consequently, that $\sin\q_{13}
\sin\d\neq 0$. Without doubt, the search for \CP-violating effects due to
the Dirac phase in $U$ represents one of the major goals of the future 
experimental studies of neutrino oscillations~\cite{Petcov:2005yy},
in experiments like T2K~\cite{Itow:2001ee} or NO$\n$A~\cite{Ayres:2004js}
as well as in the (far) future neutrino factories and/or beta
beam experiments~\cite{Albright:2004iw}. 

\subsection{Neutrinoless double-beta decay and the Majorana phases}

Theories with neutrino mass generation of the see-saw type predict 
the massive neutrinos $\n_j$ to be Majorana particles. Determining
the nature of massive neutrinos is one of the most formidable and 
pressing problems in today's neutrino physics (see, e.g., \cite{Petcov:2005yy,Aalseth:2004hb,Morales:2002zf}). Even if neutrinos are proven to be Majorana
fermions, getting information about the Majorana \CP-violating
phases in $U$, $\F_1$ and $\F_2$ [cf. Eq.~(\ref{PMNSmatrix})], will
be very difficult. Moreover, the oscillations of flavor neutrinos,
$\n_{\a}\to\n_{\a'}$ and $\bar{\n}_{\a}\to\bar{\n}_{\a'}$, $\a,\a'=e,\m,\t$,
are insensitive to the Majorana phases~\cite{Bilenky:1980cx,Langacker:1986jv}. 
On the other hand, they can affect significantly the predictions for the rates
of lepton-flavor-violating decays $\m\to e+\g,~\t\to \m+\g$, etc. in a 
large class of supersymmetric theories with type-I see-saw mechanism
(see, e.g., \cite{Pascoli:2003rq,Petcov:2005yh,Petcov:2006pc}).

In the case of three-neutrino mixing under discussion, there are, in principle,
three independent \CP~violation rephasing invariants. The first
is $J_{\rm \it CP}$, the Dirac one, associated with the Dirac phase $\d$,
which we discussed in the previous subsection. The existence of two
additional invariants, $S_1$ and $S_2$, is related to the two Majorana
\CP-violating phases in $U$. The invariants $S_1$ and $S_2$ can be chosen
as~\cite{Nieves:1987pp}
\be
S_1={\rm Im}\left(U_{\t 1}^{\star}U_{\t 2}\right),\quad S_2={\rm Im}\left(U_{\t 2}^{\star}
U_{\t 3}\right).
\ee
The rephasing invariants associated with the Majorana phases are not uniquely
determined. Instead of the flavor $\t$ involved in both $S_1$ and $S_2$, one
could have chosen $e$ or $\m$. Note also that \CP~violation due to
the Majorana phases imply that both $S_1={\rm Im}\left(U_{\t 1}^{\star}U_{\t 2}\right)\neq 0$
and ${\rm Re}\left(U_{\t 1}^{\star}U_{\t 2}\right)\neq 0$, and similarly for $S_2$.

The only feasible experiments which, at present, have the potential of 
establishing the Majorana nature of light neutrinos and of providing information
on the Majorana \CP-violating phases in $U$ are the experiments searching
for neutrinoless double-beta ($0\n\b\b$) decay, $(A,Z)\to (A,Z+2)+2e^-$. 
The  $0\n\b\b$-decay effective Majorana mass, $|\langle m\rangle |$, 
as defined in Eq.~(\ref{effmass}), which
contains all the dependence of the $0\n\b\b$-decay amplitude on the neutrino
mixing parameters, is given by the following expressions for a normal
hierarchical (NH, $m_1\ll m_2 \ll m_3$), inverted hierarchical (IH,
$m_1\ll m_2 \ll m_3$, but the $U$ matrix changes, as explained 
in Appendix~\ref{chap:param}) 
and quasi-degenerate (QD, $m_{1,2,3}\simeq m\gtrsim 0.1\ev$) neutrino
mass spectra:
\bea
|\langle m\rangle |&\simeq& \left|\sqrt{\D m^2_{\rm sol}} \sin^2 \q_{12} {\rm e}^{{\rm i}\F_2}
+\sqrt{\D m^2_{\rm atm}} \sin^2 \q_{13} e^{-2i\d}\right|,~{\rm (NH)},\\
|\langle m\rangle |&\simeq&\sqrt{\D m^2_{\rm atm}}  \left|\cos^2\q_{12}{\rm e}^{{\rm i}\F_1}+
{\rm e}^{{\rm i}\F_2}\sin^2 \q_{12} \right|,~{\rm (IH)},\\
|\langle m\rangle |&\simeq& m \left|\cos^2\q_{12}{\rm e}^{{\rm i}\F_1}+
{\rm e}^{{\rm i}\F_2}\sin^2 \q_{12} \right|,~{\rm (QD)}.
\eea
Obviously, $|\langle m \rangle |$ depends strongly on the Majorana phases:
the \CP-conserving values of $\F_2-\F_1=0,\pm \p$~\cite{Wolfenstein:1981rk}, 
for instance, determine
the range of possible values of $|\langle m \rangle |$ in the case of 
IH and QD spectra, while the \CP-conserving values of $\F_2=0,\pm \p$,
can be important in the case of NH spectrum.

The planned $0\n\b\b$-decay experiments of the next generation such as 
GERDA~\cite{Abt:2004yk}, MAJORANA~\cite{Aalseth:2004yt} or 
CUORE~\cite{Ardito:2005ar} are
aiming to probe the QD and IH ranges of $|\langle m \rangle |$ (see, e.g.,
\cite{Morales:2002zf,Aalseth:2004hb}). If the 
$0\n\b\b$-decay is observed in these experiments, the measurement
of the $0\n\b\b$-decay half-life might then allow to obtain constraints
on the combination of Majorana phases $\F_2 -\F_1$.

\section{Dirac phase leptogenesis}
\label{sec:DiracLep}

In this section we would like to investigate in detail a scenario of
leptogenesis where the \CP~violation necessary for leptogenesis
is exclusively provided by the Dirac phase $\d$, hence the name
$\d$-leptogenesis. Following closely~\cite{Anisimov:2007mw}
we start analysing the case where the heavy neutrino mass spectrum 
is hierarchical, i.e. 
$M_1\ll M_2 \ll M_3$, which, in the fully flavored regime, does not 
necessarily lead to a $N_1$-dominated scenario. Therefore, we shall
not only consider the decay of the lightest RH neutrino, but that 
of all three RH neutrinos. Then, we will assume a quasi-degenerate
mass spectrum for the heavy neutrinos, $M_1\simeq M_2 \simeq M_3$, and perform
the same analysis. 

\subsection{The hierarchical limit}
\label{sec:DiracHL}

Throughout this section, we will always assume that the fully flavored regime
holds. However, the condition of validity of the fully flavored regime,
Eq.~(\ref{M1fl}), is qualitative. One should not expect it to be exact. 
For example, this simple condition neglects the effect of $\D L=1$ scatterings
and of refractive effects, the first contributing with inverse decays
to preserve the quantum state coherence, the second, conversely,
in projecting it on the flavor basis.
Both of them can be as large as the effect from inverse decays.
Moreover, in a rigorous quantum kinetic description,
it is likely that other subtle effects will contribute to the determination of the
exact value of $M_1$ below which the fully flavored regime can be assumed.
Only making a precise study of the transition region between the unflavored 
and the fully flavored regimes solving the full system of equations, which includes
the density matrix equation (\ref{rhoL}), will allow to derive a better 
condition of validity.

In the plots showing the lower bound on $M_1$, we will then distinguish 
four regions. All plots will be cut at $M_1 = 10^{12}\,{\rm GeV}$,
since above this value, according to the condition~(\ref{unflavored}),
the unflavored regime is recovered and the asymmetry production
has to switch off, since only the phases in the $\O$ matrix can make the
total \CP~asymmetries $\ve_i$ non-zero.
On the other hand, when the condition~(\ref{M1fl}) is satisfied,
one can expect the fully flavored regime to hold. There is an intermediate
regime where a transition between the fully flavored regime
and the unflavored regime takes place.
This regime will be indicated in all plots with a squared region.
This signals that, even though we still
show the results obtained in the fully flavored regime, important
corrections are expected, especially when $M_1$ gets close to
$\sim 10^{12}\,{\rm GeV}$. Since this region describes
a transition towards the unflavored regime, where the asymmetry
production has to switch off, these corrections
are expected to reduce the final asymmetry,
making more stringent the lower bounds shown in the plots.
Furthermore, since large corrections
to the condition~(\ref{M1fl}) cannot be excluded,
we will also indicate, with a hatched region, the area where
 the condition~(\ref{M1fl}) holds but a very conservative condition,
\be\label{conservative}
M_1\lesssim {10^{11}\,{\rm GeV}\over W_1^{\rm ID}(z_{\rm B}(K_{1\a}))} \,
\ee
does not.
In this region some corrections to the results presented
cannot be excluded, but the fully flavored regime should
represent a good approximation.

We anticipate that, in the $N_1$-dominated scenario,
successful leptogenesis always requires $M_1\gtrsim 10^9\,{\rm GeV}$,
where the two-flavor regime applies. Therefore, considering
that we are assuming $\ve_1=\ve_{1\t}+\ve_{1,e+\m}=0$,
Eq.~(\ref{N1DS}) can be specialized into
\be\label{final}
\left. N_{B-L}^{\rm f} \right|_{N_1}
\simeq (\k_{1\,\t}^{\rm f}-\k_{1,e+\m}^{\rm f})\,\ve_{1\t} \, ,
\ee
showing that, in order to have a non-vanishing final asymmetry
it is necessary to have that $P_{1\t}^0 \neq P_{1,e+\m}^0$. Indeed,
the efficiency factors crucially depend on $K_{1\a}$. Useful 
analytical expressions for the flavored efficiency factors, both 
for a vanishing and a thermal 
initial $N_1$-abundance can be found in Section~\ref{sec:dependence}.
As for the flavored decay parameters in the usual orthogonal 
parametrization, Eq.~(\ref{CI}), they were given in Eq.~(\ref{Kialpha}).

The parameter $r_{1\a}$ defined in Eq.~(\ref{r1a}) is convenient to
describe the behavior of the \CP~asymmetry $\ve_{1\a}$.
Using Eq.~(\ref{e1alOm}) for real $\O$, one obtains~\cite{Abada:2006ea}
\be
r_{1\a}=-\sum_{h< l}\,
{\sqrt{m_l\,m_h}\,(m_l-m_h)\over \mt\,m_{\rm atm}}\,
\O_{h1}\,\O_{l1}\,{\rm Im}[U_{\a h}\,U_{\a l}^{\star}] \, .
\ee
Taking $\a=\t$ and specifying  the matrix elements $U_{\a j}$
from Eq.~(\ref{PMNSmatrix}), one has
\be\label{r1tau}
r_{1\t}=-{m_{\rm atm}\over \mt}\,[A_{12}+A_{13}+A_{23}],
\ee
where, in the case of normal hierarchy for the light neutrinos\footnote{
For an inverted hierarchy, the elements $A_{ij}$ must be computed 
using a modified $U$ matrix, as explained at the end of Appendix~\ref{chap:param}.},
\bea\nonumber
A_{12} & = & - {\sqrt{m_1\,m_2}\,(m_2-m_1) \over m_{\rm atm}^2}
\,\O_{11}\,\O_{21}\,
{\rm Im}[(s_{12}\,s_{23}-c_{12}\,c_{23}\,s_{13}\,{\rm e}^{{\rm i}\,\d}) \\ \nonumber
& & \;\;\;\;\;\;\;\;\;\;\;\;\;\;\;\;\;\;\;\;\;\;\;\;
\times \,(c_{12}\,s_{23}+s_{12}\,c_{23}\,s_{13}\,{\rm e}^{-{\rm i}\,\d})\,
{\rm e}^{-{{\rm i}\over 2}\,(\Phi_2-\Phi_1)}]\, , \\ \nonumber
A_{13} & = & {\sqrt{m_1\,m_3}\,(m_3-m_1)\over m_{\rm atm}^2}
\,\O_{11}\,\O_{31}\,c_{23}\,c_{13}\,
{\rm Im}[(s_{12}\,s_{23}-c_{12}\,c_{23}\,s_{13}\,{\rm e}^{{\rm i}\,\d})
{\rm e}^{{{\rm i}\over 2}\,\Phi_1}] \, ,
\\ \nonumber
A_{23} & = & - {\sqrt{m_2\,m_3}\,(m_3-m_2) \over m_{\rm atm}^2}
\,\O_{21}\,\O_{31}\,c_{23}\,c_{13}\,
{\rm Im}[(c_{12}\,s_{23}+s_{12}\,c_{23}\,s_{13}\,{\rm e}^{{\rm i}\,\d})
\,{\rm e}^{{{\rm i}\over 2}\,\Phi_2}]\, .
\eea
In the case of $\d$-leptogenesis ($\Phi_1=\Phi_2=0$),
these expressions further specialize into
\bea\nonumber
A_{12} & = & {\sqrt{m_1\,m_2}\,(m_2-m_1)\over m_{\rm atm}^2}
\,\O_{11}\,\O_{21}\,s_{23}\,c_{23}\,\D \, ,
\\ \nonumber
A_{13} & = & -{\sqrt{m_1\,m_3}\,(m_3-m_1)\over m_{\rm atm}^2}
\,\O_{11}\,\O_{31}\,c_{23}^2\,c_{12}\,c_{13}\,\D  \, ,
\\ \nonumber
A_{23} & = & - {\sqrt{m_2\,m_3}\,(m_3-m_2)\over m_{\rm atm}^2}
\,\O_{21}\,\O_{31}\,c_{23}^2\,s_{12}\,c_{13}\,\D \, ,
\eea
where we defined $\D\equiv \sin\theta_{13}\,\sin\d$.

It is now instructive to make some general considerations.
Looking at the expression Eq. (\ref{final}), one can see that
in order for the final $B\!-\!L$ asymmetry to be non-zero, two
conditions have to be simultaneously satisfied :
$\ve_{1\t}\neq 0$ and $\k_{1\t}^{\rm f} \neq \k_{1,e+\m}^{\rm f}$.
These two conditions are a specialization of two of Sakharov's
necessary conditions for baryogenesis to the case of $\d$-leptogenesis.

The first condition is the requirement to have
$C\!P$ violation and, as one could expect, from the
expressions found for the terms $A_{ij}$, one can have
$\ve_{1\t}\neq 0$ only if $\D\neq 0$.

The second condition is a specialization of the condition of
departure from thermal equilibrium in quite a non-trivial way.
Indeed, in the case of $\d$-leptogenesis, in a fully out-of-equilibrium
situation where only decays are active, no final asymmetry is generated
since $\ve_1=0$, implying that there is an equal number
of decays into leptons and antileptons. However, the presence of inverse
processes can remove this balance, yielding a different washout rate
for the $\t$ asymmetry
and for the $e+\m$ asymmetry, so that, if $K_{1\t}\neq K_{1,e+\m}$,
one has a dynamical net lepton number generation. From the flavored decay
parameters Eq.~(\ref{Kialpha}), it can be seen that this is possible independently of the
value of the Dirac phase, which is therefore directly responsible only
for $C\!P$ violation and not for lepton number violation,
exactly as in neutrino mixing, where lepton number is conserved.

It should also be noticed that the $\ve_{1\a}$'s  are expressed through
quantities ${\rm Im}[U_{\a h}\,U^{\star}_{\a l}]$ that are invariant
under a change of the PMNS matrix parametrization \cite{Nieves:1987pp,Pascoli:2006ie}.
Actually, the \CP~asymmetry itself is an invariant 
quantity~\cite{King:2006hn}. Therefore, the final $B\!-\!L$ asymmetry depends 
only on physical quantities, as it should be. 

Maximizing the asymmetry over all involved parameters
at fixed $M_1$ and $K_1$ and
imposing $\eta_B^{\rm max} \geq \eta_B^{\rm CMB}$ [cf. Eqs.~(\ref{asymgeneral}) and
(\ref{WMAP})], a lower bound on $M_1$ is obtained, which can be conveniently
expressed as in Eq.~(\ref{lbM1}).
Notice that $r_{1\t}\propto \D$, implying $N_{B-L}^{\rm f}\propto \D$ as well.
Therefore, the maximum asymmetry is obtained for $|\d|=\pi/2$
and $s_{13}=0.20$.


As we explained in Section~\ref{sec:N2}, accounting for flavor
effects it is not clear whether
the contribution from the lightest RH neutrino, $N_1$, is the only one that 
matters. Since we want our
analysis to be as general as possible, we will always estimate the asymmetry
produced by the two heavier neutrinos $N_2$ and $N_3$.  

The calculation of the contribution to the asymmetry from $N_2$-decays
proceeds in an analogous way. It can be again calculated in the
two-flavor regime, since, in the HL, successful leptogenesis always implies
\mbox{$M_2\gtrsim 10^{9}\,{\rm GeV}$}.
Therefore, one can write an expression similar to Eq.~(\ref{final})
for the contribution to the final asymmetry from $N_2$-decays,
\be\label{final2}
\left.N_{B-L}^{\rm f}\right|_{N_2}\simeq
(\k_{2\,\t}^{\rm f}-\k_{2,e+\m}^{\rm f})\,\ve_{2\t} \, .
\ee
The difference is now in the calculation of the efficiency factors
as they are suppressed by the washout from the $N_1$ inverse
processes. In the HL this additional washout factorizes and
\cite{Blanchet:2006dq,Blanchet:2006be,Vives:2005ra}
\be
\k_{2\alpha}^{\rm f}\simeq \k(K_{2\alpha})\,
\exp\left(-{3\,\pi\over 8}\,K_{1\a}\right)\, ,
\ee
where $K_{2\a}\equiv P_{2\a}^0\,K_2$. The tree-level projectors
 $P_{2\a}^0$ can be readily evaluated using the general
expression~(\ref{proj}).


The calculation of the contribution to the final asymmetry
from $N_3$-decays proceeds in a similar way and analogous
expressions hold. The only non-trivial difference is that
now, in the calculation of the efficiency factors,
one has also to include the washout from $N_2$ inverse processes,
so that
\be\label{k3a}
\k_{3\alpha}^{\rm f}\simeq \k(K_{3\alpha})\,
\exp\left[-{3\,\pi\over 8}\,(K_{1\a}+K_{2\a})\right]\, .
\ee
Notice that in the calculation of $\k_{2\a}^{\rm f}$
($\k_{3\a}^{\rm f}$) we are not including a possible effect where
part of the
asymmetry in the flavor $\a=e+\m$ produced in $N_2$- or $N_3$-decays is
orthogonal to $N_1$ inverse decays \cite{Barbieri:1999ma,Engelhard:2006yg} and is
not washed out. This washout avoidance does 
not apply to the asymmetry in the $\t$-flavor. Since in all cases we
shall consider, a $\t$-dominated scenario will be realized, we shall simply 
neglect this effect. For a short discussion about these effects, see
Section~\ref{sec:N2}.


Let us now calculate the final asymmetry in some interesting cases.

\subsubsection*{$\O=R_{13}$}

The first case we consider is $\O=R_{13}$ [cf.~Eq.~(\ref{R13})], implying $A_{12}=A_{23}=0$
in Eq.~(\ref{r1tau}). It is easy to check from Eq.~(\ref{ve2a}) that 
$\ve_{2\t}=0$, and therefore there is no
asymmetry production from $N_2$-decays even if $M_2\lesssim 10^{12}\,{\rm GeV}$.
On the other hand, one obtains
\be
r_{3\tau}=
-{2\over 3}\,{\sqrt{m_1\,m_3}\,(m_3-m_1) \over \mttt\,m_{\rm atm}}\,
\o_{31}\,\sqrt{1-\o_{31}^2}\,c_{12}\,c_{23}^2\,c_{13}\,\D \, ,
\ee
essentially the same expression as
for $r_{1\t}$ but with $\mt$ replaced by $\mttt$.
Therefore, for $M_3\lesssim 10^{12}\,{\rm GeV}$,
one has to worry about a potential non-negligible contribution
from $N_3$-decays. However, when the washout
from $N_1$ and $N_2$ inverse processes is taken into account [cf.~Eq.~(\ref{k3a})],
 we always find that the
contribution from $N_3$-decays is negligible and
the $N_1$-dominated scenario holds.

\begin{figure}
\centering
\includegraphics[width=0.32\textwidth]{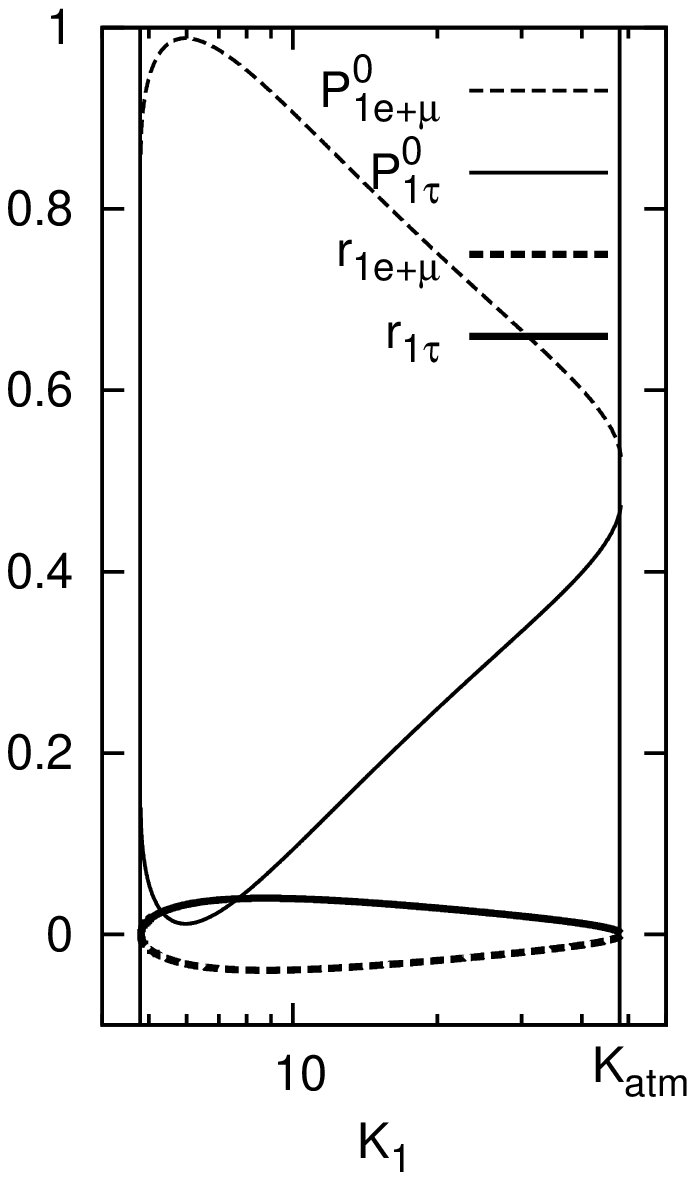}
\includegraphics[width=0.32\textwidth]{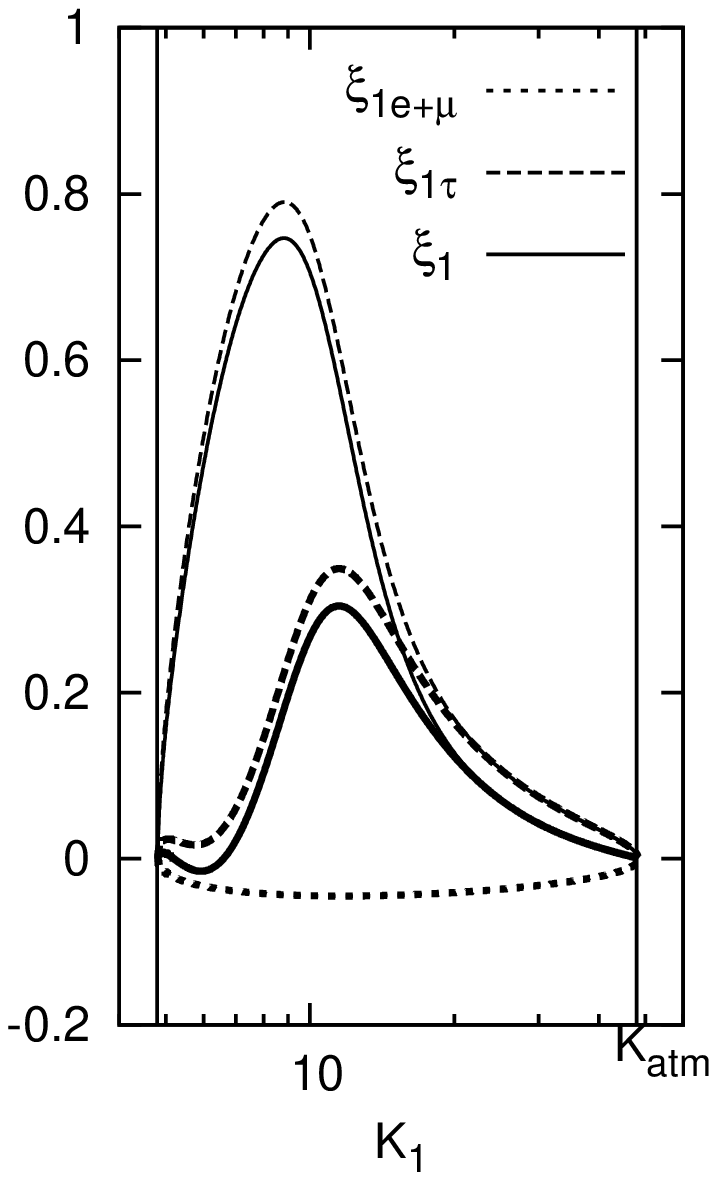}
\includegraphics[width=0.32\textwidth]{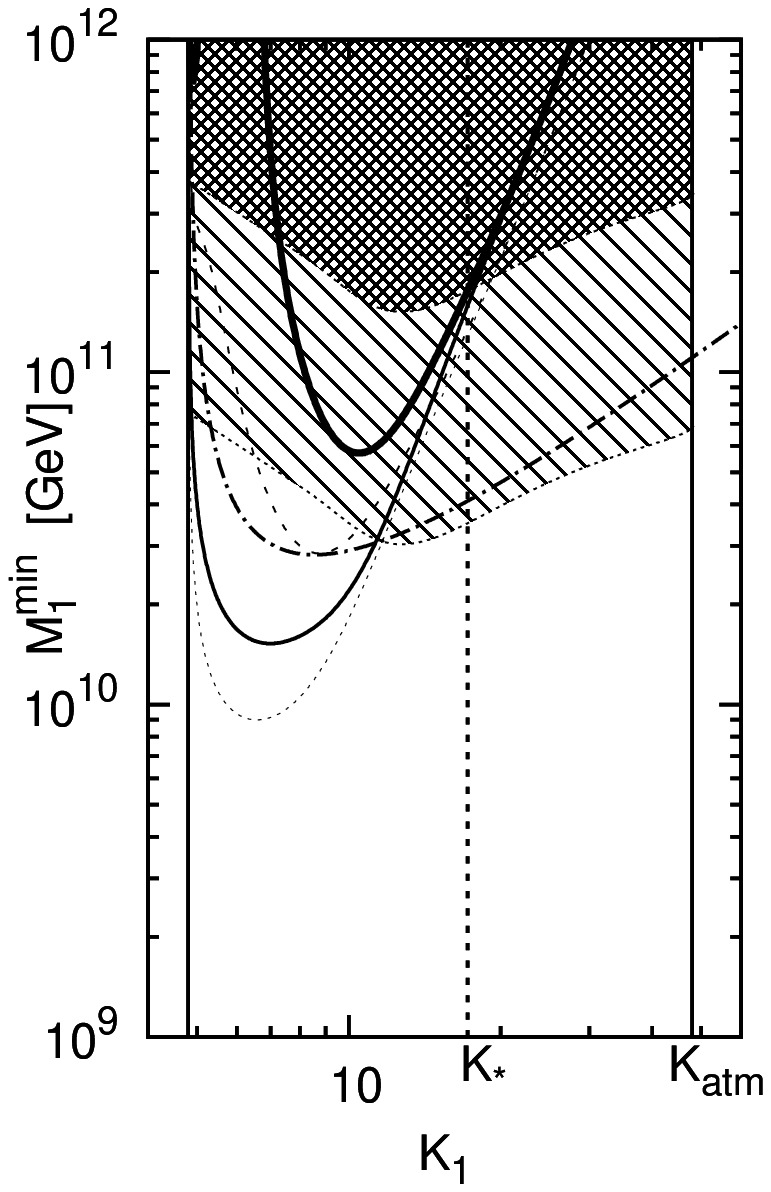}
\caption{Dependence of different quantities on $K_1$ for
$m_1/m_{\rm atm}=0.1$, $s_{13}=0.2$, $\d=-\pi/2$ and real $\O=R_{13}$
with $\o_{31}<0$.
Left panel: projectors $P^0_{1\a}$ and normalized \CP~asymmetries $r_{1\a}$;
central panel: $\xi_{1\a}$ and $\xi_1$ as
defined in Eq.~(\ref{xi1a}) for thermal (thin)
and vanishing (thick) initial $N_1$-abundance;
right panel: lower bound on $M_1$ for thermal (thin solid) and
vanishing (thick solid) initial $N_1$-abundance compared with the unflavored
result (dash-dotted line) obtained for complex $\O=R_{13}$.
In the squared region the condition~(\ref{M1fl}) is not
satisfied, and in the hatched region even the more
conservative condition~(\ref{conservative}) is not satisfied.
The dotted lines (thick for vanishing and thin
for thermal initial $N_1$-abundance) correspond still to a real
$\O=R_{13}$ but this time $\d=0$ while the only non-vanishing low-energy
phase is the Majorana phase \mbox{$\Phi_1=-\pi/2$}.}
\label{fig:HLR13}
\end{figure}

The results are shown in Fig.~\ref{fig:HLR13} for $s_{13}=0.20$, $\d=-\pi/2$
and $m_1/m_{\rm atm}=0.1$, a choice of values that approximately
maximizes the final asymmetry and yields the lower bound 
$M_1^{\rm min}(K_1)$.
In the left panel we show the tree-level projectors
$P^0_{1\a}$ and the normalized \CP~asymmetries $r_{1\a}$ [cf.~Eq.~(\ref{r1a})].
It can be seen how for $K_1\gg 10$ one has
$P^0_{1\t}\simeq P^0_{1,e+\m}\simeq  1/2$,
while for $K_1\sim 10$ one has $P^0_{1\t}\ll P^0_{1,e+\m}$.
In the central panel $\xi_1$ and the $\xi_{1\a}$'s  are plotted [cf.~Eq.~(\ref{xi1a})], 
and one can
see how for $K_1\simeq 10$ a $\t$-dominance is realized. Finally, in the
right panel, we show $M_1^{\rm min}(K_1)$ and we compare it with the
lower bound from an unflavored calculation where $\o_{31}^2$ is taken
purely imaginary \cite{DiBari:2005st}.
One can see how, at $K_1\gg 10$, the asymmetry production
rapidly dies, so that $\xi_1\ra 0$ and $M_1^{\rm min}(K_1)\ra \infty$.
 Notice that we have plotted the lower bound both
for thermal and vanishing initial $N_1$-abundances. 

We also indicated $K_{\star}$,
defined as that value of $K_1$ such that for $K_1\gtrsim K_{\star}$
the dependence on the
initial conditions can be neglected and the strong washout regime
holds. One can notice that the intermediate regime between a fully flavored
regime and the unflavored regime, the squared area, is quite extended.
In this regime corrections to the results we are showing, obtained in the
 fully flavored regime, are expected in a way that the unflavored regime
should be recovered for $M_1\rightarrow 10^{12}\,{\rm GeV}$. In this limit
the asymmetry production has to switch off and thus one expects
that the lower bound on $M_1$ becomes more restrictive
and eventually, for $M_1\rightarrow 10^{12}\,{\rm GeV}$,
the allowed region has to close up. One can see that there is essentially no 
allowed region in the strong washout regime outside the squared area.
The hatched area, where corrections cannot be excluded within current
theoretical uncertainties, cuts away almost completely any allowed region
even in the weak washout regime. 

In conclusion, the allowed
region where one can safely rely on the fully flavored regime
according to current calculations is very restricted and confined
only to a small region in the weak washout regime.


\subsubsection*{$M_3\gg 10^{14}\,{\rm GeV}$}

The second case we consider is the limit $M_3\gg 10^{14}\,{\rm GeV}$.
This corresponds to an effective 2-RH neutrino model, where the third RH
neutrino is decoupled. In this limit one has necessarily $m_1\ll m_{\rm sol}$,
implying $m_3\simeq m_{\rm atm}$, and the $\O$ matrix takes the special 
form~\cite{Frampton:2002qc,Ibarra:2003xp,Chankowski:2003rr}
\be\label{ss}
\O=
\left(
\begin{array}{ccc}
     0            &   0                 &  1 \\
\sqrt{1-\O^2_{31}}& -\O_{31}            &  0 \\
       \O_{31}    & \sqrt{1-\O^2_{31}}  &  0
\end{array}
\right)
\, .
\ee
Notice that this form of $\O$ corresponds to set
$\o_{32}=1$ and $\o_{21}=1$ in Eq.~(\ref{second}).
In the expression~(\ref{r1tau}) for $r_{1\t}$,
one has now $A_{12}=A_{13}=0$, and hence
\be
r_{1\t}\simeq
{m_{\rm atm}\over \mt}\,
\sqrt{m_2 \over m_{\rm atm}}\,
\left(1-{m_2\over m_{\rm atm}}\right)\,
\,\O_{31}\,\sqrt{1-\O^2_{31}}\,c_{23}^2\,c_{13}\,s_{12}\,\D \, .
\ee
If  $M_2\gtrsim 10^{12}\,{\rm GeV}$,
there is no contribution from the next-to-lightest RH neutrino decays, 
since these occur in the unflavored regime where $\ve_2\simeq 0$.
On the other hand, if $M_2\lesssim 10^{12}\,{\rm GeV}$, then one has
to worry about a (flavored) asymmetry generation from $N_2$-decays.
A calculation of $\ve_{2\a}$ shows that
the first term in Eq. (\ref{ve2a}) vanishes while the second term gives
\be
r_{2\t}=-{2\over 3}\,{m_{\rm atm}\over \mtt}\,
              \sqrt{m_2\over m_{\rm atm}}\,
               \left(1-{m_2\over m_{\rm atm}}\right)\,\O_{31}\,
               \sqrt{1-\O^2_{31}}\,c_{23}^2\,s_{12}\,\D \,.
\ee
This is an example of how the second term in Eq. (\ref{ve2a})
is not suppressed in the HL like the first term.
However, like for the contribution from $N_3$-decays
in the case $\O=R_{13}$, when the washout from $N_1$ inverse processes
is taken into account, one finds
$\left. N_{B-L}^{\rm f}\right|_{N_2}\ll \left.N_{B-L}^{\rm f}\right|_{N_1}$
and a $N_1$-dominated scenario is realized.

\begin{figure}
\centering
\includegraphics[width=0.32\textwidth]{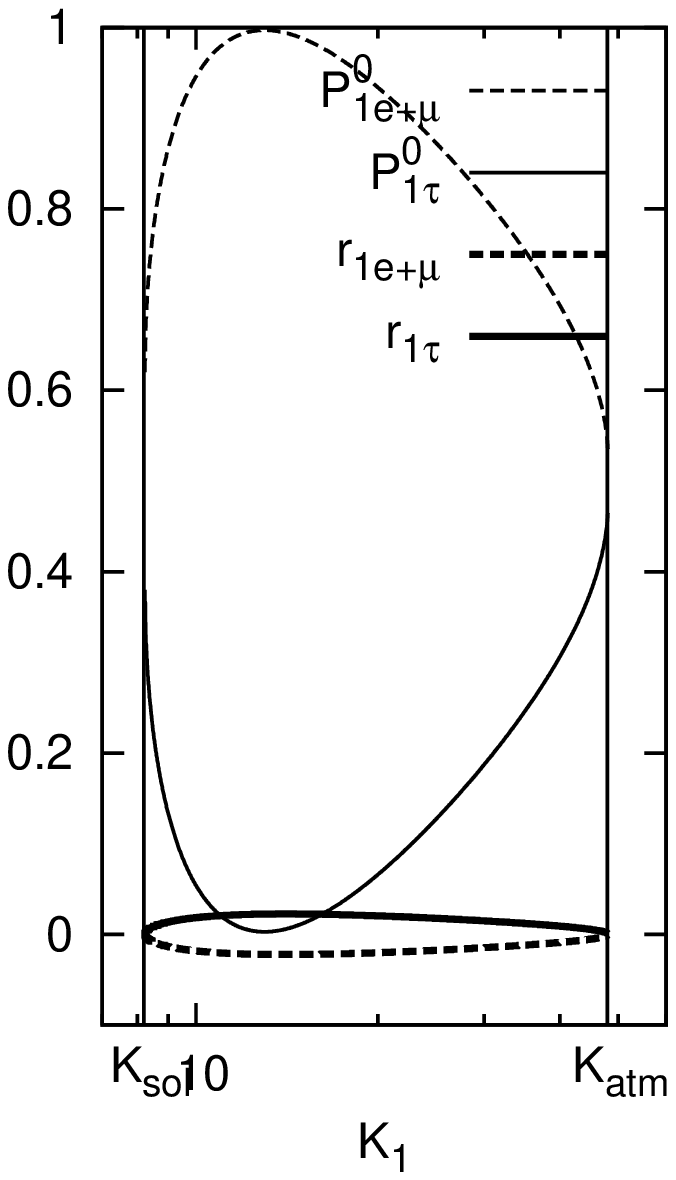}
\includegraphics[width=0.32\textwidth]{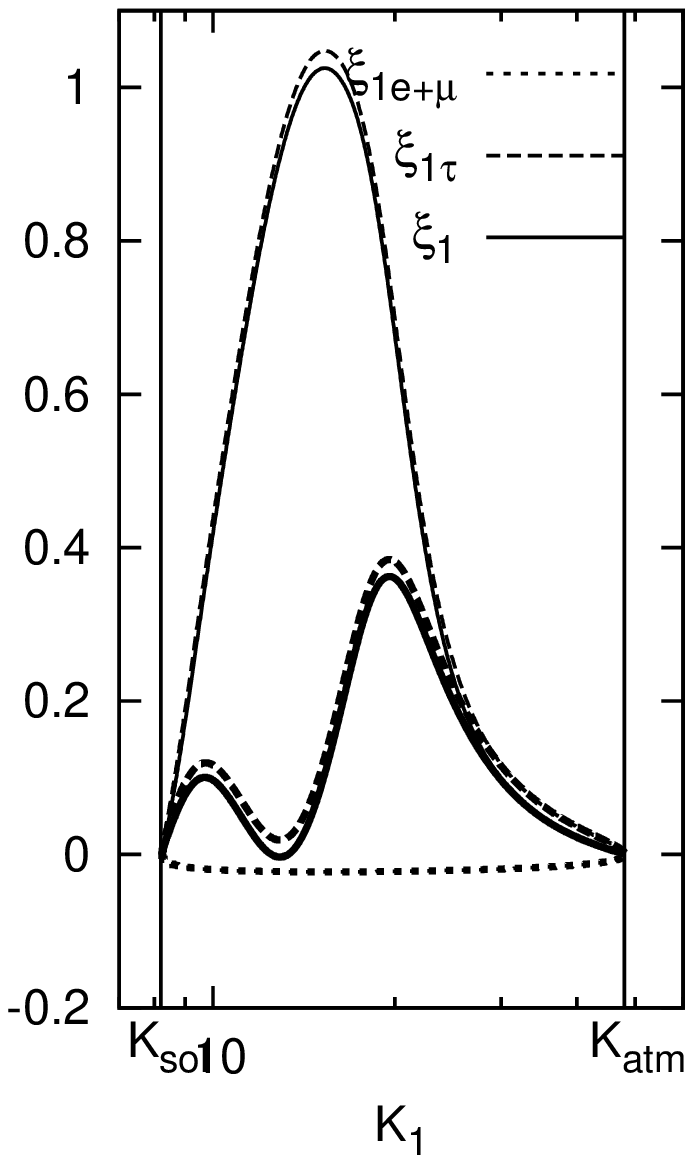}
\includegraphics[width=0.32\textwidth]{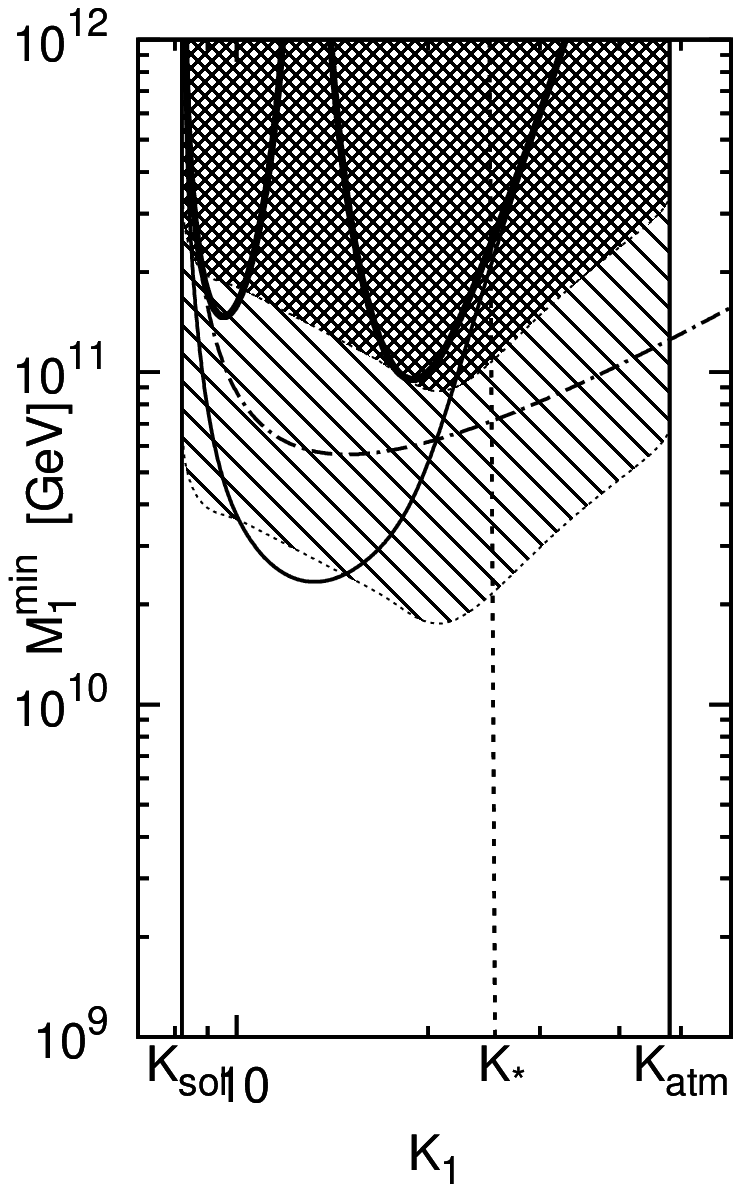}
\caption{Same quantities as in Fig.~\ref{fig:HLR13} but for the case
$M_3\gg 10^{14}\,{\rm GeV}$, corresponding to the special
form of $\O$ in Eq.~(\ref{ss}). Here we are moreover
assuming $M_2\gtrsim 10^{12}\,{\rm GeV}$ and a normal hierarchy
for the light neutrinos.
The lower bound $M_1^{\rm min}(K_1)$ is obtained for
$\o_{31}>0$ and $\d=\pi/2$.}
\label{fig:HLM3}
\end{figure}

Notice that there is a strong dependence whether one assumes
a normal or an inverted hierarchy. For normal hierarchy
the results are shown in Fig.~\ref{fig:HLM3} for $\o_{31}>0$ and $\d=\pi/2$.
For inverted hierarchy the asymmetry is so suppressed that
there is no allowed region. This means that for any choice of
the parameters one always obtains
$M_1^{\rm min}\gtrsim 10^{12}\,{\rm GeV}$.

Notice that some results for $\d$-leptogenesis in the particular
case where $M_3\gg 10^{14}\,{\rm GeV}$ have been recently
presented in~\cite{Pascoli:2006ci} for vanishing initial $N_1$-abundance.
For example the authors obtain a lower bound
$\sin\theta_{13}\gtrsim 0.09$ imposing the existence of an allowed region
for $M_1\lesssim 5\times 10^{11}\,{\rm GeV}$, while
we would obtain $\sin\theta_{13}\gtrsim 0.05$. Different reasons might 
explain this discrepancy. First, we are using a ($\sim 30\%$)
more conservative lower bound on $\overline{M}_1$
[cf.~Eq.~(\ref{barM1})]. Second, we do not include the effects of 
$\D L=1$ scatterings on the efficiency factors, which might produce 
some difference when $K_{1\a}\sim 1$, even though in the strong washout
regime this effect should be negligible.
Finally, another likely minor source of difference
is that we are not accounting for the effect of spectator processes
encoded in the matrix $C^{\ell}$ [cf.~Eq.~(\ref{Cl})] that relates
the $B/3-L_{\alpha}$ asymmetries to the $L_{\alpha}$ asymmetries~\cite{Barbieri:1999ma}.
However, notice that we do not want here to emphasize too much
a precise value of this lower bound on $\sin\theta_{13}$, since we believe that
it is anyway affected by much larger theoretical uncertainties on the validity
of the fully flavored regime. 


\subsubsection*{$\O=R_{12}$}

The third case we consider is $\O=R_{12}$ [cf.~Eq.~(\ref{R12})].
This time one has $A_{13}=A_{23}=0$ in Eq.~(\ref{r1tau}).
In the case of normal hierarchy, the $C\!P$ asymmetry, compared
to the case $\O=R_{13}$, is
suppressed by a factor $(m_{\rm sol}/m_{\rm atm})^{3/2}$,
while it is essentially the same for inverted hierarchy.
The projectors present very similar features to the case $\O=R_{13}$.
One can also again calculate, for
$M_2\lesssim 10^{12}\,{\rm GeV}$, the contribution from $N_2$-decays
to the final asymmetry, and one finds again that
the first term in Eq.~(\ref{ve2a})
vanishes, while the second produces a term $\propto M_1$, so that
\be
r_{2\t}=
{2\over 3}\,{\sqrt{m_1\,m_2}\,(m_2-m_1) \over \mtt\,m_{\rm atm}}\,
\o_{21}\,\sqrt{1-\o_{21}^2}\,s_{23}\,c_{23}\,\D \, .
\ee
When the efficiency factors are taken into account, one finds that
only in the case of normal hierarchy the contribution to
the final asymmetry from $N_2$-decays can be comparable
to the one from $N_1$-decays. However, in this case
both productions are suppressed and there
is no allowed region in the end. In the case of inverted hierarchy, 
the contribution from $N_2$-decays is always negligible compared 
to the one from $N_1$-decays. Notice, moreover, that
$\ve_{3\a}=0$ for $\O=R_{12}$, and hence there is no contribution from
$N_3$-decays.

In conclusion, for $\O=R_{12}$, the lower bound on $M_1$ for normal hierarchy
is much more restrictive than in the case $\O=R_{13}$, while it is very similar
for inverted hierarchy. A production from the two heavier RH
neutrinos can be neglected and the $N_1$-dominated scenario always
applies when the asymmetry is maximized.


\subsubsection*{$\O=R_{23}$}

The last interesting case is $\O=R_{23}$ [cf.~Eq.~(\ref{R23})].
From Eq.~(\ref{e1alOm}) one can easily check
that $\ve_{1\a}=0$. One can also check that, contrarily to the
case $\O=R_{12}$, the second term in Eq.~(\ref{ve2a}) vanishes while
the first term does not and yields
\be
r'_{2\t} \equiv {\ve_{2\t}\over \bar{\ve}(M_2)} =
{\sqrt{m_2\,m_3}\,(m_3-m_2)\over \mtt\,m_{\rm atm}}\,
\o_{32}\,\sqrt{1-\o_{32}^2}\,s_{12}\,c_{23}^2\,c_{13}\,\D \, .
\ee
Notice that this time $\ve_{2\t}\propto M_2$ and actually, more generally,
one can see that this expression is obtained from Eq.~(\ref{r1tau})
for $r_{1\t}$ in the case $\O=R_{13}$, just with the replacement
$(M_1,\mt)\rightarrow (M_2,\mtt)$.
At the same time, one has $K_1=m_1/m_{\star}$, so that
the washout from $N_1$ inverse processes vanishes for $m_1\rightarrow 0$.
For $M_3\lesssim 10^{12}\,{\rm GeV}$, one has to worry about
a possible contribution to the asymmetry from $N_3$-decays.
A straightforward  calculation shows that
\mbox{$\ve_{3\a}=(2/3)\ve_{2\a}$}, and therefore an asymmetry is
produced at $T\sim M_3$. However, we verified once more that
the washout from $N_2$ inverse processes is always strong enough
for the contribution to the final asymmetry
from $N_3$-decays to be negligible.

\begin{figure}
\centering
\includegraphics[width=0.32\textwidth]{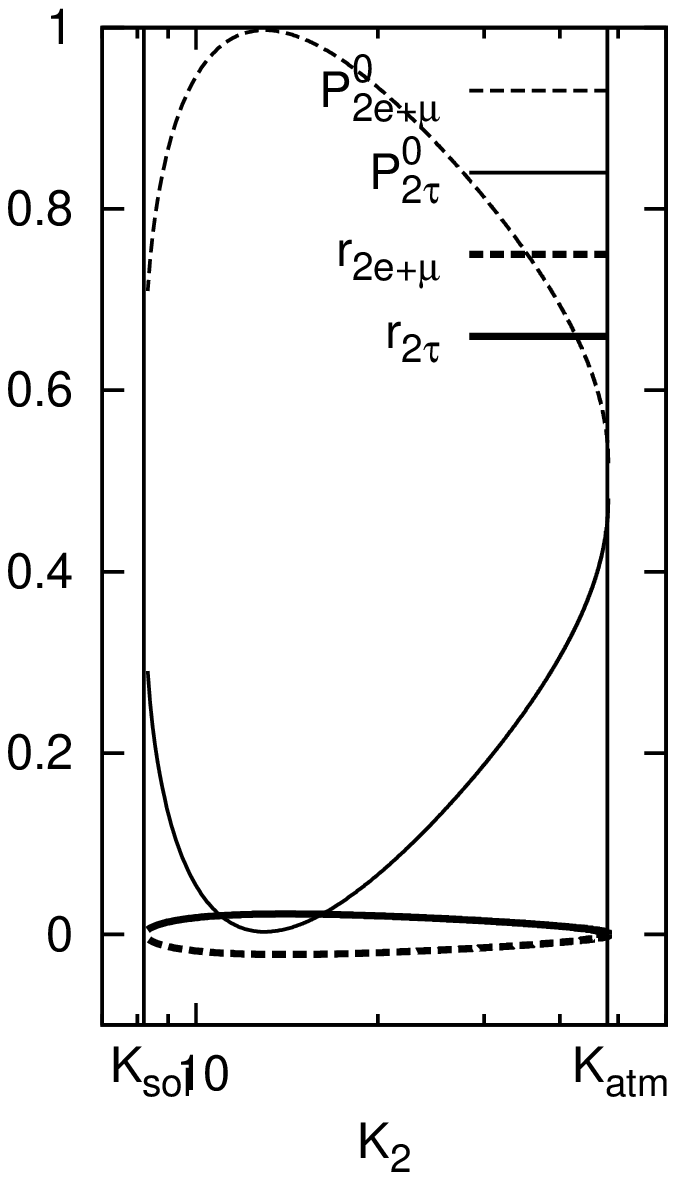}
\includegraphics[width=0.32\textwidth]{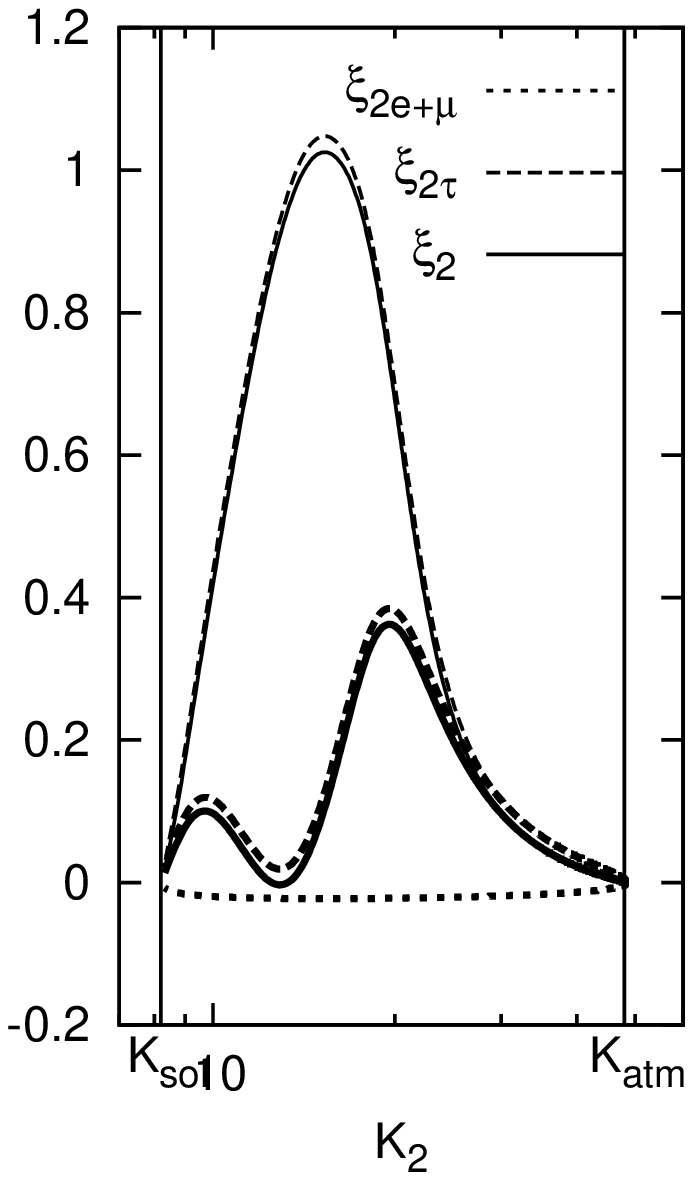}
\includegraphics[width=0.32\textwidth]{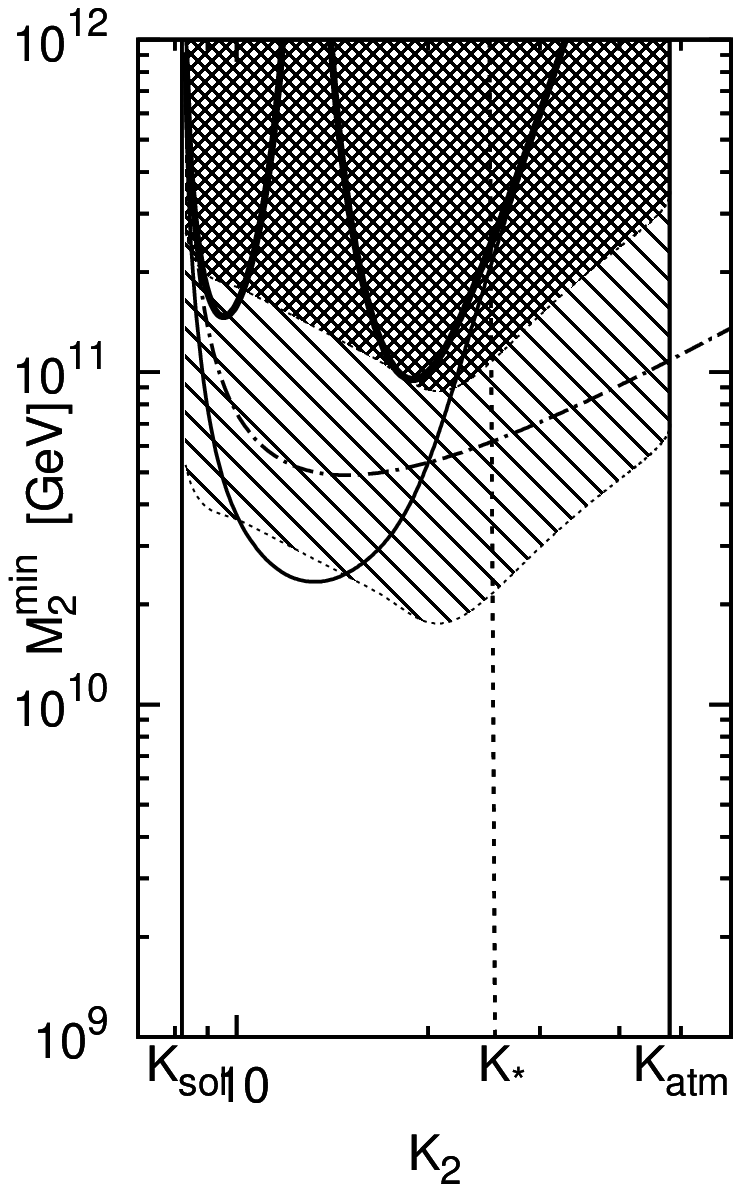}
\caption{Dependence of different quantities on $K_2$ for
$m_1=0$, $s_{13}=0.2$, $\d=\pi/2$ and real $\O=R_{23}$ with $\o_{32}>0$.
Left panel: projectors $P^0_{2\a}$ and quantities $r'_{2\a}$;
central panel: $\xi_{2\a}$ and $\xi_2$ for thermal (thin)
and vanishing (thick) initial $N_1$-abundance;
right panel: lower bound on $M_2$ for thermal (thin solid) and
vanishing (thick solid) abundance compared with the unflavored
result (dash-dotted line) as obtained in \cite{DiBari:2005st}.}
\label{fig:HLR23}
\end{figure}

In complete analogy with the unflavored
case \cite{DiBari:2005st}, one has that the lower bound $M_1^{\rm min}(K_1)$
is replaced by a lower bound $M_2^{\rm min}(K_2)$ obtained
for $\o_{32}>0$ and shown in the right panel of Fig.~\ref{fig:HLR23}.
One can see that also in this case, within
the validity of the condition~(\ref{M1fl}),
the allowed region is constrained to a small portion
falling in the weak washout regime. Assuming the very conservative 
condition of validity for the fully flavored regime outside the squared and hatched
regions, there is essentially no allowed region even in the weak washout regime.


One can wonder whether there is some choice of $\O$
beyond the special cases we analyzed where the final asymmetry
is much higher and the lower bound on $M_1$ much more
relaxed, especially in the strong washout regime. We have checked
different intermediate cases, and we can exclude such a possibility.
Therefore, the lower bound shown in Fig.~\ref{fig:HLR13} has to be considered, 
with good approximation, the lowest bound for any choice of real $\O$.

Another legitimate doubt is whether going beyond the approximations we made
the lower bound in Fig.~\ref{fig:HLR13} can be considerably relaxed.
The inclusion of non-resonant
$\D L=2$ or $\D L=1$ scatterings is not expected to produce large corrections.
Recently the effect of the off-diagonal terms in the matrix $C^{\ell}$ 
[cf.~Eq.~(\ref{Cl})] has been considered, but it
has been shown that it does not produce any relevant change
in the final asymmetry \cite{JosseMichaux:2007zj}.

Relevant corrections, as already pointed out, can only come from 
a full quantum kinetic treatment, which should
describe accurately the transition between the unflavored regime
and the fully flavored regime.

The same kind of considerations holds for the $N_2$-dominated scenario,
realized for $\O=R_{23}$. As soon as $\O$ deviates
from $R_{23}$, the washout from $N_1$ inverse processes comes into play
suppressing the final asymmetry and, at the same time, $\ve_{2\t}$
gets also suppressed. Therefore, the lower bound on $M_2$ is necessarily
obtained for $\O=R_{23}$ in complete analogy with the unflavored
treatment~\cite{DiBari:2005st}.

In conclusion, $\d$-leptogenesis in the HL is severely constrained,
confirming the conclusions of \cite{Blanchet:2006be} and \cite{Antusch:2006gy}.
In particular, imposing independence from the initial conditions,
then not even a marginally allowed region seems to survive.
Notice moreover that all plots have been obtained for $s_{13}=0.2$,
the current $3\s$ upper limit. Assuming that for
values of $M_1$ above the condition~(\ref{M1fl})  the
unflavored regime is quickly recovered and therefore that
the asymmetry production quickly switches off, then
a one-order-of-magnitude improvement of the upper limit on $\sin\theta_{13}$
would essentially rule out $\d$-leptogenesis in the HL,
even the marginally allowed regions falling in the weak washout regime.

In the next section we consider the effect
of close heavy neutrino masses in enhancing
the $C\!P$ asymmetries and relaxing the lower bounds
on $M_1, M_2$ as well as the related one on $T_{\rm reh}$.
But before concluding this section, we want to mention that
in the more general case of real $\O$ with non-vanishing Majorana phases,
an upper bound $m_1\lesssim 0.1\,{\rm eV}$
has been obtained in the fully flavored regime \cite{Branco:2006ce}
and for a hierarchical heavy neutrino mass spectrum.
This bound clearly applies also to $\d$-leptogenesis, but in this case,
considering the results we have obtained and  the expected quantum
kinetic corrections to the fully flavored regime,
the issue is actually whether an allowed region exists at all in the HL,
even for $m_1=0$. Therefore, we do not even try to place an upper bound
on $m_1$ in the HL. In the next section we show
that an upper bound on $m_1$ from successful $\d$-leptogenesis actually
holds even in the resonant limit, where the $C\!P$ asymmetries are
maximally enhanced.

\subsection{The degenerate limit}
\label{sec:DL}

We would like to show that going beyond the HL, the lower
bound on $M_1$ (or on $M_2$) can be considerably relaxed.
Nevertheless, we shall see that some interesting constraints
on the involved parameters still apply.
For simplicity, we assume from the beginning that $M_1$ (or $M_2$)
 $\ll 10^{9}\,{\rm GeV}$, so that the three-flavor regime applies,
where the muon-Yukawa interactions are also faster than inverse decays.
Therefore, when we show the flavor index $\a$,
we mean  $\a=e,\mu,\tau$. This assumption
simplifies the calculation, since we do not have
to describe a transition between the two and the three-flavor
regime, and because we can completely neglect the effect envisaged 
in~\cite{Barbieri:1999ma,Engelhard:2006yg} where
part of the asymmetry produced from $N_2$-decays is not touched
by $N_1$ inverse decays.

In order to go beyond the HL, it is convenient to use the quantity
$\d_{ji}$ defined in Eq.~(\ref{deltaji}). We are interested in the 
degenerate limit (DL),
where at least one $\d_{ji}$ is small enough that
both the asymmetry production from $N_{i,j}$-decays and the washout from 
the corresponding inverse processes can be approximately treated as
if they occurred at the same temperature, so that they can be simply added up.
The DL is a good approximation for $|\delta_{ji}|\lesssim 0.01$~\cite{Blanchet:2006dq}.

If $i,j\neq 3$ and $M_1\simeq M_2 \ll M_3$, then
one has a partial DL, where the efficiency factors can be approximated
as in Eq.~(\ref{partialDL}) for thermal initial $N_1$-abundance.
In all the cases we shall consider, we will always have 
\mbox{$K_{i\a}+K_{j\a}\gg 1$}, so that the strong washout regime
applies, and there is no need to consider the case of vanishing 
initial $N_1$-abundance. 

Another possibility
is to have a partial DL with $i,j\neq 1$, so that \mbox{$M_1\ll M_2\simeq M_3$}.
In this case one has to take into account the washout
from the lightest RH neutrino, implying
\be\label{k2}
\k_{i\a}^{\rm f}\simeq \k_{j\a}^{\rm f}\simeq
\k(K_{i\a}+K_{j\a})\exp\left(-{3\pi\over 8}K_{1\a}\right) .
\ee
Finally, in the full DL, $M_1\simeq M_2 \simeq M_3$, and 
the efficiency factor is given by Eq.~(\ref{fullDL}).

Let us now calculate the flavored $C\!P$ asymmetries.
In the case of real $\O$, which implies real $(h^{\dagger}h)_{ij}=(h^{\dagger}h)_{ji}$,
the expression~(\ref{veiaDL}) becomes
\be
\ve_{i\a}\simeq
\frac{1}{8\,\p (h^{\dag}h)_{ii}}
\sum_{j\neq i}\,(h^{\dag}h)_{i j}\,
{\rm Im}\left[h_{\a i}^{\star}\,h_{\a j}\right]\,\d_{ji}^{-1}.
\ee
We can again express the neutrino Yukawa coupling matrix through the
orthogonal representation. This time the
presence of the factor $\d_{ji}^{-1}$ does not allow to remove the sum on $j$,
as it has been possible in the HL in order to derive Eq.~(\ref{e1alOm}).
However, considering the same special cases as in the HL,
only one term $j\neq i$ survives, and we can write
\be
\hspace{-0.2cm}
\ve_{i\a}\simeq {2\,\bar{\ve}(M_i)\over 3\,\delta_{ji}}\,
\sum_{n,h<l}\,{m_n\,\sqrt{m_h\,m_l}\over\mti\,m_{\rm atm}}\,
\O_{ni}\,\O_{nj}
\,[\O_{hi}\,\O_{lj}-\O_{li}\,\O_{hj}]
\,{\rm Im}[U^{\star}_{\a h}\,U_{\a l}] \, .
\ee
The same expression holds for $\ve_{j\alpha}$
simply exchanging the $i$ and $j$ indexes.
We can always choose $j>i$ such that $M_j\geq M_i$.
In all the particular cases we shall consider,
we will have that $\ve_{k\a}=0$ for $k\neq i,j$, and moreover
the following simplifications apply:
\bea
\sum_n\,m_n\,\O_{ni}\,\O_{nj}&=&(m_q-m_p)\,\O_{ji}\,\O_{jj}\\
\sum_{h<l}\,{\sqrt{m_h\,m_l}}\,[\O_{hi}\,\O_{lj}-\O_{li}\,\O_{hj}]
&=&\sqrt{m_q\,m_p} \, ,
\eea
with $q>p$. Except for the case $M_3\gg 10^{14}\,{\rm GeV}$, we will
always have $q=j$ and $p=i$. The final $B\!-\!L$ asymmetry 
can then be expressed as
\be
N_{B-L}^{\rm f} \simeq
\sum_{\a}\,(\ve_{i\a}+\ve_{j\a})\,
\k_{\a}^{\rm f}(K_{i\a}+K_{j\a},K_{k\a})=
{\bar{\ve}(M_i)\over 3\,\d_{ji}}\,g(m_1,\O_{ji},\theta_{13},\d)\,\D\, ,
\ee
where
\bea \nonumber
g(m_1,\O_{ji},\theta_{13},\d) & \equiv &
{2\,K_{\rm atm}\,(K_i+K_j)\over K_i\,K_j}\,
\,{(m_q-m_p)\,\sqrt{m_q\,m_p}\over m_{\rm atm}^2}\,
\,
\O_{ji}\,\sqrt{1-\O_{ji}^2} \\ &  & \times   \label{g}
\,\sum_{\alpha}\,\k_{\alpha}^{\rm f}(K_{i\a}+K_{j\a},K_{k\a})\,
{{\rm Im}[U^{\star}_{\a p}\,U_{\a q}]\over\D}\,
\eea
and where
$\k_{\a}^{\rm f}(K_{i\a}+K_{j\a},K_{k\a})=\k_{i\a}^{\rm f}=\k_{j\a}^{\rm f}$
is given by one of the three expressions Eq.~(\ref{fullDL}), Eq.~(\ref{partialDL})
or Eq.~(\ref{k2}) according to the particular case.

It is interesting to notice that because of the unitarity of $U$, 
the sum over $i$ of the flavored decay parameters,
Eq.~(\ref{sumKia}), tends to $m/m_{\star}$ in the degenerate limit
for the light neutrinos ($m$ is the common mass scale) 
independently of the flavor. Therefore, the sum over the flavor in 
Eq.~(\ref{g}) tends to
vanish. This will contribute, as we shall see, to place a stringent
upper bound on the absolute neutrino mass scale in the full DL.

It is also worthwhile to notice that the sign of $\D$ cannot be predicted
from the sign of the
observed baryon asymmetry, since the sign of $g(m_1,\O_{ji},\theta_{13},\d)$
depends on the sign of $\O_{ji}$, which is undetermined. Notice also that
${{\rm Im}[U^{\star}_{\a h}\,U_{\a l}]/\D}$ does not depend on $\D$, but
nevertheless there is a dependence of $g(m_1,\O_{ji},\theta_{13},\d)$
on $\d$ and on $\theta_{13}$ coming from the tree-level projectors
$P^0_{i\a}$ in the sum $K_{i\a}+K_{j\a}$.
However, in any case, for $\D\rightarrow 0$
one has that $g(m_1,\O_{ji},\theta_{13},\d)\,\D\rightarrow 0$,
since the final asymmetry has to vanish when
$\sin\theta_{13}$ or $\sin\d$ vanishes.

The function $|g(m_1,\O_{ji},\theta_{13},\d)|$ can be
maximized over $\O_{ji}$. Indeed for $m_1=0$, since $\k< 1$ and
$K_i+K_j \leq K_{\rm atm}$, one has $g(m_1=0,K_i,\theta_{13},\d)<4$.
Increasing $m_1$ there is a
suppression due to the fact that $K_i\geq m_1/m_{\star}$, and
$g_{\rm max}(m_1,\theta_{13},\d)$ decreases monotonically. Therefore, for
any $m_1$, there is a lower bound on $M_1$ given by
\be\label{lbM1gmax}
M_1 \geq M_1^{\rm min}(m_1,\theta_{13},\d)\equiv
{3\,\overline{M}_1\over g_{\rm max}(m_1,\theta_{13},\d)}
\,{\d_{j1}\over |\D|}\, .
\ee
The $C\!P$ asymmetries, and consequently the
final baryon asymmetry, are maximally enhanced in the extreme case
of resonant leptogenesis \cite{Pilaftsis:1997jf,Pilaftsis:1998pd,Pilaftsis:2003gt}, 
when the heavy neutrino mass degeneracy is comparable to the decay
widths. An approximate resonance condition was given in Eq.~(\ref{condRL}),
where the uncertainty parameter $d=1\div 10$ was introduced because of
a discrepancy in the literature~\cite{Anisimov:2005hr,Pilaftsis:1997jf}.
When the resonance condition is satisfied, one has $\ve_1=1/d$ in the unflavored case
with maximal phase. This can be taken as a conservative
limit that implies, maximizing over $\d$, a lower bound
\be\label{lbsint}
\sin \theta_{13} \geq
\sin\theta_{13}^{\rm min} = {d\,\eta_B^{\rm CMB}\,N_{\gamma}^{\rm rec}
\over a_{\rm sph}\,
{\rm max}_{\d}[g_{\rm max}(m_1,\theta_{13}^{\rm min},\d)\,\sin\d)]} \, .
\ee

Let us now specialize the expressions for the four
special cases we have already analyzed in the HL.

\subsubsection*{$M_3\gg 10^{14}\,{\rm GeV}$}

Remember that in this case one has $(h^{\dagger}h)_{3j}=0$, implying
$\ve_{3\a}=0$, a consequence of the fact that the heaviest
RH neutrino decouples. Moreover, one has $m_1\ll m_{\rm sol}$,
so that terms $\propto m_1$ can be neglected, $m_3\simeq m_{\rm atm}$
and $m_2\simeq m_{\rm sol}$ for normal hierarchy or
$m_2\simeq m_{\rm atm}\sqrt{1-m_{\rm sol}^2/m_{\rm atm}^2}$
for inverted hierarchy. Therefore,
there is actually no dependence on $m_1$ in
$g(m_1,\O_{ji},\theta_{13},\d)$, which is given by Eq.~(\ref{g}) 
with $(i,j)=(1,2)$ and $(p,q)=(2,3)$,
\bea \nonumber
g(\O_{21},\theta_{13},\d) & \simeq &
{2\,(K_1+K_2)\,K_{\rm atm}\over K_1\,K_2}\,
\,\left(1-{m_2\over m_{\rm atm}}\right)\,
\sqrt{m_2\over m_{\rm atm}}\,
\O_{21}\,\sqrt{1-\O^2_{21}}\,   \\ \label{gM3}
& & \times
 \sum_{\a}\,\k(K_{1\a}+K_{2\a})\,
{{\rm Im}[U^{\star}_{\a2}\,U_{\a 3}]\over \D }.
\eea
\begin{figure}
\centerline{\psfig{file=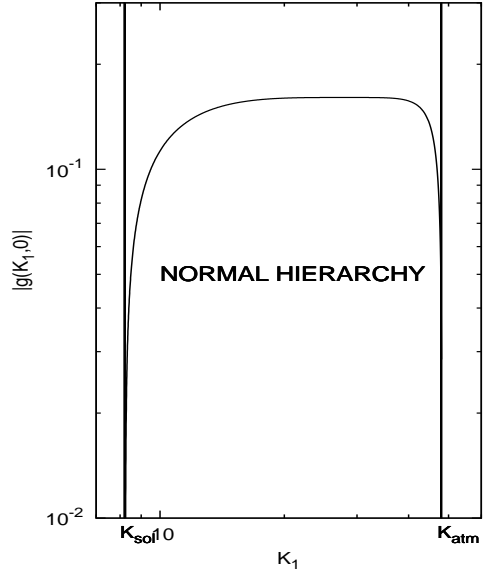,height=80mm,width=70mm,angle=0}}
\caption{Case $M_3\gg 10^{14}\,{\rm GeV}$ for normal hierarchy in the DL.
Plot of the function $|g(K_1,\theta_{13},\d)|$
in the limit $\D\rightarrow 0$.
The maximum gives the lower bound on $M_1$ [cf.~Eq.~(\ref{lbM1Minf})]
and on $\sin\theta_{13}$ [cf.~Eq.~(\ref{lbsintMinf})].}
\label{fig:gnor}
\end{figure}
In the case of normal hierarchy, $|g(\O_{21},\theta_{13},\d)|$
slightly decreases when $\D$ increases, and so the maximum 
is found for $\D=0$ and the dependence on $\theta_{13}$ and $\d$ 
disappears. Replacing the dependence on $\O_{21}$ with a 
dependence on $K_1$, we have plotted $|g(K_1,\D=0)|$ in 
Fig.~\ref{fig:gnor} for central values of $m_{\rm sol}$
and $m_{\rm atm}$. Including the errors,
one finds $g_{\rm max}\simeq 0.160 \pm 0.005$.

The ($3\s$) lower bound on $M_1$ for normal hierarchy,
from the general expression (\ref{lbM1gmax}), is then given by
\be\label{lbM1Minf}
M_1 \geq 0.9 \times \,{10^{10}}\,{\rm GeV}\,{\delta_{21}\over |\D|} \,.
\ee

In the case of inverted hierarchy, the situation is somehow opposite,
since for $\theta_{13}=0$ the electron flavor contribution vanishes
in Eq.~(\ref{gM3}) and there is an exact cancellation between the
$\tau$ and $\mu$ contributions. Consequently, the asymmetry
increases for increasing values of $\theta_{13}$ and the maximum
is found for $\sin\theta_{13}=0.2$ and $\d\simeq\pi/4$. In this case
one has that \mbox{
${\rm max}_{\theta_{13},\d}[g_{\rm max}(m_1=0,\theta_{13},\d)\,\D]
\simeq (9\pm 2)\times 10^{-8}$},
which, plugged into Eq.~(\ref{lbM1gmax}), gives at $3\s$
\be
M_1\geq 6\times 10^{15}\,{\rm GeV}\,{\delta_{21}}\,.
\ee
It should be remembered that these conditions have been
obtained in the three-flavor regime and in the DL, i.e.
they are valid for $M_{1,2}\lesssim 10^9\,{\rm GeV}$.
This implies that $\delta_{21}\lesssim 10^{-1}\,|\D|$ for normal hierarchy
and $\delta_{21}\lesssim 10^{-7}$ for inverted hierarchy.

Analogously, the general expression~(\ref{lbsint}) gives
the following ($3\s$) lower bounds on $\sin\theta_{13}$
for normal and inverted hierarchy, respectively:
\be\label{lbsintMinf}
\sin\theta_{13} \gtrsim  3.3\times 10^{-7}\,d
\hspace{5mm}
{\rm and}
\hspace{5mm}
\sin\theta_{13}\gtrsim 0.06\,d .
\ee

\subsubsection*{$\O=R_{13}$}

\begin{figure}
\centering
\includegraphics[width=0.4\textwidth]{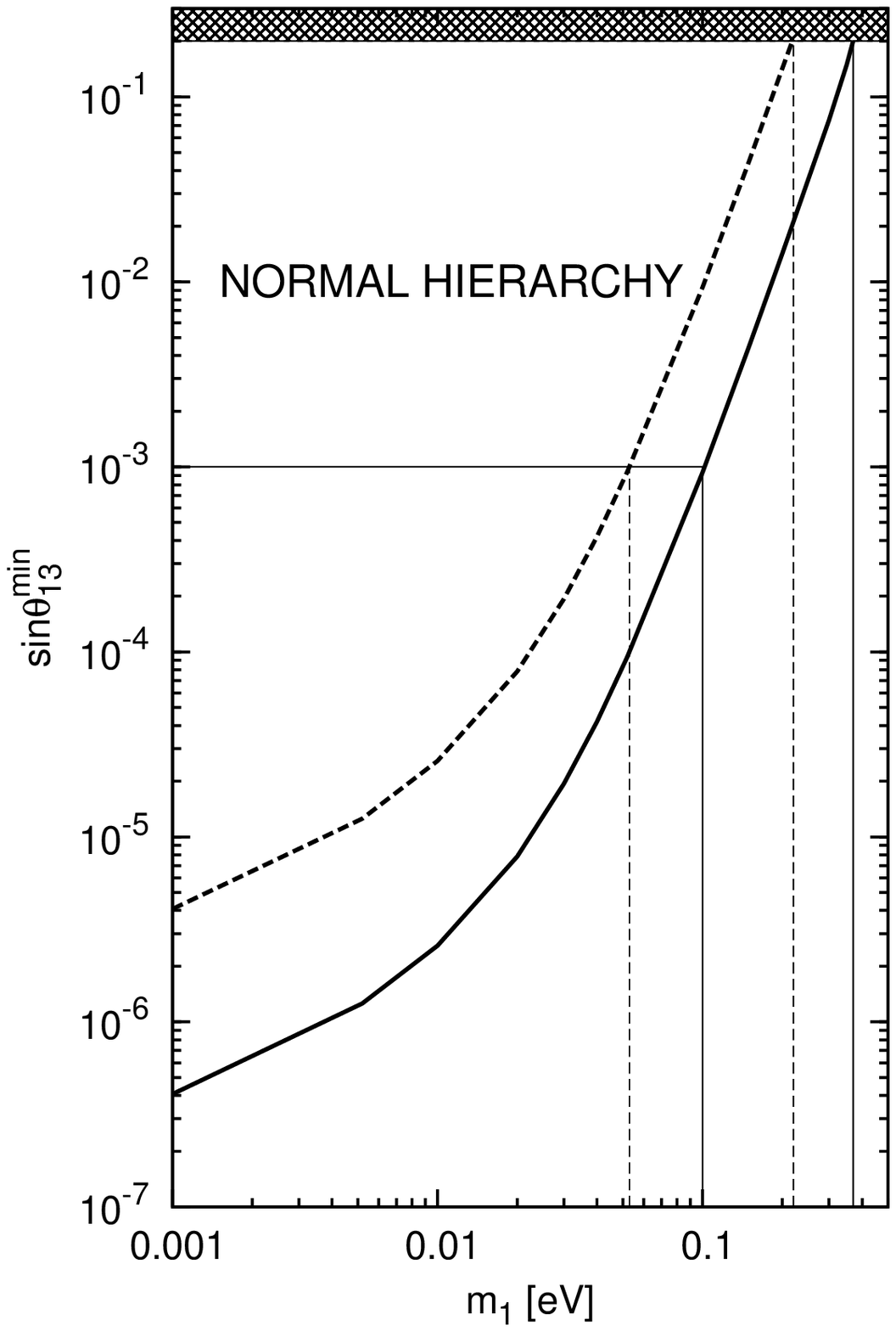}
\hspace{1cm}
\includegraphics[width=0.4\textwidth]{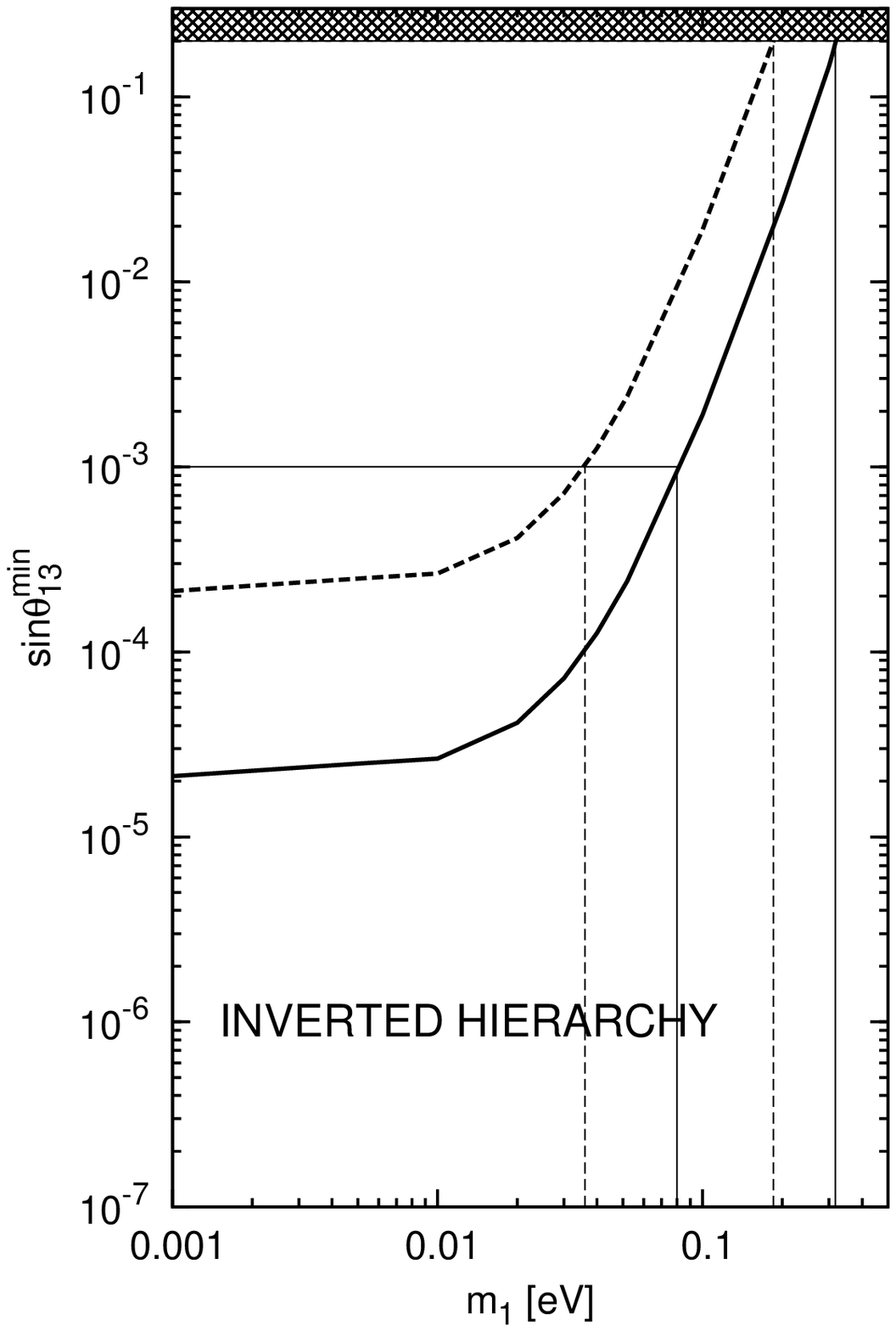}
\caption{Case $\O=R_{13}$ in the full DL. Lower bound on
$\sin\theta_{13}$ versus $m_1$ obtained in resonant leptogenesis
for $d=1$ (solid line) and $d=10$ (short-dashed line).
Values $\sin\theta_{13} > 0.20$ are excluded at $3\s$ by current
experimental data.}
\label{fig:BoundR13}
\end{figure}

In this particular case,
the next-to-lightest RH neutrino is decoupled from the other
two heavy neutrinos, which implies that $\ve_{2\a}=0$ for any $\a$ and that
$\ve_{1\a}$ does not depend on $M_2$; in particular, it
does not get enhanced if $\d_{21}\rightarrow 0$. Therefore, one
has necessarily to consider $\delta_{31}\lesssim 0.01$,
implying a full DL with all three RH neutrino masses quasi-degenerate.
The function $g(m_1,\O_{ji},\theta_{13},\d)$ is now obtained from
Eq.~(\ref{g}) with $j=q=3$ and $i=p=1$ and 
$\k_{\a}^{\rm f}=\k(K_{1\a}+K_{2\a}+K_{3\a})$,
or explicitly
\bea \nonumber
g(m_1,\O_{31},\theta_{13},\d) & \equiv &
{2\,K_{\rm atm}\,(K_1+K_3)\over K_1\,K_3}\,
\,{(m_3-m_1)\,\sqrt{m_3\,m_1}\over m_{\rm atm}^2}\,
\,
\O_{31}\,\sqrt{1-\O_{31}^2} \\ &  & \times   \label{gR13}
\,\sum_{\alpha}\,\k(K_{1\a}+K_{2\a}+K_{3\a})\,
{{\rm Im}[U^{\star}_{\a 1}\,U_{\a 3}]\over\D}\, .
\eea
It is interesting to notice that in this case an
$e$-dominance is realized. Moreover,
one has that the dependence of
$|g(m_1,\O_{31},\theta_{13},\d)|$ on $\theta_{13}$ and $\d$
is slight, and the maximum is for $\D=0$ and $m_1=0$. We
find $g_{\rm max}(0)=0.24\pm 0.01$ for normal hierarchy
and $g_{\rm max}(0)=(3.1\pm 0.2)\times 10^{-3}$ for inverted hierarchy.
The lower bound on $M_1$ in Eq.~(\ref{lbM1gmax}) yields then 
at $3\s$ for normal and inverted hierarchy,
respectively,
\be
M_1 \gtrsim 5.5\times 10^{9}\,{\rm GeV}\,{\delta_{31}\over |\D|}
\hspace{5mm}
{\rm and}
\hspace{5mm}
M_1 \gtrsim 5 \times 10^{11}\,{\rm GeV}\,{\delta_{31}\over |\D|}
\, ,
\ee
while in the case of resonant leptogenesis Eq.~(\ref{lbsint}) yields
\be
\sin\theta_{13}\gtrsim 2.3\times 10^{-7}\,d
\hspace{5mm}
{\rm and}
\hspace{5mm}
\sin\theta_{13}\gtrsim 1.5\times 10^{-5}\,d \, .
\ee
Increasing $m_1$, the value of $g_{\rm max}(m_1)$ decreases, and
the lower bound on $\sin\theta_{13}$ in resonant leptogenesis
becomes more and more
restrictive. This dependence is shown in Fig.~\ref{fig:BoundR13} 
both for normal
(left panel) and inverted (right panel) hierarchies and for
$d=1$ (solid line) and $d=10$ (short-dashed line).
Interestingly, imposing
the experimental ($3\s$) upper limit $\sin\theta_{13}\lesssim 0.20$,
one obtain the upper bound $m_1 \lesssim \,0.2\textrm{--}0.4\,{\rm eV}$,
depending on the value of $d$.
This upper bound will become more stringent if no signal for a non-vanishing
mixing angle $\q_{13}$ is seen in future neutrino oscillation experiments, since the
experimental
limit on $\sin\theta_{13}$ will then go down. Assuming no discovery, the 
most stringent experimental upper limit is expected to be reached in
neutrino factories, where one could obtain 
$\sin\theta_{13}< 10^{-3}$~\cite{Huber:2002mx}. This asymptotical
upper limit is also shown in Fig.~\ref{fig:BoundR13} and would imply
an upper bound $m_1 \lesssim \, 0.05\textrm{--}0.1\,{\rm eV}$
for normal hierarchy and $m_1 \lesssim \,0.03\textrm{--}0.08\,{\rm eV}$
for inverted hierarchy. Therefore,
an interesting interplay between two measurable quantities
is realized, making $\d$-leptogenesis falsifiable independently
of the RH neutrino mass spectrum.

In the more conservative case of normal hierarchy (see 
left panel of Fig.~\ref{fig:BoundR13}), a good approximation
for the upper bound on $m_1$ ($d=1$) is given by the fit
\be
m_1\lesssim 0.6\,
\left({\sin\theta_{13}- 2.3\times 10^{-7}} \right)^{0.25} \,{\rm eV} \, .
\ee
It is interesting that this upper bound
holds in the extreme case of resonant leptogenesis and
therefore holds for any RH neutrino spectrum. However,
we still have to verify if it holds also for a different
choice of $\O$.

\subsubsection*{$\O=R_{12}$}

The situation for $\O=R_{12}$ is quite different
compared to the previous two cases. One has now $i=p=1$
and $j=q=2$, and it is possible to have both a partial
DL with $M_1\simeq M_2 \ll M_3 \lesssim 10^{14}\,{\rm GeV}$
and a full DL. In the first case, the general expression~(\ref{g}) 
becomes
\bea \nonumber
g(m_1,\O_{21},\theta_{13},\d) & \equiv &
{2\,K_{\rm atm}\,(K_1+K_2)\over K_1\,K_2}\,
\,{(m_2-m_1)\,\sqrt{m_2\,m_1}\over m_{\rm atm}^2}\,
\,
\O_{21}\,\sqrt{1-\O_{21}^2} \\ &  & \times   \label{gR21}
\,\sum_{\alpha}\,\k(K_{1\a}+K_{2\a})\,
{{\rm Im}[U^{\star}_{\a 1}\,U_{\a 2}]\over\D}\, .
\eea
\begin{figure}
\centering
\includegraphics[width=0.4\textwidth]{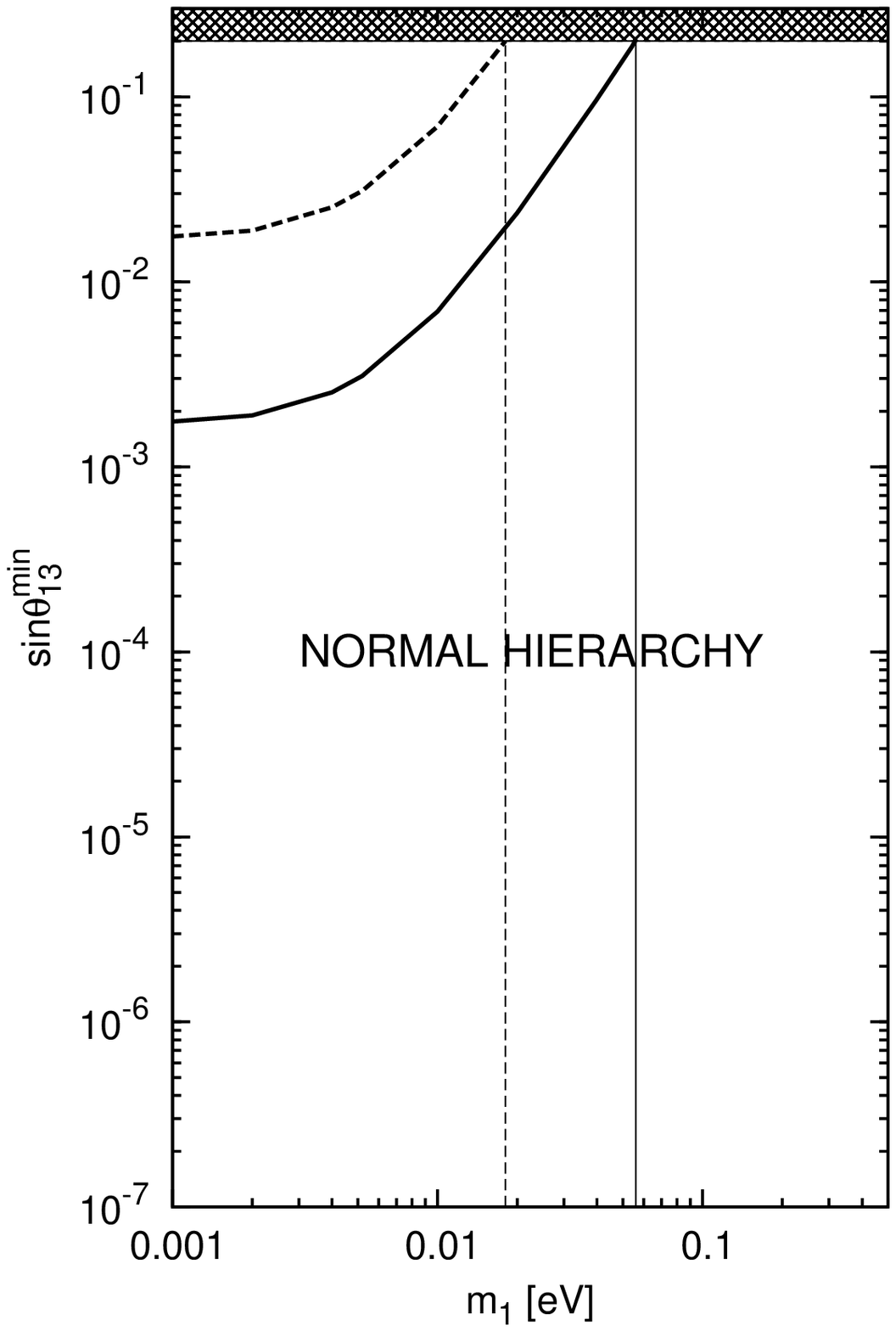}
\hspace{1cm}
\includegraphics[width=0.4\textwidth]{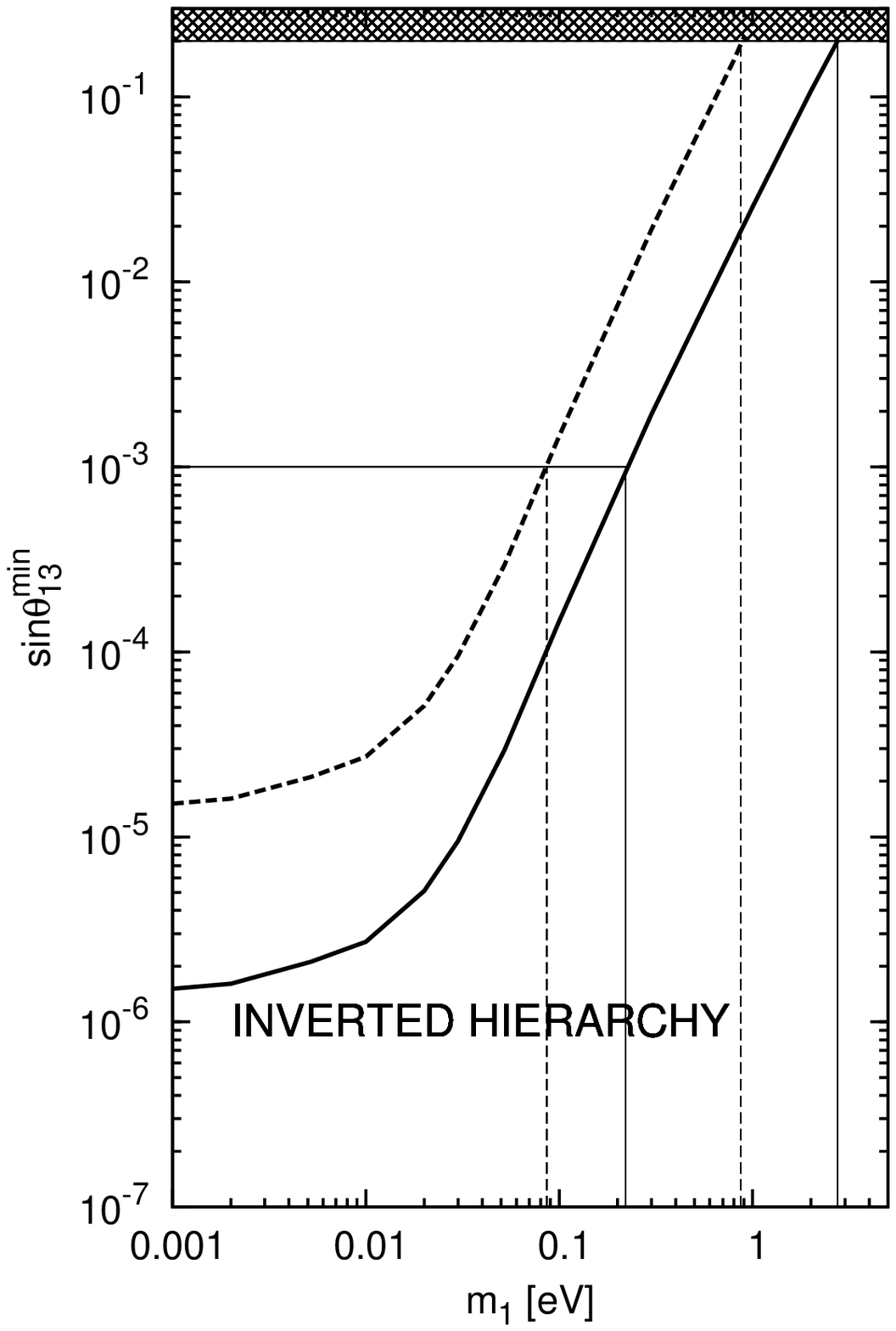}
\caption{Case $\O=R_{12}$ in the partial DL. Lower bound on
$\sin\theta_{13}$ versus $m_1$ obtained in resonant leptogenesis.
Same conventions as in the previous figure.}
\label{fig:BoundR122RHN}
\end{figure}
This time the contribution from the electron flavor vanishes.
Furthermore, for normal hierarchy, there is an almost perfect cancellation
between the $\m$ and the $\t$ contributions. In the left panel
of Fig.~\ref{fig:BoundR122RHN}, we show the lower bound on $\sin\theta_{13}$ 
versus $m_1$, and one can see that it is more restrictive than in 
the previous case, $\O=R_{13}$. 
In particular, imposing $\sin\theta_{13}<0.2$,
one obtains now a much more stringent upper bound $m_1\lesssim 0.06\,{\rm eV}$.
On the other hand, for inverted hierarchy,
the cancellation between the $\m$ and the $\tau$ flavors does not occur,
and one has a lower bound on $\sin\theta_{13}$ for $m_1\ll 0.01\,{\rm eV}$
that is similar to what has been obtained in the case $\O=R_{13}$ (see the 
right panel of Fig.~\ref{fig:BoundR122RHN}). However, here there is no
flavor cancellation for increasing values of $m_1$
because $K_{1\a}+K_{2\a}$ does not tend to a common value like
$K_{1\a}+K_{2\a}+K_{3\a}$. Therefore, one can see in Fig.~\ref{fig:BoundR122RHN} 
that this time
the upper bound on $m_1$ is much looser, both compared to normal
hierarchy and compared to $\O=R_{13}$.

\begin{figure}
\centering
\includegraphics[width=0.4\textwidth]{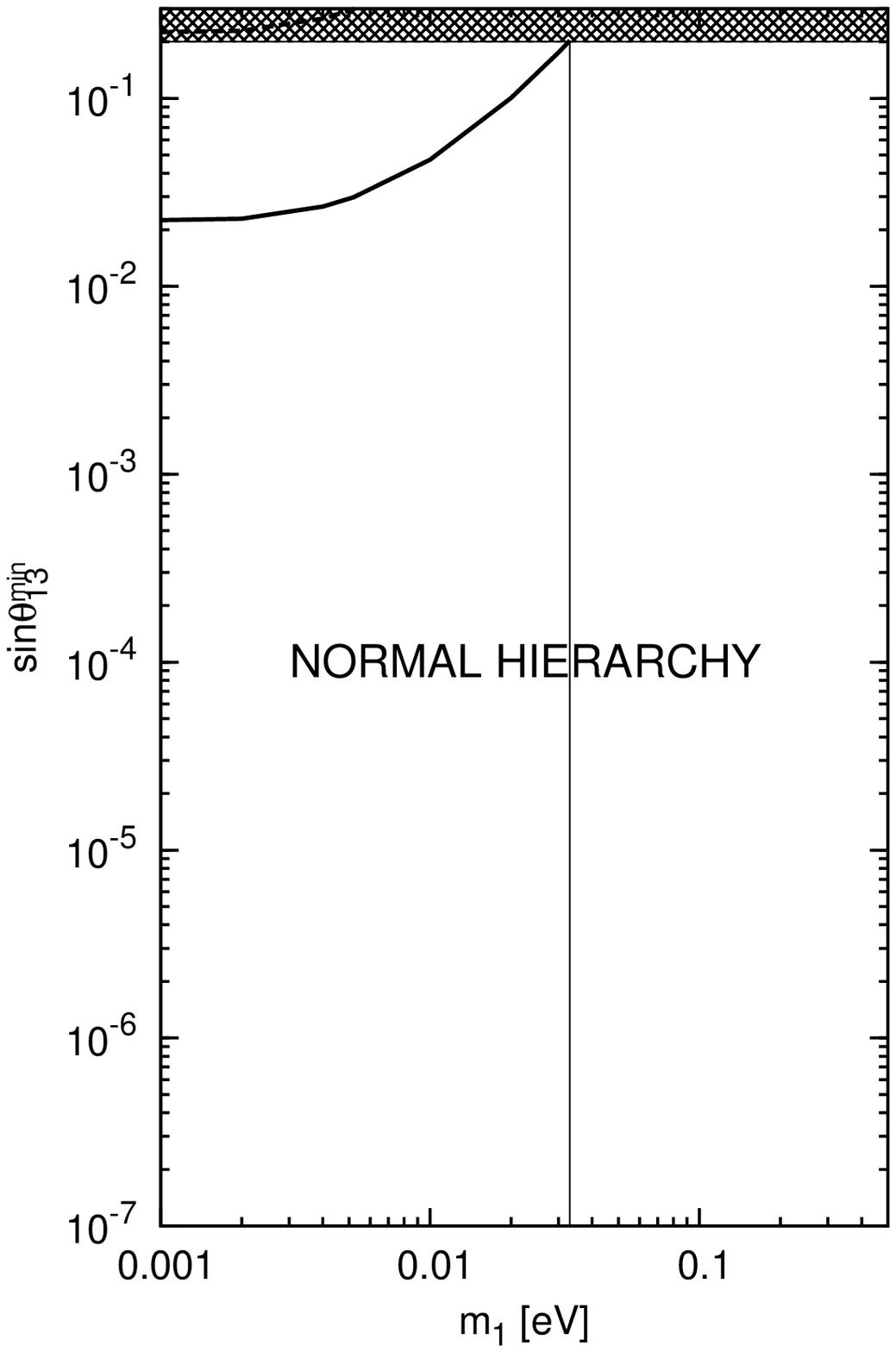}
\hspace{1cm}
\includegraphics[width=0.4\textwidth]{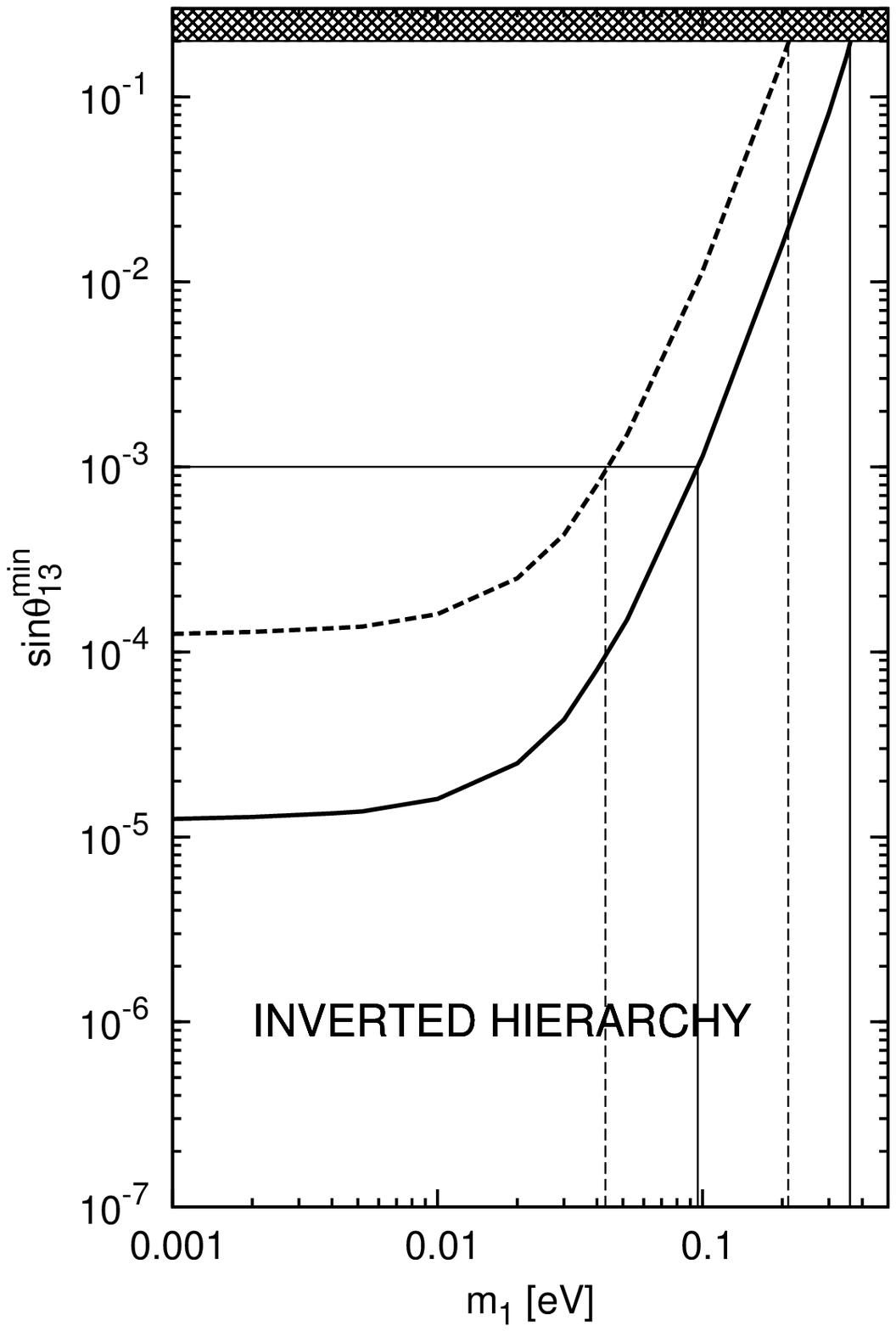}
\caption{Case $\O=R_{12}$ in the full DL. Lower bound on
$\sin\theta_{13}$ versus $m_1$ obtained in resonant leptogenesis.
Same conventions as in the previous figures.}
\label{fig:BoundR123RHN}
\end{figure}

In the full DL the flavor cancellation at large $m_1$ occurs. The
results are shown in Fig.~\ref{fig:BoundR123RHN}. One can notice that the upper bound
on $m_1$ is very restrictive for normal hierarchy, and one has
a situation similar to the case $\O=R_{13}$ for inverted hierarchy.

\subsubsection*{$\O=R_{23}$}

When $\O=R_{23}$, the lightest RH neutrino decouples and
$\ve_{1\a}=0$ independently of $M_1$.
Therefore, there is no contribution to the final asymmetry
from $N_1$-decays. On the other hand, $\ve_{2\a}$ and $\ve_{3\a}$
do not vanish and hence there is a contribution from the
decays of the two heavier RH neutrinos. Still $N_1$ inverse
processes have to be taken into account since they
contribute to the washout.
There are two different possibilities.

\begin{figure}
\centering
\includegraphics[width=0.4\textwidth]{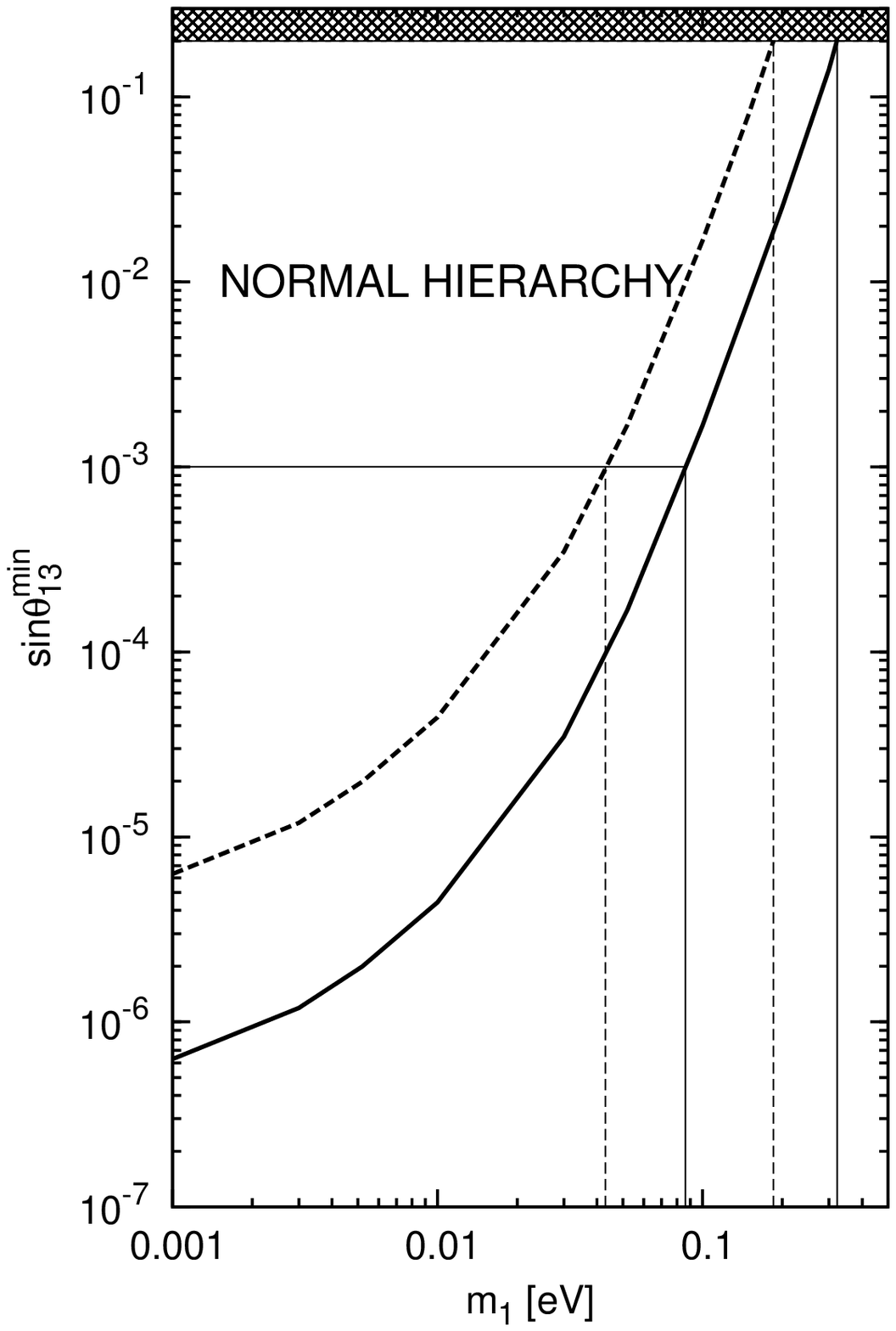}
\hspace{1cm}
\includegraphics[width=0.4\textwidth]{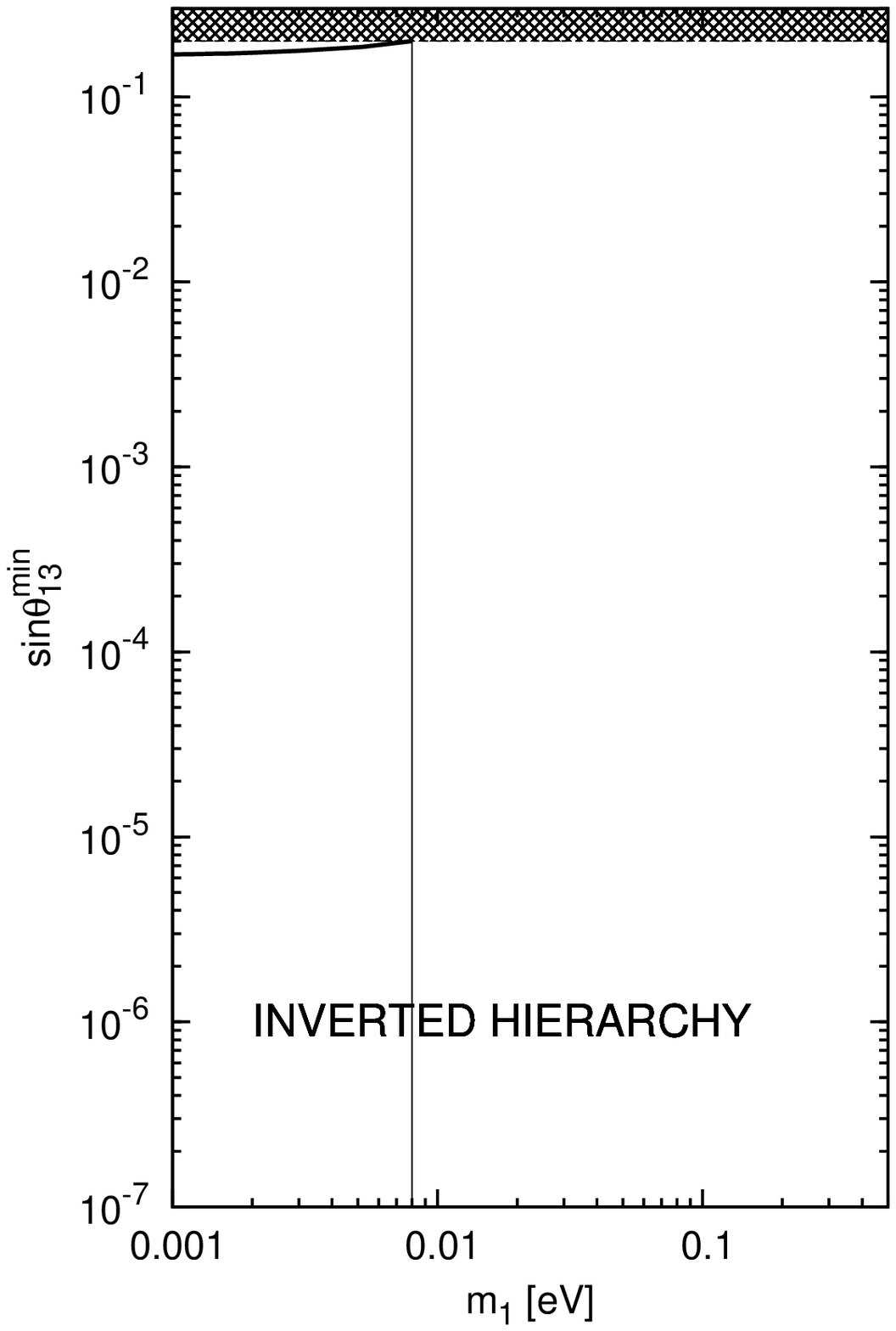}
\caption{Case $\O=R_{23}$ in the full DL.
Lower bound on $\sin\theta_{13}$ versus
$m_1$ obtained in resonant leptogenesis.
Same conventions as in the previous figures.}
\label{fig:BoundR233RHN}
\end{figure}

In the full DL the washout from $N_1$ inverse decays just cumulates
with the washout from the two heavier. Therefore, 
using expression~(\ref{g}) with $i=p=2$ and $j=q=3$
and $\k_{\a}^{\rm f}=\k(K_{1\a}+K_{2\a}+K_{3\a})$, one obtains
\bea
g(m_1,\O_{32},\theta_{13},\d) & \equiv& 
{2\,K_{\rm atm}\,(K_2+K_3)\over K_2\,K_3}\,
\,{(m_3-m_2)\,\sqrt{m_3\,m_2}\over m_{\rm atm}^2}\,
\,
\O_{32}\,\sqrt{1-\O_{32}^2} \NO\\*
&& \times 
\,\sum_{\alpha}\,\k(K_{1\a}+K_{2\a}+K_{3\a})\,
{{\rm Im}[U^{\star}_{\a 2}\,U_{\a 3}]\over\D}\, .\label{g32a}
\eea
In Fig.~\ref{fig:BoundR233RHN} we show the lower bound on
$\sin\theta_{13}$ versus $m_1$ for successful resonant leptogenesis. 
This time there is a bigger suppression than in the
case $\O=R_{13}$, especially for inverted hierarchy.

\begin{figure}
\centering
\includegraphics[width=0.4\textwidth]{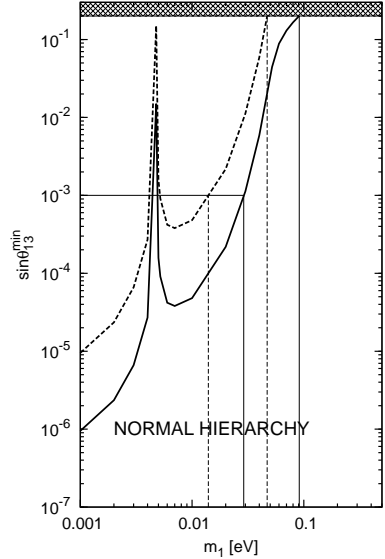}
\caption{Case $\O=R_{23}$ in the partial DL.
Lower bound on $\sin\theta_{13}$ versus $m_1$ obtained in resonant
leptogenesis. Same conventions as in the previous figures.}
\label{fig:BoundR232RHN}
\end{figure}

In the case $M_1\ll M_2\simeq M_3$, one has
\bea 
\hspace{-0.5cm}
\nonumber
g(m_1,\O_{32},\theta_{13},\d) & \equiv &
{2\,K_{\rm atm}\,(K_2+K_3)\over K_2\,K_3}\,
\,{(m_3-m_2)\,\sqrt{m_3\,m_2}\over m_{\rm atm}^2}\,
\,
\O_{32}\,\sqrt{1-\O_{32}^2} \\ 
&\times&   \label{g32b}
\,\sum_{\alpha}\,\k(K_{2\a}+K_{3\a})
\exp\left(-{3\,\pi\over 8}\,K_{1\a}\right)
{{\rm Im}[U^{\star}_{\a 2}\,U_{\a 3}]\over\D}\, .
\eea 
The lower bound on $\sin\theta_{13}$ versus $m_1$
is shown in Fig.~\ref{fig:BoundR232RHN} for normal hierarchy. One notices
that the upper bound on $m_1$ is slightly less stringent than
in the full DL.
For inverted hierarchy the asymmetry production is so
suppressed that there is no allowed region.

We can conclude this section noticing that our results
show that $\d$-leptogenesis can be falsified. In the case
of normal hierarchy, the current upper limit
$\sin\theta_{13}\lesssim 0.2$ implies
$m_1\lesssim 0.1\textrm{--}0.3\,{\rm eV}$, while, in future, a
potential upper limit
$\sin\theta_{13}\lesssim 10^{-3}$ would imply
$m_1\lesssim 0.01\textrm{--}0.1\,{\rm eV}$, with
a more precise determination depending on the possibility
of improving the current estimation of the parameter $d$ in
resonant leptogenesis.

\section{Leptogenesis from the Majorana phases}

It is apparent from the elements $A_{ij}$ in Eq.~(\ref{r1tau})
that the Majorana phases in $U$ also contribute to the 
\CP~violation necessary for leptogenesis. Actually, they can
also play the role of unique source of \CP~violation, 
as illustrated in Fig.~\ref{fig:OmReal} for $\O=R_{13}$
real and non-zero Majorana phase $\F_1$ ($\F_1=\p/2$). 
One can even say quite generally that
the Majorana phases can be more easily responsible for enough
\CP~violation than the Dirac phase,
simply because they are not associated with the small 
mixing angle $\q_{13}$. To illustrate this, we have plotted in the
right panel of Fig.~\ref{fig:HLR13} with dotted lines the case
of $\F_1=-\p/2$ and $\d=0$, compared with $\d=-\p/2$ and $\F_1=0$
for $\sin \q_{13}=0.2$, i.e. the maximal allowed value. One notices
that the lower bounds on $M_1$ for the case of non-vanishing
Majorana phase is about a factor of 2 lower than in the case
of non-vanishing Dirac phase, even for the maximal value of 
$\q_{13}$.

We do not aim here at making a thorough study of the role of the 
Majorana phases as the sole source of \CP~violation for
leptogenesis. We think that the case of the Dirac phase, 
which we analysed in detail in the previous section, represents 
a more conservative situation, and the prospects for a measurement 
seem to be more encouraging. However, for completeness, we would like
to report some results obtained in~\cite{Pascoli:2006ci} and
\cite{Molinaro:2007uv} concerning
exclusively the Majorana phase and where only the limit of hierarchical
heavy neutrinos and a vanishing initial $N_1$-abundance were considered.

When a fully hierarchical light neutrino spectrum ($m_1\ll m_{\rm sol}$) is 
considered, the authors in~\cite{Pascoli:2006ci} obtain two main results: 
for a normal hierarchy (real 
$\O$ matrix), they find the lower bound $M_1\gtrsim 3.6\times 10^{10}\gev$
for successful leptogenesis, whereas for inverted hierarchy (purely 
imaginary $\O_{12}\O_{13}$), they find $M_1\gtrsim 5.3\times 10^{10}\gev$.
It should be noted that for inverted hierarchy and real $\O$, there is
no allowed region below $10^{12}\gev$.

In~\cite{Molinaro:2007uv} the study was extended to arbitrary $m_1$, and some
new effects were found. In particular, even in the case of inverted hierarchy
and real $\O$, an allowed range was found, with the lower bound
$M_1\gtrsim 3\times 10^{10}\gev$. In the case of normal hierarchy, no such
relaxation occurs and the bounds quoted above still hold.

\section{Discussion}

We have discussed situations where the ``observable'' \CP-violating
phases, $\d$, $\F_1$ and $\F_2$, act as the only source of \CP~violation
responsible for the matter-antimatter asymmetry of the Universe. Such
possibilities, especially for the Dirac phase, which can be realistically
discovered in the future, represent by themselves a strong motivation. We may
indeed soon be in the situation to probe the second of Sakharov's necessary
conditions for baryogenesis (see Section~\ref{sec:matter-antimatter}).

As we have seen, successful leptogenesis from low-energy phases is only
marginally possible in the HL, \HL, and with dependence on the initial 
conditions. This is especially true when the only source of \CP~violation
is the Dirac phase $\d$. We have also argued that a definite conclusion on the
existence of such a marginally allowed region requires a quantum kinetic 
treatment, which is expected to shrink the already quite restricted allowed
region.

Therefore, $\d$-leptogenesis and more generally leptogenesis from low-energy
phases motivate models with quasi-degenerate RH neutrino masses, the extreme limit
being resonant leptogenesis. Even in this extreme limit, imposing
successful $\d$-leptogenesis we could derive interesting conditions on
quantities accessible in low-energy neutrino experiment: $\sin \q_{13}$, the
absolute neutrino mass scale, normal or inverted hierarchy, the Dirac phase
itself. An interesting aspect of $\d$-leptogenesis is then that it is 
falsifiable independently of the heavy neutrino mass spectrum.

There are however some objections to the scenario of leptogenesis exclusively
from low-energy phases. At the moment, it still lacks a strong theoretical
motivation. In~\cite{Branco:2006ce} a model where such a situation naturally arises
was shortly discussed. There, it was said that the simplest way of restricting
the number of \CP-violating phases is through the assumption that \CP~is a good
symmetry of the Lagrangian, only broken by the vacuum. For example, one can
add to the standard type-I see-saw framework three Higgs doublets, 
together with a $Z_3$ symmetry under which the left-handed fermion doublets
$\psi_{L j}$ transform as $\psi_{Lj} \to {\rm e}^{{\rm i}\, 2\p j/3} \psi_{Lj}$ and the Higgs
doublets as $\f_j \to {\rm e}^{-{\rm i}\, 2\p j/3} \f_j$, while all other fields transform
trivially. It can be readily shown that there is a region of parameters where
the vacuum violates \CP~through complex vacuum expectation values 
$\langle \f_i^0\rangle=v_i {\rm e}^{{\rm i}\,\q_i}$. Due to the $Z_3$ restrictions on the 
Yukawa couplings, the combination $h^{\dg} h$ is real, thus implying a real
$\O$ matrix, but keeping $U$ complex. Of course, even though such a model
might work, it is not the simplest and most economical one. 

There has been a recent claim that sequential dominance models~\cite{King:1999mb}
(see~\cite{King:2003jb} for a more complete discussion)
could represent a theoretical framework for leptogenesis from low-energy phases. 
In~\cite{King:2006hn} it was shown that these models correspond to have
an $\O$ matrix that slightly deviates from the unit matrix or from all the other
five that can be obtained  from the unit matrix exchanging rows or columns.
However, it has been noticed in~\cite{Buchmuller:2003gz,DiBari:2005st} that in
the limit ${\rm Im}[\O] \to 0$, the total \CP~asymmetries $\ve_i$ do not
necessarily vanish. Writing $\O_{ij}^2=|\O_{ij}^2|{\rm e}^{{\rm i}\,\f_{ij}}$, the correct
condition to enforce $\ve_i\to 0$ is to take the limit $\f_{ij}\to 0$. This is
a more demanding limit than ${\rm Im}[\O] \to 0$, and it is not currently 
motivated by generic sequential dominance models. This limit is not motivated
either by radiative leptogenesis~\cite{GonzalezFelipe:2003fi,Branco:2005ye} 
within the context of minimal flavor violation~\cite{Cirigliano:2005ck},
as recently considered in~\cite{Cirigliano:2006nu,Branco:2006hz,Uhlig:2007xe}.
It must however be said that the limit ${\rm Im}[\O] \to 0$, when assuming 
a vanishing initial abundance of RH neutrinos, as done in 
the works cited above, can effectively
mimic the condition $\ve_i\to 0$ because the efficiency factor $\k_i \to 0$,
when ${\rm Im}[\O] \to 0$ and ${\rm Re}[\O] \to 0$, implying $K_i \to 0$.

Another possible objection to leptogenesis from low-energy phases is that it
cannot be distinguished from the general scenario where both high- and low-energy
phases are present. In particular, the Dirac phase will likely give in this case
only a subdominant contribution, since its effect is always suppressed by the
small $\q_{13}$ mixing angle. Following this approach, one can even say that the
baryon asymmetry produced through leptogenesis is not sensitive to the phases
in $U$~\cite{Davidson:2007va}, in the sense that the baryon asymmetry can be 
accounted for with the phases of $U$ having any value. Conversely, if the phases
in $U$ are measured, the baryon asymmetry is still not constrained.
The hope is then that the theoretical framework
supporting leptogenesis from low-energy phases has some other testable predictions.
An experimental support for this model would then represent a support for 
leptogenesis from low-energy phases.

However, even if a model supports leptogenesis from low-energy phases,
how can one know if the source of \CP~violation comes from the Dirac phase
or from the Majorana phases? Actually, it was noticed that the contribution
to the final asymmetry from Majorana phases is in general dominant compared to
the one coming from the Dirac phase (see right panel of Fig.~\ref{fig:HLR13}).
Thus, it will be only possible to tell
if $\d$-leptogenesis really occurs once the Majorana phases are constrained
from neutrinoless double-beta decay experiments to give small contributions. 

One can even imagine a situation where there is an exact cancellation 
between the Dirac and Majorana 
contributions. It would be however strange to think that nature disposes 
a sufficient source of \CP~violation, sets up a second source that exactly
cancels the first one, and the observed asymmetry is explained by yet a 
third one, e.g. the phases in $\O$.


\chapter{Conclusion}
\label{chap:conclusion}


The amount of baryonic matter in the Universe which we infer, 
for instance, from the CMB temperature anisotropies represents one 
of the most important puzzles of modern cosmology. In order to
explain this number, one needs a \emph{baryogenesis} mechanism 
which generates dynamically a small baryon asymmetry in the early Universe. 
We discussed in the introduction that a solution to this problem 
necessarily leads to physics beyond the SM. It is very 
exciting that this puzzle of cosmology may actually be related
to the existence of tiny but non-zero neutrino masses, which are now
established. As a matter of fact, a simple extension of the SM 
naturally leads to small neutrino masses via the see-saw mechanism, and its 
cosmological consequence is \emph{leptogenesis}, which 
elegantly yields the required baryon asymmetry.

In the present thesis, we have thoroughly discussed the mechanism of 
leptogenesis, where a lepton asymmetry is produced 
by the decays of heavy right-handed (RH) neutrinos and then 
transferred to a baryon asymmetry by the non-perturbative sphaleron processes. 
Let us now summarize the main points discussed throughout the thesis.

First, we have presented the unflavored treatment of leptogenesis,
where the leptons produced in the decays of the heavy neutrinos have
no flavor structure. It was shown that in the hierarchical limit for
the heavy neutrino mass spectrum, \HL, it is typically enough 
to consider only the decay of the lightest RH neutrino 
($N_1$-dominated scenario), with 
a reduction of the number of parameters relevant for the computation. 
In this minimal scenario, which we refer to as the ``vanilla'' scenario,
the contribution from the lightest
RH neutrino washes out all previous asymmetry and, for the theoretically
favored values of the effective neutrino mass [cf.~Eq.~(\ref{effneut})]
$m_{\rm sol}\lesssim \mt\lesssim m_{\rm atm}$, where $\msol$ and $\matm$ stand
for the solar and atmospheric neutrino mass scales, respectively, the strong 
washout regime is obtained, with no dependence of the final asymmetry 
on the initial conditions. Additionally, the strong washout regime implies 
that the
simple picture with only decays and inverse decays, i.e. neglecting
all scattering processes, and without including thermal corrections yields 
a very good estimation of the final asymmetry. 

Interestingly, in the $N_1$-dominated scenario,
general constraints on a few parameters of the model can be derived.
For successful leptogenesis, the mass of the lightest RH neutrino, $M_1$, 
cannot be smaller than about $4\times 10^{9}\gev$ for the strong washout to be
obtained~\cite{Davidson:2002qv,Buchmuller:2002rq}. This 
lower bound leads to a related lower bound on the initial
temperature of leptogenesis, $T_{\rm in}>1.5\times 10^{9}\gev$, which
can be identified with a lower bound on the reheat temperature $T_{\rm reh}$
within inflation.
Such a high value of the reheat temperature may be in conflict with
locally supersymmetric theories due to an overproduction of gravitinos. 

The baryon asymmetry produced through
leptogenesis is also sensitive to the absolute neutrino mass scale. In
the context of vanilla leptogenesis, the stringent upper bound 
$m_1\lesssim 0.12\ev$ was obtained~\cite{Buchmuller:2002jk,Buchmuller:2003gz,Giudice:2003jh}.

Then, we introduced the ``flavored'' picture of leptogenesis, which has
been understood only recently to be the correct one for a large fraction of 
the parameter space (roughly when $M_1\lesssim 10^{12}\gev$)~\cite{Abada:2006fw,Nardi:2006fx}. 
Flavor effects introduce a dependence of the final asymmetry on 
essentially all 
parameters of the model. On the one hand, this makes the computation more 
involved, but, on the other hand, some interesting parameters, such as the
\CP-violating phases in the PMNS matrix, become accessible. About flavor effects, one
can say in general:
\begin{itemize}
\item
In most of the parameter space they lead to modifications of the predictions 
by a factor 2--3 compared to the unflavored analysis, due to a reduction of
the washout by this factor. Incidentally, the region of independence
from the initial conditions shrinks by the same amount~\cite{Blanchet:2006be}.

\item
Large modifications are possible in two cases: i) when the high-energy 
phases in the $\O$ matrix are zero or close to zero; ii) 
when a one-flavor dominance is obtained.

\item
The usually quoted values of the lower bounds on $M_1$ and $T_{\rm reh}$ (see above)
do not change when flavor effects are included~\cite{Blanchet:2006be}.


\end{itemize}

Next, we re-analysed the conditions on the temperature $T$ and on $M_1$ 
for flavor effects to be important. We found that there should be a region
in the parameter space where classical Boltzmann equations (``fully flavored
regime'') are not enough to describe the generation of asymmetry~\cite{Blanchet:2006ch}. 
Correlations in flavor space and partial losses of coherence might indeed be 
relevant there, so that a quantum kinetic equation in the form of a density matrix 
equation should be used. Interestingly, the region concerned is exactly the one where the 
upper bound on the absolute neutrino mass scale seems to be evaded. Therefore,
at the moment it is not clear whether the upper bound on $m_1$ from successful 
vanilla leptogenesis quoted above disappears, is simply relaxed, or still holds.

Then, we went beyond the minimal picture where only the lightest
RH neutrino is considered. In the quasi-degenerate limit,
\DL, the contributions from all RH neutrinos have to be included. Due to 
the enhancement of the \CP~asymmetry in this limit~\cite{Covi:1996wh}, the 
lower bounds on $M_1$ and $T_{\rm reh}$ can be lowered down to the TeV scale 
in the extreme case of resonant leptogenesis~\cite{Pilaftsis:1997jf,Pilaftsis:2003gt}. 
Accounting for flavor effects, this 
might even be possible without having all Yukawa couplings unnaturally 
small~\cite{Pilaftsis:2004xx,Pilaftsis:2005rv}, even though it remains to be
proven that the condition of validity of the fully flavored equations is
satisfied in this case.

Actually, the production of asymmetry from the heavier RH neutrinos $N_2$ and $N_3$ 
is not only important when considering quasi-degenerate heavy neutrinos. 
Even for hierarchical heavy neutrinos, it was noticed 
in~\cite{DiBari:2005st} that a particular choice of the $\O$ matrix 
leads to production of asymmetry by $N_2$ instead of $N_1$. This implies
that the lower bound on $M_1$ does not apply anymore and is replaced
by a lower bound on $M_2$, which still yields a lower bound on 
$T_{\rm reh}$~\cite{DiBari:2005st}. 
When flavor effects are included, the contributions from $N_2$ and $N_3$ are 
potentially more important, due to the reduced washout
from $N_1$~\cite{Vives:2005ra,Engelhard:2006yg}. Specifically, the domain
of applicability of the $N_2$-dominated scenario is expected to be enlarged, 
and $N_3$ might be important as well, even though it seems complicated to 
avoid the washout both from $N_1$ and $N_2$.

The last possibility of going beyond the typical 
scenario that we discussed is when one element of the $\O$ matrix, namely
$\O_{22}$, has a non-trivial value different from 0 or 1. It turns out that making
$|\O_{22}|$ large opens the possibility of relaxing the lower bound on $M_1$
by 3 orders of magnitude even for \mbox{$M_1\ll M_2$}~\cite{Raidal:2004vt}. 
Moreover, with moderate values of $|\O_{22}|$
and moderate degeneracies $M_1\simeq M_2$, the upper bound on the neutrino
mass can be evaded~\cite{Hambye:2003rt}.

Finally, we studied the special case where the source of \CP~violation required
for leptogenesis stems exclusively from the phases in the PMNS matrix. We focused 
on the Dirac
phase, for which the prospects of measurement are the most promising. We found that
for hierarchical heavy neutrino masses leptogenesis from the Dirac phase ($\d$-leptogenesis) 
is only
marginally allowed and in the weak washout regime. For quasi-degenerate heavy neutrinos,
the strong washout 
is recovered, and we could even derive a upper bound on $m_1$ dependent on $\q_{13}$ for
successful resonant leptogenesis, which is the most favorable case one can 
imagine. Roughly speaking, for the $3\s$ upper limit $\sin \q_{13}=0.2$, 
resonant $\d$-leptogenesis only works for 
$m_1\lesssim 0.2\ev$~\cite{Anisimov:2007mw}. If the experimental upper limit on 
$\sin \q_{13}$ decreases in the future, so does the upper bound on $m_1$ for
successful $\d$-leptogenesis.

The see-saw mechanism and leptogenesis have very appealing 
features. However, in their vanilla form, they seem very difficult to prove
or disprove. The scale of the heavy neutrinos necessary for leptogenesis, as well
as to have a ``natural'' see-saw mechanism, is too large to be accessible at
future colliders. The only possibility to produce heavy neutrinos at colliders would
be to have at the same time TeV masses and large Yukawa couplings. Even though
such a possibility relies on a cancellation mechanism rather than a see-saw 
mechanism~\cite{Kersten:2007vk,deGouvea:2007uz}, the question whether it is
imaginable to have signals from heavy RH neutrinos at colliders is very
exciting by itself~\cite{delAguila:2007em,Kersten:2007vk,deGouvea:2007uz}. 
The inclusion of successful leptogenesis in the picture certainly deserves 
investigation.

Another way of probing directly leptogenesis would be to measure a primordial
lepton asymmetry of the order of the baryon asymmetry. The relic neutrino
background would carry such an information. Although the sole detection
of relic neutrinos is in principle possible~\cite{Cocco:2007qv,Lazauskas:2007da,Blennow:2008fh}, 
measuring an asymmetry of order $10^{-10}$ in it seems hopeless.

If a direct test of leptogenesis seems to be out of reach, then one
has to wait for an accumulation of indirect hints in favor of this scenario.

First, the observation of neutrinoless double-beta decay
would establish the Majorana nature of light neutrinos, hence supporting
the see-saw mechanism and leptogenesis. The next-generation experiments
such as GERDA~\cite{Abt:2004yk} or CUORE~\cite{Ardito:2005ar} are likely to
observe a signal if the light neutrino mass hierarchy is quasi-degenerate
or inverted. Furthermore, the discovery of the Majorana nature of neutrinos
would immediately tell that lepton number is violated, hence verifying
the first Sakharov's condition, thanks to the presence of sphalerons.

Second, the discovery of \CP~violation in the neutrino sector in future
long-baseline neutrino experiments such as T2K~\cite{Itow:2001ee} or 
NO$\n$A~\cite{Ayres:2004js} will certainly strengthen the case for 
leptogenesis. It will indeed tell that the second Sakharov's condition
is verified. Moreover, 
the source of \CP~violation in neutrino mixing might be sufficient to 
explain the origin of the matter-antimatter asymmetry of the Universe 
without resorting to the unobservable high-energy phases in the $\O$
matrix.

Conversely, it is interesting to ask oneself if there are ways to 
disprove leptogenesis. A non-observation of neutrinoless double-beta 
decay in the next-generation experiments would not disprove
by itself the see-saw mechanism and leptogenesis. The mass hierarchy might
simply be normal, hence out of reach of the next-generation 
experiments. However, if neutrino oscillation experiments determine
the mass hierarchy to be inverted~\cite{Hagiwara:2005pe,Mena:2005ri}, 
but no signal is found in neutrinoless double beta decay experiments,
then one would conclude that the see-saw mechanism is not at the
origin of the neutrino masses, and that leptogenesis is not responsible
for the baryon asymmetry of the Universe.

As concerns \CP~violation, the non-observation of leptonic \CP~violation
in future neutrino experiments will not weaken the case for leptogenesis
in a significant way. Instead, it would mean that the Dirac phase and/or
the angle $\q_{13}$ are too small to give an noticeable contribution to the
final asymmetry. The source
of \CP~violation responsible for leptogenesis can be still given by
the remaining 5 \CP-violating phases present in the model.

The absolute neutrino mass scale may also provide a way to contrain 
significantly leptogenesis. As a matter of fact, we have seen that the
latter scenario leads to an upper bound on the absolute neutrino
mass scale of about $0.1\ev$, which is however subject to modification
due to flavor effects. Assuming that the upper bound remains below 
roughly $0.3\ev$, implying $m_{\n_e}\lesssim 0.3\ev$, a signal in the future 
KATRIN experiment~\cite{Osipowicz:2001sq}, which claims a discovery
potential down to about $0.35\ev$, would severely constrain leptogenesis in its
minimal version. On the other hand, other versions, such as the
one with quasi-degenerate heavy neutrinos, would still remain viable.

Even though the production of heavy neutrinos at future colliders seems 
to be unlikely, the LHC experiment at CERN will still have an indirect impact on 
leptogenesis. Indeed, if supersymmetry is discovered and the parameters
measured are consistent with successful electroweak baryogenesis 
(e.g. light stop and light Higgs~\cite{Balazs:2004ae,Menon:2004wv}), 
then the case for leptogenesis
will become weaker. On the other hand, if electroweak baryogenesis is
ruled out at the LHC, then leptogenesis, as one of the remaining 
possibilities to explain the baryon asymmetry of the Universe, will 
become stronger.

In conclusion, since leptogenesis is unavoidable when considering the see-saw
mechanism for the generation of neutrino masses, 
it provides a very elegant explanation to 
one of the outstanding problems of modern cosmology, the origin
of the matter-antimatter asymmetry of the Universe. Even though a direct
test is challenging, there is no doubt that in the next few years more 
experimental evidence 
will become available to weaken or strengthen the case for leptogenesis.






\appendix
\renewcommand{\chaptermark}[1]{\markboth{App. \thechapter:\ #1}{}}
\chapter{Neutrino mixing parameters}
\label{chap:param}

The currently existing data from neutrino oscillation experiments can
be well described by the Lagrangian,
\be
\mathcal{L}=-{g\over \sqrt{2}}\overline{\ell_{L}}\g^{\m}\n_{L}W_{\m}
-{1\over 2}\overline{(\n_L)^c}m_{\n}\n_L -\overline{\ell_R}m_{\ell}
\ell_L+ h.c. \, ,
\ee
which includes the weak charged current interaction in the lepton
sector, a Majorana mass term for neutrinos, and a Dirac mass
term for charged leptons. When diagonalizing the neutrino
and charged lepton mass matrices via $m_{\n}=U_{\n}^{\star}m_{\n}^{\rm diag}
U_{\n}^{\dagger}$ and $m_{\ell}=V_{\ell}m_{\ell}^{\rm diag}U_{\ell}^{\dg}$,
one obtains the lepton mixing matrix, known also as the 
Pontecorvo-Maki-Nakagawa-Sakata (PMNS) mixing 
matrix~\cite{Pontecorvo:1957qd,Pontecorvo:1957cp,Maki:1962mu}, in the
weak charged lepton current
\be
U=U_{\ell}^{\dg}U_{\n}.
\ee
The charged current interaction can then be written in the mass basis
as
\be
-{g\over \sqrt{2}}\overline{\ell_{L\a}}\g^{\m}U_{\a i}\n_{Li}W_{\m}.
\ee
Note that it is conventional to choose the basis where the mass matrix
for the charged leptons is diagonal, in which case $U=U_{\n}$, i.e.
the PMNS matrix is the matrix that diagonalizes the neutrino mass
matrix. Neutrinos with a given flavor are then related to neutrinos
with a given mass through the PMNS matrix,\footnote{
If one wants the relation between the flavor eigenstates and the mass
eigenstates, it is given by $|\n_{\a}\rangle=\sum_{j=1}^3 U^{\star}_{\a j}
 |\n_{j}\rangle, \quad \a=e,\m,\t $. At first sight, this may seem
surprising, but it is simply due to the fact that, by convention, field
operators create antiparticles, or annihilate particles!~\cite{Strumia:2006db}}
\be
\n_{\a L}=\sum_{j=1}^3 U_{\a j} \n_{jL} \, ,\quad \a=e,\m,\t \, .
\ee
Note also that the number of neutrino flavor states with mass below $M_Z/2$ 
is known to be three from the precise measurement of the $Z$-width at 
LEP~I~\cite{PDBook}.

The standard parametrization of the PMNS matrix was given 
in Eq.~(\ref{PMNSmatrix2}). Carrying out the matrix product, one obtains
\be\label{PMNSmatrix}
U=V \times {\rm diag}({\rm e}^{{\rm i}\,{\Phi_1\over 2}}, {\rm e}^{{\rm i}\,{\Phi_2\over 2}}, 1),
\ee
\begin{displaymath}
V=\left( \begin{array}{ccc}
c_{12}\,c_{13} & s_{12}\,c_{13} & s_{13}\,{\rm e}^{-{\rm i}\,\d} \\
-s_{12}\,c_{23}-c_{12}\,s_{23}\,s_{13}\,{\rm e}^{{\rm i}\,\d} &
c_{12}\,c_{23}-s_{12}\,s_{23}\,s_{13}\,{\rm e}^{{\rm i}\,\d} & s_{23}\,c_{13} \\
s_{12}\,s_{23}-c_{12}\,c_{23}\,s_{13}\,{\rm e}^{{\rm i}\,\d}
& -c_{12}\,s_{23}-s_{12}\,c_{23}\,s_{13}\,{\rm e}^{{\rm i}\,\d}  &
c_{23}\,c_{13}
\end{array}\right).
\end{displaymath}
We shall use in all calculations
$\theta_{12}=\pi/6$ and $\theta_{23}=\pi/4$, compatible
with the results from neutrino oscillation experiments
discussed in the introduction, Eqs.~(\ref{atmparam}) and 
(\ref{solparam}). For the third angle,
we shall use the $3\s$ range \mbox{$s_{13}=0\textrm{--}0.2$} 
obtained from Eq.~(\ref{theta13}).

Concerning the mass spectrum of light neutrinos, we shall use
the convention that $m_1\leq m_2 \leq m_3$, whatever the hierarchy
is. This means that when a normal hierarchy is considered, one
has $m_3^2-m_2^2=\D m_{\rm atm}^2$ and $m_2^2-m_1^2=\D m_{\rm sol}^2$ 
[cf.~Eqs.~(\ref{atmmass}) and (\ref{solmass})],
whereas for an inverted hierarchy, one has $m_3^2-m_2^2=\D m_{\rm sol}^2$ 
and $m_2^2-m_1^2=\D m_{\rm atm}^2$. Defining the two convenient
quantities
\be
m_{\rm atm}\equiv
\sqrt{\D m_{\rm atm}^2+\D m_{\rm sol}^2}=(0.052\pm 0.002)~{\rm eV},
\ee
and
\be
m_{\rm sol}\equiv
\sqrt{\D m_{\rm sol}^2}=(0.0089\pm 0.0002)~{\rm eV},
\ee
one has that the light neutrino spectrum is quasi-degenerate when
\mbox{$m_1\gg m_{\rm atm}$}, while for $m_1\ll m_{\rm sol}$
it is fully hierarchical.

Within the convention for the light neutrino masses we are using,
it must be pointed out that the case of inverted hierarchy
is obtained by performing a cyclic permutation of the columns in
the PMNS matrix Eq.~(\ref{PMNSmatrix}), such that the 
$i$-th column becomes the $(i+1)$-th~\cite{Anisimov:2007mw}.

Finally, let us note that we shall always neglect the effect 
of the running
of neutrino parameters from high energy to low 
energy~\cite{Babu:1993qv,Antusch:2003kp}.


\chapter[The see-saw mechanism with three RH neutrinos]{The see-saw 
mechanism with three RH neutrinos}
\label{chap:see-saw}

The (type-I) see-saw mechanism is based on the following extension of the SM: 
\begin{equation}\label{lagrangian}
\mathcal{L}= \mathcal{L}_{\rm SM} +i \overline{N_{Ri}}\g_{\m}\partial^{\m} N_{Ri} - 
h_{\a i} \overline{\ell_{L\a}} N_{Ri} \tilde{\F} -{1\over 2}
M_{{\rm M} i} \overline{(N_{Ri})^c}N_{Ri} +h.c.~,
\end{equation}
where $\a=e,\m,\t$, three new fields $N_{Ri},~i=1,2,3$,\footnote{In 
principle, it would be possible to consider
the addition of only two RH neutrinos, since only two light neutrinos
are known to be massive. But, in the following, we want to be slightly
more general and therefore keep three RH neutrinos.}
a Majorana mass matrix $M_{\rm M}$ and a Yukawa coupling matrix $h$ have 
been introduced. We chose here the basis where the charged-lepton 
Yukawa matrix and the Majorana mass matrix are diagonal.
The superscript $c$ denotes charge 
conjugation, defined as $\phi^c \equiv C \bar{\phi}^T$, where
the charge conjugation matrix $C$ satisfies
\be
C^{-1}\g_{\m}C=-\g_{\m}^T,~C^T=-C,~C^{\dg}=C^{-1}.
\ee
The subscripts $L$ and $R$ denote the left-handed and right-handed
chiral projections, respectively: $P_{L,R}\equiv (1\pm\g_5)/2$. 
Finally, the $SU(2)_L\times U(1)_Y$ charge
assignments are as follows:
\begin{eqnarray}
\tilde{\F}\equiv i\s_2 \F^{\star}=\left(\begin{array}{c}\phi_0^{\star} \\ 
\phi_+^{\star} 
\end{array}\right),&
\quad (2,-1) \, ,\NO \\
\ell_{L\a}=\left(\begin{array}{c} \n_{\a} \\  
\a^- \end{array}\right)_L,&
\quad (2,-1) \, ,\NO \\
N_{Ri},&\quad (1,0)\,.\NO 
\end{eqnarray}
It should be stressed that the new fields $N_{Ri}$, often called
right-handed (RH) neutrinos, are singlets
under $SU(2)_L\times U(1)_Y$.

In Eq.~(\ref{lagrangian}) the Yukawa-type term is simply the analog of
the other mass terms for fermions in the SM. Concerning
the Majorana mass term, it is new and unique, but in full generality
there is no reason why it should be absent. This particularity
for neutrinos is due to the fact that they are the only neutral
fermions in the SM.

After
spontaneous symmetry breaking, a Dirac mass term $m_{\rm D}=h v$,
is generated by the vacuum expectation value of the Higgs field, $v=174$~GeV. 
Using the identity $\overline{\n_L}N_R = \overline{(N_R)^c}(\n_L)^c$, one
can then rewrite the mass term in a more compact form:
\begin{equation}\label{massmatrix}
\mathcal{L}_{\rm mass}=-{1\over 2} \left(\begin{array}{cc}
\overline{\n_L} & 
\overline{(N_R)^c}\end{array}\right)
\left(\begin{array}{cc} 0& m_{\rm D}\\
m_{\rm D}^T & M_{\rm M} \end{array}\right)\left(\begin{array}{c}
(\n_L)^c \\ N_R\end{array}\right)+h.c.
\end{equation}
The see-saw mechanism then assumes that the entries of the Majorana
mass matrix matrix are much larger than all Dirac matrix elements,
i.e. \mbox{$M_{\rm M}\gg m_{\rm D}$}. Under this assumption, the mass matrix
in Eq.~(\ref{massmatrix}) can be block-diagonalized with a
$6\times 6$ unitary matrix $V$:
\be \label{diagmat}
V \left(\begin{array}{cc} 0& m_{\rm D}\\
m_{\rm D}^T & M_{\rm M} \end{array}\right) V^T\simeq
\left(\begin{array}{cc} m_{\n}& 0\\
0 & M \end{array}\right)K,
\ee
where
\be\label{mnu}
m_{\nu}=  m_{\rm D} {1\over M_{\rm M}} m_{\rm D}^T \,,
\ee
\be
V=\left(\begin{array}{cc} 1& -m_{\rm D}M_{\rm M}^{-1}\\
M_{\rm M}^{-1}m_{\rm D}^T & 1 \end{array}\right),
\ee
and 
\be
K=\left(\begin{array}{cc} -1& 0\\
0& 1 \end{array}\right).
\ee
The $K$ matrix only makes sure that the mass eigenvalues
are positive.

The matrix in Eq.~(\ref{mnu}) corresponds to the light neutrino
mass matrix, which has naturally suppressed entries due to the heavy
scale in the denominator. This is the reason why this mechanism is 
called the see-saw mechanism. As explained in Appendix~\ref{chap:param},
the matrix that diagonalizes the mass matrix $m_{\n}$ for the light
neutrinos is the PMNS mixing matrix [cf. Eq.~(\ref{PMNSmatrix})], 
so that
\be
U^{\dagger}m_{\n}U^{\star}={\rm diag}(m_1,m_2,m_3),
\ee
which are the masses of the three light neutrinos.
The corresponding mass eigenstates are given by
\be
\n_i
\simeq 
\sum_{\a}\left(U^T\right)_{i\a}\left\{\left[\n_{L\a}-(\n_{L\a})^c\right]-
m_{{\rm D}\a i}M^{-1}_i \left[(N_{Ri})^c-N_{Ri}\right]\right\}.
\ee
It can be easily checked that $\n_i=-\n^c_i$, which means
that the light neutrinos are Majorana particles.

On the other hand, the lower right block on the right-hand side of 
Eq.~(\ref{diagmat}) was diagonal before the 
diagonalization and, to leading order, it remains diagonal, i.e. $M\simeq M_{\rm M}$. 
The entries are $M_1\leq M_2 \leq M_3$, corresponding to three 
heavy neutrinos. The corresponding mass eigenstates are given by
\be
N_i\simeq \sum_{\a} M^{-1}_i (m_{\rm D}^T)_{i\a}\left[\n_{L\a}+(\n_{L\a})^c\right]+
\left[(N_{Ri})^c+N_{Ri}\right].
\ee
It can be checked again that $N^c_i=N_i$, which means that
the heavy neutrinos are Majorana particles.

Finally, let us note that the extension of the SM in Eq.~(\ref{lagrangian})
introduces 18 new parameters: 6 masses, 6 mixing angles and 6 
\CP-violating phases. The number of parameters in principle accessible 
in low-energy neutrino experiments is 9: the masses of the 3 light neutrinos 
and the 6 parameters (3 mixing angles and 3  
\CP-violating phases) in the PMNS matrix [cf.~Eq.~(\ref{PMNSmatrix})].



\providecommand{\href}[2]{#2}\begingroup\raggedright\endgroup

\end{document}